\documentclass[12pt]{article}
\usepackage{amsfonts,amsmath,amssymb}
\usepackage{graphicx}

\usepackage[makeindex]{imakeidx}
\makeindex

\usepackage{hyperref}
\hypersetup{
    colorlinks,
    citecolor=green,
    filecolor=green,
    linkcolor=blue,
    urlcolor=green
}

\newfont{\twelvemsb}{msbm10 scaled\magstep1}
\newfont{\eightmsb}{msbm8}
\newfam\msbfam
\textfont\msbfam=\twelvemsb \scriptfont\msbfam=\eightmsb
\catcode`\@=11
\def\Bbb{\ifmmode\let\next\Bbb@\else
\def\next{\errmessage{Use \string\Bbb\space only in math mode}}\fi\next}
\def\Bbb@#1{{\fam\msbfam{{#1}}}}

\newcommand{\be}{\begin{equation}}
\newcommand{\ee}{\end{equation}}
\newcommand{\ba}{\begin{eqnarray}}
\newcommand{\ea}{\end{eqnarray}}
\newcommand{\h}{\hat}
\newcommand{\m}{\mathcal}
\newcommand{\ti}{\tilde}
\newcommand{\nn}{\nonumber}
\newcommand{\q}{\theta}

\begin{document}

\sloppy
\renewcommand{\thefootnote}{\fnsymbol{footnote}}
\newpage
\setcounter{page}{1} \vspace{0.7cm}
\vspace*{1cm}
\begin{center}
{\bf Asymptotic Bethe Ansatz on the GKP vacuum as a defect spin chain: }\\
{\bf scattering, particles and minimal area Wilson loops} \\
\vspace{.6cm} {Davide Fioravanti $^a$, Simone Piscaglia $^{a,\,b}$, Marco Rossi $^c$}
\footnote{E-mail: fioravanti@bo.infn.it, piscagli@bo.infn.it, rossi@cs.infn.it} \\
\vspace{.3cm} $^a$ {\em Sezione INFN di Bologna, Dipartimento di Fisica e Astronomia,
Universit\`a di Bologna} \\
{\em Via Irnerio 46, Bologna, Italy}\\
\vspace{.3cm} $^b$ {\em Dipartimento di Fisica and INFN, Universit\`a
di Torino} \\
{\em Via P. Giuria 1, Torino, Italy}\\
\vspace{.3cm} $^c${\em Dipartimento di Fisica dell'Universit\`a
della Calabria and INFN, Gruppo collegato di Cosenza} \\
{\em Arcavacata di Rende, Cosenza, Italy} \\
\end{center}
\renewcommand{\thefootnote}{\arabic{footnote}}
\setcounter{footnote}{0}
\begin{abstract}
{\noindent Moving from Beisert-Staudacher equations, the complete set of Asymptotic Bethe Ansatz equations and $S$-matrix for the excitations over the GKP vacuum is found. The resulting model on this new vacuum is an integrable spin chain of length $R=2\ln s$ ($s=$ spin) with particle rapidities as inhomogeneities, two (purely transmitting) defects and $SU(4)$ (residual R-)symmetry. The non-trivial dynamics of ${\cal N}=4$ SYM appears in elaborated dressing factors of the 2D two-particle scattering factors, all depending on the 'fundamental' one between two scalar excitations. From scattering factors we determine bound states. In particular, we study the strong coupling limit, in the non-perturbative, perturbative and giant hole regimes. Eventually, from these scattering data we construct the $4D$ pentagon transition amplitudes (perturbative regime). In this manner, we detail the multi-particle contributions (flux tube) to the MHV gluon scattering amplitudes/Wilson loops (OPE or BSV series) and re-sum them to the Thermodynamic Bubble Ansatz.}

\end{abstract}
\vspace{6cm}

\newpage


\tableofcontents

\index{}
\printindex
\newpage


\section{Introduction}
\setcounter{equation}{0}

The study of the energy of the excitations on a suitably chosen vacuum state is a problem which is common to very many physical theories.
It often happens that most intriguing excitations arise over a vacuum state which is an intricate superposition of
'basic' states, {\it i.e.} a sort of Fermi sea of interacting 'pseudoparticles'. In general, this vacuum may be dubbed antiferromagnetic as the prototypical example in the realm of integrable models is the antiferromagnetic
vacuum state of the Heisenberg spin chain. In their turn, important excitations on it are called spinons or solitons, whilst magnons are the (pseudo)particles forming the see on the ferromagnetic vacuum. In an easy Bethe Ansatz perspective \cite{Bethe}, spinons may appear as holes in a distribution of a large number of real Bethe roots. As a consequence, these holes are constrained by quantisation conditions for their rapidities, which may anew be seen as Bethe(-Yang) equations for these new 'fundamental' particles. Of course, we expect this phenomenon to be of non-perturbative nature, so that integrability is the right realm to exploit it.

A similar, but obviously much richer situation, arises
in the framework of Beisert-Staudacher Asymptotic Bethe Ansatz (ABA) equations determining, via a specific root configuration, the anomalous dimension (energy) of the single trace fields in planar ${\cal N}=4$ SYM \cite{BS}. In this context one can choose as 'antiferromagnetic' vacuum the configuration which contains a large number, $s$, of type-4 roots
and which describes, up to wrapping corrections \cite {WRA}, high spin ($=s$) twist two operators, namely, sketchily,
\be
\mathcal{O}=Tr\, Z D_+^{s} Z + \dots \quad , \label{tw-op}
\ee
where $Z$ is one of the three (complex) scalars of the theory. In fact, this is likely the 'simplest' example of Wilson twist operator. It belongs to the paradigmatic $sl(2)$ sector of scalar operators, which are made up of only one (out of three) complex scalar $Z$ and the (light-cone) covariant derivative $D_+$, so enjoying the sketchy form
\begin{equation}
{\mbox {Tr}} (D_+^s Z^L)+.... \, , \label {sl2op}
\end{equation}
where dots stand for permutations. Built up in this selected way they result to be perturbatively closed under renormalisation, so forming a sector. These composite single trace operator have of course Lorentz spin $s$ and twist (or length in the ferromagnetic/half-BPS vacuum perspective) $L$, with minimum value $L=2$ for which (a descendant\footnote{In this case, the spin may be shifted by a finite amount which does not affect our analysis and results at high spin.} of) the Gubser-Klebanov-Polyakov (GKP) 'vacuum' solution is realised \cite{GKP}. Also, the AdS/CFT correspondence \cite{MGKPW} relates an operator (\ref {sl2op}) to a spinning folded closed strings on $\text{AdS}_5\times\text{S}^5$ spacetime, with angular momenta $s/\sqrt{\lambda}$ and $L/\sqrt{\lambda}$ on each space respectively, the 't Hooft coupling in the multi-color $N_c\rightarrow  \infty$ (planar) regime
\be
\lambda \equiv N_c g^2_{YM}\,\,;\,\,\, g^2\equiv \frac{\lambda}{8\pi^2} \, ,
\label{lambda}
\ee
being connected to the string tension $T=\frac{\sqrt{\lambda}}{2\pi}$ \cite{GKP, FTT}. On the other hand we may think of the operators (\ref{sl2op}) as obtained from the GKP vacuum (\ref{tw-op}) by adding scalar excitations on top of it. Of course, when $L>2$ we can realise states with different energies, at fixed $L$, and typically the minimal energy has been more extensively studied for 'large size' $s\rightarrow +\infty$. In particular, the minimal anomalous dimension of (\ref {sl2op}) has been proven to enjoy at one loop the same leading behaviour $\sim \ln s$ at high spin (and fixed $L$) \cite {BGK}, as in the opposite string regime (strong coupling) \cite{GKP}. Later on, the coefficient of this term\footnote{This is the so-called {\it universal scaling function}, $f(g)$, which does not depend on $L$ and equals twice the {\it cusp anomalous dimension} (renormalisation divergence \cite{Poly-cusp}) of a light-like Wilson cusp, as in QCD\cite{Korchemsky:1987wg}.} was obtained at all loops from the solution of a linear integral equation directly derived from the Beisert-Staudacher equations via the root density approach \cite{BES}. In very brief summary, as computed in \cite{FGR5}, the high spin (asymptotic) expansion (at fixed $g$ and $L$) enjoys the peculiar form
\be
\gamma(g,s,L)= f(g) \ln s + f_{sl}(g,L) + \sum_{n=1}^\infty \gamma^{(n)}(g,L)  \, (\ln s)^{-n} + O\left ( (\ln s)/s \right ) \, ,
\label{subl}
\ee
in inverse integer powers of the size\footnote{In fact, it is consistent with the length of the long classical string $R\sim \ln (s/\sqrt{\lambda})$ \cite{GKP, FTT}.} $R\sim \ln s$, except the sub-leading $(\ln s)^0$ contribution $f_{sl}(g,L)$ (defect contribution). The latter, which reduces to the so-called virtual scaling function for $L=2$, has been captured in \cite{FRS} by a Non-Linear Integral Equation (NLIE) and in \cite{BFR} by a linear integral equation (by means of which explicit strong coupling expansions can be performed \cite{FZ}, along the lines of those for the cusp \cite{BKK}). Up to this order, we can be sure that this expansion enjoys the same form at all perturbative orders in QCD, or its Mellin transform, {\it i.e.} the evolution kernels \cite{KM}. Moreover, in the supersymmetric case similar linear integral equations hold for all the coefficients in (\ref{subl}) \cite{FGR5} and also for the next order $O\left ( (\ln s)/s \right )$ \cite{reciprocity}, and all these, -- importantly the first two $f(g)$ and $f_{sl}(g,L)$, -- are now believed to be exactly given by the ABA without wrapping\footnote{For instance in \cite {BJL} wrapping corrections to ABA start to contribute at order $e^{-R}=e^{-2\ln s}=1/s^2$, inducing to think of a factor $2$ in the size of the folded string $R=2\ln s+\dots$.}, also thanks to these recent studies.

The latter were focussed on the same scalar, $Z$, added to (\ref {tw-op}), but we can generalise to the other fields: indeed, elementary one-particle excitations may correspond to inserting one of the other fields {\it i.e.}, besides the other two scalars, a gauge field (gluon) or a Fermi field (gaugino)\footnote{Notice that in the half-BPS vacuum description this state would belong to a longer spin chain of length $L=3$.}. In other words, they are the lowest twist ($=$three) operators/states with the form
\be
\mathcal{O}_{1-particle}=Tr\, Z D_+^{s-s'} \varphi D_+^{s'} Z + \dots \quad , \label{exc-tw-op}
\ee
where $\varphi=Z, W, X$, the scalars, or $\varphi=F_{+\bot}, \bar{F}_{+\bot}$, the two components of the gauge field, or $\varphi= \Psi_+$, $\bar{\Psi}_+$, the $4+4$ (anti-)fermions, respectively. Besides the energy, one can determine also the momentum of an operator through the Beisert-Staudacher ABA equations. Along this line, the one-particle dispersion relations of the excitations (\ref{exc-tw-op}) have been receiving  much attention in the different coupling regimes ({\it cf.} for instance \cite{Freyhult:2009fc} and references therein); but recently they have been summarised, corrected and put forward in an illuminating work by Basso \cite{Basso} (also reference therein).

On the same footing, we start wondering in \cite{FRO6} about the scattering $S$-matrix which may be attached to the two-particle states (of, at least, twist-$4$)
\be
\mathcal{O}_{2-particles}=Tr\, Z D_+^{s-s_1-s_2} \varphi_1 D_+^{s_1} \varphi_2 D_+^{s_1} Z+ \dots \quad ,
\ee
where $\varphi_1$ and $\varphi_2$ may be any general elementary local field as $\varphi$ in (\ref{exc-tw-op}), whereas in \cite{FRO6} we confined our attention to the peculiar ({\it cf.} below) case $\varphi_1=\varphi_2=Z$. In fact, as argued above, we expect the Beisert-Staudacher quantisation conditions to give correct results at leading $\ln s$ and next to leading order $(\ln s)^0$. And then, regarding $R\sim \ln s$ as the size of the system, these orders are exactly the ones we need to write down $2D$ (many-particle) scattering amplitudes, {\it i.e.} (on-shell) quantisation conditions, for rapidities of excitations on the GKP vacuum. Generalising to all the other scalars, \cite{BREJ} have deduced the entire $SO(6)$ scattering, while we have computed in \cite{FPR} all the $g$-depending scalar factors of the different scattering channels, neglecting the $SU(4)$ representation structure.

Moving from this lack, we shall make here our analysis deeper, by computing explicitly the matrix structures of the different $SU(4)$ representations carried by the 'elementary' particles and by their bound states. We will not only consider the two-body scattering, but also in general the multi-particle 2D scattering amplitudes. As a byproduct we will see a well know characterisation of integrable theories, namely the elasticity and factorisation, {\it i.e.} the determination of many-particle scattering by the two-particle one. Besides the traditional name of Bethe-Yang equations, we can call these quantisation conditions Asymptotic Bethe Ansatz equations as well, but now the term 'Asymptotic' refers to the new length $\sim \ln s$, which measures the validity of the equations (and to the 'new' vacuum). More precisely, from the BMN (ferromagnetic) vacuum \cite{BMN} (no roots) we will switch on, in the Beisert-Staudacher equations, the configurations corresponding to the GKP (antiferromagnetic) vacuum and to all possible 'elementary' excitations over the GKP vacuum; to accomplish this, we will be using the idea of converting many (Bethe) algebraic equations describing an excited state into few non-linear integral equations (NLIEs) \cite{DDV,FMQR, FR, FRS, BFR}. In this way, we will obtain the quantisation conditions of all the 'elementary' excitations over the GKP vacuum and show that the structure of these equations coincides with Bethe equations of a inhomogeneous spin chain of length $R=2\ln s$ with two identical (purely transmitting) defects and a $SU(4)$ symmetry in different representations (where the particle rapidities represent the inhomogeneities). Of course, the scalar pre-factors in front of the above $SU(4)$ matrix structure are dependent on $g$ and characteristic of the theory (and GKP vacuum). Nevertheless, we can express all in terms of the scalar-scalar one \cite{FPR}.  Moreover, we will discuss in many details the consequences of switching to a different vacuum which basically means that any elementary particle interacts with the sea of covariant derivatives namely the type-4 roots. For instance, the poles of the new 2D scattering factors of these particle imply the entrance of bound states thereof into the spectrum and then the existence of new scattering amplitudes for the latter particles. As anticipated, not only the 2D scattering amplitudes, but also many physical quantities assume novel expressions, as for instance the energy, momentum \cite{Basso} and all the other conserved charges carried by a single elementary or composite excitation ({\it cf.} below). Furthermore, the scattering of any particle onto two defects arises, as anticipated in \cite{BREJ, FPR}, though they were absent in the ferromagnetic setup, and they are likely to be associated to the two external holes (or tips of the GKP string). Together with the change of length, these are somehow unprecedented features in the theory of (quantum) integrable systems though their common origin can be traced back to the $sl(2)$ spin chain (describing the one-scalar sector (\ref{sl2op}) at one-loop): nevertheless, we are used to insert the defect {\it ab initio} in the theory on the ferromagnetic vacuum and then finding the anti-ferromagnetic dynamics with defect (possibly characterised by a different scattering factor) and the same length.

As a consequence of this new ABA, also the exact Thermodynamic Bethe Ansatz (TBA) \cite{TBA-Z1, TBA-Y} for the spectrum of anomalous dimensions as derived from its mirror version should ostensibly look very differently\footnote{The attentive reader may guess many aspects of it from the form of {\it string/stack solutions} as reported in section \ref{sez9}.} than the usual one on the BMN vacuum \cite{TBA}, although they should give the same spectrum after all. Even more interestingly, a recent series of papers by Basso, Sever and Vieira (BSV) extended to all terms the operator product expansion of \cite{TBA-amp2} and thus proposed a non-perturbative approach to 4D gluon scattering amplitudes/null polygonal Wilson loops (which are allegedly the same \cite{AM-amp, DKS, BHT}) in ${\cal N}=4$ SYM, which relies on these 2D scattering factors as input data or building blocks \cite{BSV1,BSV2,BSV3,BSV4,BSV5}. In this perspective, light-like polygonal Wilson loops (WLs) can thought of as an infinite sum over more fundamental polygons, namely square and pentagonal WLs, whose knowledge relies on the GKP scattering factors. By virtue of the AdS/CFT strong/weak duality, this superposition of pentagons and squares should lead, at large coupling $g$, to the classical string regime, namely the minimisation of the supersymmetric string action \cite{AM-amp}. In general, this is a complicated problem of minimal area (string action) subtending a polygon living on the boundary of $AdS_5$, and results in a set of non-linear coupled integral equations \cite{TBA-amp1, TBA-amp3, TBA-amp2, Hatsuda:2010cc}. For some still hidden reason, their form resembles that of a relativistic Thermodynamic Bethe (or Bubble, in this case!) Ansatz system whose free energy yields the area \cite{TBA-Y, TBA-Z1}\footnote{In the particular case of the hexagon WL \cite{TBA-amp1} the system does coincide with the usual TBA one \cite{TBA-Z2}}. Instead, we wish in this paper to construct this TBA set-up by summing the infinite BSV series and performing a saddle point evaluation. For this aim, we will perform a propaedeutic analysis of all the different strong coupling regimes.

The article is organised according to the following plan. In section \ref{sez2} we derive the ABA equations, first at one-loop as exemplifying case so to highlight all the relevant features, then for any value of the coupling. In section \ref{sez3} the conserved charges of the excitations (on the GKP vacuum) are computed. In section \ref{sez4} the strong coupling limit of the scattering factors is considered, in the different dynamical regimes, {\it i.e.} non-perturbative, perturbative and giant hole regimes. Section \ref{sez5} contains equivalent forms for the momentum associated to any elementary particle excitation, in particular that elaborated in \cite{Basso}. Section \ref{sez6} is a study of the strong coupling behaviour of (the scattering factor for) the spin chain defects. In section \ref{sez7} the properties of the different kinds of particles under the SU(4) symmetry are taken into exam, so that in section \ref{sez8} we are able to describe the structure of the overall S-matrix. In section \ref{sez9} the so-called {\it string hypothesis} is used on the GKP ABA, in order to survey the on-shell states and, in particular, the bound states of elementary particles; for later purposes, an accurate study is devoted to the behaviour of bound states of gluons at any coupling and fermion-antifermion$=$meson  and its bound states which, instead, do appear only at the leading order of perturbative strong coupling regime, {\it i.e.} in the classical string theory. Section \ref{sez10} computes all the string perturbative expressions of the pentagonal amplitudes (contributing to the BSV series). Finally, these infinite contributions are summed up exactly in section \ref{sez11} as for the hexagonal Wilson loop; the result is remarkably coinciding with the Yang-Yang functional and Thermodynamic Bubble Ansatz (TBA) equations for minimal area solution \cite{TBA-amp1, TBA-amp2, TBA-amp3}. After the conclusions (section \ref{sez12}), several appendices follow. In appendices \ref{functions} and \ref{formuli} the definition of the functions employed throughout the text, as well as some useful formulae and integrals are listed. Appendix \ref{scatt-fact} is a synopsis of scattering factors, displayed at arbitrary coupling, one-loop, strong coupling (mirror transformed too), while appendix \ref{some-calc} gives some details about their derivation. Finally, in appendix \ref{Bethe-eqs} all the ABA equations are listed.

\section{General equations}\label{sez2}
\setcounter{equation}{0}

\subsection{Excitations}

The main aim of this section is to write Bethe equations describing 'elementary' excitations over the long GKP string,
in the more general case when $H$ scalars ($u_h, h=1,...,H$), $N_F$ large fermions, $N_{\bar F}$ large antifermions, $N_f$ small fermions, $N_{\bar f}$ small antifermions, $N_g$ gauge fields $F_{+\perp}$ and $N_{\bar g}$ gauge fields $\bar F_{+\perp}$ are present.

In the notation of Beisert-Staudacher equations \cite {BS} the GKP vacuum is described by a large even number $s$ of type-4 roots filling the interval $[-b,b]$ of the real axis, together with two external holes \cite {KOR, BGK}. In the large $s$ limit $b$ is  approximated with $s/2$ and the positions of the two external holes with $\pm s/\sqrt{2}$: corrections to those estimates give rise to $O(1/s^2)$ terms in the final Bethe equations, that we will neglect.

It is a general fact that, in order to deal with a large number of Bethe roots, it is convenient to use their counting function $Z_4(v)$, which satisfies a nonlinear integral equation \cite {DDV, FMQR}. We found then natural to apply that strategy to the study of GKP vacuum and its excitations. In this approach scalar excitations, which are represented by holes in the distribution of type-4 roots in $[-b,b]$, are classified by quantisation conditions for $Z_4(v)$. The same function $Z_4(v)$ governs the interaction between roots with different flavour and scalars.

Coming in specific to the classification of the various excitations \cite {Basso}, we have already said that scalars are represented by holes in the distribution of type-4 roots. Large (small) fermions are described by $u_3$-type ($u_1$-type) roots and large (small) antifermions are described by $u_5$-type ($u_7$-type) roots. Rapidity of large fermions is the function $x_F(u_3)=x(u_3)$, where ($g$ is related to the 't Hooft coupling $\lambda$ by $\lambda =8\pi^2 g^2$)
\be
x(u)=\frac{u}{2} \left [ 1+ \sqrt{1-\frac{2g^2}{u^2}} \right ] \, , \quad u^2 \geq 2g^2 \, ,  \label {xudef}
\ee
with the arithmetic definition of the square root. Therefore, rapidity of large fermions satisfies the inequality $|x_F|\geq g/\sqrt{2}$. On the other hand rapidity of small fermions is the function $x_f(u_1)=\frac{g^2}{2x(u_1)}$,
with definition (\ref {xudef}) for $x(u)$ and, consequently, it is constrained by the inequality
$|x_f|\leq g/\sqrt{2}$. Changing $u_3 \rightarrow u_5$ and $u_1 \rightarrow u_7$ allows to describe large and small antifermions, respectively.

Gauge fields $F_{+\perp}$ with rapidity $u_j^g$ correspond to stacks,
\be
u_{2,j}=u^{g}_j \, , \quad u_{3,j}=u^g_j \pm i/2 \, , \quad j=1,...,N_g \, ,
\ee
with real centres $u^g_j$, while gauge fields $\bar F_{+\perp}$ with rapidity $u_j^{\bar g}$ are described by stacks,
\be
u_{6,j}=u^{\bar g}_j \, , \quad u_{5,j}=u^{\bar g}_j \pm i/2 \, , \quad j=1,...,N_{\bar g} \, ,
\ee
with real centres $u^{\bar g}_j$.

We consider also the presence of isotopic roots, which do not carry momentum and energy, but take into account internal degrees of freedom, i.e. the residual $SU(4)$ symmetry of the GKP vacuum. In specific, we have
$K_a$ roots $u_{a,j}$ of type $u_2$,
\be
u_{a,j}=u_{2,j} \, , \quad j=1,...,K_a  \, , \label {uais}
\ee
$K_c$ roots  $u_{c,j}$ of type $u_6$
\be
u_{c,j}=u_{6,j} \, , \quad  j=1,...,K_c \, ,  \label {ucis}
\ee
and $K_b$ stacks,
\ba
u_{4,j}=u_{b,j}\pm \frac{i}{2} \, , \quad
u_{b,j}=u_{3,j}=u_{5,j} \, , \quad j=1,...,K_b \, , \label {stack}
\ea
with centers $u_{b,j}$.

\medskip

We are now going to present our derivation of the full set of Bethe-Yang equations for excitations on the GKP vacuum.
For excitations with rapidity $u_{m}$ belonging to the representation $\rho $ of the symmetry group $SU(4)$ of the GKP vacuum such equations will appear in the following form
\ba\label{Lievac}
&& \prod_{m}
\frac{u_{q,k}-u_{m}+i\vec\alpha_q\cdot\vec w_{\rho}}{u_{q,k}-u_{m}-i\vec\alpha_q\cdot\vec w_{\rho}} =
\prod_{j\neq k}^{K_{q}}\frac{u_{q,k}-u_{q,j}+i\vec\alpha_q\cdot\vec\alpha_{q}}{u_{q,k}-u_{q,j}-i\vec\alpha_q\cdot \vec\alpha_{q}}\,
\prod_{q'\neq q}\prod_{j=1}^{K_{q'}}
\frac{u_{q,k}-u_{q',j}+i\vec\alpha_q\cdot\vec\alpha_{q'}}{u_{q,k}-u_{q',j}-i\vec\alpha_q\cdot\vec\alpha_{q'}}\ , \\
1&=&e^{iRP(u_{m})+2iD(u_m)} \prod _{q}\prod _{k=1}^{K_q}\frac{u_{m}-u_{q,k}+i\vec\alpha_q\cdot\vec w_{\rho}}{u_{m}-u_{q,k}-i\vec\alpha_q\cdot\vec w_{\rho}}\prod _{m ' \not= m} S (u_{m}, u_{m'}) \label {qua} \, ,
\ea
where $\{\alpha_q\}$ are the set of simple roots of $SU(4)$, $u_{q,k}$ are the isotopic roots associated and $\vec w_{\rho}$ the highest weight of the representation $\rho $. The structure of these equations agrees with the general pattern shown in \cite {OW}. While the first equation (\ref {Lievac}) comes from the symmetry properties of the vacuum, the second
one (\ref {qua}) is a quantisation condition for the rapidity $u_{m}$ of an excitation moving in a one dimensional chain. Within this interpretation, $R$ is given the meaning of the physical length of the chain, $P (u_{m})$ that of the momentum
of an excitation with rapidity $u_{m}$. The extra term $2D(u_m)$ in the exponent is interpreted as the effect of two {\it purely transmitting} ({\it i.e.} without reflection \cite{Baj-G}) defects related to the tips of the GKP string. The rational factor in the right hand side of (\ref {qua}) takes into account the internal degrees of freedom: solving (\ref {Lievac}) one obtains $u_{q,k}$ in terms of $u_{m}$: plugging this result in the rational term in (\ref {qua}) one obtains, together with the products over the various $S (u_{m}, u_{m '})$, the phase change due to the scattering between an excitation with rapidity $u_{m}$ and the other excitations with rapidity $u_{m'}$.

We will start with the one loop case, where all factors entering equations (\ref {qua}) are written in an explicit form, specifically in terms of products of Euler Gamma functions.
The general all loops case will appear as a technical complication of the one loop, since the building blocks
of the various equations (\ref {qua}) will be obtained after solving a linear integral equation.

\subsection{Equations at one loop}

{\bf Scalars}

\medskip

In order to show how our strategy works, we first concentrate on the one loop case.
We start from the fourth of the Beisert-Staudacher equations, in the presence of a large number
$s$ of real type-4 roots, together with the general pattern of excitations and isotopic roots described in previous section. We remark that in the one loop case only large fermions and large antifermions are present: for uniformity of notations in this subsection we will denote large fermions rapidity $x_F$ with $u_F$ and large antifermions rapidity $x_{\bar F}$ with $u_{\bar F}$.

We introduce the counting function
\ba
Z_4(v)&=& iL \ln \frac{\frac{i}{2}-v}{\frac{i}{2}+v}+i \sum _{j=1}^{K_4} \ln \frac{i-v+u_{4,j}}{i+v-u_{4,j}} +2i \sum _{j=1}^{K_b}
\ln \frac{\frac{i}{2}+v-u_{b,j}}{\frac{i}{2}-v+u_{b,j}} + i \sum _{j=1}^{N_F} \ln \frac{\frac{i}{2}+v-u_{F,j}}{\frac{i}{2}-v+u_{F,j}} + \nonumber \\
&+& i \sum _{j=1}^{N_{\bar F}} \ln \frac{\frac{i}{2}+v-u_{\bar F,j}}{\frac{i}{2}-v+u_{\bar F,j}}+i \sum _{j=1}^{N_g} \ln \frac{i+v-u^g_j}{i-v+u^g_j} +
i \sum _{j=1}^{N_{\bar g}} \ln \frac{i+v-u^{\bar g}_j}{i-v+u^{\bar g}_j} \, , \label {Z40}
\ea
where the sum up to $K_4=s+2K_b$ is a sum over $s$ real type-4 roots and over the $2K_b$ complex type-4 roots contained in the stack (\ref {stack}).
In terms of $Z_4$ the fourth of Beisert-Staudacher equations reads as
\be
e^{-iZ_4(u_{4,k})}=(-1)^{L+K_4+2K_b+N_F+N_{\bar F}+N_g+N_{\bar g}-1} \label {Z4-qua} \, .
\ee
In addition, in the large $s$ limit the behaviour of $Z_4(v)$ is dominated by the second term in the right hand side, which implies that for $v$ real $Z_4'(v) <0$. With this information we can prove that
the length $L$ is not independent of the total number of excitations. Indeed,
it is widely known \cite{KOR, BGK} that condition (\ref {Z4-qua}) is satisfied on the real axis not only by the type-4 real roots, but also by $H+2$ real numbers, called holes. $H$ holes are inside the interval $[-b,b]$ ($b =s/2 +O(1/s^2))$ and, consequently, are often called 'internal' or 'small'; the remaining two holes have positions $\pm s/\sqrt{2}+O(1/s^2)$ and then are dubbed 'external' or 'large'.
Since $Z_4'(v)<0$ for $v$ real, the difference between the extremal values on the real axis $Z_4(+\infty)-Z_4(-\infty)$ has to count the total number of real roots and holes, i.e.
\be
Z_4(+\infty)-Z_4(-\infty)=-2\pi (s+H+2) \, , \label {diff-Z}
\ee
On the other hand, the definition (\ref  {Z40}) implies the asymptotic behaviours
\be
Z_4(\pm \infty)=\mp \pi (L+K_4-2K_b-N_F-N_{\bar F}-N_g-N_{\bar g}) =
\mp \pi (L+s-N_F-N_{\bar F}-N_g-N_{\bar g})\, . \label {diff-Z2}
\ee
Comparison between (\ref {diff-Z}) and (\ref  {diff-Z2}) gives the desired connexion
\be
L=H+2+N_F+N_{\bar F}+N_g+N_{\bar g} \, , \label {L-ecc}
\ee
between the length $L$ and the total number of excitations.

Relation (\ref {L-ecc}) once plugged in the exponent of the right hand side of (\ref {Z4-qua}), provides a simplification of quantisation condition for type-4 roots and real holes. Restricting to holes, whose position we call $u_h, h=1,...,H$, we get the compact formula (remember that $s$ is always even)
\be
e^{-iZ_4(u_h)}=(-1)^{H-1} \, . \label {quant}
\ee
After fixing these preliminary aspects, we come back to equation (\ref {Z40}). In order to get manageable expressions, we convert the sum over real type-4 roots into an integral by means of the master equation \cite {FMQR}:
\be
u_{4,j} \quad \textrm{real} \quad \Rightarrow \quad \sum _{j=1}^s f(u_{4,j})=-f\left (\frac{s}{\sqrt{2}} \right )
-f\left (-\frac{s}{\sqrt{2}} \right )-\sum _{h=1}^H f(u_h) - \int _{-\infty}^{+\infty} \frac{dv}{2\pi} f(v)\frac{d}{dv} [ Z_4(v)-2L_4(v) ] \, ,  \label {summaster}
\ee
where $L_4(v)= {\mbox {Im}} \ln [1+(-1)^{H} \, e^{iZ_4(v-i0^+)}]$.
We specialise formula (\ref {summaster}) to our case and get
\ba
&& \sum _{j=1}^{K_4}i \ln \frac{i-v+u_{4,j}}{i+v-u_{4,j}}= \sum _{j=1}^{K_b}i \ln \frac{(\frac{i}{2}-v+u_{b,j})(\frac{3i}{2}-v+u_{b,j})}{(\frac{i}{2}+v-u_{b,j})(\frac{3i}{2}+v-u_{b,j})}
-i \ln \frac{(i-v+\frac{s}{\sqrt{2}})(i-v-\frac{s}{\sqrt{2}})}{(i+v-\frac{s}{\sqrt{2}})(i+v+\frac{s}{\sqrt{2}})}- \nonumber \\
&-& i \sum _{h=1}^H \ln \frac{i-v+u_h}{i+v-u_h} + \int _{-\infty}^{+\infty}\frac{dw}{\pi} \frac{1}{1+(v-w)^2}  [Z_4(w)-2L_4(w)] \label {sumK4} \, .
\ea
Eventually, putting together (\ref {Z40}) and (\ref {sumK4}), we find that
\be
Z_4(v)= F(v)+2 \int _{-\infty}^{+\infty} dw \, G(v-w) \, L_4(w) \, , \label {Z40eq}
\ee
where $F(v), G(v)$ satisfy the linear integral equations
\ba\label {F0u}
F(v)&=&iL \ln \frac{\frac{i}{2}-v}{\frac{i}{2}+v}+i\sum _{j=1}^{K_b}\ln \frac{\frac{i}{2}+v-u_{b,j}}{\frac{i}{2}-v+u_{b,j}}\frac{\frac{3i}{2}-v+u_{b,j}}{\frac{3i}{2}+v-u_{b,j}}+ i\sum _{j=1}^{N_g} \ln \frac{i+v-u^g_j}{i-v+u^g_j}+\nonumber \\
&+& i\sum _{j=1}^{N_{\bar g}} \ln \frac{i+v-u^{\bar g}_j}{i-v+u^{\bar g}_j}+ i \sum _{j=1}^{N_F} \ln \frac{\frac{i}{2}+v-u_{F,j}}{\frac{i}{2}-v+u_{F,j}} + i \sum _{j=1}^{N_{\bar F}} \ln \frac{\frac{i}{2}+v-u_{\bar F,j}}{\frac{i}{2}-v+u_{\bar F,j}} + \\
 &+& i \sum _{h=1}^H \ln \frac{i+v-u_h}{i-v+u_h} +i \ln \frac{i+v-\frac{s}{\sqrt{2}}}{i-v+\frac{s}{\sqrt{2}}}
\frac{i+v+\frac{s}{\sqrt{2}}}{i-v-\frac{s}{\sqrt{2}}}+\int _{-\infty}^{+\infty}\frac{dw}{\pi} \frac{1}{1+(v-w)^2} F(w)
 \, ,  \nn\\
G(u-v) &=& -\frac{1}{\pi}\frac{1}{1+(u-v)^2} + \int _{-\infty}^{+\infty} \frac{dw}{\pi}  \frac{1}{1+(u-w)^2} G(w-v)     \, .  \label {G0eq}
\ea
Equation (\ref {F0u}) is solved exactly; however, we remember that the pattern of excitations discussed before holds
only in the large spin limit. To be consistent with that, we have to use the large $s$ asymptotic behaviour
\be
 i \ln \frac{\Gamma \left (1+iv-i \frac{s}{\sqrt{2}}\right )}{\Gamma \left (1-iv+i \frac{s}{\sqrt{2}}\right )}
+ i \ln \frac{\Gamma \left (1+iv+i \frac{s}{\sqrt{2}}\right )}{\Gamma \left (1-iv-i \frac{s}{\sqrt{2}} \right )} \rightarrow
-4v \ln \frac{s}{\sqrt{2}} +O(1/s^2) \, ,
\ee
and write the final result for $F(v)$ as
\ba
F(v)&=&-iL \ln \frac{\Gamma \left (\frac{1}{2}+iv\right )}{\Gamma \left (\frac{1}{2}-iv\right )}-4v \ln \frac{s}{\sqrt{2}}  + i\sum _{h=1}^H \ln \frac{\Gamma (1+iv-iu_h)}{\Gamma (1-iv+iu_h)}+ \nonumber \\
&+& i \sum _{j=1}^{K_b} \ln \frac{\frac{i}{2}+v-u_{b,j}}{\frac{i}{2}-v+u_{b,j}} + i\sum _{j=1}^{N_g} \ln \frac{\Gamma (1+iv-iu_j^g)}{\Gamma (1-iv+iu_j^g)}+  i\sum _{j=1}^{N_{\bar g}} \ln \frac{\Gamma (1+iv-iu_j^{\bar g})}{\Gamma (1-iv+iu_j^{\bar g})} +\label {1loopF} \\
&+& i\sum _{j=1}^{N_F} \ln \frac{\Gamma (\frac{1}{2}+iv-iu_{F,j})}{\Gamma (\frac{1}{2}-iv+iu_{F,j})}+
i\sum _{j=1}^{N_{\bar F}} \ln \frac{\Gamma (\frac{1}{2}+iv-iu_{\bar F,j})}{\Gamma (\frac{1}{2}-iv+iu_{\bar F,j})} +O(1/s^2) \, . \nonumber
\ea
On the other hand, the solution of (\ref {G0eq}) for $G$ reads as
\be
G(v-w)= \frac{1}{2\pi}[\psi (1+iv-iw)+ \psi (1-iv+iw) ]\, . \quad \label {1loopG}
\ee
Then, we notice that in the large $s$ limit $Z_4(v)\sim F(v) \sim -4v \ln (s/\sqrt{2}) +O(s^0)$. This means that the leading behaviour of the nonlinear term in (\ref {Z40eq}) is the same as that of the analogous term for the GKP vacuum: therefore, we can use results of \cite {FRS} and approximate the non linear term in (\ref {Z40eq}) as
\be
2 \int  _{-\infty}^{+\infty} dw \, G(v-w) \, L_4(w) =-2v \ln 2 + O(1/s^2) \, .  \label {1loopnl}
\ee
Plugging (\ref {1loopF}) and (\ref {1loopnl}) into (\ref {Z40eq}), we eventually get
\ba
Z_4(v)&=&-iL \ln \frac{\Gamma \left (\frac{1}{2}+iv\right )}{\Gamma \left (\frac{1}{2}-iv\right )}-4v \ln s + i\sum _{h=1}^H \ln \frac{\Gamma (1+iv-iu_h)}{\Gamma (1-iv+iu_h)}+ \nonumber \\
&+& i \sum _{j=1}^{K_b} \ln \frac{\frac{i}{2}+v-u_{b,j}}{\frac{i}{2}-v+u_{b,j}} + i\sum _{j=1}^{N_g} \ln \frac{\Gamma (1+iv-iu_j^g)}{\Gamma (1-iv+iu_j^g)}+  i\sum _{j=1}^{N_{\bar g}} \ln \frac{\Gamma (1+iv-iu_j^{\bar g})}{\Gamma (1-iv+iu_j^{\bar g})} +\label {1loopZ} \\
&+& i\sum _{j=1}^{N_F} \ln \frac{\Gamma (\frac{1}{2}+iv-iu_{F,j})}{\Gamma (\frac{1}{2}-iv+iu_{F,j})}+
i\sum _{j=1}^{N_{\bar F}} \ln \frac{\Gamma (\frac{1}{2}+iv-iu_{\bar F,j})}{\Gamma (\frac{1}{2}-iv+iu_{\bar F,j})} +O(1/s^2) \, . \nonumber
\ea
Now, it is clear that imposing quantisation condition (\ref {quant}) on (\ref {1loopZ}) provides a constraint between the rapidity $u_h$ of a scalar and the rapidities of all the other excitations. As in all integrable models, this constraint has the general form (\ref {qua}): therefore we could be tempted to use (\ref {quant}) to define momenta and scattering factors of excitations, as well as the effective length of the chain. Such procedure, however, will provide scattering factors $i\ln S$ which diverge as $u_{\ast}\ln u_{\ast}$ when the rapidity $u_{\ast}$ of a generic excitation becomes very large. Fortunately, it happens that this problem can be avoided if we make use of the zero momentum condition, which is a selection rule to extract physical states out of the Beisert-Staudacher equations.
To be specific, all physical states have to satisfy the condition $e^{iP}=1$, where
\be
P=i\sum _{j=1}^{K_4} \ln \frac{u_{4,j}+\frac{i}{2}}{u_{4,j}-\frac{i}{2}} \, .
\ee
Since $K_4=s+2K_b$ is even, we can also write
\be
P=i\sum _{j=1}^{K_4} \ln \frac{\frac{i}{2}+u_{4,j}}{\frac{i}{2}-u_{4,j}}+ 2\pi {\mathbb{Z}} \, .
\ee
This expression is regular for $u_{4,j}=0$ and, therefore, it is more convenient for our calculations:
\ba
P&=&i\sum _{j=1}^{K_4} \ln \frac{\frac{i}{2}+u_{4,j}}{\frac{i}{2}-u_{4,j}}= i \sum _{h=1}^H \ln \frac{\frac{i}{2}-u_h}{\frac{i}{2}+u_h}+
i \sum _{j=1}^{K_b} \ln \frac{i+u_{b,j}}{i-u_{b,j}}+ \nonumber \\
&+& \pi K_b-i \int  _{-\infty}^{+\infty} \frac{dv}{2\pi} \ln \frac{\frac{i}{2}+v}{\frac{i}{2}-v} \frac{d}{dv} [ Z_4(v)-2L_4(v)]= \pi K_b -i \sum _{h=1}^H \ln \frac{\Gamma \left (\frac{1}{2}+iu_h \right )}{\Gamma \left (\frac{1}{2}-iu_h \right )}- \nonumber \\
&-& i \sum _{j=1}^{N_g} \ln \frac{\Gamma \left (\frac{3}{2}+iu_j^{g} \right )}{\Gamma \left (\frac{3}{2}-iu_j^{g} \right )}
-i \sum _{j=1}^{N_{\bar g}} \ln \frac{\Gamma \left (\frac{3}{2}+iu_j^{\bar g} \right )}{\Gamma \left (\frac{3}{2}-iu_j^{\bar g} \right )}-  \nonumber \\
&-& i \sum _{j=1}^{N_F} \ln \frac{\Gamma \left (1+iu_j^{F} \right )}{\Gamma \left (1-iu_j^{F} \right )} -
 i \sum _{j=1}^{N_{\bar F}} \ln \frac{\Gamma \left (1+iu_j^{\bar F} \right )}{\Gamma \left (1-iu_j^{\bar F} \right )}+
 O(1/s^2) \, .\label {P-1}
\ea
As a technical remark, we notice that nonlinear terms give no contributions at the orders $\ln s$ and $(\ln s )^0$.

Putting together (\ref {1loopZ}) and (\ref  {P-1}) we obtain the equality
\ba
Z_4(v)-P&=&i  \sum _{j=1}^{K_b} \ln \frac{v-u_{b,j}+\frac{i}{2}}{v-u_{b,j}-\frac{i}{2}}+ \nonumber \\
&+& i \sum  _{h=1}^H \ln \frac{\Gamma \left (\frac{1}{2}-iv\right )\Gamma \left (\frac{1}{2}+iu_h \right ) \Gamma (1+iv-iu_h)}{\Gamma \left (\frac{1}{2}+iv\right )\Gamma \left (\frac{1}{2}-iu_h\right ) \Gamma (1-iv+iu_h)} + \nonumber \\
&+& i \sum  _{j=1}^{N_g} \ln \frac{\Gamma \left (\frac{1}{2}-iv\right )\Gamma \left (\frac{3}{2}+iu_j^g \right ) \Gamma (1+iv-iu_j^g)}{\Gamma \left (\frac{1}{2}+iv\right )\Gamma \left (\frac{3}{2}-iu_j^g \right ) \Gamma (1-iv+iu_j^g)} + (g \rightarrow \bar g) + \label {Fv}\\
&+& i \sum  _{j=1}^{N_F} \ln \frac{\Gamma \left (\frac{1}{2}-iv\right )\Gamma (1+iu_j^F ) \Gamma \left (\frac{1}{2}+iv-iu_j^F \right )}{\Gamma \left (\frac{1}{2}+iv\right )\Gamma (1-iu_j^F) \Gamma \left (\frac{1}{2}-iv+iu_j^F \right)} + (F \rightarrow \bar F) - \nonumber \\
&-& 2i \ln \frac{\Gamma \left ( \frac{1}{2}+iv \right )}{\Gamma \left ( \frac{1}{2}-iv \right )}
-4v \ln s + O(1/s^2) \, .  \nonumber
\ea
Therefore, we have gained the possibility to write the condition $e^{-i[Z_4(u_h)-P]}=(-1)^{H-1}$ as a convenient alternative to (\ref{quant}):
\ba
1&=&e^{4iu_h \ln s } \left(\frac{\Gamma(\frac{1}{2}-iu_h)}{\Gamma(\frac{1}{2}+iu_h)}\right)^2
 \prod _{\stackrel {h'=1}{h'\not=h}}^{H} (-)\frac{\Gamma \left (\frac{1}{2}-iu_h \right ) \Gamma \left (\frac{1}{2}+iu_{h'} \right ) \Gamma (1+iu_h-iu_{h'})}{\Gamma \left (\frac{1}{2}+iu_h \right )
\Gamma \left (\frac{1}{2}-iu_{h'} \right )\Gamma (1-iu_h+iu_{h'})}
 \prod _{j=1}^{K_b} \frac{u_h-u_{b,j}+\frac{i}{2}}{u_h-u_{b,j}-\frac{i}{2}} \cdot  \nonumber\\
&\cdot & \prod _{j=1}^{N_{g}}
\frac{\Gamma\left(1+i(u_h-u^g_j)\right)\Gamma\left(\frac{1}{2}-iu_h\right)\Gamma\left(\frac{3}{2}+iu^g_j\right)}
{\Gamma\left(1-i(u_h-u^g_j)\right)\Gamma\left(\frac{1}{2}+iu_h\right)\Gamma\left(\frac{3}{2}-iu_j^g\right)}
\prod _{j=1}^{N_{\bar g}} \frac{\Gamma\left(1+i(u_h-u^{\bar g}_j)\right)\Gamma\left(\frac{1}{2}-iu_h\right)\Gamma\left(\frac{3}{2}+iu^{\bar g}_j\right)}{\Gamma\left(1-i(u_h-u^{\bar g}_j)\right)\Gamma\left(\frac{1}{2}+iu_h\right)\Gamma\left(\frac{3}{2}-iu_j^{\bar g}\right)}
   \nonumber\\
&\cdot &\prod _{j=1}^{N_F} \frac{\Gamma(\frac{1}{2}+i(u_h-u_{F,j}))\Gamma(1+iu_{F,j})\Gamma(\frac{1}{2}-iu_h)}
{\Gamma(\frac{1}{2}-i(u_h-u_{F,j}))\Gamma(1-iu_{F,j})\Gamma(\frac{1}{2}+iu_h)}
\prod _{j=1}^{N_{\bar F}} \frac{\Gamma(\frac{1}{2}+i(u_h-u_{{\bar F},j}))\Gamma(1+iu_{{\bar F},j})\Gamma(\frac{1}{2}-iu_h)}{\Gamma(\frac{1}{2}-i(u_h-u_{{\bar F},j}))\Gamma(1-iu_{{\bar F},j})\Gamma(\frac{1}{2}+iu_h)} \nonumber\\
\label {scal-by}
\ea
We take (\ref {scal-by}) as Bethe-Yang equations for scalars. In the spirit of (\ref {qua}) we make the following identifications:

$\bullet$ Length of the chain $R=2\ln s $

$\bullet$ Momentum of a scalar $P^{(s)}_0(u_h)=2 u_h$

The terms in (\ref {scal-by}) depending on two rapidities have the natural interpretation of scattering factors between scalars and other excitations. Using notations given in Appendix \ref {scatt-fact}, we write
\ba
1&=&e^{iRP^{(s)}_0(u_h)} \left[\frac{\Gamma(\frac{1}{2}-iu_h)}{\Gamma(\frac{1}{2}+iu_h)}\right]^2
 \prod _{j=1}^{K_b} \frac{u_h-u_{b,j}+\frac{i}{2}}{u_h-u_{b,j}-\frac{i}{2}}  \prod _{\stackrel {h'=1}{h'\not=h}}^{H} S_0^{(ss)}(u_h,u_{h'}) \cdot  \nonumber\\
&\cdot & \prod _{j=1}^{N_{g}} S^{(sg)}_0(u_h,u_j^g)  \prod _{j=1}^{N_{\bar g}} S^{(s\bar g)}_0(u_h,u_j^{\bar g})
\prod _{j=1}^{N_F} S^{(sF)}_0(u_h,u_{F,j})   \prod _{j=1}^{N_{\bar F}} S^{(s\bar F)}_0 (u_h, u_{\bar F,j})
\label {scal-by2} \, ,
\ea
where $S^{(s \ast)}_0$ denotes the scattering factors between a scalar and a generic excitation.
We remark that $i\ln S^{(s \ast)}_0$ behaves like $\ln u_{\ast}$ when the rapidity of an excitation becomes large.
Eventually, the last term
\be
\left[ \frac{\Gamma(\frac{1}{2}-iu_h)}{\Gamma(\frac{1}{2}+iu_h)}\right ] ^2
\ee
has the form of the phase delay due to two purely transmitting defects.

\medskip

Finally, in view of generalisations to all loops we find convenient to identify the various pieces entering the function $Z_4(v)$
\ba
Z_4(v)&=&\Theta '_0(v,s/\sqrt{2})+\Theta '_0(v,-s/\sqrt{2})+\sum _{h=1}^H \Theta '_0(v,u_h) + i \sum _{j=1}^{K_b} \ln \frac{\frac{i}{2}+v-u_{b,j}}{\frac{i}{2}-v+u_{b,j}} + \nonumber \\
&+& \sum _{j=1}^{N_g} F^G_0(v,u^g_j) + \sum _{j=1}^{N_{\bar g}} F^G_0(v,u^{\bar g}_j)+\sum _{j=1}^{N_F} F^F_0(v,u_{F,j}) +\sum _{j=1}^{N_{\bar F}} F^F_0(v,u_{\bar F,j})-2v \ln 2 \, , \label {Z4abs}
\ea
as solutions of integral equations
\ba
&& \Theta '_0(v,u)=\phi _0(v-u)+\Phi _0(v) - \int _{-\infty}^{+\infty} dw   \varphi _0(v-w) \Theta '_0(w,u)=\nonumber \\
&&= i \ln \frac{\Gamma \left (1+iv-iu\right )\Gamma (1/2-iv)}{\Gamma \left (1-iv+iu\right )\Gamma (1/2+iv)} \, ,  \\
&& F^F_0(v,u)=\chi _0(v-u|1)+\Phi _0(v) - \int  _{-\infty}^{+\infty} dw   \varphi _0(v-w) F^F_0(w,u) = \nonumber \\
&&= i \ln \frac{\Gamma \left (1/2+iv-iu\right )\Gamma (1/2-iv)}{\Gamma \left (1/2-iv+iu\right )\Gamma (1/2+iv)}
\label {F-F0} \, , \\
&& F^G_0(v)=\chi _0(v-u|2)+\Phi _0(v) - \int  _{-\infty}^{+\infty} dw   \varphi _0(v-w)  F^G_0(w,u) =  \nonumber \\
&& = i \ln \frac{\Gamma \left (1+iv-iu\right )\Gamma (1/2-iv)}{\Gamma \left (1-iv+iu\right )\Gamma (1/2+iv)}\, ,
\label {F-G0}
\ea
where $\Phi _0$, $\phi _0$, $\chi _0$ are defined in Appendix \ref {functions}. In (\ref {Z4abs}) the large $s$ limit has to be taken in the first two terms in the right hand side. This limit gives
\be
\Theta '_0(v,s/\sqrt{2})+\Theta '_0(v,-s/\sqrt{2}) \rightarrow -4v \ln \frac{s}{\sqrt{2}} -2i \ln \frac{\Gamma \left (\frac{1}{2}+iv \right )}{\Gamma \left (\frac{1}{2}-iv \right )} \, .
\ee

\medskip

{\bf Fermions}

\medskip

The equations for (large) fermions with rapidity $x_{F,k}=u_{F,k}$ come from the (inverse of the) third of the Beisert-Staudacher equations.
We have
\ba
1&=&\prod _{j=1}^{K_a} \frac{u_{F,k}-u_{a,j}+\frac{i}{2}}{u_{F,k}-u_{a,j}-\frac{i}{2}}\prod _{j=1}^{N_g} \frac{u_{F,k}-u^g_{j}+\frac{i}{2}}{u_{F,k}-u^g_{j}-\frac{i}{2}}(-1)^{K_4}\prod _{j=1}^{K_4} \frac{\frac{i}{2}+u_{4,j}-u_{F,k}}{\frac{i}{2}+u_{F,k}-u_{4,j}}= \nonumber \\
&=& \prod _{j=1}^{K_a} \frac{u_{F,k}-u_{a,j}+\frac{i}{2}}{u_{F,k}-u_{a,j}-\frac{i}{2}}
\prod _{j=1}^{N_g} \frac{u_{F,k}-u^g_{j}+\frac{i}{2}}{u_{F,k}-u^g_{j}-\frac{i}{2}}
(-1)^{K_b}\prod _{j=1}^{K_b} \frac{i-u_{F,k}+u_{b,j}}{i+u_{F,k}-u_{b,j}} \prod _{h=1}^H \frac{\frac{i}{2}+u_{F,k}-u_h}{\frac{i}{2}+u_h-u_{F,k}} \cdot \nonumber \\
&\cdot & \textrm{exp} \Bigl [ -\int  _{-\infty}^{+\infty} \frac{dv}{2\pi}  \ln \frac{\frac{i}{2}+v-u_{F,k}}{\frac{i}{2}-v+u_{F,k}}\frac{d}{dv}\left (Z_4(v)-2L_4(v) \right )
\Bigr ] \label {fer-1} \, ,
\ea
since $K_4=s+2K_b$ is even. Then, we evaluate the integral ($L_4$ contributes with subleading $O(1/s^2)$ terms)
\ba
&& -\int  _{-\infty}^{+\infty} \frac{dv}{2\pi}  \ln \frac{\frac{i}{2}+v-u_{F,k}}{\frac{i}{2}-v+u_{F,k}}\frac{d}{dv}\left (Z_4(v)-2L_4(v) \right )= L \ln \frac{\Gamma (1-iu_{F,k})}{\Gamma (1+iu_{F,k})} +4i u_{F,k} \ln s - \nonumber \\
&-& \sum _{h=1}^H \ln \frac{\Gamma (\frac{3}{2}+iu_h-iu_{F,k})}{\Gamma (\frac{3}{2}-iu_h+iu_{F,k})}
-  \sum _{j=1}^{N_g}\ln \frac{\Gamma \left (\frac{3}{2}+iu_j^g-iu_{F,k}\right )}{\Gamma \left (\frac{3}{2}-iu_j^g+iu_{F,k}\right )}
-  \sum _{j=1}^{N_{\bar g}}\ln \frac{\Gamma \left(\frac{3}{2}+iu_j^{\bar g}-iu_{F,k}\right )}{\Gamma \left (\frac{3}{2}-iu_j^{\bar g}+iu_{F,k}\right )} - \nonumber \\
&-& \sum  _{j=1}^{N_F} \ln \frac{\Gamma (1+iu_{F,j}-iu_{F,k})}{\Gamma (1-iu_{F,j}+iu_{F,k})}
- \sum  _{j=1}^{N_{\bar F}} \ln \frac{\Gamma (1+iu_{\bar F,j}-iu_{F,k})}{\Gamma (1-iu_{\bar F,j}+iu_{F,k})}
+\sum _{j=1}^{K_b}\ln \frac{i+u_{F,k}-u_{b,j}}{i-u_{F,k}+u_{b,j}} \, .
\ea
Now, in order to reproduce scattering factors already appearing in equations for scalars (\ref  {scal-by}, \ref  {scal-by2}), we use the zero momentum condition: we multiply (\ref {fer-1}) with $1=e^{iP}$, with $P$ given by (\ref {P-1}). Using notations defined in Appendix \ref {scatt-fact} we write the final Bethe equations for fermionic excitations as
\ba
1&=& e^{i R P^{(F)}_0 (u_{F,k}) }  \left [ \frac{\Gamma (1-iu_{F,k})}{\Gamma (1+iu_{F,k})} \right ]^2
\prod _{j=1}^{K_a} \frac{u_{F,k}-u_{a,j}+i/2}{u_{F,k}-u_{a,j}-i/2}
\prod _{h=1}^H  S^{(Fs)}_0(u_{F,k},u_h)
 \cdot \nonumber \\
&\cdot & \prod _{j=1}^{N_g} S^{(Fg)}_0 (u_{F,k},u_j^g)
\prod _{j=1}^{N_{\bar g}} S^{(F\bar g)}_0 (u_{F,k},u_j^{\bar g})
\prod _{j=1}^{N_F} S^{(FF)}_0(u_{F,k},u_{F,j}) \prod _{j=1}^{N_{\bar F}} S^{(F\bar F)}_0 (u_{F,k},u_{\bar F,j})
\label {fer-by} \, ,
\ea
where we introduced the length $R=2\ln s$ of the chain and the momentum $P^{(F)}_0(u_{F,k})=2 u_{F,k}$ of a fermionic excitation. As for scalars, the term
\be
\left [ \frac{\Gamma (1-iu_{F,k})}{\Gamma (1+iu_{F,k})} \right ]^2
\ee
stands for phase delay due to purely transmitting defects.

\medskip

Equations for large antifermions come from the (inverse of the) fifth of the Beisert-Staudacher equations and are obtained in a completely similar way as in
the fermions case. The final result is:
\ba
1&=& e^{i R P^{(F)}_0( u_{\bar F,k})}  \left [ \frac{\Gamma (1-iu_{\bar F,k})}{\Gamma (1+iu_{\bar F,k})} \right ]^2
\prod _{j=1}^{K_c} \frac{u_{\bar F,k}-u_{c,j}+\frac{i}{2}}{u_{\bar F,k}-u_{c,j}-\frac{i}{2}}
\cdot \nonumber \\
&\cdot & \prod _{j=1}^{N_g} S^{(\bar Fg)}_0 (u_{\bar F,k},u_j^g)
\prod _{j=1}^{N_{\bar g}} S^{(\bar F\bar g)}_0 (u_{\bar F,k},u_j^{\bar g})
\prod _{j=1}^{N_F} S^{(\bar FF)}_0(u_{\bar F,k},u_{F,j}) \prod _{j=1}^{N_{\bar F}} S^{(\bar F\bar F)}_0 (u_{\bar F,k},u_{\bar F,j})
\label {antifer-by} \, ,
\ea
where we introduced the length $R=2\ln s$ of the chain, the momentum $P^{(\bar F)}_0(u_{\bar F,k})=2 u_{\bar F,k}$ of a antifermionic excitation and the 'defect' term
\be
\left [ \frac{\Gamma (1-iu_{\bar F,k})}{\Gamma (1+iu_{\bar F,k})} \right ]^2 \, .
\ee

\medskip

{\bf Gluons}

\medskip

In the presence of a large number $s$ of real type-4 roots, a gluon with rapidity $u_k^g$ is described by a stack composed of a single type-2 root $u_{2,k}=u_k^g$ and a two-string formed by two type-3 roots $u_{3,k}=u_k^g \pm i/2$.
Rapidity $u_k^g$ is then constrained by the equation obtained by multiplying together
the (inverse of the) second of the Beisert-Staudacher equations with $u_{2,k}=u_k^g$ with
the (inverse of the) third for $u_{3,k}=u_k^g \pm i/2$, i.e.
\be
1=\prod _{\stackrel {j=1}{j\not= k}}^{N_g} \frac{u_k^g-u_j^g+i}{u_k^g-u_j^g-i} \prod _{j=1}^{K_b} \frac{u_k^g-u_{b,j}+\frac{i}{2}}{u_k^g-u_{b,j}-\frac{i}{2}} \prod _{j=1}^{N_F} \frac{u_k^g-u_{F,j}+\frac{i}{2}}{u_k^g-u_{F,j}-\frac{i}{2}}
\prod _{j=1}^{K_4} \frac{u^{g}_k-u_{4,j}-i}{u^g_k-u_{4,j}+i} \label {gl5} \, .
\ee
We concentrate on the last term in (\ref {gl5}), which we rewrite as (remember that $K_4$ is even)
\ba
&& \prod _{j=1}^{K_4} \frac{u^{g}_k-u_{4,j}-i}{u^g_k-u_{4,j}+i}= \prod _{j=1}^{K_4} \frac{i-u^{g}_k+u_{4,j}}{i+u^g_k-u_{4,j}}= \prod _{j=1}^{K_b} \left [ \frac{\frac{3i}{2}-u_k^g+u_{b,j}}{\frac{3i}{2}+u_k^g-u_{b,j}} \
\frac{\frac{i}{2}-u_k^g+u_{b,j}}{\frac{i}{2}+u_k^g-u_{b,j}} \right ] \cdot \nonumber \\
&\cdot & \prod _{h=1}^H \frac{i+u_k^g-u_h}{i-u_k^g+u_h}
\exp \left [-\int  _{-\infty}^{+\infty} \frac{dv}{2\pi}  \ln \frac{i-u_k^g+v}{i+u_k^g-v}\frac{d}{dv}\left (Z_4(v)-2L_4(v) \right ) \right ] \, .
\ea
The integral term equals
\ba
&& -\int \frac{dv}{2\pi}  \ln \frac{i-u_k^g+v}{i+u_k^g-v}\frac{d}{dv}\left (Z_4(v)-2L_4(v) \right )=
4iu_k^g \ln s-L\ln \frac{\Gamma\left (\frac{3}{2}+iu_k^g \right )}{\Gamma \left (\frac{3}{2}-iu_k^g \right)} + \nonumber \\
&+& \sum _{h=1}^H \ln \frac{\Gamma (2-iu_h+iu_k^g)}{\Gamma (2+iu_h-iu_k^g)}+
 \sum _{j=1}^{N_g} \ln \frac{\Gamma (2-iu_j^g+iu_k^g)}{\Gamma (2+iu_j^g-iu_k^g)} +
 \sum _{j=1}^{N_{\bar g}} \ln \frac{\Gamma (2-iu_j^{\bar g}+iu_k^g)}{\Gamma (2+iu_j^{\bar g}-iu_k^g)}+ \nonumber \\
 &+&   \sum _{j=1}^{N_F} \ln \frac{\Gamma \left (\frac{3}{2}-iu_{F,j}+iu_k^g \right )}{\Gamma \left (\frac{3}{2}+iu_{F,j}-iu_k^g \right )}
 +  \sum _{j=1}^{N_{\bar F}} \ln \frac{\Gamma \left (\frac{3}{2}-iu_{\bar F,j}+iu_k^g \right )}{\Gamma \left (\frac{3}{2}+iu_{\bar F,j}-iu_k^g \right )}
 -\sum _{j=1}^{K_b} \ln \frac{\frac{3i}{2}-u_k^g+u_{b,j}}{\frac{3i}{2}+u_k^g-u_{b,j}} \, .
\ea
Putting the last two formul{\ae} together, we have
\ba
&& \prod _{j=1}^{K_4} \frac{u^{g}_k-u_{4,j}-i}{u^g_k-u_{4,j}+i}= e^{4iu_k^g \ln s}
\left [ \frac{\Gamma \left (\frac{3}{2} -i u_k^g \right )}{\Gamma \left (\frac{3}{2} + i u_k^g \right )} \right ]^L
\prod _{j=1}^{K_b} \frac{\frac{i}{2}-u_k^g+u_{b,j}}{\frac{i}{2}+u_k^g-u_{b,j}}  \prod _{h=1}^H \frac{\Gamma (1-iu_h+iu_k^g)}{\Gamma (1+iu_h-iu_k^g)}  \cdot \nonumber \\
&\cdot & \prod _{j=1}^{N_g} \frac{\Gamma (2-iu_j^g+iu_k^g)}{\Gamma (2+iu_j^g-iu_k^g)}
\prod _{j=1}^{N_{\bar g}} \frac{\Gamma (2-iu_j^{\bar g}+iu_k^g)}{\Gamma (2+iu_j^{\bar g}-iu_k^g)}
\prod _{j=1}^{N_F} \frac{\Gamma \left (\frac{3}{2}-iu_{F,j}+iu_k^g \right )}{\Gamma \left (\frac{3}{2}+iu_{F,j}-iu_k^g \right )}
\prod _{j=1}^{N_{\bar F}} \frac{\Gamma \left (\frac{3}{2}-iu_{\bar F,j}+iu_k^g \right )}{\Gamma \left (\frac{3}{2}+iu_{\bar F,j}-iu_k^g \right )} \nonumber \, .
\ea
We now plug such expression in (\ref {gl5}) and multiply the resulting expression by $1=e^{iP}$, where $P$ is given by (\ref {P-1}). We observe the exact cancelation of the term depending on type-$b$ isotopic roots and get the final set of equations, written in terms of scattering factors listed in Appendix \ref {scatt-fact}:
\ba
1&=&e^{iR P^{(g)}(u_k^g)}
\left [ \frac{\Gamma \left (\frac{3}{2} -i u_k^g \right )}{\Gamma \left (\frac{3}{2} + i u_k^g \right )} \right ]^2
\prod _{h=1}^H S_0^{(gs)}(u_k^g, u_h)  \cdot \nonumber \\
&\cdot &  \prod _{\stackrel {j=1}{j\not= k}}^{N_g} S_0^{(gg)}(u_k^g,u_j^g)
\prod _{j=1}^{N_{\bar g}} S_0^{(g\bar g)}(u_k^g,u_j^{\bar g}) \prod _{j=1}^{N_F} S_0^{(gF)}(u_k^g,u_{F,j}) \prod _{j=1}^{N_{\bar F}} S_0^{(g\bar F)}(u_k^g,u_{\bar F , j}) \, , \label {glu-by}
\ea
where we introduced the length $R=2\ln s$ of the chain and the momentum $P^{(g)}_0(u_k^g)=2 u_k^g$ of a gluon $F_{+\bot}$. In this case, the effect of the two transmitting defects on gluons is
\be
\left [ \frac{\Gamma \left (\frac{3}{2} -i u_k^g \right )}{\Gamma \left (\frac{3}{2} + i u_k^g \right )} \right ]^2 \, .
\ee

In analogous fashion, we obtain the equation for the gluon field $\bar F_{+\bot}$:
\ba
1&=&e^{iR P^{(g)}(u_k^{\bar g})}
\left [\frac{\Gamma \left (\frac{3}{2}-iu_k^{\bar g} \right )}{\Gamma \left (\frac{3}{2}+iu_k^{\bar g}\right )}\right ]^2 \prod _{h=1}^H S_0^{(\bar gs)}(u_k^{\bar g}, u_h)  \cdot \nonumber \\
&\cdot &  \prod _{\stackrel {j=1}{j\not= k}}^{N_g} S_0^{(\bar gg)}(u_k^{\bar g},u_j^g)
\prod _{j=1}^{N_{\bar g}} S_0^{(\bar g\bar g)}(u_k^{\bar g},u_j^{\bar g}) \prod _{j=1}^{N_F} S_0^{(\bar g F)}(u_k^{\bar g},u_{F,j}) \prod _{j=1}^{N_{\bar F}} S_0^{(\bar g \bar F)}(u_k^{\bar g},u_{\bar F , j}) \, , \label {barglu-by}
\ea
where again $R=2\ln s$ is the length of the chain and $P^{(g)}_0(u_k^{\bar g})=2 u_k^{\bar g}$ is the momentum of the gluon excitation $\bar F_{+\bot}$.

\medskip

{\bf Isotopic roots}

\medskip

We remember (see (\ref {uais}, \ref {ucis}, \ref {stack}) the definition of the three sets of isotopic roots, which do not carry momentum and energy, but take into account the $su(4)$ symmetry of the GKP vacuum.

We have
$K_a$ roots $u_{a,j}$ of type $u_2$, $K_c$ roots  $u_{c,j}$ of type $u_6$ and $K_b$ stacks,
$ u_{4,j}=u_{b,j}\pm \frac{i}{2} \, , \quad
u_{3,j}=u_{5,j}=u_{b,j}$ with centers $u_{b,j}$.

The equations for the isotopic roots
$u_a$ and $u_c$ come directly from the second and the sixth of the Beisert-Staudacher equations: we observe the cancelation of the contributions coming from gauge field stacks:
\ba
1&=&\prod _{j\not=k}^{K_a} \frac{u_{a,k}-u_{a,j}+i}{u_{a,k}-u_{a,j}-i} \prod _{j=1}^{N_F}
\frac{u_{a,k}-u_{F,j}-\frac{i}{2}}{u_{a,k}-u_{F,j}+\frac{i}{2}}  \prod _{j=1}^{K_b}
\frac{u_{a,k}-u_{b,j}-\frac{i}{2}}{u_{a,k}-u_{b,j}+\frac{i}{2}} \label {1lfineq1} \\
1&=&\prod _{j\not=k}^{K_c} \frac{u_{c,k}-u_{c,j}+i}{u_{c,k}-u_{c,j}-i} \prod _{j=1}^{N_{\bar F}}
\frac{u_{c,k}-u_{\bar F,j}-\frac{i}{2}}{u_{c,k}-u_{\bar F,j}+\frac{i}{2}} \prod _{j=1}^{K_b}
\frac{u_{c,k}-u_{b,j}-\frac{i}{2}}{u_{c,k}-u_{b,j}+\frac{i}{2}} \label {1lfineq2} \, .
\ea
For what concerns the isotopic roots $u_b$, we consider the product of the third Beisert-Staudacher equation for $u_{3,k}=u_{b,k}$ with the fifth for $u_{5,k}=u_{b,k}$, the fourth for $u_{4,k}=u_{b,k}+i/2$ and the fourth for $u_{4,k}=u_{b,k}-i/2$. We arrive at the following equation,
\ba
1&=& \left (\frac{u_{b,k}-i}{u_{b,k}+i}\right )^L \prod _{j=1}^{K_a} \frac{u_{b,k}-u_{a,j}-\frac{i}{2}}{u_{b,k}-u_{a,j}+\frac{i}{2}} \prod _{j=1}^{K_c} \frac{u_{b,k}-u_{c,j}-\frac{i}{2}}{u_{b,k}-u_{c,j}+\frac{i}{2}} \cdot \nonumber \\
&\cdot & \prod _{j=1}^{K_4} \frac{u_{b,k}-u_{4,j}+\frac{i}{2}}{u_{b,k}-u_{4,j}-\frac{i}{2}}
\prod _{j=1}^{K_4} \frac{u_{b,k}-u_{4,j}-\frac{3i}{2}}{u_{b,k}-u_{4,j}+\frac{3i}{2}}
\prod _{j=1}^{K_b} \left (  \frac{u_{b,k}-u_{b,j}+i}{u_{b,k}-u_{b,j}-i} \right )^2 \cdot \label  {1lequinte}\\
& \cdot & \prod _{j=1}^{N_g} \frac{u_{b,k}-u^g_{j}+\frac{3i}{2}}{u_{b,k}-u^g_{j}-\frac{3i}{2}}
\prod _{j=1}^{N_{\bar g}} \frac{u_{b,k}-u^{\bar g}_{j}+\frac{3i}{2}}{u_{b,k}-u^{\bar g}_{j}-\frac{3i}{2}}
\prod _{j=1}^{N_F} \frac{u_{b,k}-u_{F,j}+i}{u_{b,k}-u_{F,j}-i}
\prod _{j=1}^{N_{\bar F}} \frac{u_{b,k}-u_{\bar F,j}+i}{u_{b,k}-u_{\bar F,j}-i}
\, . \nonumber
\ea
We have
\ba
&& \prod _{j=1}^{K_4} \frac{u_{b,k}-u_{4,j}+\frac{i}{2}}{u_{b,k}-u_{4,j}-\frac{i}{2}}=
\prod _{j=1}^{K_4} \frac{\frac{i}{2}+u_{b,k}-u_{4,j}}{\frac{i}{2}-u_{b,k}+u_{4,j}} =
\prod _{\stackrel {j=1}{j\not= k}}^{K_b} \frac{u_{b,k}-u_{b,j}+i}{u_{b,k}-u_{b,j}-i} \prod _{h=1}^{H} \frac{\frac{i}{2}-u_{b,k}+u_h}{\frac{i}{2}+u_{b,k}-u_h} \nonumber \cdot \\
&\cdot & \textrm{exp} \left [ - \int _{-\infty}^{+\infty}\frac{dv}{2\pi} \ln \frac{\frac{i}{2}+u_{b,k}-v}{\frac{i}{2}-u_{b,k}+v} (Z_4'(v)-2L_4'(v)) \right ]
\left [ 1+O(1/s^2) \right ]
\label  {1lequ2} \, .
\ea
Plugging (\ref {Z40}) into the integral in the last term of (\ref  {1lequ2}), we find that
\ba
&& \textrm{exp} \left [ - \int _{-\infty}^{+\infty} \frac{dv}{2\pi} \ln \frac{\frac{i}{2}+u_{b,k}-v}{\frac{i}{2}-u_{b,k}+v} (Z_4'(v)-2L_4'(v)) \right ]=
\left (\frac{i+u_{b,k}}{i-u_{b,k}}\right )^L \prod _{j=1}^{K_b} \left (  \frac{i-u_{b,k}+u_{b,j}}{i+u_{b,k}-u_{b,j}} \right )^2
\cdot \nonumber \\
&\cdot & \prod _{j=1}^{K_4} \frac{\frac{3i}{2}+u_{b,k}-u_{4,j}}{\frac{3i}{2}-u_{b,k}+u_{4,j}}
\prod _{j=1}^{N_g} \frac{\frac{3i}{2}-u_{b,k}+u^g_{j}}{\frac{3i}{2}+u_{b,k}-u^g_{j}}
\prod _{j=1}^{N_{\bar g}} \frac{\frac{3i}{2}-u_{b,k}+u^{\bar g}_{j}}{\frac{3i}{2}+u_{b,k}-u^{\bar g}_{j}}
\prod _{j=1}^{N_F} \frac{i-u_{b,k}+u_{F,j}}{i+u_{b,k}-u_{F,j}}
\prod _{j=1}^{N_{\bar F}} \frac{i-u_{b,k}+u_{\bar F,j}}{i+u_{b,k}-u_{\bar F,j}} \nonumber \, .
\ea
Putting all together, we eventually get the following equation, for the third isotopic root $u_b$:
\be
1=\prod _{j=1}^{K_a} \frac{u_{b,k}-u_{a,j}-\frac{i}{2}}{u_{b,k}-u_{a,j}+\frac{i}{2}} \prod _{j=1}^{K_c} \frac{u_{b,k}-u_{c,j}-\frac{i}{2}}{u_{b,k}-u_{c,j}+\frac{i}{2}} \prod _{\stackrel {j=1}{j\not= k}}^{K_b} \frac{u_{b,k}-u_{b,j}+i}{u_{b,k}-u_{b,j}-i}
\prod _{h=1}^{H} \frac{u_{b,k}-u_h-\frac{i}{2}}{u_{b,k}-u_h+\frac{i}{2}}  \, , \label {1lfineq3}
\ee
which does not depend on the roots associated to gluons.

\medskip

\subsection{The general (all loops) case}

\medskip

We now generalise all the results discussed in the one loop case to the most general all loops case.
For the sake of clarity, the complete set of equations is summarised in Appendix \ref {Bethe-eqs}.

As we did in the one loop case, we start from scalar excitations.

\medskip

{\bf Scalars}

\medskip

Let us introduce the counting function for the type-4 roots
\ba
Z_4(v)&=& iL \ln \left (-\frac{x^-(v)}{x^+(v)} \right ) + i \sum _{j\not=k}^{K_4} \ln \left (- \frac{x^{-}(v)-x^{+}_{4,j}}{x^{+}(v)-x^{-}_{4,j}} \frac{1-\frac{g^2}{2x^{+}(v)x^{-}_{4,j}}}{1-\frac{g^2}{2x^{-}(v)x^{+}_{4,j}}} \sigma ^2 (v,u_{4,j}) \right ) + \label {Zeta4} \\
&+& 2i \sum _{j=1}^{K_b}\ln \frac{x^+(v)-x_{b,j}}{x_{b,j}-x^-(v)}
+i \sum _{j=1}^{N_F}\ln \frac{x^+(v)-x_{F,j}}{x_{F,j}-x^-(v)}
+i \sum _{j=1}^{N_{\bar F}}\ln \frac{x^+(v)-x_{\bar F,j}}{x_{\bar F,j}-x^-(v)} + \nonumber \\
&+&i \sum _{j=1}^{N_f}\ln \frac{1-\frac{x_{f,j}}{x^+(v)}}{1-\frac{x_{f,j}}{x^-(v)}}
+i \sum _{j=1}^{N_{\bar f}} \ln \frac{1-\frac{x_{\bar f,j}}{x^+(v)}}{1-\frac{x_{\bar f,j}}{x^-(v)}} + \nonumber \\
&+& i \sum _{j=1}^{N_g}\ln \frac{x^+(v)-x_{j}^{g+}}{x^-(v)-x_{j}^{g+}}\frac{x_{j}^{g-}-x^+(v)}{x^-(v)-x_{j}^{g-}}
+i \sum _{j=1}^{N_{\bar g}}\ln \frac{x^+(v)-x_{j}^{\bar g+}}{x^-(v)-x_{j}^{\bar g+}}
\frac{x_{j}^{\bar g-}-x^+(v)}{x^-(v)-x_{j}^{\bar g-}} \, ,
\nonumber
\ea
where $\sigma ^2 (v,u)$ is the so-called dressing factor \cite {DRE,BES}, $x^{\pm}(v)=x(v\pm i/2)$ and we are using the
notations
\ba
&& x_j^{g\pm}=x\left(u_j^g \pm \frac{i}{2}\right ) \, , \quad x_j^{\bar g\pm}=x\left(u_j^{\bar g} \pm \frac{i}{2}\right ) \nonumber \\
&& x_{4,j}^{\pm}=x\left (u_{4,j}\pm \frac{i}{2}\right ) \, , \quad x_{b,j}=x(u_{b,j})\, .  \label {x-not}
\ea
The property $e^{iZ_4(u_{4,k})}=(-1)^{H-1}$ follows from the definition (\ref {Zeta4}) and from the relation (\ref {L-ecc}) between $L$ and the number of the various excitations: the condition $e^{iZ_4(u_h)}=(-1)^{H-1}$ identifies the $H$ internal holes, i.e. the scalar excitations.

It is convenient to write (\ref {Zeta4}) in terms of functions $\Phi$, $\phi $, $\chi$, introduced in the Appendix \ref {functions}
\ba
Z_4(v)&=&L\Phi (v) -\sum _{j=1}^{K_4} \phi (v,u_j) +2i \sum _{j=1}^{K_b} \ln \frac{x^+(v)-x_{b,j}}{x_{b,j}-x^-(v)}+ \nonumber \\
&+&\sum _{j=1}^{N_g} \chi (v,u^g_j|1) + \sum _{j=1}^{N_{\bar g}} \chi (v,u^{\bar g}_j|1)+\sum _{j=1}^{N_F} \chi_F(v,u_{F,j}) +\sum _{j=1}^{N_{\bar F}} \chi_F(v,u_{\bar F,j}) - \label {count-scal1} \\
&-& \sum _{j=1}^{N_f} \chi_H(v,u_{f,j}) - \sum _{j=1}^{N_{\bar f}} \chi_H(v,u_{\bar f,j}) \, , \nonumber
\ea
where $s$ of the type-4 roots involved in the sum are real, while $2K_b$ are part of the stack defining the isotopic root $u_b$.
We concentrate on the real type-4 roots and write the sum over them as an integral by means of (\ref {summaster}), getting
\ba
Z_4(v)&=&L\Phi (v) +2i \sum _{j=1}^{K_b} \ln \frac{x^+(v)-x_{b,j}}{x_{b,j}-x^-(v)}-
\sum _{j=1}^{K_b} [\phi (v,u_{b,j}+i/2)+ \phi (v,u_{b,j}-i/2)]+ \nonumber \\
&+&\sum _{j=1}^{N_g} \chi (v,u^g_j|1) + \sum _{j=1}^{N_{\bar g}} \chi (v,u^{\bar g}_j|1)+\sum _{j=1}^{N_F} \chi_F(v,u_{F,j}) +\sum _{j=1}^{N_{\bar F}} \chi_F(v,u_{\bar F,j}) - \label {count-scal2} \\
&-& \sum _{j=1}^{N_f} \chi_H(v,u_{f,j}) - \sum _{j=1}^{N_{\bar f}} \chi_H(v,u_{\bar f,j})+\sum _{h=1}^H \phi (v,u_h) + \nonumber \\
&+& \phi (v, s/\sqrt{2}) + \phi (v, -s/\sqrt{2}) -\int  _{-\infty}^{+\infty} dw \varphi (v,w) [Z_4(w)-2L_4(w)] + O(1/s^2) \, , \nonumber
\ea
where $\varphi$ is defined in (\ref {fphidef}).
Then, we can write
\be\label{scalar_counting}
Z_4(v)= F(v)+2 \int  _{-\infty}^{+\infty} dw G(v,w) L_4(w) \, ,
\ee
where $F(v)$ satisfies the linear integral equation
\ba
F(v)&=&L\Phi (v) +2i \sum _{j=1}^{K_b} \ln \frac{x^+(v)-x_{b,j}}{x_{b,j}-x^-(v)}-
\sum _{j=1}^{K_b} [\phi (v,u_{b,j}+i/2)+ \phi (v,u_{b,j}-i/2)]+ \nonumber \\
&+&\sum _{j=1}^{N_g} \chi (v,u^g_j|1) + \sum _{j=1}^{N_{\bar g}} \chi (v,u^{\bar g}_j|1)+\sum _{j=1}^{N_F} \chi_F(v,u_{F,j}) +\sum _{j=1}^{N_{\bar F}} \chi_F(v,u_{\bar F,j}) - \label {Fu} \\
&-& \sum _{j=1}^{N_f} \chi_H(v,u_{f,j}) - \sum _{j=1}^{N_{\bar f}} \chi_H(v,u_{\bar f,j})+\sum _{h=1}^H \phi (v,u_h) + \nonumber \\
&+& \phi (v, s/\sqrt{2}) + \phi (v, -s/\sqrt{2})
 -\int  _{-\infty}^{+\infty} dw \varphi (v,w) F(w) +O(1/s^2) \, , \nonumber
\ea
and
\be
G(v,w) = \varphi (v,w) - \int  _{-\infty}^{+\infty} dw' \varphi (v,w') G(w',w)     \, .  \label {Geq}
\ee
We now work out the solution to (\ref {Fu}). The part depending on the isotopic roots is written in an explicit form. For the remaining parts we remember that $L=H+2+N_g+N_{\bar g}+N_F+N_{\bar F}$: this allows to put $L\Phi (v)$ together with the other functions in the right hand side of (\ref {Fu}). Eventually, the solution to (\ref {Fu}) is written in terms of solutions of linear integral equations. In specific, we have
\ba
F(v)&=&\Theta '(v,s/\sqrt{2})+\Theta '(v,-s/\sqrt{2})+\sum _{h=1}^H  \Theta '(v,u_h) + i \sum _{j=1}^{K_b} \ln \frac{i/2+v-u_{b,j}}{i/2-v+u_{b,j}} + \nonumber \\
&+& \sum _{j=1}^{N_g} F^G(v,u^g_j) + \sum _{j=1}^{N_{\bar g}} F^G(v,u^{\bar g}_j)+\sum _{j=1}^{N_F} F^F(v,u_{F,j}) +
\sum _{j=1}^{N_{\bar F}} F^F(v,u_{\bar F,j}) \label {forc-scal} \\
&+& \sum _{j=1}^{N_f} F^f(v,u_{f,j}) + \sum _{j=1}^{N_{\bar f}} F^f(v,u_{\bar f,j}) + O(1/s^2) \, , \nonumber
\ea
where
\ba
&& \Theta '(v,u)=\phi (v,u)+\Phi (v) - \int _{-\infty}^{+\infty} dw   \varphi (v,w) \Theta '(w,u)\, ,  \label {Theta-prime} \\
&& F^F(v,u)=\chi _F(v,u)+\Phi (v) - \int  _{-\infty}^{+\infty} dw   \varphi (v,w) F^F(w,u)
\label {F-F} \, , \\
&& F^f(v,u)=-\chi _H(v,u) - \int  _{-\infty}^{+\infty} dw \varphi (v,w) F^f(w,u) \, , \label {F-f} \\
&& F^G(v,u)=\chi (v,u|1)+\Phi (v) - \int  _{-\infty}^{+\infty} dw   \varphi (v,w)  F^G(w,u)
\label {F-G} \, .
\ea
We first analyse the $s$-depending terms.
A tedious calculation\footnote {Formula (\ref {Thetasol}) clarifies the origin of the length $2\ln s$ and the two 'defects': they are both due to the interaction with the two heavy large holes.} shows that in the large $s$ limit
\be
\Theta '(v,s/\sqrt{2})+\Theta '(v,-s/\sqrt{2})= \ln \frac{s}{\sqrt{2}} \left [ -4v+Z_{BES} (v) \right ] - 2 \tilde P(v) + O(1/s^2) \, , \label {Thetasol}
\ee
where $Z_{BES}(v)=-Z_{BES}(-v)$ and $\frac{d}{dv} Z_{BES}(v)=\sigma _{BES}(v)$, $\sigma _{BES}(v)$ being the famous BES density \cite {BES} and $\tilde P(v)$ is the solution of the integral equation\footnote {We write the kernel of equation (\ref {tildePeq}) in an explicitly antisymmetric form in order to avoid one loop divergencies.}
\be
\tilde P(v)=-\Phi (v) - \int _{-\infty}^{+\infty} \frac{dw}{2} [ \varphi (v,w)- \varphi (v,-w) ] \tilde P(w) \, .
\label {tildePeq}
\ee
Then, we pass to study the nonlinear term $NL(v)=2\int  _{-\infty}^{+\infty} dw G(v,w) L_4(w) $.
The same term was computed in \cite {FRO6}, where only real type-4 roots were present.
Here we can use the same results, since in the large $s$ limit the leading behaviour of $Z_4(v)$ does not depend on
the presence of excitations. We have
\be
NL(v)=-2v\ln 2 + \frac{\ln 2}{2}Z_{BES}(v)+O(1/s^2) \, .
\ee
Putting everything together, we arrive at
\ba
Z_4(v)&=&\ln s \left [ -4v+Z_{BES} (v) \right ] - 2 \tilde P(v)+\sum _{h=1}^H  \Theta '(v,u_h) + i \sum _{j=1}^{K_b} \ln \frac{i/2+v-u_{b,j}}{i/2-v+u_{b,j}} + \nonumber \\
&+& \sum _{j=1}^{N_g} F^G(v,u^g_j) + \sum _{j=1}^{N_{\bar g}} F^G(v,u^{\bar g}_j)+\sum _{j=1}^{N_F} F^F(v,u_{F,j}) +
\sum _{j=1}^{N_{\bar F}} F^F(v,u_{\bar F,j}) \label {count-scal-all} \\
&+& \sum _{j=1}^{N_f} F^f(v,u_{f,j}) + \sum _{j=1}^{N_{\bar f}} F^f(v,u_{\bar f,j}) + O(1/s^2) \, , \nonumber
\ea
Now, as in the one loop case, before imposing the quantisation condition for holes, we introduce the
momentum of the chain
\ba
P&=&i \sum _{j=1}^{K_4} \ln \frac{x_{4,j}^+}{x_{4,j}^-}=- \sum _{j=1}^{K_4} \Phi (u_{4,j}) =
\sum _{h=1}^{H} \Phi (u_h)  +
i\sum _{j=1}^{K_b} \ln \left ( -\frac{x_{b,j}^{++}}{x_{b,j}^{--}} \right ) + \nonumber \\
&+& \int  _{-\infty}^{+\infty} \frac{dv}{2\pi} \Phi (v) \frac{d}{dv} [ Z_4(v)-2L_4(v)] \label {imp} \, ,
\ea
where $x_{b,j}^{\pm \pm}=x(u_{b,j}\pm i)$.
Terms containing $L_4$ give no contributions at the orders $\ln s $ and $(\ln s )^0$. Terms containing $u_b$ produce only a term $\pi K_b$. The dependence on excitations is worked out after inserting for $Z_4(v)$ expression (\ref {count-scal-all}). However, for our convenience we prefer to work directly on the expression $Z_4(v)-P$: after some calculation (see Appendix \ref {some-calc} for details) we arrive at the expression
\ba
Z_4(v)-P&=& \ln s \left [ -4v + Z_{BES}(v) \right ] - 2 \tilde P(v) + i \sum _{j=1}^{K_b} \ln \frac{v-u_{b,j}+\frac{i}{2}}{v-u_{b,j}-\frac{i}{2}} + \sum _{h=1}^H \Theta (v,u_h) + \nonumber \\
&+& \sum _{j=1}^{N_F} i\ln S^{(sF)}(v,u_{F,j}) +  \sum _{j=1}^{N_{\bar F}} i\ln S^{(s\bar F)}(v,u_{\bar F,j})
+ \sum _{j=1}^{N_f} i\ln S^{(sf)}(v,u_{f,j}) +  \label {Z4-imp}\\
&+& \sum _{j=1}^{N_{\bar f}} i\ln S^{(s\bar f)}(v,u_{\bar f,j})
+ \sum _{j=1}^{N_g} i\ln S^{(sg)}(v,u_j^g) +  \sum _{j=1}^{N_{\bar g}} i\ln S^{(s\bar g)}(v,u_j^{\bar g})
\nonumber \, ,
\ea
where we introduced the scalar-scalar phase
\be
\Theta (v,u)=\Theta ' (v,u)+\tilde P(u) = i \ln [-S^{(ss)}(v,u) ] \label {sca-sca}
\ee
and the scattering factors $S^{(s \ast)}(v,u_j^{\ast})$ between scalars and other excitations, which are listed in Appendix \ref {scatt-fact}.
Imposing the quantisation condition
$e^{-i[Z_4(u_h)-P]}=(-1)^{H-1}$, we get the final Bethe equation for scalars
\ba
1&=&e^{iRP^{(s)}(u_h)+2iD^{(s)}(u_h)}
 \prod _{j=1}^{K_b} \frac{u_h-u_{b,j}+\frac{i}{2}}{u_h-u_{b,j}-\frac{i}{2}}  \prod _{\stackrel {h'=1}{h'\not=h}}^{H} S^{(ss)}(u_h,u_{h'})\prod _{j=1}^{N_{g}} S^{(sg)}(u_h,u_j^g)  \prod _{j=1}^{N_{\bar g}} S^{(s\bar g)}(u_h,u_j^{\bar g}) \cdot  \nonumber\\
&\cdot & \prod _{j=1}^{N_F} S^{(sF)}(u_h,u_{F,j})   \prod _{j=1}^{N_{\bar F}} S^{(s\bar F)}(u_h, u_{\bar F,j})
\prod _{j=1}^{N_f} S^{(sf)}(u_h,u_{f,j})   \prod _{j=1}^{N_{\bar f}} S^{(s\bar f)}(u_h, u_{\bar f,j})
\label {scal-byall} \, ,
\ea
where we introduced the length of the chain $R=2\ln s$, the momentum of a scalar excitation with rapidity $u$
\be
P^{(s)}(u)=2u-\frac{1}{2}Z_{BES}(u) \label {scal-mom}
\ee
and the effect of the two purely transmitting defects
\be
2D^{(s)}(u)=2\tilde P(u). \label {scal-def}
\ee
Important properties of the scalar-scalar phase (\ref {sca-sca}), which can be proven using equations
(\ref {Theta-prime}, \ref {tildePeq}) are
\be
\Theta (u,v)=-\Theta (-u,-v) \, , \quad \Theta (u,v)=-\Theta (v,u) \, . \label {Theta-pro}
\ee
Eventually, we remember an efficient way proposed in \cite {FPR} to compute the scalar-scalar phase.
We found that
\be
\Theta (u,v)=M(u,v)-M(v,u) \, , \label {ThetaMrel}
\ee
where $M(u,v)=Z^{(1)}(u)+Z(u;v)$ and $Z^{(1)}(u), Z(u;v)$ are univocally defined by the conditions
\be
\frac{d}{du}Z^{(1)}(u)=\sigma ^{(1)}(u) \, , \ \ \frac{d}{du} Z(u;v)= \sigma (u;v) \, , \ \ Z^{(1)}(u)=-Z^{(1)}(-u) \, , \ \ Z(u;v) = -Z(-u;v) \, , \label  {Z-cond}
\ee
with the functions $\sigma ^{(1)}(u), \sigma (u;v)$ solutions of equations (\ref {sigma(1)}, \ref {tilSigeqA}), respectively.

This procedure provides an alternative (with respect to solving equation (\ref {tildePeq})) way to determine the function $\tilde P(u)$, once $Z_{BES}(u)$, $Z^{(1)}(u)$ and $Z(u;v)$ are known.
Indeed using (\ref {Thetasol}) we have
\ba
2\tilde P(v)&=&\lim _{s\rightarrow +\infty} \left [ \ln \frac{s}{\sqrt{2}}\left [ -4v+Z_{BES} (v) \right ]-\Theta (v,s/\sqrt{2})-\Theta (v,-s/\sqrt{2})\right ] = \nonumber \\
&=& \lim _{s\rightarrow +\infty} \left [ \ln \frac{s}{\sqrt{2}}\left [ -4v+Z_{BES} (v) \right ]-2Z^{(1)}(v)-Z(v;s/\sqrt{2})-Z(v;-s/\sqrt{2}) \right ] \, . \label {Thetasol2}
\ea
A final alternative to compute $\tilde P(u)$ is to look at equation (\ref {count-scal-all}) when no excitations nor isotopic roots are present. Then we see that $-2\tilde P(u)$ represent the contribution $O(\ln s ^0)$ to the twist two counting function of the pure $sl(2)$ sector. This function has been analysed in \cite {BFR,FZ} (in notations of the second of \cite {FZ} it is connected to the function $S^{extra}$).

\medskip

{\bf Fermions}

\medskip

The equations for large fermions come from the (inverse) of the third of the Beisert-Staudacher equations.
We have
\ba
1&=&\prod _{j=1}^{K_a} \frac{u_{F,k}-u_{a,j}+i/2}{u_{F,k}-u_{a,j}-i/2}\prod _{j=1}^{N_g} \frac{u_{F,k}-u^g_{j}+i/2}{u_{F,k}-u^g_{j}-i/2}(-1)^{K_4}\prod _{j=1}^{K_4} \frac{x_{4,j}^+-x_{F,k}}{x_{F,k}-x_{4,j}^-}= \nonumber \\
&=& \prod _{j=1}^{K_a} \frac{u_{F,k}-u_{a,j}+i/2}{u_{F,k}-u_{a,j}-i/2}
\prod _{j=1}^{N_g} \frac{u_{F,k}-u^g_{j}+i/2}{u_{F,k}-u^g_{j}-i/2}
\prod _{j=1}^{K_b} \frac{x_{b,j}^{++}-x_{F,k}}{x_{F,k}-x_{b,j}^{--}} \prod _{h=1}^H \frac{x_{F,k}-x_h^-}{x_h^+-x_{F,k}} \cdot \nonumber \\
&\cdot & \textrm{exp} \left [ i \int  _{-\infty}^{+\infty} \frac{dv}{2\pi} \chi _F(v,u_{F,k}) \frac{d}{dv}\Bigl (Z_4(v)-2L_4(v) \Bigr ) \right ] \label
{ferm1} \, ,
\ea
since $K_4=s+2K_b$ is even, where $x_h^{\pm}=x \left (u_h \pm \frac{i}{2} \right )$.
As in the one loop case, we multiply such expression by $1=e^{iP}$. Then we use expression (\ref {count-scal-all}) for $Z_4(v)$ and remember that the term containing $L_4(v)$ gives subleading $O(1/s^2)$ contributions. We get
\ba\label{Bethe_Ferm}
1&=& e^{iRP^{(F)}(u_{F,k}) + 2i D^{(F)}(u_{F,k})}\prod _{j=1}^{K_a} \frac{u_{F,k}-u_{a,j}+i/2}{u_{F,k}-u_{a,j}-i/2} \prod _{h=1}^{H} S^{(F s)} (u_{F,k},u_h) \cdot \label  {fineq4} \\
& \cdot &  \prod _{j=1}^{N_{F}} {S}^{(FF)}(u_{F,k},u_{F,j})  \prod _{j=1}^{N_{\bar F}} {S}^{(F\bar F)}(u_{ F,k},u_{\bar F,j})
\prod _{j=1}^{N_{f}} {S}^{(Ff)}(u_{F,k},u_{f,j})  \prod _{j=1}^{N_{\bar f}} {S}^{(F\bar f)}(u_{ F,k},u_{\bar f,j})
\cdot \nonumber \\
&\cdot & \prod _{j=1}^{N_g}  {S}^{(Fg)}(u_{F,k},u_{j}^g) \prod _{j=1}^{N_{\bar g}}
{S}^{(F\bar g)}(u_{F,k},u_{j}^{\bar g}) \, ,
\nonumber
\ea
where $R=2\ln s$ is the length of the chain and
\ba
P^{(F)}(u)&=&-\int _{-\infty}^{+\infty} \frac{dv}{2\pi} \left [ \chi _F(v,u)+\chi _F(-v,u) \right ] \left [ 1-\frac{\sigma _{BES}(v)}{4} \right ]
\, , \label {P-fer} \\
2 D^{(F)}(u)&=& -\int _{-\infty}^{+\infty} \frac{dv}{2\pi} \left [ \chi _F(v,u)+\chi _F(-v,u) \right ] \frac{d}{dv} \tilde P(v) \label {D-fer} \,
\ea
are the momentum of a fermion and the effect on it of the two defects.

\medskip

The equations for large antifermions come from the (inverse of the) fifth of the Beisert-Staudacher equations. Their derivation is analogous to the fermionic case:
\ba\label{Bethe_antiFerm}
1&=& e^{iRP^{(F)}(u_{\bar F,k}) + 2 i D^{(F)}(u_{\bar F,k})} \prod _{j=1}^{K_c} \frac{u_{\bar F,k}-u_{c,j}+i/2}{u_{\bar F,k}-u_{c,j}-i/2} \prod _{h=1}^{H} S^{(\bar F s)} (u_{\bar F,k},u_h) \cdot \label  {fineq5} \\
& \cdot &  \prod _{j=1}^{N_{F}} {S}^{(\bar FF)}(u_{\bar F,k},u_{F,j})  \prod _{j=1}^{N_{\bar F}} {S}^{(\bar F\bar F)}(u_{\bar F,k},u_{\bar F,j})
\prod _{j=1}^{N_{f}} {S}^{(\bar Ff)}(u_{\bar F,k},u_{f,j})  \prod _{j=1}^{N_{\bar f}} {S}^{(\bar F\bar f)}(u_{\bar F,k},u_{\bar f,j}) \nonumber \\
&\cdot & \prod _{j=1}^{N_g}  {S}^{(\bar Fg)}(u_{\bar F,k},u_{j}^g) \prod _{j=1}^{N_{\bar g}}
{S}^{(\bar F\bar g)}(u_{\bar F,k},u_{j}^{\bar g}) \, .
\nonumber
\ea
\medskip
Equations for small fermions are obtained starting from the (inverse of the) first of the Beisert-Staudacher equations. We have
\ba
1&=& \prod _{j=1}^{K_2} \frac{u_{f,k}-u_{2,j}+i/2}{u_{f,k}-u_{2,j}-i/2} \prod _{j=1}^{K_4} \frac{1-\frac{x_{f,k}}{x_{4,j}^{+}}}{1-\frac{x_{f,k}}{x_{4,j}^{-}}}=
\prod _{j=1}^{K_a} \frac{u_{f,k}-u_{a,j}+i/2}{u_{f,k}-u_{a,j}-i/2}
\prod _{j=1}^{N_g} \frac{u_{f,k}-u_{j}^g+i/2}{u_{f,k}-u_{j}^g-i/2} \cdot \nonumber \\
&\cdot &
\prod _{j=1}^{K_b} \frac{1-\frac{x_{f,k}}{x_{b,j}^{++}}}{1-\frac{x_{f,k}}{x_{b,j}^{--}}}
 \prod _{h=1}^{H} \frac{1-\frac{x_{f,k}}{x_{h}^{-}}}{1-\frac{x_{f,k}}{x_{h}^{+}}}
 \textrm{exp} \left [ -i \int  _{-\infty}^{+\infty} \frac{dv}{2\pi} \chi _H(v,u_{f,k}) \frac{d}{dv}\Bigl (Z_4(v)-2L_4(v) \Bigr ) \right ]
  \label  {int-eq6} \, .
\ea
In contrast with the fermionic case, we do not multiply this equality by $1=e^{iP}$.
We use expression (\ref {count-scal-all}) for $Z_4(v)$. Working out the various terms
we get
\ba
1&=& e^{iRP^{(f)}(u_{f,k}) + 2 i D^{(f)}(u_{f,k})} \prod _{j=1}^{K_a} \frac{u_{f,k}-u_{a,j}+i/2}{u_{f,k}-u_{a,j}-i/2} \prod _{h=1}^{H} S^{(f s)} (u_{f,k},u_h) \cdot \label  {fineq6} \\
& \cdot &  \prod _{j=1}^{N_{F}} {S}^{(fF)}(u_{f,k},u_{F,j})  \prod _{j=1}^{N_{\bar F}} {S}^{(f\bar F)}(u_{f,k},u_{\bar F,j}) \prod _{j=1}^{N_f}  {S}^{(ff)}(u_{f,k},u_{f,j}) \prod _{j=1}^{N_{\bar f}}
{S}^{(f\bar f)}(u_{f,k},u_{\bar f,j}) \nonumber \\
&\cdot & \prod _{j=1}^{N_g}  {S}^{(fg)}(u_{f,k},u_{j}^g) \prod _{j=1}^{N_{\bar g}}
{S}^{(f\bar g)}(u_{f,k},u_{j}^{\bar g})\, ,
\nonumber
\ea
where
\ba
P^{(f)}(u)&=&\int _{-\infty}^{+\infty} \frac{dv}{2\pi} \left [ \chi _H(v,u)+\chi _H(-v,u) \right ] \left [ 1-\frac{\sigma _{BES}(v)}{4} \right ]
\, , \label {P-smfer} \\
2 D^{(f)}(u)&=& \int _{-\infty}^{+\infty} \frac{dv}{2\pi} \left [ \chi _H(v,u)+\chi _H(-v,u) \right ] \frac{d}{dv} \tilde P(v) \label {D-smfer} \, .
\ea

\medskip
In a completely analogous way we work on the (inverse of the) seventh of the Beisert-Staudacher equations, which gives the quantisation condition for
small antifermions:
\ba
1&=&  e^{iRP^{(f)}(u_{\bar f,k}) + 2 i D^{(f)}(u_{\bar f,k})}  \prod _{j=1}^{K_c} \frac{u_{\bar f,k}-u_{c,j}+i/2}{u_{\bar f,k}-u_{c,j}-i/2} \prod _{h=1}^{H} S^{(\bar f s)} (u_{\bar f,k},u_h) \cdot \label  {fineq7} \\
& \cdot &  \prod _{j=1}^{N_{F}} {S}^{(\bar f F)}(u_{\bar f,k},u_{F,j})  \prod _{j=1}^{N_{\bar F}} {S}^{(\bar f \bar F)}(u_{\bar f,k},u_{\bar F,j})
\prod _{j=1}^{N_{f}} {S}^{(\bar f f)}(u_{\bar f,k},u_{f,j})  \prod _{j=1}^{N_{\bar f}} {S}^{(\bar f \bar f)}(u_{\bar f,k},u_{\bar f,j}) \nonumber \\
&\cdot & \prod _{j=1}^{N_g}  {S}^{(\bar f g)}(u_{\bar f,k},u_{j}^g) \prod _{j=1}^{N_{\bar g}}
{S}^{(\bar f\bar g)}(u_{\bar f,k},u_{j}^{\bar g})
\nonumber \, .
\ea

\medskip

{\bf Gluons}

\medskip

As in the one loop case, we multiply (the inverse of) the second of the Beisert-Staudacher equations
for $u_{2,k}=u_k^g$ with (the inverse of) the third for $u_{3,k}=u_k^g + i/2$ and (the inverse of) the third
for $u_{3,k}=u_k^g - i/2$. We then get the following equations for the center of the gluonic string $u_k^g$:
\ba
1&=&\prod _{\stackrel {j=1}{j\not= k}}^{N_g}\frac{u_k^g-u_j^g+i}{u_k^g-u_j^g-i} \prod _{j=1}^{K_4} \frac{(x_k^{g+}-x_{4,j}^+)(x_k^{g-}-x_{4,j}^+)}{(x_k^{g+}-x_{4,j}^-)(x_k^{g-}-x_{4,j}^-)}
\prod _{j=1}^{N_F} \frac{u_k^g-u_{F,j}+i/2}{u_k^g-u_{F,j}-i/2}
\prod _{j=1}^{N_f} \frac{u_k^g-u_{f,j}+i/2}{u_k^g-u_{f,j}-i/2} \cdot \nonumber \\
&\cdot & \prod _{j=1}^{K_b} \frac{u_k^g-u_{b,j}+i/2}{u_k^g-u_{b,j}-i/2} \, .
\ea
Making explicit the type-4 roots, we arrive at
\ba
1&=& \prod _{\stackrel {j=1}{j\not= k}}^{N_g}\frac{u_k^g-u_j^g+i}{u_k^g-u_j^g-i}
\prod _{j=1}^{N_F} \frac{u_k^g-u_{F,j}+i/2}{u_k^g-u_{F,j}-i/2}
\prod _{j=1}^{N_f} \frac{u_k^g-u_{f,j}+i/2}{u_k^g-u_{f,j}-i/2}
\prod _{j=1}^{K_b} \frac{u_k^g-u_{b,j}+i/2}{u_k^g-u_{b,j}-i/2}  \cdot \nonumber \\
&\cdot & \prod _{h=1}^H \frac{(x_k^{g-}-x_h^-)(x_k^{g+}-x_h^{-})}{(x_k^{g+}-x_h^+)(x_h^+-x_k^{g-})}
\prod _{j=1}^{K_b} (-1) \frac{(x_k^{g+}-x_{b,j}^{++})(x_{b,j}^{++}-x_k^{g-})}{(x_k^{g-}-x_{b,j}^{--})(x_k^{g+}-x_{b,j}^{--})} \cdot \nonumber \\
&\cdot &  \textrm{exp} \left [ i \int  _{-\infty}^{+\infty} \frac{dv}{2\pi} \chi (v,u_k^g|1) \frac{d}{dv}\Bigl (Z_4(v)-2L_4(v) \Bigr ) \right ] \, .
\ea
Following what we did for (large) fermions, we multiply such expression by $1=e^{iP}$. Then we use expression (\ref {count-scal-all}) for $Z_4(v)$. We observe the exact cancelation of terms involving the isotopic root $u_b$ and eventually for the field $F_{+\bot}$ we obtain the equations
\ba
1&=&  e^{iRP^{(g)}(u_{k}^g) + 2 i D^{(g)}(u_{k}^g)} \prod _{j=1, j\not=k}^{N_{g}}
 {S}^{(gg)}(u_{k}^{g},u_{j}^{g}) \prod _{j=1}^{N_{\bar g}} {S}^{(g\bar g)}(u_{k}^{g},u_{j}^{\bar g})
  \prod _{h=1}^H S^{(gs)}(u_k^g,u_h) \cdot \label  {fineq8} \\
&\cdot &  \prod _{j=1}^{N_F} {S}^{(gF)}(u_{k}^{g},u_{F,j})
 \prod _{j=1}^{N_{\bar F}} {S}^{(g\bar F)}(u_{k}^{g},u_{\bar F,j})  \prod _{j=1}^{N_f} {S}^{(gf)}(u_{k}^{g},u_{f,j})
 \prod _{j=1}^{N_{\bar f}} {S}^{(g\bar f)}(u_{k}^{g},u_{\bar f,j})
\nonumber \, ,
\ea
where
\ba
P^{(g)}(u)&=&-\int _{-\infty}^{+\infty} \frac{dv}{2\pi} \left [ \chi (v,u|1)+\chi (-v,u|1) \right ] \left [ 1-\frac{\sigma _{BES}(v)}{4} \right ]
\, , \label {P-glu} \\
2 D^{(g)}(u)&=& -\int _{-\infty}^{+\infty} \frac{dv}{2\pi} \left [ \chi (v,u|1)+\chi (-v,u|1) \right ] \frac{d}{dv} \tilde P(v) \label {D-glu} \, .
\ea
The procedure for the field $\bar F_{+\bot}$ is completely analogous, hence we give only the final equations
\ba
1&=& e^{iRP^{(g)}(u_{k}^{\bar g}) + 2 i D^{(g)}(u_{k}^{\bar g})} \prod _{j=1}^{N_{g}}
 {S}^{(\bar gg)}(u_{k}^{{\bar g}},u_{j}^{g}) \prod _{j=1, j\not=k}^{N_{\bar g}} {S}^{(\bar g\bar g)}(u_{k}^{{\bar g}},u_{j}^{\bar g}) \prod _{h=1}^H S^{(\bar gs)}(u_k^{\bar g},u_h) \cdot \label {fineq9} \\
&\cdot &  \prod _{j=1}^{N_F} {S}^{(\bar gF)}(u_{k}^{{\bar g}},u_{F,j})
 \prod _{j=1}^{N_{\bar F}} {S}^{(\bar g\bar F)}(u_{k}^{{\bar g}},u_{\bar F,j})  \prod _{j=1}^{N_f} {S}^{(\bar gf)}(u_{k}^{{\bar g}},u_{f,j})
 \prod _{j=1}^{N_{\bar f}} {S}^{(\bar g\bar f)}(u_{k}^{{\bar g}},u_{\bar f,j}) \, .
\nonumber \, .
\ea

{\bf Isotopic roots}

The equations for the isotopic roots $u_a$ and $u_c$ come directly from the second and the sixth of the Beisert-Staudacher equations and their derivation is completely analogous to the one loop case: the only difference is that in the general all loops case also small fermions are present.
\ba
1&=&\prod _{j\not=k}^{K_a} \frac{u_{a,k}-u_{a,j}+i}{u_{a,k}-u_{a,j}-i} \prod _{j=1}^{N_F}
\frac{u_{a,k}-u_{F,j}-i/2}{u_{a,k}-u_{F,j}+i/2} \prod _{j=1}^{N_f}
\frac{u_{a,k}-u_{f,j}-i/2}{u_{a,k}-u_{f,j}+i/2}  \prod _{j=1}^{K_b}
\frac{u_{a,k}-u_{b,j}-i/2}{u_{a,k}-u_{b,j}+i/2} \label {fineq1} \\
1&=&\prod _{j\not=k}^{K_c} \frac{u_{c,k}-u_{c,j}+i}{u_{c,k}-u_{c,j}-i} \prod _{j=1}^{N_{\bar F}}
\frac{u_{c,k}-u_{\bar F,j}-i/2}{u_{c,k}-u_{\bar F,j}+i/2} \prod _{j=1}^{n_{\bar f}}
\frac{u_{c,k}-u_{\bar f,j}-i/2}{u_{c,k}-u_{\bar f,j}+i/2}  \prod _{j=1}^{K_b}
\frac{u_{c,k}-u_{b,j}-i/2}{u_{c,k}-u_{b,j}+i/2} \label {fineq2}
\ea
Then, we consider the product of the third equation for $u_{3,k}=u_{b,k}$ with the fifth for $u_{5,k}=u_{b,k}$, the fourth for $u_{4,k}=u_{b,k}+i/2$ and the fourth for $u_{4,k}=u_{b,k}-i/2$. We arrive at the following equation
\ba
1&=& \prod _{j=1}^{K_a} \frac{u_{b,k}-u_{a,j}-i/2}{u_{b,k}-u_{a,j}+i/2} \prod _{j=1}^{K_c} \frac{u_{b,k}-u_{c,j}-i/2}{u_{b,k}-u_{c,j}+i/2} \prod _{j=1}^{K_4} \frac{u_{b,k}-u_{4,j}+i/2}{u_{b,k}-u_{4,j}-i/2}
 \prod _{j=1}^{N_g} \frac{u_{b,k}-u^g_{j}-i/2}{u_{b,k}-u^g_{j}+i/2}
\cdot \nonumber  \\
&& \prod _{j=1}^{N_{\bar g}} \frac{u_{b,k}-u^{\bar g}_{j}-i/2}{u_{b,k}-u^{\bar g}_{j}+i/2}
\left ( \frac{x^{--}_{b,k}}{x^{++}_{b,k}} \right )^L \prod _{j\not=k}^{K_4} \frac{x^{--}_{b,k}-x^{+}_{4,j}}{x^{++}_{b,k}-x^{-}_{4,j}} \frac{1-\frac{g^2}{2x^{++}_{b,k}x^{-}_{4,j}}}{1-\frac{g^2}{2x^{--}_{b,k}x^{+}_{4,j}}} \sigma ^2 (u_{b,k}+i/2,u_{4,j})
\sigma ^2 (u_{b,k}-i/2,u_{4,j}) \nonumber \\
&&  \cdot\prod _{j=1}^{K_b} \left ( \frac{x^{++}_{b,k}-x_{b,j}}{x^{--}_{b,k}-x_{b,j}} \right )^2  \prod _{j=1}^{N_F} \frac{x_{b,k}^{++}-x_{F,j}}{x_{b,k}^{--}-x_{F,j}}
\prod _{j=1}^{N_{\bar F}} \frac{x_{b,k}^{++}-x_{\bar F,j}}{x_{b,k}^{--}-x_{\bar F,j}}
\prod _{j=1}^{N_f} \frac{1-\frac{x_{f,j}}{x_{b,k}^{++}}}{1-\frac{x_{f,j}}{x_{b,k}^{--}}}
\prod _{j=1}^{N_{\bar f}} \frac{1-\frac{x_{\bar f,j}}{x_{b,k}^{++}}}{1-\frac{x_{\bar f,j}}{x_{b,k}^{--}}}
\nonumber \\
&& \cdot\prod _{j=1}^{N_g} \frac{x_{b,k}^{++}-x_{j}^{g+}}{x_{b,k}^{--}-x_{j}^{g+}}   \frac{x_{b,k}^{++}-x_{j}^{g-}}{x_{b,k}^{--}-x_{j}^{g-}} \prod _{j=1}^{N_{\bar g}} \frac{x_{b,k}^{++}-x_{j}^{\bar g+}}{x_{b,k}^{--}-x_{j}^{\bar g+}}   \frac{x_{b,k}^{++}-x_{j}^{\bar g-}}{x_{b,k}^{--}-x_{j}^{\bar g-}} \, , \label  {equinte}
\ea
where
\be
L=H+2+N_F+N_{\bar F}+N_g+N_{\bar g} \, .
\ee
We have
\ba
&& \prod _{j=1}^{K_4} \frac{u_{b,k}-u_{4,j}+i/2}{u_{b,k}-u_{4,j}-i/2}= \prod _{\stackrel {j=1}{j\not= k}}^{K_b} \frac{u_{b,k}-u_{b,j}+i}{u_{b,k}-u_{b,j}-i} \prod _{h=1}^{H} \frac{u_{b,k}-u_h-i/2}{u_{b,k}-u_h+i/2} \left (1+O(1/s^2)
\right )  \nonumber \cdot \\
&\cdot & \textrm{exp} \left [ - \int _{-\infty}^{+\infty}\frac{dv}{2\pi} \ln \frac{u_{b,k}-v+i/2}{u_{b,k}-v-i/2} \frac{d}{dv}(Z_4(v)-2L_4(v)) \right ]
\label  {equ2} \, ,
\ea
where for $Z_4(v)$ it is convenient to use form (\ref {Zeta4}).
It is remarkable that, plugging (\ref {Zeta4}) into the integral in the last term of (\ref  {equ2}), we find that
\be
\textrm{exp} \left [ - \int _{-\infty}^{+\infty} \frac{dv}{2\pi} \ln \frac{u_{b,k}-v+i/2}{u_{b,k}-v-i/2} \frac{d}{dv} Z_4(v) \right ]
\ee
produces \footnote {This cancelation was already noticed and proven by Basso in Appendix C.2 of \cite {Basso}.} massive cancelations in (\ref {equinte}). On the other hand, the nonlinear term containing $L_4'(v)$ gives a negligible $O(1/s^2) $ contribution. Eventually, for the third isotopic root $u_b$ we obtain the same equation as in the one loop:
\be
1=\prod _{j=1}^{K_a} \frac{u_{b,k}-u_{a,j}-i/2}{u_{b,k}-u_{a,j}+i/2} \prod _{j=1}^{K_c} \frac{u_{b,k}-u_{c,j}-i/2}{u_{b,k}-u_{c,j}+i/2} \prod _{\stackrel {j=1}{j\not= k}}^{K_b} \frac{u_{b,k}-u_{b,j}+i}{u_{b,k}-u_{b,j}-i}
\prod _{h=1}^{H} \frac{u_{b,k}-u_h-i/2}{u_{b,k}-u_h+i/2} \label {fineq3} \, .
\ee

\section {Conserved observables} \label{sez3}
\setcounter{equation}{0}

Momentum was already obtained in previous sections: therefore, we concentrate on higher charges $Q_r$ and in particular on anomalous dimensions $\gamma =Q_2$.
Let us introduce the function
\be
q_r(u)= \frac{ig^2}{r-1}  \left [ \left (\frac{1}{x^+(u)}\right )^{r-1}- \left (\frac{1}{x^-(u)} \right )^{r-1} \right ] \, ,
\quad r \geq 2 \, , \label {qr}
\ee
whose Fourier transform reads
\be
\hat q_r(k)=2\pi i g^2 \left ( \frac{\sqrt{2}}{ig} \right )^{r-1} e^{-\frac{|k|}{2}} \frac{J_{r-1}(\sqrt{2}gk)}{k} \, . \label {qrk}
\ee
The $r$-th charge of an excited state over the GKP vacuum enjoys the expression
\ba
Q_r&=&\sum _{j=1}^{K_4} q_r (u_{4,j})=\sum  _{j=1}^{K_b} [ q_r(u_{b,j}+i/2) + q_r(u_{b,j}-i/2) ] -\sum _{h=1}^H q_r (u_h) - \nonumber \\
&-& \int _{-\infty} ^{+\infty} \frac{dv}{2\pi} q_r (v) \frac{d}{dv} [ Z_4 (v)-2L_4(v) ] + O\left ( \frac{1}{s^2} \right ) \, ,
\ea
where for $Z_4(v)$ we use expression (\ref  {count-scal-all}). Doing this, we observe the exact cancelation of the dependence on the isotopic root $u_b$ and we are left with the formula
\ba
Q_r&=& -\sum _{h=1}^{H} q_r(u_h) +
\ln s \int \frac{dv}{2\pi} q_r(v) \frac{d}{dv} [4v-Z_{BES}(v)]+
 \int \frac{dv}{2\pi} q_r(v) \frac{d}{dv} 2 \tilde P(v)
 - \nonumber \\
 &-& \int \frac{dv}{2\pi} q_r(v) \frac{d}{dv} \Bigl [ \sum _{h=1}^H \Theta '(v,u_h) +  \sum _{j=1}^{N_g} F_G(v,u^g_j) + \sum _{j=1}^{N_{\bar g}} F_G(v,u^{\bar g}_j)+
 \sum _{j=1}^{N_F} F_F(v,u_{F,j})+ \nonumber \\
&+&  \sum _{j=1}^{N_{\bar F}} F_F(v,u_{\bar F,j})+ \sum _{j=1}^{N_f} F_f(v,u_{f,j}) +
\sum _{j=1}^{N_{\bar f}} F_f(v,u_{\bar f,j})+ 2\int dw G(v,w) L_4(w)-2L_4(v) \Bigr ] = \nonumber \\
&=& \ln s \int \frac{dv}{2\pi} q_r(v) \frac{d}{dv} [4v-Z_{BES}(v)]+
 \int \frac{dv}{2\pi} q_r(v) \frac{d}{dv} 2 \tilde P(v) + \nonumber \\
&+& \sum _{h=1}^{H} Q_r^{(s)}(u_h) + \sum _{j=1}^{N_g} Q_r^{(g)}(u_j^g) + \sum _{j=1}^{N_{\bar g}} Q_r^{(g)}(u^{\bar g}_j) + \sum _{j=1}^{N_F} Q_r^{(F)}(u_{F,j}) + \sum _{j=1}^{N_{\bar F}} Q_r^{(F)}(u_{\bar F,j})+ \nonumber \\
&+& \sum _{j=1}^{N_f} Q_r^{(f)}(u_{f,j}) + \sum _{j=1}^{N_{\bar f}} Q_r^{(f)}(u_{\bar f,j}) + O(1/s^2) \, .
\label {Qr}
\ea
The first two terms in the right hand side of (\ref {Qr}) are contributions from the GKP background.
The remaining terms in (\ref{Qr}) are the contributions that any single particle brings to the overall value of the $r$-th charge.\\
$\bullet $ For scalars we have
\be
Q^{(s)}_r(u)=-q_r(u)
-\int \frac{dv}{2\pi} q_r(v) \frac{d}{dv} \Theta '(v,u)  \, . \label {Qrscal}
\ee
Restricting to $r$ even, we use relation (2.21) of \cite {FPR} to write
\be
Q^{(s)}_r(u)=-q_r (u) - \int _{-\infty}^{+\infty} \frac{dk}{4\pi^2} \hat q_r (k) [ \hat \sigma ^{(1)}(k)+\hat \sigma (k;u) ] \label {Qrscal-2} \, ,
\ee
where the functions $\hat \sigma ^{(1)}(k)$, $\hat \sigma (k;u)$ satisfy equations (2.19), (2.20) of \cite {FPR}, respectively. It is convenient to introduce the functions, defined for $k>0$,
\ba
S^{(1)}(k)&=& \frac{\sinh \frac{k}{2}}{\pi k} \left [ \hat \sigma ^{(1)}(k) + \frac{\pi}{\sinh \frac{k}{2}} \left (1-e^{-\frac{k}{2}} \right ) \right ] \label {S1-def} \, , \\
S(k;u)&=& \frac{\sinh \frac{k}{2}}{\pi k} \left [ \hat \sigma (k;u)- \frac{2\pi e^{-k}}{1-e^{-k}} (\cos ku-1) \right ]
\label {Su-def} \, ,
\ea
and to expand them in Neumann series
\be
S^{(1)}(k)=\sum _{p=1}^{+\infty} S_p^{(1)} \frac{J_p(\sqrt{2}gk)}{k} \, , \quad
S(k;u)=\sum _{p=1}^{+\infty} S_p^{\prime}(u) \frac{J_p(\sqrt{2}gk)}{k} \, . \label {S-Neu}
\ee
We use defining equations (3.10) of \cite {FGR1} for $S_p^{(1)}$ and (3.15) of \cite {FRO6} for $S_p^{\prime}(u)$
to simplify (\ref {Qrscal-2}) as follows
\be
Q^{(s)}_r(u)=\frac{ig^2}{r-1} \left ( \frac{\sqrt{2}}{ig} \right )^{r-1}\left [ S_{r-1}^{(1)}+S_{r-1}^{\prime}(u) \right ] \label {Qrscal-3} \, .
\ee
We could not find a formula analogous to (\ref {Qrscal-3}) in the case $r$ odd.

When $r=2$, this simple expression can be connected with the first of (4.6) of \cite {Basso}.
We indeed remember formula (4.36) of \cite {FGR3} and that
\be
S_1^{\prime}(u)=-2\pi \sum _{n=1}^{+\infty} \frac{(-1)^n u^{2n}}{(2n)!} \tilde S_1^{(n)} \, ,
\ee
where for $\tilde S_1^{(n)}$ we use formula (4.35) of \cite {FGR3} to connect with the solution of the BES equation. Operating in this way we get, after some algebra,
\be
Q^{(s)}_2(u)=\gamma^{(s)}(u)=\int _{0}^{+\infty} \frac{dt}{t} \left [ \frac{e^{-\frac{t}{2}}-\cos tu}{e^{\frac{t}{2}}-e^{-\frac{t}{2}}}
\gamma _-^{\textrm{{\o}}}(\sqrt{2}gt) + \frac{\cos tu -e^{\frac{t}{2}}}{e^{\frac{t}{2}}-e^{-\frac{t}{2}}}
\gamma _+^{\textrm{{\o}}}(\sqrt{2}gt) \right ] \, ,
\ee
i.e. the first of (4.6) of \cite {Basso}. Functions $\gamma _{\pm}^{\textrm{{\o}}}$ are defined in (\ref {BES-BAS}, \ref {BES-BAS2}).

It can be of interest to express $\gamma^{(s)}(u)$ in the $O(6)$ limit \cite {AM}.
We introduce
\be
m(g)=\frac{2^{\frac{5}{8}}\pi ^{\frac{1}{4}}}{\Gamma \left (\frac{5}{4} \right )} g^{\frac{1}{4}}e^{-\frac{\pi g}{\sqrt{2}}}\left [ 1+O\left (\frac{1}{g} \right ) \right ] \, . \label {m(g)}
\ee
In the $O(6)$ limit
$\sqrt{2}g S_1^{(1)}=m(g)-1$ and $\sqrt{2}g S_1^{\prime}(u)=m(g) \left (\cosh \frac{\pi}{2}u
-1 \right )$. Substituting in (\ref {Qrscal-3}) we get
\be
\gamma^{(s)}(u)=m(g) \cosh \frac{\pi}{2}u -1 \, ,
\ee
and for the complete anomalous dimension in presence only of scalar excitations
\be
\gamma = \ln s f(g) + f_{sl}(g)+  \sum _{h=1}^{H} \left ( m(g) \cosh \frac{\pi}{2}u_h -1 \right ) \, .
\ee

$\bullet $ For gluons we have
\be
Q^{(g)}_r(u)=-\int \frac{dv}{2\pi} q_r(v) \frac{d}{dv} F^G (v,u) \, .
\ee
Using equations (\ref {Theta-prime}, \ref {F-F}), we arrive at the formula
\be
Q^{(g)}_r(u)=- \int \frac{dv}{2\pi} \left [ \chi (v,u |1)+\Phi (v) \right ] \frac{d}{dv} Q_r ^{(s)} (v) \, .
\ee
When $r=2$ we have
\ba
&& Q^{(g)}_2(u)=\gamma ^{(g)}(u) = \int_{-\infty}^{\infty}\frac{dk}{4\pi^2}\,\frac{i\pi}{e^{\frac{k}{2}}-e^{-\frac{k}{2}}}\,
(\gamma _+^{\textrm{{\o}}}(\sqrt{2}gk)- \textrm{sgn}(k)\gamma _-^{\textrm{{\o}}}(\sqrt{2}gk))
\left[\frac{2\pi}{ik}\,e^{-|k|}e^{-iku}- \right. \\
&&\ \ -\frac{2\pi}{ik}\left. \,e^{-\frac{|k|}{2}}\sum_{n=1}^{\infty}\left(\left(\frac{g}{\sqrt{2}i x(u+ \frac{i}{2})}\right)^n+\left(\frac{g}{\sqrt{2}i x(u- \frac{i}{2})}\right)^n\right)\,J_n(\sqrt{2}gk)
-\frac{2\pi}{ik}\,J_0(\sqrt{2}gk)e^{-\frac{|k|}{2}}
\right]  \nonumber \, .
\ea
By means of the relation (3.40) in \cite{Basso}, and making use of the identity
\be
\int_0^\infty \frac{dk}{k}\,e^{-k\left(\frac{1}{2}\pm iu\right)}\, J_n(\sqrt{2}gk)=
\frac{(\pm 1)^n}{n}\left(\frac{g}{\sqrt{2}i x(u\mp \frac{i}{2})}\right)^n  \qquad ,
\ee
the expression above becomes
\be
\gamma ^{(g)}(u)=\int_0^\infty\frac{dk}{k}\,\frac{\gamma _+^{\textrm{{\o}}}(\sqrt{2}gk)}
{1-e^{-k}}\,[\cos ku \,e^{-\frac{k}{2}}-1]
-\int_0^\infty\frac{dk}{k}\,\frac{\gamma _-^{\textrm{{\o}}}(\sqrt{2}gk)}{e^{k}-1}\,[\cos ku \,e^{-\frac{k}{2}}-1] \, .
\ee

$\bullet $ Large fermions and antifermions with rapidity $u$ carry an amount of $r$-charge equal to
\be\label{Qferm}
Q^{(F)}_r(u)=- \int \frac{dv}{2\pi} q_r(v)  \frac{d}{dv} F_F (v,u)  \, .
\ee
Using equations (\ref {Theta-prime}, \ref {F-F}), we arrive at the formula
\be
Q^{(F)}_r(u)=- \int \frac{dv}{2\pi} \left [ \chi _F (v,u)+\Phi (v) \right ] \frac{d}{dv} Q_r ^{(s)} (v) \, .
\ee
For small fermions we have
\be
Q^{(f)}_r(u)=\int \frac{dv}{2\pi} \chi _H (v,u) \frac{d}{dv} Q_r ^{(s)} (v) \, .
\ee
When $r=2$,
the very same reasonings outlined above apply to large fermions, so that ($Q^{(F)}_2(u)=\gamma ^{(F)}(u)$)
\be
\gamma ^{(F)}(u) =\int_0^\infty\frac{dk}{k}\,\frac{\gamma _+^{\textrm{{\o}}}(\sqrt{2}gk)-\gamma _-^{\textrm{{\o}}}(\sqrt{2}gk)}{e^{k}-1}\,[\cos ku -1]
+\int_0^\infty\frac{dk}{k}\,\gamma _+^{\textrm{{\o}}}(\sqrt{2}gk)
\,[\frac{1}{2}\cos ku-1] \, .
\ee
Analogously for $r=2$ and small fermions
\be
Q^{(f)}_2(u)=\gamma ^{(f)}(u)=-\frac{1}{2}\int_0^\infty\frac{dk}{k}\,\gamma _+^{\textrm{{\o}}}(\sqrt{2}gk) \,\cos ku
\, .
\ee

\subsection {One loop}

All the previous expression are explicitly computed at one loop,
upon introducing the following notation for the derivatives of the digamma function
\ba
&&\psi^{(n)}(z)\equiv \left(\frac{d}{dz}\right)^n \psi(z) \, , \\
&&\psi^{(0)}(z)\equiv \psi(z) \, .
\ea
$\bullet $ Scalars
\be
Q^{(s)}_r(u)=-\frac{i^r g^2}{(r-1)!} \left[\psi^{(r-2)} (1/2-iu) + (-1)^r \psi^{(r-2)} (1/2+iu) -\psi^{(r-2)}(1)(1+(-1)^r) \right]  \, ;
\ee
$\bullet $ Fermions and antifermions
\be
Q^{(F)}_r(u)=-\frac{i^r g^2}{(r-1)!} \left [\psi^{(r-2)} (1-iu) + (-1)^r \psi^{(r-2)} (1+iu) -\psi^{(r-2)}(1)(1+(-1)^r) \right ] \, ;
\ee
$\bullet $ Gluons
\be
Q^{(g)}_r(u)=-\frac{i^r g^2}{(r-1)!} \left [\psi^{(r-2)} (3/2-iu) + (-1)^r \psi^{(r-2)} (3/2+iu) -\psi^{(r-2)}(1)(1+(-1)^r) \right ] \, .
\ee

\section{Strong coupling regimes of 2D scattering factors} \label{sez4}
\setcounter{equation}{0}

In this section we want to give a detailed analysis of the different  strong coupling limits of the 2D scattering factors  $S^{\ast \ast '}(u,v)$ of sub-section 2.3. In fact there are different ways of performing the $g\rightarrow +\infty$ limit as these give rise to different results or regimes, so paralleling what happens to the energy/momentum dispersion relations \cite{Basso}.

First, we shall discuss the regime, relevant only for scalars (as the other excitations decouples towards very high energy), where we keep their rapidities fixed, namely the so-called non-perturbative regime. In this case integrations inside the expressions for the various scattering factors receive the leading contributions from the region where the integration variables are fixed (while sending $g\rightarrow +\infty$). This regime is dominated by scalars which are the only ones to have a non-trivial (finite) $S$-factor, whilst the other $S$-factors involving other excitations
reduce to one. Here we find out the (usual) $O(6)$ non-linear sigma model scattering theory as low energy string theory \cite{AM, FGR1, BK} \footnote{At next approximation it would be perturbed by irrelevant fields as suggested by the expansion of the energy in inverse powers of the size ($R$) \cite{FRO6} ({\it cf.} also the dispersion relation in \cite{Basso} and the effective field theory of \cite{Zarembo:2011ag}).}. Alternatively, we can first rescale the external rapidities $u=\sqrt{2}g \bar u$, $v=\sqrt{2} g \bar v$ and then send $g\rightarrow +\infty$. If the rescaled variables, $\bar u$ and $\bar v$, have modulus smaller than one we are (with the exception of scalars, see discussion below) in the perturbative string regime (where the irrelevant and relevant perturbations of the $O(6)$ non-linear sigma prevail on it putting at zero its mass); while if their modulus is greater than one we are in the so-called giant hole (semiclassical soliton) regime. In both cases, in order to have the maximum contribution to the integrals, after rescaling external rapidities, we have to perform the same rescaling of the integration variables $u_i=\sqrt{2}g \bar u_i$ and eventually take the limit $g\rightarrow +\infty$.

\subsection{Scalars}

{\bf Scalars in the non-perturbative regime}

We report the strong coupling limit of the scalar-scalar scattering factor in the non-perturbative regime, i.e. $g\rightarrow +\infty$, with $u,v$ fixed (details on the calculation can be found in \cite {FPR}, see also \cite{FRO6} and \cite{BREJ}):
\be\label{theta-non-perturb}
g\rightarrow +\infty \quad \Rightarrow \quad \Theta (u,v)\rightarrow \Theta _{np}(u-v) = -i \ln \frac{\Gamma \left (1-i\frac{u-v}{4} \right ) \Gamma \left (\frac{1}{2}+i\frac{u-v}{4} \right )}{\Gamma \left (1+i\frac{u-v}{4} \right ) \Gamma \left (\frac{1}{2}-i\frac{u-v}{4} \right )}-\textrm{gd} \left ( \frac{\pi (u-v)}{2} \right ) \, ,
\ee
which depends only on the difference of the rapidities and coincide with the pre-factor of the $S$-matrix, as derived in \cite{Zam^2}, of the $O(6)$ non-linear sigma model upon the identification (of the hyperbolic rapidities) $\theta=\pi u/2$ and $\theta'=\pi v/2$. This definitely supports the proposal of the latter model by \cite{AM} as that describing the string at low energy (see also subsequent studies \cite{FGR1,BK}).\\
\medskip

{\bf Scalars in the perturbative regime}

As in \cite {Basso} the perturbative regime for scalars is recovered by introducing a new rapidity $z$
as
\be
u=\frac{2}{\pi}\ln \frac{z}{m(g)} \, , \ \textrm{for} \ u>0 \ \ \textrm{(right mover)}; \quad u=\frac{2}{\pi}\ln \frac{m(g)}{z} \, , \ \textrm{for} \ u<0 \ \ \textrm{(left mover)} \label {uz} \, ,
\ee
$m(g)$ being the non-perturbative mass (\ref {m(g)}).
The rapidity $z$ is kept fixed in the region $m(g)<z<1$ as $g \rightarrow +\infty$. If $m(g)<z<1$ formula (\ref {theta-non-perturb}) is valid, therefore the function $\Theta$ in the perturbative regime is obtained by plugging
(\ref {uz}) into (\ref {theta-non-perturb}).
\medskip

\noindent
{\bf Scalars in the scaling regimes}

We rescale the rapidities $u=\sqrt{2}g \bar u$, $v=\sqrt{2}g \bar v$ and then send $g\rightarrow +\infty$, with $\bar u$, $\bar v$ fixed and $|\bar u| >1$, $|\bar v|>1$.  This is the so-called giant hole regime. Details on the calculation can be found in \cite {FPR}. We give the final result for the double derivative, which will be useful for next computations
\be
\frac{d}{d\bar u} \frac{d}{d\bar v} \Theta (\sqrt{2} g \bar u , \sqrt{2} g \bar v)= \sqrt{2}g
\frac{\left ( \frac{\bar u+1}{\bar u-1}\right )^{1/4} \left ( \frac{\bar v-1}{\bar v+1}\right )^{1/4} +
\left ( \frac{\bar u-1}{\bar u+1}\right )^{1/4} \left ( \frac{\bar v+1}{\bar v-1}\right )^{1/4} }{\bar u-\bar v }
+O(g^0) \, .
\label {der2theta}
\ee
This result (\ref {der2theta}) agrees with corresponding formula coming from using the scattering phase (2.34) of \cite {DoreyZhao}.

\medskip

Another possibility is to define rescaled rapidities (with a bar) $u=\sqrt{2}g \bar u$, $v=\sqrt{2}g \bar v$ and then send $g\rightarrow +\infty$, with $\bar u$, $\bar v$ fixed and $|\bar u| <1$, $|\bar v|<1$. Although for the other particles this second possibility gives rise to the perturbative string regime (giving for understood an obvious modification $u\rightarrow x(u)$  for the (small) fermion, {\it cf.} below), it does not in the case of scalars as given in \cite{Basso}, because of their non-perturbative, dynamically generated mass. Yet, we need to consider the scalar $\Theta$ in this regime at least to access the other $S$-matrix elements (depending on it). In fact, we may write the limiting value
\be
\frac{d^2}{d\bar ud\bar v}\Theta (u,v)=2\pi  \frac{d}{d\bar u }\delta  (\bar u -\bar v) + \frac{1}{\sqrt{2}g} \frac{d^2}{d \bar v^2} P\frac{1}{\bar v -\bar u}+ O(1/g^2) \, . \label {der2theta-per}
\ee
Importantly, formula (\ref {der2theta-per}) is valid also in the domains $|\bar u| <1$, $|\bar v|>1$ and $|\bar u| >1$, $|\bar v|<1$. We will make
frequent use of (\ref {der2theta-per}) in this Section.

From (\ref {der2theta-per}) we can infer
\be
\Theta (u,v)=-\pi \textrm{sgn} (u-v)- \frac{1}{\sqrt{2}g}\frac{1}{\bar v - \bar u} +O(1/g^2) \Rightarrow
S^{(ss)}(u,v)=e^{\frac{i}{\sqrt{2}g}\frac{1}{\bar v -\bar u}+O(1/g^2)}
\ee

\subsection{Gluons}

{\bf  Gluons in the perturbative regime}

We want to study the gluon-gluon scattering factor (\ref {Sgg}) in the limit $g\rightarrow +\infty$, with
$u=\bar u \sqrt{2}g$, $v=\bar v\sqrt{2}g$, $\bar u$, $\bar v $ fixed and
$ \bar u ^2 <1$,  $\bar v^2 <1$ (perturbative regime).

We have
\be
i\ln \left ( - S^{(gg)} (u,v) \right )= \mathcal{I}_1+\mathcal{I}_2+\mathcal{I}_3 \, ,
\ee
where
\ba
 \mathcal{I}_1&=&\tilde \chi (u,v|1,1)=
 -2 \arctan \sqrt{2}g (\bar v-\bar u) = \nonumber \\
 &=& -\pi \textrm{sgn}(\bar v-\bar u) + \frac{\sqrt{2}}{g(\bar v-\bar u)}+ O(1/g^3)  \label {i1} \, .
\ea
Passing to study $\mathcal{I}_2$ and $\mathcal{I}_3$, we first remark that in the perturbative regime
\be
x^{\mp} (u) =\left [\frac{g}{\sqrt{2}}+\frac{1}{4\sqrt{1-\bar u^2}} \right ] ( \bar u \mp i \sqrt{1-\bar u ^2} ) + O(1/g) \, . \label {xpmlimit}
\ee
Since we have to work out $\chi (w, u|1) +\Phi (w)$, in addition to (\ref {xpmlimit}) we need also to know the behaviour of $x^{\pm}(w)$ when $w=\bar w\sqrt{2}g$
and $g\rightarrow +\infty$. When $|\bar w|<1$ we can use (\ref {xpmlimit}). On the other hand, for $|\bar w|>1$
we have
\be
x^{\pm} (w)= \sqrt{2}g \bar x (\bar w) \pm \frac{i}{4} \frac{1+\sqrt{1-\frac{1}{\bar w ^2}}}{\sqrt{1-\frac{1}{\bar w^2}}} + O(1/g) \, , \quad \bar x (\bar w)= \frac{\bar w}{2} \left [ 1+\sqrt{1-\frac{1}{\bar w ^2}} \right ] \, .
\label {xpmlimit2}
\ee
Using results (\ref {xpmlimit}, \ref {xpmlimit2}), we arrive at the relations, valid for $w=\bar w\sqrt{2}g$
and $g\rightarrow +\infty$:
\ba
\chi (w, u|1) +\Phi (w) &=& \frac{1}{\sqrt{2}g (\bar u -\bar w)} +O(1/g^2) \, , \quad \textrm{when} \ \ |\bar w|>1 \, , \\
\chi (w, u|1) +\Phi (w) &=& O(1/g)  \, , \quad \textrm{when} \ \ |\bar w|<1 \, . \label {chiphiper}
\ea
Therefore, we have
\ba
\mathcal{I}_2=- \int_{-\infty}^{+\infty} \frac{dw}{2\pi}\, [ \chi(w,u|1)
+\Phi (w)]\, \frac{d}{dw}[ \chi(w,v|1)+\Phi (w) ]=
O(1/g^2) \label {i2} \, .
\ea
For what concerns the last term $\mathcal{I}_3$ in the rhs of (\ref {Sgg}), we find convenient to perform the change of variables $w=\sqrt{2} g \bar w$, $z=\sqrt{2} g \bar z$:
\be
\mathcal{I}_3= \int_{-\infty} ^{+\infty}\frac{d\bar w}{2\pi}\int_{-\infty}^{+\infty}\frac{d\bar z}{2\pi}\, \,  [ \chi(w,u|1) +\Phi (w)] \left [ \frac{d}{d\bar w} \frac{d}{d\bar z} \Theta ( \sqrt{2}g \bar w, \sqrt{2}g \bar z) \right ] \,  [ \chi(z,v|1) +\Phi (z)] \, .
\ee
Now, from formul{\ae} (\ref {der2theta}, \ref {der2theta-per}), we deduce that the leading behaviour of the double derivative of the scalar-scalar phase is realised in the giant hole regime $|\bar w|>1$, $|\bar z|>1$.
Therefore, we can write
\be
\mathcal{I}_3 \cong \int_{|\bar w|>1} \frac{d\bar w}{2\pi} \int_{|\bar z|>1} \frac{d\bar z}{2\pi}
\frac{1}{\sqrt{2}g \bar w-\sqrt{2}g\bar u } \frac{1}{\sqrt{2}g \bar z-\sqrt{2}g \bar v}
\frac{d}{d\bar w} \frac{d}{d\bar z} \Theta (\sqrt{2} g \bar w , \sqrt{2} g \bar z)  \label {i3} \, .
\ee
Plugging (\ref {der2theta}) into (\ref {i3}) and performing the integrations we arrive at
\be
\mathcal{I}_3= \frac{1}{2\sqrt{2}g (\bar u - \bar v )} \left [ 2 - \left ( \frac{1+\bar u}{1- \bar u}\right )^{1/4} \left ( \frac{1-\bar v}{1+\bar v }\right )^{1/4}  - \left ( \frac{1-\bar u}{1+ \bar u}\right )^{1/4} \left ( \frac{1+\bar v}{1-\bar v }\right )^{1/4} \right ] \label {i3fin} \, .
\ee
Now, summing up (\ref {i1}, \ref {i2}, \ref {i3fin}) we obtain the final result for the gluon-gluon scattering phase at the order $O(1/g)$:
\be
S^{(gg)}(u,v) = \textrm{exp} \left [ \frac {i}{\sqrt{2}g (\bar u - \bar v)} \left ( 1+
\frac{1}{2}\left ( \frac{1+\bar u}{1- \bar u}\right )^{1/4} \left ( \frac{1-\bar v}{1+\bar v }\right )^{1/4}  + \frac{1}{2}\left ( \frac{1-\bar u}{1+ \bar u}\right )^{1/4} \left ( \frac{1+\bar v}{1-\bar v }\right )^{1/4} \right ) \right ] \label {Sglu-str} \, .
\ee
The expression above agrees with the correspondent result of Basso, Sever, Vieira \cite {BSV1}.

\medskip

{\bf Gluons in the giant hole regime}

We now want to compute the gluon-gluon scattering factor (\ref {Sgg}) in the limit $g\rightarrow +\infty$, with
$u=\bar u \sqrt{2}g$, $v=\bar v \sqrt{2}g$, $\bar u$, $\bar v$ fixed and
$ \bar u ^2 >1$,  $\bar v^2 >1$ (giant hole regime).

As a preliminary calculation we consider the quantity $\chi (v,u|1)+\Phi (v)$. For its expression we refer to (\ref {chiphi}). Computing the scaling limit $u=\bar u \sqrt{2}g$, $v=\bar v \sqrt{2}g$, $g\rightarrow +\infty$, $\bar u$, $\bar v$ fixed and $ \bar u ^2 >1$,  $\bar v^2 >1$, we find that
\be
\chi (v,u|1)+\Phi (v)= \pi \textrm{sgn} (v-u) -\pi \textrm{sgn} v +O(1/g) \label {chiphilea} \, .
\ee
Since $\chi (v,u|1)+\Phi (v)$ is at most $O(g^0)$, properties (\ref {der2theta}, \ref {der2theta-per}) imply that the following part of the integral $\mathcal{I}_3$,
\be
\mathcal{I}_3^>= \int_{|\bar w|>1} \frac{d\bar w}{2\pi}\int_{|\bar z|>1}\frac{d\bar z}{2\pi}\, \,  [ \chi(w,u|1) +\Phi (w)] \left [ \frac{d}{d\bar w} \frac{d}{d\bar z} \Theta ( \sqrt{2}g \bar w, \sqrt{2}g \bar z) \right ] \,  [ \chi(z,v|1) +\Phi (z)] \label {I3up} \, ,
\ee
gives actually the dominant contribution (proportional to $g$) to $i\ln \left ( - S^{(gg)} (u,v) \right )$. The integrations in (\ref {I3up}) are easily performed and the final result\footnote {In order to get (\ref  {Sgggh}) we use the properties $\Theta (u,\pm \sqrt{2}g)=\Theta (\pm \sqrt{2}g,v) =0$ which are proven using expressions given in \cite {DoreyZhao}.} is
\be
i\ln \left ( - S^{(gg)} (u,v) \right )= i\ln \left ( - S^{(g\bar g )} (u,v) \right )=\Theta (u,v) + O(g^0) = i\ln \left ( - S^{(ss)} (u,v) \right )+O(g^0) \, .
\label {Sgggh}
\ee

\medskip

{\bf Gluons in the non-scaling regime}

In this regime we send $g\rightarrow +\infty$ keeping the excitations rapidities fixed.
For what concerns gluons, if we send $g\rightarrow +\infty$, with gluons and scalar rapidities, $u$, $v$ respectively, fixed, we get that
\be
\chi (v, u|1)+\Phi (v)=O(1/g^2) \, . \label {chiphinp}
\ee
In order to get (\ref {chiphinp}), relation
\be
g\rightarrow +\infty \, , \quad u \ \textrm{fixed}   \quad \Rightarrow x^{\pm}(u)=\pm \frac {ig}{\sqrt {2}} + \frac {u\pm \frac {i}{2}}{2}\mp i \frac {\left (u\pm \frac {i}{2}\right )^2}{4\sqrt {2} g}+O(1/g^3) \, ,  \label {xpmnplimit}
\ee
is useful. Result (\ref {chiphinp}) means that in this regime
the gluon-gluon scattering phase $S^{(gg)} (u,v)$ reduces to $\frac{u-v+i}{u-v-i}$.

\subsection{Fermions}

{\bf Fermions in the perturbative regime}

We want to find the strong coupling limit of the fermion-fermion scattering factor in the perturbative regime.
As we will show in a moment, this regime fits in the small fermion case. We start from
\be
{S}^{(ff)}(u,v)= \textrm{exp} \Bigl \{ i \int _{-\infty}^{+\infty}\frac{dw}{2\pi} \chi _H (w,u)  \frac{d}{dw} \chi _H (w,v)- i \int \frac{dw}{2\pi} \frac{dz}{2\pi} \chi _H (w,u) \frac{d^2}{dwdz}\Theta (w,z) \chi _H(z,v)  \Bigr \}   \label {Sff} \, ,
\ee
where
\be
\chi _H (w,u)= -i \ln \frac{1-\frac{x_f(u)}{x^+(w)}}{1-\frac{x_f(u)}{x^-(w)}} \, , \quad x_f(u) =\frac{u}{2}\left [ 1-\sqrt{1-\frac{2g^2}{u^2}} \right ] \, .
\ee
In the perturbative regime the fermion rapidity scales as $x_f(u)= \sqrt{2}g \bar x _f(\bar u)$, $u=\sqrt{2}g \bar u$, with
\be
\bar x_f(\bar u)= \frac{\bar u}{2}\left [ 1-\sqrt{1-\frac{1}{\bar u^2}} \right ]\, , \quad |\bar u| \geq 1 \, , \quad |\bar x _f(\bar u)| \leq \frac{1}{2} \, .
\ee
It is then clear that we are in the small fermion case.

For what concerns scalar rapidity, we make the rescaling $w=\sqrt{2}g \bar w$ and we develop at strong coupling.
We have to distinguish two cases.

$\bullet$ If $|\bar w |>1$, then
\be
x^{\pm} (w)= \sqrt{2}g \bar x (\bar w) \pm \frac{i}{4} \frac{1+\sqrt{1-\frac{1}{\bar w ^2}}}{\sqrt{1-\frac{1}{\bar w^2}}} + O(1/g) \, , \quad \bar x (\bar w)= \frac{\bar w}{2} \left [ 1+\sqrt{1-\frac{1}{\bar w ^2}} \right ] \, .
\ee
In this case we have
\be
\chi _H (w,u)= -i \ln \frac{1-\frac{x_f(u)}{x^+(w)}}{1-\frac{x_f(u)}{x^-(w)}}\cong - \frac{\bar x _f(\bar u)}{\sqrt{2}g} \frac{1}{\bar w \sqrt{1-\frac{1}{\bar w ^2}}} \frac{1}{\bar x _f(\bar u) -\bar x(\bar w)} +O(1/g^2) \, . \label {chiH1}
\ee
$\bullet$ If $|\bar w |<1$, then
\be
x^{\pm} (w)= \frac{g}{\sqrt{2}} [ \bar w \pm i \sqrt{1-\bar w^2} ] + O(g^0) \, ,
\ee
In this second case
\be
\chi _H (w,u)= -i \ln \frac{1-2\bar x_f (\bar w -i \sqrt{1-\bar w^2})}{1-2\bar x_f (\bar w +i \sqrt{1-\bar w^2})}
+O(1/g) \, . \label {chiH2}
\ee
In general we write $-i \ln S ^{(ff)}=I_1+I_2$, where
\be
I_1= \int _{-\infty}^{+\infty}\frac{dw}{2\pi} \chi _H (w,u)  \frac{d}{dw} \chi _H (w,v)=I_1^{>}+I_1^{<} \, ,
\ee
where
\ba
I_1^{>}&=& \int _{|\bar w|>1}\frac{dw}{2\pi} \chi _H (w,u)  \frac{d}{dw} \chi _H (w,v) \sim O(1/g^2) \, ,  \\
I_1^{<}&=& \int _{|\bar w|<1}\frac{dw}{2\pi} \chi _H (w,u)  \frac{d}{dw} \chi _H (w,v) \, ,
\ea
and
\be
I_2=-\int \frac{dw}{2\pi} \frac{dz}{2\pi} \chi _H (w,u) \frac{d^2}{dwdz}\Theta (w,z) \chi _H(z,v)=I_2^{>}+I_2^{rest}
\, , \ee
where
\ba
I_2^{>}&=&-\int _{|\bar w|, |\bar z| \geq 1}\frac{d\bar w}{2\pi} \frac{d\bar z}{2\pi} \chi _H (w,u) \frac{d^2}{d\bar wd\bar z}\Theta (w,z) \chi _H(z,v) \, , \\
I_2^{rest}&=&-\int _{|\bar w| or |\bar z| \leq 1}\frac{d\bar w}{2\pi} \frac{d\bar z}{2\pi} \chi _H (w,u) \frac{d^2}{d\bar wd\bar z}\Theta (w,z) \chi _H(z,v) \, .
\ea
In order to evaluate $I_2$ we need first to compute (the second derivative of) $\Theta (w,z)$. This happens to depend on the domain of $w,z$. When $|\bar w|$ and $|\bar z|$ are both greater than one, we can use formula (2.33) of the letter \cite {FPR}: in particular, in this domain $\frac{d^2}{d\bar wd\bar z}\Theta (w,z)=O(g)$.  In the remaining domains (i.e $|\bar w|$ and $|\bar z|$ are not both greater than one), we have formula (\ref {der2theta-per}):
\be
\frac{d^2}{d\bar wd\bar z}\Theta (w,z)=2\pi \frac{d}{d\bar w }\delta  (\bar w -\bar z) + \frac{1}{\sqrt{2}g} \frac{d^2}{d \bar z^2} P\frac{1}{\bar z -\bar w}+ O(1/g^2) \, .
\ee
Using this formula we can estimate $I_2^{rest}$. We have
\ba
I_2^{rest}&=&-\int _{|\bar w|<1}\frac{dw}{2\pi} \chi _H (w,u)  \frac{d}{dw} \chi _H (w,v) - \nonumber \\
&-& \frac{1}{\sqrt{2}g} \int _{|\bar w|, |\bar z| \leq 1}\frac{d\bar w}{2\pi} \frac{d\bar z}{2\pi} \chi _H (w,u)
\frac{d^2}{d\bar w^2} P \frac{1}{\bar z -\bar w} \chi _H (z,v) + O(1/g^2) \label {I2rest} \, ,
\ea
where we used the fact that $ \chi _H (w,u)$ is $O(1/g)$ when $|\bar w|>1$. The first term in (\ref {I2rest}) cancels
$I_1^{<}$. The second term equals
\ba
&-&\frac{1}{\sqrt{2}g} \int _{|\bar w|, |\bar z| \leq 1}\frac{d\bar w}{2\pi} \frac{d\bar z}{2\pi} \chi _H (w,u)
\frac{d^2}{d\bar w^2} P \frac{1}{\bar z -\bar w} \chi _H (z,v) = \nonumber\\
&=& -\frac{2\sqrt{2}}{g} (\bar x_f (\bar u) - \bar x_f (\bar v) ) \frac{1+4 \bar x_f (\bar u) \bar x_f (\bar v)}{1-4 \bar x_f (\bar u) \bar x_f (\bar v)} \frac{\bar x_f (\bar u) \bar x_f (\bar v)}{(1-4 \bar x_f (\bar u) ^2 ) (1-4 \bar x_f (\bar v) ^2 )} \label {Irestfinal} \, .
\ea
In order to get this result, we made use of the approximation
\be
\frac{d}{d\bar w}\chi_H(w,u)=\frac{\bar w- 2 \bar x_f(\bar u)}{\sqrt{1-\bar w^2}(\bar w-\bar u)}
-\frac{1}{2\sqrt{2}g(\bar w-\bar u)^2}+O\left(\frac{1}{g^2}\right)
\ee
and of the integral (\ref {integ2bis}).
For what concerns $I_2^{>}$, we have
\ba
I_2^{>}&=&-  \frac{1}{\sqrt{2}g} \int _{|\bar w|\geq 1}\frac{d\bar w}{2\pi} \int _{|\bar z|\geq 1}\frac{d\bar z}{2\pi} \frac{1}{\bar w \sqrt{1-\frac{1}{\bar w ^2}}} \frac{\bar x _f(\bar u)}{\bar x _f (\bar u)-\bar x(\bar w)} \cdot  \nonumber \\ &\cdot &
\frac{\left ( \frac{\bar w -1}{\bar w +1}\right ) ^{\frac{1}{4}} \left ( \frac{\bar z+1}{\bar z -1}\right ) ^{\frac{1}{4}} +
\left ( \frac{\bar w +1}{\bar w -1}\right ) ^{\frac{1}{4}} \left ( \frac{\bar z-1}{\bar z +1}\right ) ^{\frac{1}{4}}}{\bar w - \bar z}  \frac{1}{\bar z \sqrt{1-\frac{1}{{\bar z}^2}}} \frac{\bar x _f(\bar v)}{\bar x _f(\bar v)-\bar x(\bar z)} \, .
\ea
Now, we use the identity
\be
\frac{\left ( \frac{\bar w -1}{\bar w +1}\right ) ^{\frac{1}{4}} \left ( \frac{\bar z+1}{\bar z -1}\right ) ^{\frac{1}{4}} + \left ( \frac{\bar w +1}{\bar w -1}\right ) ^{\frac{1}{4}} \left ( \frac{\bar z-1}{\bar z +1}\right ) ^{\frac{1}{4}}}{\bar w - \bar z} = \frac{1}{\bar x (\bar w)-\bar x (\bar z)} \frac{\sqrt{1+\sqrt{1-\frac{1}{\bar w^2}} } \sqrt{1+\sqrt{1-\frac{1}{\bar z^2}}}}{\left (1-\frac{1}{\bar w ^2} \right )^{\frac{1}{4}} \left (1-\frac{1}{\bar z ^2} \right )^{\frac{1}{4}} }
\ee
and arrive at
\ba
I_2^{>}&= & -  \frac{1}{\sqrt{2}g} \int _{|\bar w|\geq 1}\frac{d\bar w}{2\pi} \int _{|\bar z|\geq 1}\frac{d\bar z}{2\pi} \frac{1}{\bar w \sqrt{1-\frac{1}{\bar w ^2}}} \frac{\bar x _f(\bar u)}{\bar x _f(\bar u)-\bar x(\bar w)} \cdot \nonumber \\ &\cdot & \frac{1}{\bar x (\bar w)-\bar x (\bar z)} \frac{\sqrt{1+\sqrt{1-\frac{1}{\bar w^2}} } \sqrt{1+\sqrt{1-\frac{1}{\bar z^2}}}}{\left (1-\frac{1}{\bar w ^2} \right )^{\frac{1}{4}} \left (1-\frac{1}{\bar z ^2} \right )^{\frac{1}{4}} }
\frac{1}{\bar z \sqrt{1-\frac{1}{{\bar z}^2}}} \frac{\bar x _f(\bar v)}{\bar x _f(\bar v)-\bar x(\bar z)} \, .
\ea
We now use the symmetry properties of the integrand under the exchange $\bar w$ with $\bar z$ and factorise the integral as
\be
I_2 ^{>} =  - \frac{1}{2\sqrt{2}g} \bar x _f(\bar u) \bar x _f(\bar v) [ \bar x _f(\bar v)-\bar x _f(\bar u) ] \gimel (\bar u, \bar v) ^2 \, ,
\ee
where
\be
\gimel (\bar u, \bar v)= \int _{|\bar w|\geq 1}\frac{d\bar w}{2\pi} \frac {\sqrt{1+\sqrt{1-\frac{1}{\bar w^2}} } }
{\bar w \left (1-\frac{1}{\bar w ^2} \right )^{\frac{3}{4}}} \frac{1}{(\bar x _f(\bar u)-\bar x(\bar w)) (\bar x _f(\bar v)-\bar x(\bar w)) } \, .
\ee
We change variable of integration from $\bar w$ to $\bar x(\bar w)=y$. We get
\be
\gimel (\bar u, \bar v)= \int _{|y|\geq 1/2} \frac{dy}{2\pi y} \sqrt{\frac{2}{1-\frac{1}{4y ^2}}}
\frac{1}{\bar x _f(\bar u)-y} \frac{1}{\bar x _f(\bar v)-y} \, ,
\ee
which can be exactly computed by means of (\ref {integ2bis}):
\be
\gimel (\bar u, \bar v)=\frac{\sqrt{2}}{\bar x _f(\bar u)-\bar x _f(\bar v)} \left [ \frac{1}{\sqrt{1-4\bar x _f(\bar u)^2}}- \frac{1}{\sqrt{1-4\bar x _f(\bar v)^2}} \right] \, .
\ee
Therefore, we obtain
\be
I_2^{>}= - \frac{1}{\sqrt{2}g} \frac{\bar x _f(\bar u) \bar x _f(\bar v)}{\bar x _f(\bar v)-\bar x _f(\bar u)} \left [ \frac{1}{\sqrt{1-4\bar x _f(\bar v)^2}} - \frac{1}{\sqrt{1-4\bar x _f(\bar u)^2}} \right ]^2 \, . \label {I2>final}
\ee
Adding (\ref {Irestfinal}, \ref {I2>final}) we arrive at the final formula
\ba
S^{(ff)}(u,v)&=& \textrm{exp} \Bigl \{-\frac{2i\sqrt{2}}{g} (\bar x_f (\bar u) - \bar x_f (\bar v) ) \frac{1+4 \bar x_f (\bar u) \bar x_f (\bar v)}{1-4 \bar x_f (\bar u) \bar x_f (\bar v)} \frac{\bar x_f (\bar u) \bar x_f (\bar v)}{(1-4 \bar x_f (\bar u) ^2 ) (1-4 \bar x_f (\bar v) ^2 )}  - \nonumber \\
&-& \frac{i}{\sqrt{2}g} \frac{\bar x _f(\bar u) \bar x _f(\bar v)}{\bar x _f(\bar v)-\bar x _f(\bar u)} \left [ \frac{1}{\sqrt{1-4\bar x _f(\bar v)^2}} - \frac{1}{\sqrt{1-4\bar x _f(\bar u)^2}} \right ]^2 +O(1/g^2) \Bigr \} \, .
\label {Sff-final}
\ea

{\bf Fermions in the giant hole regime}

\medskip

In the giant hole regime which fits into the large fermion case the fermion rapidity scales as $x_F(u)= \sqrt{2}g \bar x (\bar u)$, $u=\sqrt{2}g \bar u$, with
\be
\bar x (\bar u)= \frac{\bar u}{2}\left [ 1+\sqrt{1-\frac{1}{\bar u^2}} \right ]\, , \quad |\bar u| \geq 1 \, , \quad |\bar x (\bar u)| \geq \frac{1}{2} \, ,
\ee
where $\bar x (\bar u)$ was already defined in the second of (\ref {xpmlimit2}).
Referring then to formula (\ref {SFF}) for large fermions,
we first show that if $w=\bar w \sqrt{2}g$, $u=\bar u \sqrt{2} g$ and $g\rightarrow +\infty$, with $|\bar w|>1$, then
\be
\chi _F(w,u)+\Phi (w) = \pi \textrm{sgn} (w-u) - \pi \textrm{sgn} (w) + O(1/g) \label {chiFlim} \, .
\ee
In order to prove (\ref {chiFlim}), it is convenient to start from (\ref {chiF2}) and then use (\ref {xpmlimit2}).
Therefore, the situation is completely analogous to the gluon case: relation (\ref  {chiFlim}) implies that in (\ref {SFF}) the dominant contribution comes from
integrations in the second term in the region $|\bar w|>1$, $|\bar x|>1$, where the scalar-scalar factor $\Theta $ is proportional to $g$. The final result is
\be
i\ln S^{(FF)}(u,v)=\Theta (u,v)+O(g^0) =i\ln (-S^{(ss)}(u,v))+O(g^0) = i\ln (-S^{(gg)}(u,v))+O(g^0)
\label {SFFgh} \, .
\ee

{\bf Fermions in the non-scaling regime}

\medskip

We send $g\rightarrow +\infty$ by keeping fixed all the rapidities. For fermions, rapidities are the variables $x$: therefore
if we keep $x$ fixed, we are necessarily in the small fermion case, i.e. $|x_f|<g/\sqrt {2}$. We can then show that
\be\label{chiH-non-perturb}
\lim _{g\rightarrow +\infty} \chi _H(u,v)=x_f(v) \left [ -\frac {2\sqrt {2}}{g}+\frac {1}{g^2} +O(1/g^3) \right ] \, .
\ee
This means that in this regime $i\ln S^{(ff)}(u,v)= O(1/g^2) $.

\subsection{Mixed factors}

{\bf Scalar-gluon}

\medskip

$\bullet $ Perturbative regime

\medskip

We start from the exact expression (\ref {Sgs})
\be
i\ln [S^{(sg)}(u,v)]=\chi (u,v|1) +\Phi (u) - \int \frac{dw}{2\pi} \frac{d\Theta}{dw}(u,w) [\chi (w,v|1)+\Phi (w)] \, , \ee
where both the scalar and the gluon are in the perturbative regime. This means that the $u$ rapidity is parametrised as
(\ref {uz}) and the $v$ rapidity is scaled as $v=\sqrt{2}g \bar v$. In these hypothesis we have that
\be
\chi (u,v|1) +\Phi (u)= \frac{1}{\sqrt{2}g} \frac{1}{\bar v- \textrm{sgn}(u)} \, .
\ee
In addition, in first approximation, we can integrate in the region $|w|<\frac{2}{\pi}\ln m$ and use for $\Theta $ the expression (\ref {theta-non-perturb}) in which rapidities are parametrised as (\ref {uz}).
The final result is
\be
i\ln [S^{(sg)}(u,v)]= \frac{1}{2\pi \sqrt{2} g} \frac{1}{\bar v - \textrm{sgn} (u)} \left [\pi-\Theta _{np} \left (\frac{2}{\pi} \ln z \right ) \right ] \, , \label {Sgs-pert}
\ee
where $|u|=\frac{2}{\pi}\ln \frac{z}{m(g)}$, $m(g)<z<1$ being the scalar rapidity in the perturbative regime.

\medskip

$\bullet $ Giant hole regime

\medskip

In the giant hole regime we use formula (\ref {chiphilea}) for the limiting expression of $\chi (v,u|1)+\Phi (v)$ when
both $|\bar v|$ and $|\bar u|$ are greater than one. Then the leading (i.e. $O(g)$) contribution to $i\ln S ^{(sg)}(u,v)$ comes from integration in the second term of (\ref {Sgs}) in the region $|\bar w|>1$. This integration is easily done and the result is
\be
i\ln S ^{(sg)}(u,v)= i\ln S^{(s\bar g)}(u,v)=\Theta (u,v) + O(g^0) \, . \label {Sgsgh}
\ee

\medskip

$\bullet $ Non-scaling regime

\medskip

In order to compute the scattering phase $S^{(sg)}(u,v)$ in the non-scaling regime, we have to plug  the expressions (\ref{der2theta-per}) and (\ref{chiphinp}) into (\ref {Sgs}). Since $\chi(u,v|1)+\Phi(u)$ is of order $O(1/g^2)$, we claim that
\be
i\ln S^{(sg)}(u,v)=[\chi(u,v|1)+\Phi(u)]-\int \frac{dw}{2\pi}\,\frac{d}{dw}\Theta(u,w)\,[\chi(w,v|1)+\Phi(w)]
=O\left(\frac{1}{g^2}\right) \ .
\ee

{\bf Gluons-fermions}

\medskip

$\bullet$ Perturbative regime

\medskip

We study the scattering factor between gluons with rapidity $u$ and (small) fermions with rapidity $x_f(v)$ in perturbative regime of the strong coupling limit, i.e. $u=\sqrt{2}g \bar u$, with $|\bar u|\leq 1$ and $x_f(v)=\sqrt{2}g \bar x_f(\bar v)$, $v=\sqrt{2}g \bar v$,
with $|\bar v| \geq 1$, $|\bar x_f(\bar v)| \leq 1/2$.

We start from
\be
i\ln \left (- S^{(gf)}(u,v) \right ) = I_1+I_2+I_3 \, ,
\ee
where
\be
I_1=2 \arctan 2(u-v) \, ,
\ee
\be
I_2=\int _{-\infty}^{+\infty} \frac{dw}{2\pi} [\chi  (w,u|1)+\Phi (w)] \frac{d}{dw} \chi _H (w,v) \, ,
\ee
\be
I_3=-  \int \frac{dw}{2\pi}\frac{dz}{2\pi} [\chi  (w,u|1)+\Phi (w)] \frac{d^2}{dwdz}\Theta (w,z) \chi _H(z,v) \, .
\ee
In the perturbative regime
\be
I_1 = \pi \textrm{sgn} (\bar u -\bar v) - \frac{1}{\sqrt{2}g (\bar u -\bar v)} + O(1/g^2) \, ,
\ee
\be
I_2=\int _{|\bar w|<1}\frac{dw}{2\pi} [\chi  (w,u|1)+\Phi (w)] \frac{d}{dw} \chi _H (w,v) +O(1/g^2) \, ,
\ee
\ba
I_3&=&\int _{|\bar w|, |\bar z| \geq 1}\frac{d\bar w}{2\pi}\frac{d\bar z}{2\pi} \frac{1}{\bar u -\bar w}
\frac{\left (\frac{\bar w +1}{\bar w-1} \right )^{\frac{1}{4}} \left (\frac{\bar z -1}{\bar z+1} \right )^{\frac{1}{4}} + \left (\frac{\bar w -1}{\bar w+1} \right )^{\frac{1}{4}} \left (\frac{\bar z +1}{\bar z-1} \right )^{\frac{1}{4}}}
{\bar w-\bar z} \frac{\bar x_f(\bar v)}{\sqrt{2}g \bar z \sqrt{1-\frac{1}{\bar z ^2}}} \frac{1}{\bar x_f(\bar v)-
\bar x(\bar z)} - \nonumber \\
&-& \int _{|\bar w|<1}\frac{dw}{2\pi} [\chi  (w,u|1)+\Phi (w)] \frac{d}{dw} \chi _H (w,v) +O(1/g^2) \, .
\ea
We evaluate the sum $I_2+I_3$ by first performing integration in $\bar w$ with the help of (\ref  {integ3}), getting
\be
I_2+I_3=-\frac{\bar x_f(\bar v)}{2g}\int _{|\bar z|\geq 1}\frac{d\bar z}{2\pi} \frac{1}{\bar z \sqrt{1-\frac{1}{\bar z ^2}}}
\frac{1}{\bar x_f(\bar v)-\bar x(\bar z)} \frac{\left (\frac{1+\bar u}{1-\bar u} \right )^{\frac{1}{4}} \left (\frac{\bar z -1}{\bar z+1} \right )^{\frac{1}{4}} - \left (\frac{1-\bar u}{1+\bar u} \right )^{\frac{1}{4}} \left (\frac{\bar z +1}{\bar z-1} \right )^{\frac{1}{4}}}{\bar u-\bar z}  +O(1/g^2) \, .
\ee
Then, we integrate in $\bar z$, using (\ref  {integ4}). We obtain
\be
I_2+I_3=\frac{1}{4g} \frac{\sqrt{\frac{1-2\bar x_f(\bar v)}{1+2\bar x_f(\bar v)}} \left (\frac{1+\bar u}{1-\bar u} \right )^{\frac{1}{4}} + \sqrt{\frac{1+2\bar x_f(\bar v)}{1-2\bar x_f(\bar v)}} \left (\frac{1-\bar u}{1+\bar u} \right )^{\frac{1}{4}} - \sqrt{2}}{\bar v -\bar u}+O(1/g^2) \, .
\ee
Summing up $I_1+I_2+I_3$ we get the final result
\be
S^{(gf)} (u,v)=\textrm{exp} \left [ \frac{i}{4g} \frac{\sqrt{2}+ \sqrt{\frac{1-2\bar x _f}{1+2\bar x _f}} \left (\frac{1+\bar u}{1-\bar u} \right )^{\frac{1}{4}} + \sqrt{\frac{1+2\bar x _f}{1-2\bar x _f}} \left (\frac{1-\bar u}{1+\bar u} \right )^{\frac{1}{4}}}{\bar u -\bar v} +O(1/g^2) \right ] \, .
\ee
For what concerns $i\ln S^{(\bar g f)}(u,v)=I_2+I_3$, we have
\be
S^{(\bar gf)} (u,v)=\textrm{exp} \left [ \frac{i}{4g} \frac{ \sqrt{\frac{1-2\bar x _f}{1+2\bar x _f}} \left (\frac{1+\bar u}{1-\bar u} \right )^{\frac{1}{4}} + \sqrt{\frac{1+2\bar x _f}{1-2\bar x _f}} \left (\frac{1-\bar u}{1+\bar u} \right )^{\frac{1}{4}}-\sqrt{2}}{\bar u -\bar v} +O(1/g^2) \right ] \, .
\ee

\medskip

$\bullet$ Giant hole regime

\medskip

Since both $\chi (v,u|1)+\Phi (v)$ and $\chi _F (v,u)+\Phi (v)$ have the same limiting nonzero expression (\ref {chiphilea}) when $|\bar v|>1$, $|\bar u|>1$, the leading expressions for $i\ln (-S ^{(gF)}(u,v))$ and
$i\ln (S ^{(\bar gF)}(u,v))$ coincide with the one for $i\ln S ^{(FF)}(u,v)$. Therefore,
\be
i\ln (-S ^{(gF)}(u,v))= i\ln (S ^{(\bar gF)}(u,v))+O(g^0)=\Theta (u,v) +O(g^0) \, . \label {SgFgh}
\ee

\medskip

$\bullet$ Non-scaling regime

\medskip

As written before, in this regime fermions are necessarily small. Then,
since $\chi (w,u|1)+\Phi (w)$ is $O(1/g^2)$ and $\chi _H(w,v)$ is $O(1/g)$, integrals $I_2, I_3$ are both $O(1/g^3)$.
For what concerns $I_1$, since fermionic rapidities $u_{f,k}$ are bounded by the inequality $u_{f,k}^2>2g^2$, we can safely approximate $-e^{iI_1}=1+O(1/g)$. Therefore, in the non-scaling regime
$S^{(gf)}(u,v)=S^{(\bar g \bar f)}(u,v)=1+O(1/g)$ and $S^{(\bar gf)}(u,v)=S^{(g \bar f)}(u,v)=1+O(1/g^3)$.

\medskip

{\bf Scalars-fermions}

\medskip

$\bullet$ Perturbative regime

\medskip

We start from the exact expression (\ref {SsF}),
\be
-i\ln  S ^{(sf)}(u,v)=\chi _H (u,v) - \int _{-\infty}^{+\infty} \frac{d\bar w}{2\pi} \frac{d\Theta}{d\bar w}(u,w) \chi _H(w,v) \, ,
\ee
and make the parametrisations (\ref {uz}) for $u$ and $v=\sqrt{2}g \bar v$. At leading order we have
\be
\chi _H (u,v)= 2^{\frac{3}{4}}\sqrt{\frac{\ln \frac{g}{\sqrt{2}}}{\pi g}} \, \frac{1}{\bar v-1 + \sqrt{\bar v^2-1}} \, .
\ee
Then, as for the scalar-gluon case, in first approximation, we can integrate in the region $|w|<\frac{2}{\pi}\ln m (g)$ and use for $\Theta $ the expression (\ref {theta-non-perturb}) in which rapidities are parametrised as (\ref {uz}).
The final result is
\be
-i \ln S ^{(sf)}(u,v)= i \ln S ^{(fs)}(v,u) = 2^{-\frac{1}{4}}\sqrt{\frac{\ln \frac{g}{\sqrt{2}}}{\pi g}} \, \frac{1}{\bar v-1 + \sqrt{\bar v^2-1}}
\left [ 1- \frac{1}{\pi} \Theta _{np} \left ( \frac{2}{\pi}\ln z \right ) \right ] \label {Ssfper} \, ,
\ee
with $|u|=\frac{2}{\pi}\ln \frac{z}{m(g)}$.
\medskip

$\bullet$ Giant hole regime

\medskip

Since $\chi _F(v,u)+\Phi (v)$ and $\chi (v,u|1)+\Phi (v)$ go to the same limit, i.e. $\pi \textrm{sgn} (v-u) -\pi
\textrm{sgn} (v)$, by comparing (\ref {SsF}) and (\ref {Sgs}) we get the equality $i\ln S^{(sF)}(u,v)=-i\ln S^{(gs)}(v,u)+O(g^0)$, which, together with (\ref {Sgsgh}), gives
\be
i \ln S ^{(sF)}(u,v)=\Theta (u,v) + O(g^0) \label {SsFgh} \, .
\ee

\medskip

$\bullet$ Non-scaling regime

\medskip

We perform the non-perturbative limit of the scalar-fermion scattering phase
\be\label{Ssferm_s}
i\ln S^{(sf)}(u,v)=-\chi_H(u,v)+\int_{-\infty}^\infty \frac{dw}{2\pi}\,\frac{d}{dw}\Theta(u,w)\,\chi_H(w,v) \, ,
\ee
by taking $g\longrightarrow\infty$ while keeping the scalar rapidities finite, whereas the modulus of the fermionic rapidities $x_f$ must be $|x_f|<g/\sqrt{2}$. Under these assumptions, we can make use of the approximations (\ref{chiH-non-perturb}) and (\ref{theta-non-perturb}) for $\chi_H(u,v)$ and $\Theta(u,v)$; eventually, we find:
\be
i\ln S^{(sf)}(u,v)=O\left(\frac{1}{g^3}\right) \ .
\ee

\medskip

\subsection{Remark on the non-scaling regime}

We showed that in the non-scaling regime all the factors $S^{\ast \ast '}(u,v)$ go as $1+O(1/g^2)$,
with the exception of the scalar-scalar one which goes as
\be
S^{(ss)}(u,v)=- \frac{\Gamma \left (1+i\frac{u-v}{4}\right ) \Gamma \left (\frac{1}{2}-i\frac{u-v}{4}\right )}
{\Gamma \left (1-i\frac{u-v}{4}\right ) \Gamma \left (\frac{1}{2}+i\frac{u-v}{4}\right )} \textrm{exp} \left [i \textrm{gd} \frac{\pi}{2}
(u-v) \right ] [1+O(1/g)] \label {S06} \, .
\ee
In addition to that, we recall that the fermionic rapidities $u_{f,k}$ satisfy the inequalities
$u_{f,k}^2>2g^2$. Therefore in this regime all the rational factors involving fermionic rapidities
(which appear in the quantisation conditions for fermions and in the equations for isotopic roots $u_a$ and $u_c$) go to one. In addition all the exponentials of momenta and defect ($=e^i(P+D)$) go to $1$, with the exception of those for scalars. Summarising, in the limit $g\rightarrow +\infty$ with rapidities fixed (and finite), the non-trivial equations are
\ba
&& 1=e^{iRP^{(s)}(u_h)+iD^{(s)}(u_h)}
 \prod _{j=1}^{K_b} \frac{u_h-u_{b,j}+\frac{i}{2}}{u_h-u_{b,j}-\frac{i}{2}}  \prod _{\stackrel {h'=1}{h'\not=h}}^{H} S^{(ss)}(u_h,u_{h'}) \, , \label {Bethe06} \\
&&1 = \prod _{j\not=k}^{K_a} \frac{u_{a,k}-u_{a,j}+i}{u_{a,k}-u_{a,j}-i} \prod _{j=1}^{K_b}
\frac{u_{a,k}-u_{b,j}-i/2}{u_{a,k}-u_{b,j}+i/2} \, , \nonumber \\
&& \prod _{h=1}^{H} \frac{u_{b,k}-u_h+i/2}{u_{b,k}-u_h-i/2}=\prod _{j=1}^{K_a} \frac{u_{b,k}-u_{a,j}-i/2}{u_{b,k}-u_{a,j}+i/2} \prod _{j=1}^{K_c} \frac{u_{b,k}-u_{c,j}-i/2}{u_{b,k}-u_{c,j}+i/2} \prod _{\stackrel {j=1}{j\not= k}}^{K_b} \frac{u_{b,k}-u_{b,j}+i}{u_{b,k}-u_{b,j}-i}
 \nonumber \, , \\
&& 1=\prod _{j\not=k}^{K_c} \frac{u_{c,k}-u_{c,j}+i}{u_{c,k}-u_{c,j}-i}   \prod _{j=1}^{K_b}
\frac{u_{c,k}-u_{b,j}-i/2}{u_{c,k}-u_{b,j}+i/2} \, , \nonumber \\
&& 1=\prod _{j\not=k}^{N_g} \frac{u^{g}_{k}-u^{g}_{j}+i}{u^{g}_{k}-u^{g}_{j}-i} \, , \nonumber \\
&& 1=\prod _{j\not=k}^{N_{\bar g}} \frac{u^{\bar g}_{k}-u^{\bar g}_{j}+i}{u^{\bar g}_{k}-u^{\bar g}_{j}-i} \nonumber  \, ,
\ea
with $S^{(ss)}$ given by (\ref {S06}). Since equations for gluons have no solutions for finite rapidities (i.e. $N_g=N_{\bar g}=0$), equations (\ref {Bethe06}) show that in the non-perturbative regime the only active excitations are the six scalars. The other excitations are obliged to assume infinite rapidities and thus decouple to very high energy from the scalars. The latter satisfy the above ABA (\ref{Bethe06}) which is the same we can derive from the $O(6)$ non-linear sigma model $S$-matrix of \cite{Zam^2}. Therefore, also the exact TBA would be that of the $O(6)$ model (if we can neglect the exchange of the $g\rightarrow +\infty$ limit with the thermodynamics).

\section{Particle momentum in different forms} \label{sez5}
\setcounter{equation}{0}

Momentum was already thoroughly discussed by Basso in \cite {Basso}. The aim of this section is to show that
the expressions for momenta of the various excitations we found (in our notations) in previous sections agree
with corresponding formul{\ae} of \cite {Basso}.

$\bullet$ Scalars

We found (\ref {scal-mom}) that the momentum of a scalar excitation is
\be
P^{(s)}(u)=2u-\frac{1}{2}Z_{BES}(u) \label {scal-imp} \, .
\ee
Now, using the mapping (\ref {BES-BAS}), valid for $k>0$,
\be
\frac{i\sinh \frac{k}{2}}{\pi}\hat Z_{BES} (k)= \frac{\gamma _+^{\textrm{\o}}(\sqrt{2}gk) + \gamma _-^{\textrm{\o}}(\sqrt{2}gk)}{k} \, ,
\ee
between our quantities and quantities used in \cite {Basso}, we immediately write (\ref {scal-imp}) in the form
reported in \cite {Basso} (second of the (4.6)).

\medskip

$\bullet$ Gluons

For a gluon with rapidity $u$ we found for the momentum the expression (\ref {P-glu}), which we can write in Fourier space as
\ba
&& P^{(g)}(u) = \int _{-\infty} ^{+\infty}\frac{dk}{4\pi^2} \Bigl [ -2\pi \frac{\sin ku}{k} e^{-|k|} + i \sum _{n=1}
\left (\frac{g}{i\sqrt{2}x^-(u)} \right )^n \frac{2\pi}{k} e^{-\frac{|k|}{2}} J_n(\sqrt{2}gk) + \nonumber \\
&& \ \ \ +i \sum _{n=1}
\left (\frac{g}{i\sqrt{2}x^+(u)} \right )^n \frac{2\pi}{k} e^{-\frac{|k|}{2}} J_n(\sqrt{2}gk) \Bigr ] (-4\pi \delta (k) + \frac{1}{2} \hat \sigma _{BES} (k) ) = \nonumber \\
&& = 2u - \left ( \frac{g^2}{x^-} + \frac{g^2}{x^+} \right ) +\int _{-\infty} ^{+\infty} \frac{dk}{8\pi^2} \Bigl [ -2\pi \frac{\sin ku}{k} e^{-|k|} +  i \sum _{\stackrel {n=1} {n \ \ \textrm{odd}}}^{+\infty}
\left (\frac{g}{i\sqrt{2}x^-(u)} \right )^n \frac{2\pi}{k} e^{-\frac{|k|}{2}} J_n(\sqrt{2}gk) + \nonumber \\
&& \ \ \ +i \sum  _{\stackrel {n=1} {n \ \ \textrm{odd}}}^{+\infty}
\left (\frac{g}{i\sqrt{2}x^+(u)} \right )^n \frac{2\pi}{k} e^{-\frac{|k|}{2}} J_n(\sqrt{2}gk) \Bigr ] \nonumber
\hat \sigma _{BES} (k) \, .
\ea
Now, we use the equalities
\be
\int _{-\infty} ^{+\infty} \frac{dk}{k} e^{-\frac{|k|}{2}} J_n(\sqrt{2}gk) \hat \sigma _{BES} (k)= 4\pi [ \sqrt{2}g \delta _{n,1}-
\gamma ^{\textrm{\o}} _n ] \, , \quad n \ \textrm{odd} , ,
\ee
and
\be
\sum _{\stackrel {n=1} {n \ \ \textrm{odd}}}^{\infty} \left [ \left ( \frac{g}{i\sqrt{2}x^-} \right )^n +
\left ( \frac{g}{i\sqrt{2}x^+} \right )^n \right ] \gamma _n^{\textrm{{\o}}}=-i
\int _{0}^{+\infty} \frac{dk}{k} \sin ku \ e ^{-\frac{k}{2}} \ \gamma _- ^{\textrm{{\o}}} (\sqrt{2}gk) \, ,
\ee
to eventually obtain
\be
P^{(g)}(u)=2u - \int _{0}^{+\infty} \frac{dk}{k} \sin ku \ e ^{-\frac{k}{2}} \left [ \frac{\gamma _-^{\textrm{{\o}}} (\sqrt{2}gk)}{1-e^{-k}} +  \frac{\gamma _+^{\textrm{{\o}}} (\sqrt{2}gk)}{e^{k}-1} \right ] \, ,
\ee
which agrees with the second of (4.9) of \cite {Basso}.

$\bullet$ Large fermions

The momentum associated to a large fermion with rapidity $u$ enjoys the expression (\ref {P-fer}). In Fourier space it reads
\ba
P^{(F)}(u) = \int _{-\infty} ^{\infty} \frac{dk}{2\pi} \left[ \frac{\sin ku}{k} e^{-\frac{|k|}{2}} +
 \sum _{n=1}
\left (\frac{g}{i\sqrt{2}x(u)} \right )^n \frac{e^{-\frac{|k|}{2}}}{ik}  J_n(\sqrt{2}gk) \right] (4\pi \delta (k) - \frac{ \hat \sigma _{BES} (k)}{2} )= \nonumber \\
= 2u - \int _{0}^{+\infty} \frac{dk}{k} \sin ku \  \frac{\gamma _+^{\textrm{{\o}}} (\sqrt{2}gk)+\gamma _-^{\textrm{{\o}}} (\sqrt{2}gk)}{e^{k}-1}-\frac{1}{2}\int_0^{+\infty} \frac{dk}{k}\, \sin ku\, \gamma _-^{\textrm{{\o}}} (\sqrt{2}gk) \quad , \nn
\ea
which recalls the second of (4.10) of \cite {Basso}.
In order to obtain the equation in the last line, we made use of the relation
\be
\int_0^{\infty} \frac{dk}{2k}\, \sin ku\, \gamma _-^{\textrm{{\o}}} (\sqrt{2}gk)=
i\sum_{n=1}^{\infty}\left(\frac{g}{\sqrt{2}ix(u)}\right)^{2n-1} \gamma _{2n-1}^{\textrm{{\o}}} \qquad ,
\ee
which holds for $u^2>2g^2$.

$\bullet$ Small fermions

The reasonings for the momentum of a small fermion with rapidity $u$ mimic very closely the large fermion case.
We start from our expression (\ref {P-smfer}) and in Fourier space we eventually get the result ($u^2>2g^2$)
\be
P^{(f)}(u)=
\frac{1}{2}\int_0^{+\infty} \frac{dk}{k}\, \sin (ku)\, \gamma _-^{\textrm{{\o}}} (\sqrt{2}gk) \, ,
\ee
therefore matching the second of (4.12) of \cite {Basso}.

\section{Strong coupling analysis of the defect term}\label{sez6}
\setcounter{equation}{0}

We now perform a quantitative analysis of the strong coupling limit of the defect which appears in the Bethe equations on the GKP vacuum.

\subsection{Scalars}

It is convenient to concentrate on the function $Z_4$ (\ref {count-scal-all}) in absence of excitations, which equals
\be
Z_4(u) |_{NE}=- 2\ln s P^{(s)}(u)-2 D^{(s)}(u) \label {count-scal-noexc} \, .
\ee
The study of this function, which relies also on previous results, provides information on both the momentum and the defect of the scalar.

\medskip

\textbf{Non-perturbative regime}

\medskip

In this regime we send $g\rightarrow +\infty$, keeping the rapidity $u$ fixed.
We can use results from \cite {FRO6} where the non perturbative regime for the pure $sl(2)$ sector is studied.
We found that the function (\ref {count-scal-noexc}) has the form
\be
Z_4(u)|_{NE} =-2 m(g) \ln \frac{2\sqrt{2}s}{g} \ \sinh \frac{\pi}{2} u + O(m(g)^3) \, ,
\ee
where $m(g)$ is given by (\ref {m(g)}).
Therefore, the contribution of the two defect is proportional to the momentum
\be
D^{(s)}(u)=  m(g) \ln \frac{2\sqrt{2}}{g} \ \sinh \frac{\pi}{2} u + O(m(g)^3) \, , \label {def-np}
\ee
so allowing us to fully re-absorb them into a re-definition of the size $R(g)$ as in \cite{BB, FRO6}

\medskip

\textbf{Perturbative regime}

\medskip

We obtain interesting formul{\ae} in this regime by plugging (\ref {uz}) in (\ref {def-np}), namely:
\be
D^{(s)}(u)= \frac{z}{2} \ln \frac{2\sqrt{2}}{g}  + \ldots \, ,
\ee
for right movers ($u>0$), while
\be
D^{(s)}(u)= - \frac{z}{2}\ln \frac{2\sqrt{2}}{g}  + \ldots \, ,
\ee
for left movers ($u<0$), where evidently the dots $\ldots$ mean sub-leading corrections.

\medskip

\textbf{Scaling regimes}

\medskip

We introduce the density $\sigma _4(u)|_{NE}=\frac{d}{du} Z_4(u)|_{NE}$ and rescale the rapidity $u=\sqrt{2}g \bar u$.
By using techniques developed in \cite{BKK}, we eventually find
\ba
\sigma _4(\sqrt{2}g \bar u)|_{NE}&=&\int _{0}^{+\infty} \frac{d\bar h}{\sqrt{2}g} \cos \bar h \bar u \left [ - \frac{e^{-\frac{\bar h}{2\sqrt{2}g}}}{\cosh \frac{\bar h}{\sqrt{2}g}} \Gamma _+ (\bar h) + \frac{e^{\frac{\bar h}{2\sqrt{2}g}}}{\cosh \frac{\bar h}{\sqrt{2}g}} \Gamma _- (\bar h) \right ] -4\ln s +
\nonumber \\
&+& \frac{2\pi}{\cosh \pi \sqrt{2}g \bar u} +
2 \psi \left ( \frac{1}{2}+i \bar u \sqrt{2}g \right ) + 2 \psi \left ( \frac{1}{2}-i \bar u \sqrt{2}g \right )+ \\
&+& \psi \left ( \frac{5}{8}+ \frac{i\sqrt{2}g\bar u}{4} \right ) + \psi \left ( \frac{5}{8}- \frac{i\sqrt{2}g\bar u}{4} \right )
- \psi \left ( \frac{7}{8}+ \frac{i\sqrt{2}g\bar u}{4} \right )- \psi \left ( \frac{7}{8}- \frac{i\sqrt{2}g\bar u}{4} \right ) \nonumber \, ,
\ea
where the functions $\Gamma _{\pm}$ satisfy the relations, valid when $|\bar u|<1$:
\ba
&& \int _{0}^{+\infty} \frac{d\bar h}{\sqrt{2}g} \sin \bar h \bar u [ \Gamma _-(\bar h)+\Gamma _+(\bar h) ]= -2i [ \psi (1-i\bar u \sqrt{2}g) -
\psi (1+i\bar u \sqrt{2}g)] \\
&& \int _{0}^{+\infty} \frac{d\bar h}{\sqrt{2}g} \cos \bar h \bar u [ \Gamma _-(\bar h)-\Gamma _+(\bar h) ]= 4\ln s - 2[ \psi (1-i\bar u \sqrt{2}g) + \psi (1+i\bar u \sqrt{2}g)] \, .
\ea
Going to the strong coupling limit $g\rightarrow +\infty$, with $\bar u$ fixed, we find
\be
\sigma _4(\sqrt{2}g \bar u)|_{NE}=\int _{0}^{+\infty} \frac{d\bar h}{\sqrt{2}g} \cos \bar h \bar u \left [ - \frac{e^{-\frac{\bar h}{2\sqrt{2}g}}}{\cosh \frac{\bar h}{\sqrt{2}g}} \Gamma _+ (\bar h) + \frac{e^{\frac{\bar h}{2\sqrt{2}g}}}{\cosh \frac{\bar h}{\sqrt{2}g}} \Gamma _- (\bar h) \right ] -4\ln \frac{s}{g} + O(g^0) \, ,
\ee
where
\ba
&& \int _{0}^{+\infty} \frac{d\bar h}{\sqrt{2}g} \sin \bar h \bar u [ \Gamma _-(\bar h)+\Gamma _+(\bar h) ]= O(g^0) \, , \quad |\bar u|<1 \, , \\
&& \int _{0}^{+\infty} \frac{d\bar h}{\sqrt{2}g} \cos \bar h \bar u [ \Gamma _-(\bar h)-\Gamma _+(\bar h) ]= 4\ln \frac{s}{g} + O(g^0) \, , \quad |\bar u|<1 \, .
\ea
Solutions to these equations go differently according to the value of $|\bar u|$.
If $|\bar u|<1$ we have
\be
 \sigma _4(u)|_{NE}=- \frac{\sqrt{2}\pi}{g} \delta (\bar u) + O(1/g^2) \, , \label {sigma-scal}
\ee
which means that $P^{(s)}(u)$ is exponentially small and that $D^{(s)}(u)= - \frac{\sqrt{2}\pi}{g} \delta (\bar u) + O(1/g^2)$.

On the other hand, if $|\bar u|>1$, we have
\be
 \sigma _4(u)|_{NE}=-2 \ln \frac{s}{g} \frac{d}{du} P^{(s)}(u)+ O(g^0) \label {sigma-scal-2} \, ,
\ee
with $P^{(s)}(u) = O(g)$
or, alternatively, the proportionality to the momentum
\be
D^{(s)}(u)=- P^{(s)}(u) \ln g + O(g) \, ,
\ee
which would allow us to re-absorb, at this order (only), fully the defect into a simple redefinition of the size $R$. Using now (5.27) of \cite {Basso} we can then express $D^{(s)}(u)$ in terms of the rapidity
$\bar x (\bar u)=\frac{1}{2}\bar u + \frac{1}{2}\bar u \sqrt{1-\frac{1}{\bar u^2}}$ (defined in the second of (\ref {xpmlimit2})) as
\be
D^{(s)}(u)=- \sqrt{2} g \ln g \left [ 2\bar x (\bar u) \sqrt{1-\frac{1}{4\bar x ^2 (\bar u)}}-\arctan \left ( 2\bar x (\bar u) \sqrt{1-\frac{1}{4 \bar x ^2 (\bar u) }}\right ) \right ]
 +O(g) \, .  \label {s-gh}
\ee

\subsection{Gluons}

\textbf{Perturbative and giant hole regimes}

\medskip

We start from the formula
\be
2\ln s \ P^{(g)}(u)+2 D^{(g)}(u)=\int \frac{dv}{2\pi} [\chi (v,u|1)+\chi (-v,u|1)]\, \frac{1}{2}\sigma _4(v)|_{NE} \, ,
\ee
Scaling $u=\sqrt{2}g \bar u$ and $v=\sqrt{2}g \bar v$, we have that
$\chi (v,u|1)+\Phi (v)$ is $O(1/g)$ if $|\bar u|<1$ and (at most) $O(g^0)$ if $|\bar u|>1$.
Referring to (\ref {sigma-scal}, \ref {sigma-scal-2}), we remark that the integration receives leading contribution from the region $|\bar v|>1$. We conclude that
\ba
&&|\bar u|<1 \quad \Rightarrow \quad 2\ln s P^{(g)}(u)+2 D^{(g)}(u)=2\ln \frac{s}{g} \ P^{(g)}(u)+ O(g^0)\, , \\
&&|\bar u|>1 \quad \Rightarrow \quad 2\ln s P^{(g)}(u)+2 D^{(g)}(u)=2\ln \frac{s}{g} \ P^{(g)}(u)+ O(g) \, .
\ea
We can now refer to formul{\ae} of \cite {Basso} for the momentum of the gauge field and arrive at the final expressions
\be
|\bar u|<1 \quad \Rightarrow \quad D^{(g)}(u)=-\ln g P^{(g)}(u)+O(g^0)= -\frac{\ln g }{\sqrt{2}} \left [ \left ( \frac{1+\bar u}{1-\bar u} \right ) ^{\frac{1}{4}} - \left ( \frac{1-\bar u}{1+\bar u} \right ) ^{\frac{1}{4}} \right ]+ O(g^0) \, ,
\ee
\ba
&& |\bar u|>1 \quad \Rightarrow \quad
D^{(g)}(u)=-\ln g P^{(g)}(u)+O(g)= \nonumber \\
&& - \sqrt{2} g \ln g \left [ 2\bar x (\bar u) \sqrt{1-\frac{1}{4\bar x ^2 (\bar u)}}-\arctan \left ( 2\bar x (\bar u) \sqrt{1-\frac{1}{4 \bar x ^2 (\bar u)}}\right ) \right ] +O(g) \, ,  \label {g-gh}
\ea
$\bar x(\bar u)$ being defined in the second of (\ref {xpmlimit2}).

\subsection{Fermions}

\textbf{Perturbative regime}

\medskip

In this regime the scaled rapidity needs to satisfy $|\bar x_f (\bar u)|<1/2$, so that it belongs to a small fermion. Therefore, we start from the formula
\be
2\ln s \ P^{(f)}(u)+2 D^{(f)}(u)=-\int \frac{dv}{2\pi} [\chi _H(v,u)+\chi _H(-v,u)]\, \frac{1}{2}\frac{d}{dv}Z_4(v)|_{NE} \, .
\ee
Scaling $v=\sqrt{2}g \bar v$,
we have that $\chi _H(v,u)$ is $O(1/g)$ if $|\bar v|>1$ and $O(g^0)$ if $|\bar v|<1$.
Referring to (\ref {sigma-scal}, \ref {sigma-scal-2}), we remark that the integration receives leading contribution from the region $|\bar v|>1$. We conclude that
\be
2\ln s P^{(f)}(u)+2 D^{(f)}(u)=2\ln \frac{s}{g} \ P^{(f)}(u)+ O(g^0)
\ee
and
\be
D^{(f)}(u)=-\ln g P^{(f)}(u)+ O(g^0)=-2 \ln g \frac{\bar x_f (\bar u)}{\sqrt{1-4\bar x_f (\bar u)^2}}+ O(g^0) \, .
\ee

\medskip

\textbf{Giant hole regime}

\medskip

On the contrary, here the scaled rapidity is $|\bar x_F (\bar u)|>1/2$, so to characterise a large fermion. Therefore, we start from the formula
\be
2\ln s \ P^{(F)}(u)+2 D^{(F)}(u)=\int \frac{dv}{2\pi} [\chi _F(v,u)+\chi _F(-v,u)]\, \frac{1}{2}\sigma_4(v)|_{NE}
\, .
\ee
Scaling $v=\sqrt{2}g \bar v$, we now have that $\chi _H(v,u)$ is $O(g^0)$ for all $|\bar v|$.
Referring to (\ref {sigma-scal}, \ref {sigma-scal-2}), we remark that the integration receives leading contribution from the region $|\bar v|>1$. We conclude that
\be
2\ln s P^{(F)}(u)+2 D^{(F)}(u)=2\ln \frac{s}{g} \ P^{(F)}(u)+ O(g)
\ee
and, consequently, that
\ba
&& D^{(F)}(u)=-\ln g P^{(F)}(u)= \nonumber \\
&=& - \sqrt{2} g \ln g \left [ 2\bar x (\bar u) \sqrt{1-\frac{1}{4\bar x ^2 (\bar u) }}-\arctan \left ( 2\bar x (\bar u) \sqrt{1-\frac{1}{4 \bar x ^2 (\bar u)}}\right ) \right ] +O(g) \, .  \label {F-gh}
\ea

\section{The SU(4) symmetry}\label{sez7}
\setcounter{equation}{0}

The particles we are addressing to (scalars, gluons, fermions and anti-fermions) belong to a specific multiplet under the $SU(4)$ symmetry (${\bf 6}$, ${\bf 1}$, ${\bf 4}$ and ${\bf \bar 4}$, respectively). In fact, the scattering matrix possess this symmetry. Starting from the scattering matrices derived in the previous section, the Bethe equations may be assembled for every sort of excitation; anyway, they are actually able to catch only a single state in each multiplet, precisely the one corresponding to the highest weight state of the representation. In this section the focus moves to a few sectors of the complete theory, which include just one type (or two at most) of excitations along with the set of isotopic roots, aiming at elucidating the behaviour of the different kinds of particle under $SU(4)$.

Following \cite{OW}, a set of Bethe equations can be formulated for any spin chain associated to a simple Lie algebra.
Therefore, given the set of simple roots of a simple Lie algebra$\{\alpha_q\}$, and chosen a representation $\rho $ by fixing its highest weight $\vec w_{\rho}$ (or equivalently a tern of positive integer Dynkin labels), the relative Bethe equations arise, with a further generalization stemming from the introduction of a set of inhomogeneities (labelled by their rapidities $u_m$) along the spin chain:
\be\label{Lie_spin_chain}
\prod_{m}
\frac{u_{q,k}-u_m+i\vec\alpha_q\cdot\vec w_{\rho}}{u_{q,k}-u_m-i\vec\alpha_q\cdot\vec w_{\rho}} =
\prod_{j\neq k}^{K_{q}}\frac{u_{q,k}-u_{q,j}+i\vec\alpha_q\cdot\vec\alpha_{q}}{u_{q,k}-u_{q,j}-i\vec\alpha_q\cdot \vec\alpha_{q}}\,
\prod_{q'\neq q}\prod_{j=1}^{K_{q'}}
\frac{u_{q,k}-u_{q',j}+i\vec\alpha_q\cdot\vec\alpha_{q'}}{u_{q,k}-u_{q',j}-i\vec\alpha_q\cdot\vec\alpha_{q'}}\ .
\ee

Turning to the $su(4)$ algebra, we perform a choice
of three simple roots $\vec\alpha_k$, along with the three simple roots $\vec\varphi_k$, resulting from the defining condition $\displaystyle\frac{2\vec\alpha_j\cdot \vec\varphi_k}{(\vec\alpha_j)^2}=\delta_{kj}$\ :
\ba\label{roots}
\vec\alpha_1 &=& \left(\frac{1}{2},\frac{\sqrt{3}}{2},0\right)
\qquad\qquad\qquad\qquad\,
\vec\varphi_1 = \left(\frac{1}{2},\frac{1}{2\sqrt{3}},\frac{1}{2\sqrt{6}}\right)
\nonumber\\
\vec\alpha_2 &=& \left(\frac{1}{2},-\frac{\sqrt{3}}{2},0\right)
\qquad\qquad\qquad\ \quad
\vec\varphi_2 = \left(\frac{1}{2},-\frac{1}{2\sqrt{3}},\frac{1}{\sqrt{6}}\right) \\
\vec\alpha_3 &=& \left(-\frac{1}{2},\frac{1}{2\sqrt{3}},\frac{2}{\sqrt{6}}\right)
\qquad\qquad\qquad
\vec\varphi_3 = \left(0,0,\frac{3}{2\sqrt{6}}\right) \qquad\ \ .\nonumber
\ea
To sum up, the Bethe equations in (\ref{Lie_spin_chain}) specialize to the $su(4)$ algebra:
\ba\label{su(4)BetheEqs}
		\prod_m \frac{u_{a,k}-u_m+i\vec\alpha_1\cdot\vec w_{\rho}}{u_{a,k}-u_m-i\vec\alpha_1\cdot\vec w_{\rho}}
&=& \prod _{j\not=k}^{K_a} \frac{u_{a,k}-u_{a,j}+i}{u_{a,k}-u_{a,j}-i}
\prod _{j=1}^{K_b} \frac{u_{a,k}-u_{b,j}-i/2}{u_{a,k}-u_{b,j}+i/2} \nonumber\\
		\prod_m\frac{u_{b,k}-u_m+i\vec\alpha_2\cdot\vec w_{\rho}}{u_{b,k}-u_m-i\vec\alpha_2\cdot\vec w_{\rho}}
&=& \prod_{j\neq k}^{K_b} \frac{u_{b,k}-u_{b,j}+i}{u_{b,k}-u_{b,j}-i}
\prod _{j=1}^{K_a} \frac{u_{b,k}-u_{a,j}-i/2}{u_{b,k}-u_{a,j}+i/2} \prod_{j=1}^{K_c} \frac{u_{b,k}-u_{c,j}-i/2}{u_{b,k}-u_{c,j}+i/2} \nonumber\\
		\prod_m\frac{u_{c,k}-u_m+i\vec\alpha_3\cdot\vec w_{\rho}}{u_{c,k}-u_m-i\vec\alpha_3\cdot\vec w_{\rho}}
&=& \prod _{j\not=k}^{K_c} \frac{u_{c,k}-u_{c,j}+i}{u_{c,k}-u_{c,j}-i}
\prod _{j=1}^{K_b}\frac{u_{c,k}-u_{b,j}-i/2}{u_{c,k}-u_{b,j}+i/2}
\ea
\medskip\\
\textbf{$\bullet$ Scalar sector:}\\
When considering a system composed only of $H$ scalar excitations with rapidities $\{u_h\}$,
together with $K_a$ roots $u_a$, $K_b$ roots $u_b$ and $K_c$ roots $u_c$, the equations for the isotopic roots (\ref{1lfineq1}, \ref{1lfineq2}, \ref{1lfineq3}) take the form :
\ba\label{scalSector}
1&=& \prod _{j\not=k}^{K_a}\frac{u_{a,k}-u_{a,j}+i}{u_{a,k}-u_{a,j}-i} \prod_{j=1}^{K_b}\frac{u_{a,k}-u_{b,j}-\frac{i}{2}}{u_{a,k}-u_{b,j}+\frac{i}{2}} \\
\prod_{h=1}^{H} \left(\frac{u_{b,k}-u_h+\frac{i}{2}}{u_{b,k}-u_h-\frac{i}{2}}\right)
&=& \prod _{j=1}^{K_b} \frac{u_{b,k}-u_{b,j}+i}{u_{b,k}-u_{b,j}-i}\prod _{j=1}^{K_a} \frac{u_{b,k}-u_{a,j}-\frac{i}{2}}{u_{b,k}-u_{a,j}+\frac{i}{2}} \prod _{j=1}^{K_c} \frac{u_{b,k}-u_{c,j}-\frac{i}{2}}{u_{b,k}-u_{c,j}+\frac{i}{2}} \nonumber\\
1&=& \prod _{j\not=k}^{K_c}\frac{u_{c,k}-u_{c,j}+i}{u_{c,k}-u_{c,j}-i} \prod _{j=1}^{K_b} \frac{u_{c,k}-u_{b,j}-\frac{i}{2}}{u_{c,k}-u_{b,j}+\frac{i}{2}} \nonumber
\ea
A comparison with (\ref{su(4)BetheEqs}) promptly reveals that the equations (\ref{scalSector}) coincide with those for a spin chain associated to the antisymmetric (\textbf{6}) representation of $su(4)$, whose highest weight is, according to our convention, $\vec{w}_6=\vec{\varphi}_2$. The $H$ hole rapidities ($u_h$, $h=1,\dots,H$) can be read as inhomogeneities along the spin chain, and their dynamics are regulated by the equations (\ref{scal-byall}), suitably adapted to the case at hand.
\medskip\\
\textbf{$\bullet$ (Large) Fermionic sector}\\
Let us stick now to a system composed of
$N_F$ large fermions $u_{F,j}$,\ $j=1,...,N_F$, together with $K_a$ roots $u_a$,
$K_b$ roots $u_b$ and $K_c$ roots $u_c$. While the fermions satisfy Bethe equations (\ref{Bethe_Ferm}),
the auxiliary roots obey relations:
\ba\label{fermSector}
\prod _{j=1}^{N_F}\left(
\frac{u_{a,k}-u_{F,j}+\frac{i}{2}}{u_{a,k}-u_{F,j}-\frac{i}{2}}\right)&=&\prod _{j\not=k}^{K_a} \frac{u_{a,k}-u_{a,j}+i}{u_{a,k}-u_{a,j}-i} \prod _{j=1}^{K_b}
\frac{u_{a,k}-u_{b,j}-\frac{i}{2}}{u_{a,k}-u_{b,j}+\frac{i}{2}} \\
1&=&\prod _{j=1}^{K_b} \frac{u_{b,k}-u_{b,j}+i}{u_{b,k}-u_{b,j}-i}\prod _{j=1}^{K_a} \frac{u_{b,k}-u_{a,j}-\frac{i}{2}}{u_{b,k}-u_{a,j}+\frac{i}{2}}\prod_{j=1}^{K_c}\frac{u_{b,k}-u_{c,j}-\frac{i}{2}}{u_{b,k}-u_{c,j}+\frac{i}{2}} \nonumber\\
1&=&\prod _{j\not=k}^{K_c} \frac{u_{c,k}-u_{c,j}+i}{u_{c,k}-u_{c,j}-i}
\prod _{j=1}^{K_b} \frac{u_{c,k}-u_{b,j}-\frac{i}{2}}{u_{c,k}-u_{b,j}+\frac{i}{2}} \nonumber
\ea
A look at (\ref{su(4)BetheEqs})
suggests the equations (\ref{fermSector}) should be associated
to a spin chain related to the fundamental representation (\textbf{4}) of $su(4)$ (with highest weight $\vec{w}_4=\vec{\varphi}_1$), where the large fermions behave as inhomogeneities, with rapidities $u_{F,j}$, $j=1,\dots,N_F$.

Otherwise, when only large antifermions (in number of $N_{\bar F}$) appear in the vacuum, again accompanied by $K_a$ isotopic roots $u_a$, $K_b$ roots $u_b$ and $K_c$ roots $u_c$, the system is described by the set of Bethe equations
(\ref{Bethe_antiFerm}) together with the isotopic roots equations:
\ba\label{antifermSector}
1&=&\prod _{j\not=k}^{K_a} \frac{u_{a,k}-u_{a,j}+i}{u_{a,k}-u_{a,j}-i} \prod _{j=1}^{K_b}
\frac{u_{a,k}-u_{b,j}-\frac{i}{2}}{u_{a,k}-u_{b,j}+\frac{i}{2}} \\
1&=&\prod _{j=1}^{K_b} \frac{u_{b,k}-u_{b,j}+i}{u_{b,k}-u_{b,j}-i}\prod _{j=1}^{K_a} \frac{u_{b,k}-u_{a,j}-\frac{i}{2}}{u_{b,k}-u_{a,j}+\frac{i}{2}}\prod_{j=1}^{K_c}\frac{u_{b,k}-u_{c,j}-\frac{i}{2}}{u_{b,k}-u_{c,j}+\frac{i}{2}} \nonumber\\
\prod _{j=1}^{N_{\bar F}}
\left(\frac{u_{c,k}-u_{\bar F,j}+\frac{i}{2}}{u_{c,k}-u_{\bar F,j}-\frac{i}{2}}\right)
&=&\prod _{j\not=k}^{K_c} \frac{u_{c,k}-u_{c,j}+i}{u_{c,k}-u_{c,j}-i}
\prod _{j=1}^{K_b} \frac{u_{c,k}-u_{b,j}-\frac{i}{2}}{u_{c,k}-u_{b,j}+\frac{i}{2}} \quad .\nonumber
\ea
The (\ref{antifermSector}) are in fact the equations for $\bar{\textbf{4}}$ spin chain (highest weight $\vec{w}_{\bar 4}=\vec{\varphi}_3$), as may be read from (\ref{su(4)BetheEqs}).

Some interest should be paid to a system including both $N_F$ (large) fermions and $N_{\bar F}$ (large) antifermions;
in this case, the isotopic roots satisfy the relations:
\ba\label{fermantifermSector}
\prod _{j=1}^{N_F}\left(
\frac{u_{a,k}-u_{F,j}+\frac{i}{2}}{u_{a,k}-u_{F,j}-\frac{i}{2}}\right)&=&\prod _{j\not=k}^{K_a} \frac{u_{a,k}-u_{a,j}+i}{u_{a,k}-u_{a,j}-i} \prod _{j=1}^{K_b}
\frac{u_{a,k}-u_{b,j}-\frac{i}{2}}{u_{a,k}-u_{b,j}+\frac{i}{2}} \\
1&=&\prod _{j=1}^{K_b} \frac{u_{b,k}-u_{b,j}+i}{u_{b,k}-u_{b,j}-i}\prod _{j=1}^{K_a} \frac{u_{b,k}-u_{a,j}-\frac{i}{2}}{u_{b,k}-u_{a,j}+\frac{i}{2}}\prod_{j=1}^{K_c}\frac{u_{b,k}-u_{c,j}-\frac{i}{2}}{u_{b,k}-u_{c,j}+\frac{i}{2}} \nonumber\\
\prod _{j=1}^{N_{\bar F}}
\left(\frac{u_{c,k}-u_{\bar F,j}+\frac{i}{2}}{u_{c,k}-u_{\bar F,j}-\frac{i}{2}}\right)
&=&\prod _{j\not=k}^{K_c} \frac{u_{c,k}-u_{c,j}+i}{u_{c,k}-u_{c,j}-i}
\prod _{j=1}^{K_b} \frac{u_{c,k}-u_{b,j}-\frac{i}{2}}{u_{c,k}-u_{b,j}+\frac{i}{2}} \quad\ \ .\nonumber
\ea
On the basis of $su(4)$ simple roots and fundamental weights (\ref{roots}), we can claim that the equations (\ref{fermantifermSector}) are associated to a spin chain, related to the representation of $su(4)$ whose Dynkin labels are $(1,0,1)$; in other terms, we found as its highest weight $\vec{w}_{15}=\vec{\varphi}_1+\vec{\varphi}_3$, and that leads to the \textbf{15}. The reason lies in the way how fermions (in the \textbf{4}) and antifermions (in the $\bar{\textbf{4}}$) scatter, since the process can be decomposed into two channels, according to the 
rule
\be
\textbf{4}\otimes\bar{\textbf{4}} = \textbf{1} \oplus \textbf{15}  \quad ;
\ee
the singlet \textbf{1} channel is not explicitly appearing in (\ref{fermantifermSector}), but
it can be revealed upon imposing some constraints on the isotopic roots (see next section).

We eventually remark that analogous considerations and expressions hold if we replace large fermions/antifermions
with small fermions/antifermions.
\medskip\\
\textbf{$\bullet$ Gauge field sector}\\
When only $N_g$ gluons (with rapidities $u^g_j$) are excited over the vacuum,
the isotopic roots decouple from them, since
\ba\label{gluSector}
1&=& \prod _{j\not=k}^{K_a}\frac{u_{a,k}-u_{a,j}+i}{u_{a,k}-u_{a,j}-i} \prod_{j=1}^{K_b}\frac{u_{a,k}-u_{b,j}-\frac{i}{2}}{u_{a,k}-u_{b,j}+\frac{i}{2}} \\
1&=& \prod _{j=1}^{K_b} \frac{u_{b,k}-u_{b,j}+i}{u_{b,k}-u_{b,j}-i}\prod _{j=1}^{K_a} \frac{u_{b,k}-u_{a,j}-\frac{i}{2}}{u_{b,k}-u_{a,j}+\frac{i}{2}} \prod _{j=1}^{K_c} \frac{u_{b,k}-u_{c,j}-\frac{i}{2}}{u_{b,k}-u_{c,j}+\frac{i}{2}} \nonumber\\
1&=& \prod _{j\not=k}^{K_c}\frac{u_{c,k}-u_{c,j}+i}{u_{c,k}-u_{c,j}-i} \prod _{j=1}^{K_b} \frac{u_{c,k}-u_{b,j}-\frac{i}{2}}{u_{c,k}-u_{b,j}+\frac{i}{2}} \nonumber \quad .
\ea
Therefore, gluon excitations behave like singlets (\textbf{1}) under $SU(4)$. The very same reasoning applies to barred-gluons.

\section{Eigenvalues}\label{sez8}
\setcounter{equation}{0}

While commenting on the equations (\ref{fermantifermSector}), we hinted the role the $SU(4)$ symmetry takes in the scattering between fermions and antifermions.
Now we are going to examine in some more detail several scattering processes involving different kinds of particles. In general, given two types of particles $\alpha$ and $\beta$, transforming under the representations of $su(4)$ $\rho_\alpha$ and $\rho_\beta$, which act respectively on the spaces $V_\alpha$ and $V_\beta$, their scattering decomposes according to the Clebsch-Gordan rule
\be
\rho_\alpha \otimes \rho_\beta = \bigoplus_\Omega \rho_\Omega \, .
\ee
Recalling \cite{OW}, the scattering matrix $\hat S^{(\alpha\beta)}$ (defined on $V_\alpha\otimes V_\beta$) between excitations $\alpha$ and $\beta$ enjoys the spectral decomposition
\be
\hat S^{(\alpha\beta)}=\sum _{\Omega} S^{(\alpha\beta)}_{\Omega} P_{\Omega} \, ,
\ee
where $S^{(\alpha\beta)}_{\Omega}$ are the eigenvalues of the matrix $\hat S^{(\alpha\beta)}$, relatives to the (normalized) eigenvectors $P_{\Omega}$, which act as projectors onto the space $V_\Omega$, \textit{i.e.} $P_\Omega(V_\alpha\otimes V_\beta)=V_\Omega$.
In this section we list the eigenvalues corresponding to the scattering between excitations on the top of the GKP string.

{\bf Scalar-scalar}\\
The scalar-scalar scattering was completely clarified in \cite {BREJ}, here we list eigenvalues and corresponding isotopic roots; since scalars belong to the $\textbf{6}$, the decomposition follows:
\be
\textbf{6}\otimes{\textbf{6}} = \textbf{1} \oplus \textbf{15} \oplus \textbf{20} \quad .
\ee
The singlet \textbf{1} channel involves two type-$b$ isotopic roots, which shall be related to the hole rapidities $u_h,\,u_{h'}$ according to $u_{b,1}= \frac{1}{2} \left ( u_h+u_{h'}- \sqrt{\frac{1+(u_h-u_{h'})^2}{3}} \right )$
and $u_{b,2}= \frac{1}{2} \left ( u_h+u_{h'}+ \sqrt{\frac{1+(u_h-u_{h'})^2}{3}} \right )$, together with $a$ and $c$ roots $u_a=u_c=\frac{u_h+u_{h'}}{2}$. These constraints on the isotopic roots lead us to the eigenvalue
\be
S^{(ss)}_{\textbf{1}}(u_h,u_{h'})=\frac{u_h-u_{h'}+2i}{u_h-u_{h'}-2i}\,\frac{u_h-u_{h'}+i}{u_h-u_{h'}-i}\, S^{(ss)}(u_h,u_{h'}) \ ,
\ee
where the scalar factor $S^{(ss)}(u_h,u_{h'})$ can be read from (\ref{Sss}).
The adjoint channel $\Omega=\textbf{15}$ requires one $b$-type root, satisfying $u_b=\frac{u_h+u_{h'}}{2}$ and no $a$ nor $c$ roots $K_a=K_c=0$. Eventually, the resulting eigenvalue follows
\be
S^{(ss)}_{\textbf{15}}(u_h,u_{h'})=
\frac{u_h-u_{h'}+i}{u_h-u_{h'}-i}\, S^{(ss)}(u_h,u_{h'}) \ .
\ee
Finally, the $\Omega=\textbf{20}$ channel request no isotopic roots ($K_a=K_b=K_c=0$), so that the eigenvalue simply coincides with (\ref{Sss})
\be
S^{(ss)}_{\textbf{20}}(u_h,u_{h'})=S^{(ss)}(u_h,u_{h'}) \ .
\ee

{\bf Fermion-fermion}\\
In the fermion-fermion scattering, we have two eigenvalues corresponding to the decomposition $\textbf{4}\otimes \textbf{4}=\textbf{10}\oplus \textbf{6}$. The first one, for $\Omega=\textbf{10}$, corresponds to no isotopic roots
and therefore it holds
$$
S^{(FF)}_{\textbf{10}}(u_{F,1},u_{F,2})=S^{(FF)}(u_{F,1},u_{F,2}) \ ,
$$
where the scalar factor corresponds to (\ref{SFF}).
The second one, for the $\Omega=\textbf{6}$ channel, is obtained from the solution with $K_a=1$, $K_b=2$, $K_c=0$, such that
$u_a=\frac{u_{F,1}+u_{F,2}}{2}$, while $u_{b,1}=u_{F,1}$ and $u_{b,2}=u_{F,2}$; consequently, we find that:
$$
S^{(FF)}_{\textbf{6}}(u_{F,1},u_{F,2})=\frac{u_{F,1}-u_{F,2}+i}{u_{F,1}-u_{F,2}-i} \ S^{(FF)}(u_{F,1},u_{F,2}) \ .
$$

{\bf Fermion-antifermion}\\
As previously mentioned, the fermion-antifermion scattering is associated to the decomposition
$\textbf{4} \otimes \mathbf{\bar 4}=\textbf{15}\oplus \textbf{1}$. Turning to the $\Omega=\textbf{15}$ channel, no isotopic roots are involved,
therefore the eigenvalue equals the scalar factor:
\be
S^{(F\bar F)}_{\textbf{15}}(u_{F,1},u_{\bar F,1})=S^{(FF)}(u_{F,1},u_{\bar F,1}) \ .
\ee
The singlet channel instead is obtained from the solution with $K_a=K_b=K_c=1$, where the isotopic roots satisfy the constraints
$u_a=\frac{3}{4}u_{F,1} + \frac{1}{4}u_{\bar F,1}$, $u_{b,1}=\frac{1}{2}u_{F,1}+\frac{1}{2}u_{\bar F,1}$ and $u_{c,1}=\frac{1}{4}u_{F,1}+\frac{3}{4}u_{\bar F,1}$. As a consequence, we obtain as the eigenvalue for the   $\Omega=\textbf{1}$ channel:
\be\label{ferm-antiferm}
S^{(F\bar F)}_{\textbf{1}}(u_{F,1},u_{\bar F,1})=\frac{u_{F,1}-u_{\bar F,1}+2i}{u_{F,1}-u_{\bar F,1}-2i} \ S^{(FF)}(u_{F,1},u_{\bar F,1}) \ .
\ee
The same result, i.e.
\be\label{ferm-antiferm-small}
S^{(f\bar f)}_{\textbf{1}}(u_{f,1},u_{\bar f,1})=\frac{u_{f,1}-u_{\bar f,1}+2i}{u_{f,1}-u_{\bar f,1}-2i} \ S^{(ff)}(u_{f,1},u_{\bar f,1}) \ ,
\ee
holds for small fermions. We highlighted (\ref {ferm-antiferm-small}), since this factor is responsible
of the appearance of a mass two particle at $g=+\infty $. This issue will be discussed in next section.

\medskip

\section{Classification of possible bound states}\label{sez9}
\setcounter{equation}{0}

\subsection{String solutions at large size}

In the large size limit, $R\rightarrow +\infty$, solutions to Bethe Ansatz equations show many (numerical and analytic) evidences that they organise into strings or stacks (generalised strings with different isospin or nested degrees of freedom). Their derivation follows as customary \cite{strings}. Let a complex rapidity $u_k^{\ast}$ exists, whose imaginary part be different from zero, but with sign so that the factor $e^{iRP(u_k^{\ast})}$ goes to zero (infinity) in the large $R$ limit: then another rapidity $u_j^{\ast '}$ must exist with the same real part, but imaginary part lowered if $\textrm{Im} u_k^{\ast} >0$  or, otherwise, raised if $\textrm{Im} u_k^{\ast} <0$, by an appropriate quantity, in order to drive rational factors in $S^{\ast \ast '}(u_k^{\ast},u_j^{\ast '})$ to infinity (zero), thus balancing the ABA equations. The process can continue by involving a finite number of extra rapidities displaced at regular distances until a string of $m$ roots disposed around a real 'center' is closed. Since the 'wave function' of a string of $m$ roots is by construction rapidly decreasing at $\pm \infty$, we naturally associate this configuration with a bound state of $m$ 'elementary' excitations.

In this section we discuss some possible bound states, with the important caveat that the list below is not meant to provide a complete classification of the particles living in the theory. This is indeed an interesting problem in itself and will be possibly dealt with in a future publication.

We also remark that, strictly speaking, the complexes of solutions we provide below are meant to be valid for finite values of the coupling constant, $i.e.\ g\neq 0$ and $g<\infty$. At $g=+\infty$ the situation ought to be different, as it can be inferred from considerations on the classical (quadratic) string theory action. Indeed, its small fluctuations in the bosonic sector consists of two mass $=\sqrt{2}$ (real) bosons and one mass $2$ (real) boson, besides the five massless bosons (of the $O(6)$ non-linear sigma model) \cite{FTT, AM}. Seemingly, this mass $2$ boson degree of freedom is missing in the gauge theory \footnote{Before us, many authors shared this concern, as, for instance, \cite{AM, Basso, Zarembo:2011ag, BSV3}.}, but, in the following, we find evidence that, with this mass, there is indeed a composite state made up of a small fermion and a small antifermion in their singlet channel. By means of the string mechanism discussed before, this bound state cannot exist at {\it finite} $g$, since divergences (zeroes) of the phase factor $e^{iRP^{(f)}(u_f)}$ of small fermions for complex rapidities go together with divergences (zeroes) of the S matrices, instead of compensating each other: thus this is a {\it resonance}. Indeed, after specialising the isotopic roots to the values defining the singlet channel of formula (\ref{ferm-antiferm-small}) and after switching off for simplicity's sake all the excitations (dots in the equation below) but a fermion and an antifermion, we have to cope with the coupled equations
\ba
1&=& e^{iRP^{(f)}(u_{f,1})+2iD^{(f)}(u_{f,1})} \frac{u_{f,1}-u_{\bar f,1}+2i}{u_{f,1}-u_{\bar f,1}-2i} \ S^{(ff)}(u_{f,1},u_{\bar f,1})\dots \, , \nonumber \\
1&=& e^{iRP^{(f)}(u_{\bar f,1})+2iD^{(f)}(u_{\bar f,1})} \frac{u_{\bar f,1}-u_{f,1}+2i}{u_{\bar f,1}-u_{f,1}-2i} \ S^{(ff)}(u_{\bar f,1},u_{f,1})\dots \label {pot-string} \, .
\ea
Thus, we can verify that the candidate string-like solutions $u_{f,1}=u_{M,1}\pm i$, $ u_{\bar f,1}=u_{M,1}\mp i$, and related isotopic rapidities of the singlet (\ref{ferm-antiferm-small}), with $u_{M,1}$ real, cannot satisfy (at large $R$) equations (\ref {pot-string}), since $ \textrm{Im}P^{(f)}(u_f) >0 $
if $ \textrm{Im} u_f <0$ and viceversa\footnote{We take up the occasion to highlight that we have found in the previous part of the paper that this opposition of signs happens only for the small fermions}. However, this remains true as long as $g$ is finite, while this bound state can appear as a new 'particle' when the value of $g$ is strictly $+\infty$. In fact, the point $g=+\infty$ is rather peculiar and singular, as complex scaled rapidities ($\bar u = u/\sqrt{2}g$) all collapse into the real axis, thus making possible a solution of ABA equations with a stack with two (small) fermion-antifermion rapidities (besides the isotopic rapidities): the two constituents possess no binding energy and no 'breathing', but they are just 'one on the other'. This is indeed a new (real) scalar, named here 'meson', coming to life only in the classical string regime $g=+\infty$ \footnote{Only at this value its rapidity, otherwise virtual \cite{Zamolodchikov:2013ama}, enters the physical domain \cite{Zarembo:2009au, Zamolodchikov:2013ama}.}. In summary, our following analysis of the ABA scattering on the GKP vacuum shows evidence for the existence of this bosonic particle with mass $2$ as long as $g\rightarrow +\infty$. And not only: the same mechanism at $g\rightarrow +\infty$ sustains the existence of a bound state of $k=1, 2, 3,\dots$ mesons with mass $m_k=2 k$ (zero static binding energy, as well as for gluon bound states). This is a new bosonic sector with respect to the classical (quadratic) string spectrum (and, {\it a fortiori}, to previous gauge theory analyses), but yet indispensable to be considered in the BSV series for 4D amplitudes, -- as we shall see --, for making checks with and reproducing the string minimal area solution (in other words the Thermodynamic Bubble Ansatz (TBA)). On the other end if this is an important way to check the validity of the series, it also confirms the pentagonal amplitude values and the 2D scattering factors entailing them. Eventually, the formation of mesons and bound states thereof shows a sort of {\it confinement} phenomenon at strong coupling as for the 4D amplitudes/Wilson loops, in that the contribution of the constituents, the fermions, to them is subtlety sub-dominant (as $g\rightarrow +\infty$, {\it cf.} \cite{BSV3} and below). In fact, this negligibility is not true for the 2D scattering amplitudes in themselves, but in their contribution to the 4D ones.

A more mathematical understanding of the small fermion-antifermion state ought to arrive \cite{35} from the collision of the poles into the integration (real) axis \cite{Nekrasov:2009rc, Meneghelli:2013tia, Bourgine:2014yha, Bettelheim:2014gma, BSV3}: this will give us the opportunity to explain the meson bound states and hence the {\it confinement} under a different light.\footnote{We are particularly grateful to I. Kostov and J.-E. Bourgine for explanatory discussions on this point.}\\
\medskip\\
{\bf $\bullet$ Gluonic strings:} \\
A first example is provided by strings made up of gluons or, alternatively, barred gluons, as equations (\ref{glu-fin}) and (\ref{barglu-fin}) suggest.
In this case one remarks the emergence of complex of solutions characterized by length $m$ and real centre $u_{k}^{g,m}$:
\ba
u_{k'}^{g,m}&=&u_{k}^{g,m}+\frac{i}{2}(m-1-2k') \, , \quad k'=0,...,m-1 \ ;
\ea
the very same structure may be built by assembling barred-gluon rapidities, too.
We will study more extensively bound states of gluons in next subsection, where we will show that they can be also obtained starting from the BMN vacuum by considering stacks of roots of type 1, 2 and 3.\\
\medskip\\
{\bf $\bullet$ More bound states:} \\
Along with gluonic strings, whose structure is quite ordinary, the ABA equations also admit the existence of more peculiar kinds of complexes whose composition and length result completely determined by the SU(4) symmetry of the vacuum. In fact, the structure of the Bethe equations for the SU(4) spin chain (\ref{Lie_spin_chain}), reflected in the equations for auxiliary roots (\ref{iso-1}, \ref{iso-3}, \ref{iso-2}), prevents these strings from including more than
two massive roots (or exceptionally three, as in (\ref{3particles}) below), intertwined with isotopic roots which are spaced by a constant distance fully fixed by (\ref{Lie_spin_chain}). The presence of isotopic roots is necessary for these strings to effectively represent solutions of the ABA equations and live in some definite scattering channel (see section \ref{sez8}), thus behaving in a broad sense like bound states which belong to some SU(4) 'isospin' multiplet. Below, such peculiar strings are listed according to their composition and SU(4) behaviour.\\
\medskip\\
{\bf $\bullet$ Bound states of large fermions in the $\mathbf{6}$ channel:}
\be\label{stringaFF}
u_{F,k,\pm} = u_{k} \pm \frac{i}{2}
\quad\quad u_{a,k} = u_{k}
\quad\quad u_{b,k,\pm} = u_{k} \pm \frac{i}{2} \, ;
\ee
the same structure occurs with antifermions too, upon substituting fermions with antifermions and the central $a$-root with a $c$-root.\\
{\bf $\bullet$ Bound states of large fermions in the singlet channel\footnote{Anyway we remark that bound states of this sort do not play any role in the strong coupling perturbative regime and, noticeably, their centres need to lie on the real axis in the region $|u_{M,k}|<\sqrt{2}g$, hence inside a square root branch cut in the large fermionic $u$-rapidity plane, so that perhaps they should not even be considered physical.
It is thus far from being obvious that any relation exist with what in the following we will refer to as 'meson' bound states, which exclusively subsist at $g=\infty$ and are made of small fermions, instead.}:}
\ba\label{prima0}
u_{F,k} &=& u_{M,k}+i  \quad\quad
u_{a,k} = u_{M,k}+\frac{i}{2}  \quad\quad
u_{b,k} = u_{M,k}  \nonumber\\
u_{\bar F,k} &=& u_{M,k}-i  \quad\quad
u_{c,k} = u_{M,k}-\frac{i}{2}
\ea
(the complex conjugate of (\ref{prima0}) is a solution, too).\\
\medskip\\
{\bf $\bullet$ Bound states of scalars in the $\mathbf{15}$ channel:}
\be
u_{h,k,\pm} = u_{k} \pm\frac{i}{2} \quad\quad u_{b,k} = u_{k}  \ .\\
\ee
\medskip\\
{\bf $\bullet$ Bound states of scalars in the singlet channel:}
\be
u_{h,k,\pm} = u_{k} \pm i  \quad\quad
u_{b,k,\pm} = u_{k} \pm\frac{i}{2}  \quad\quad
u_{a,k} =  u_{c,k} = u_{k}   \ ;
\ee
it is important to point out that these strings made of holes do not survive to the strong coupling limit in the non perturbative regime, as they are destroyed by poles of (\ref{theta-non-perturb}). Indeed, it is a well known result that in this regime the scalar dynamics is regulated by the O(6) non-linear $\sigma$-model, which lacks bound states.\\
\medskip\\
{\bf $\bullet$ Mixed bound states of large fermions and gluons in the $\mathbf{10}$ channel:}
\be\label{3particles}
u_{F,k,\pm} = u_{k} \pm\frac{i}{2} \quad\quad
u_{k}^{g} = u_{k} \ .
\ee
(the same also holds for barred-gluons and large antifermions).
\footnote{In addition to the string configurations listed above, several further complexes of solutions could be found, although strictly speaking they should not be considered actual bound states, since they are not endowed with real valued momenta; an example is offered by strings made of one single scalar and one fermion (or antifermion) bound together, whose distance gets fixed by the SU(4) symmetry to $\frac{3i}{2}$, and which could be probably related to a similar state described in \cite{BSV3}.}\\
\medskip\\
{\bf $\bullet$ (Purely) Magnonic strings:} \\
Also three distinct kinds of massless strings made of isotopic roots only, one for each type, can be found:
\ba
u_{a,k,j}^{A}&=&u_{a,k}^{A}+\frac{i}{2}(A-1-2j) \, , \quad j=0,...,A-1 \ ; \nonumber\\
u_{b,k,j}^{B}&=&u_{b,k}^{B}+\frac{i}{2}(B-1-2j) \, , \quad j=0,...,B-1 \ ; \\
u_{c,k,j}^{C}&=&u_{c,k}^{C}+\frac{i}{2}(C-1-2j) \, , \quad j=0,...,C-1 \ . \nonumber
\ea

\subsection{Bound states of gluons}

On the BMN vacuum with a sea of $u_4$ roots bound states of excitations $F_{+\perp}$ with rapidity $u_{k}^{g,m}$ can be constructed \cite {Basso} as stacks involving type 1, type 2 and type 3 roots:
\ba
u_{1,k}&=&u_{k}^{g,m}+\frac{i}{2}(m-2-2k') \, , \quad k'=0,...,m-2 \nonumber \\
u_{2,k}&=&u_{k}^{g,m}+\frac{i}{2}(m-1-2k') \, , \quad k'=0,...,m-1   \label  {glu-str} \\
u_{3,k}&=&u_{k}^{g,m}+\frac{i}{2}(m-2k') \, , \quad k'=0,...,m  \nonumber \, .
\ea
Analogously, bound states of gauge fields $\bar F_{+\perp}$ with rapidity $u_{k}^{\bar g,m}$ are obtained from (\ref {glu-str}), with $g \rightarrow \bar g$ and
$u_1,u_2,u_3 \rightarrow u_5,u_6,u_7$.
In presence of bound states of gluons Bethe equations should be modified as follows.

Bethe equations for bound states of $F_{+\perp}$ ($N_g^l$ ($N_{\bar g}^l$) is the number of bound states of $F_{+\perp}$ ($\bar F_{+\perp}$) with length $l$: their centers are indicated with $u_j^{g,l}$ ($u_j^{\bar g,l}$)) are
\ba
1&=&  e^{iRP^{(g)}_m(u_{k}^{g,m}) +2 i D^{(g)}_m(u_{k}^{g,m})} \prod _{l=1}^{+\infty} \prod _{j=1}^{N_{g}^{l}}
 {S}^{(gg)}_{ml}(u_{k}^{g,m},u_{j}^{g,l}) \prod _{l=1}^{+\infty} \prod _{j=1}^{N_{\bar g}^{l}} {S}^{(g\bar g)}_{ml}(u_{k}^{g,m},u_{j}^{\bar g,l})
  \prod _{h=1}^H S^{(gs)}_m(u_k^{g,m},u_h) \cdot  \nonumber \\
&\cdot &  \prod _{j=1}^{N_F} {S}^{(gF)}_m(u_{k}^{g,m},u_{F,j})
 \prod _{j=1}^{N_{\bar F}} {S}^{(g\bar F)}_m(u_{k}^{g,m},u_{\bar F,j})  \prod _{j=1}^{N_f} {S}^{(gf)}_m(u_{k}^{g,m},u_{f,j})
 \prod _{j=1}^{N_{\bar f}} {S}^{(g\bar f)}_m(u_{k}^{g,m},u_{\bar f,j}) \, ,
\label  {glu-finm}
\ea
where momentum and defect are given by
\ba
P^{(g)}_m(u)&=&-\int _{-\infty}^{+\infty} \frac{dv}{2\pi} \left [ \chi (v,u|m)+\chi (-v,u|m) \right ] \left [ 1-\frac{\sigma _{BES}(v)}{4} \right ]
\, , \label {P-glum} \\
2 D^{(g)}_m(u)&=& -\int _{-\infty}^{+\infty} \frac{dv}{2\pi} \left [ \chi (v,u|m)+\chi (-v,u|m) \right ] \frac{d}{dv} \tilde P(v) \label {D-glum} \,
\ea
and the various scattering factors are listed in Appendix \ref {scatt-fact}.

Exchanging $g$ with $\bar g$ we get Bethe equations for bound states of $\bar F_{+\perp}$.
The other equations can be obtained from the equations written when simple gluons are present (and collected in Appendix \ref {Bethe-eqs}) by means of the replacements:
\be
\prod _{j=1}^{N_g} S ^{\ast g} (u_{\ast}, u^{g}_j) \rightarrow \prod _{l=1}^{+\infty} \prod  _{j=1}^{N_g^{(l)}}
S^{(\ast g)}_{l}(u_{\ast}, u^{g,l}_j) \quad , \quad \prod _{j=1}^{N_{\bar g}} S ^{\ast \bar g} (u_{\ast}, u^{\bar g}_j) \rightarrow \prod _{l=1}^{+\infty} \prod  _{j=1}^{N_{\bar g}^{(l)}} S^{(\ast \bar g)}_{l}(u_{\ast}, u^{\bar g,l}_j) \, .
\ee
We now show that equations (\ref {glu-finm}) and others, which constrain centers of the string (\ref {glu-str}), are not independent of equations describing excitations on the GKP vacuum, but actually can be obtained from these by considering strings involving gluons.

In order to obtain (\ref {glu-finm}), we shall first consider equations (\ref {fineq8}) for gluons,
\ba
1&=&e^{iRP^{(g)}(u_{k'}^{g}) +2iD^{(g)}(u_{k'}^{g})}\prod _{j=1,\neq k'}^{N_g} \frac{u_{k'}^{g}-u_j^{g}+i}{u_{k'}^{g}-u_j^{g}-i}
S_{red}^{(gg)}(u_{k'}^{g},u_j^{g}) \prod _{j=1}^{N_F} \frac{u_{k'}^{g}-u_{F,j}+\frac{i}{2}}{u_{k'}^{g}-u_{F,j}-\frac{i}{2}}
S_{red}^{(gF)}(u_{k'}^{g},u_{F,j}) \cdot \nonumber \\
&\cdot &  \prod _{j=1}^{N_f} \frac{u_{k'}^{g}-u_{f,j}+\frac{i}{2}}{u_{k'}^{g}-u_{f,j}-\frac{i}{2}}
S_{red}^{(gf)}(u_{k'}^{g},u_{f,j}) \prod _{h=1}^H S^{(gs)}(u_{k'}^g, u_h)  \cdot \label {eq-g} \\
&\cdot & \prod _{j=1}^{N_{\bar g}} S^{(g\bar g)}(u_{k'}^g,u_j^{\bar g}) \prod _{j=1}^{N_{\bar F}}
S^{(g\bar F)}(u_{k'}^g,u_{\bar F,j}) \prod _{j=1}^{N_{\bar f}} S^{(g\bar f)}(u_{k'}^g,u_{\bar f,j})  \nonumber \, ,
\ea
where $S_{red}$ (\ref  {Sgg}, \ref {SgF}) stand for the $S$ factors deprived of the rational factors $-e^{\pm i\chi _0}$. The appearance of the rational factor
\be
\frac{u_{k'}^{g}-u_j^{g}+i}{u_{k'}^{g}-u_j^{g}-i}
\ee
suggests that strings of gluons rapidities with the same real part and imaginary parts separated of $i$
form. In specific,
\ba
u_{k'}^{g}&=& u_k^{g,m}+\frac{i}{2} (m-1-2k') \, , \quad k'=0,..., m-1 \, , \label {rap-str1} \\
u_{k'}^{\bar g}&=& u_k^{\bar g,m}+\frac{i}{2} (m-1-2k') \, , \quad k'=0,..., m-1 \label {rap-str2} \, ,
\ea
where the real centers of the strings are in the region $ |u_k^{g,m}| <\sqrt{2}g$, $|u_k^{\bar g,m}| <\sqrt{2}g$.

Performing the products over $k'$ we arrive at the equation
\ba
1&=&e^{iRP_m^{(g)}(u_k^{g,m}) +2iD_m^{(g)}(u_k^{g,m})}\prod _{j=1}^{N_g} \frac{u_k^{g,m}-u_j^{g}+\frac{i}{2}(m+1)}{u_k^{g,m}-u_j^{g}-\frac{i}{2}(m+1)}
\frac{u_k^{g,m}-u_j^{g}+\frac{i}{2}(m-1)}{u_k^{g,m}-u_j^{g}-\frac{i}{2}(m-1)} S_{red,m}^{(gg)}(u_k^{g,m},u_j^{g}) \cdot \nonumber \\
&\cdot & \prod _{j=1}^{N_F} \frac{u_k^{g,m}-u_{F,j}+\frac{im}{2}}{u_k^{g,m}-u_{F,j}-\frac{im}{2}}
S_{red,m}^{(gF)}(u_k^{g,m},u_{F,j})
\prod _{j=1}^{N_f} \frac{u_k^{g,m}-u_{f,j}+\frac{im}{2}}{u_k^{g,m}-u_{f,j}-\frac{im}{2}}
S_{red,m}^{(gf)}(u_k^{g,m},u_{f,j})
\prod _{h=1}^H S^{(gs)}_m(u_k^{g,m}, u_h)  \nonumber \\
&\cdot & \prod _{j=1}^{N_{\bar g}} S^{(g\bar g)}_m(u_k^{g,m},u_j^{\bar g}) \prod _{j=1}^{N_{\bar F}}
S^{(g\bar F)}_m(u_k^{g,m},u_{\bar F,j}) \prod _{j=1}^{N_{\bar f}} S^{(g\bar f)}_m(u_k^{g,m},u_{\bar f,j})
\, , \label {eq-gm}
\ea
where
\be
 S_{red,m}^{(g\ast)}(u,v)=\prod _{l=-\frac{m-1}{2}}^{\frac{m-1}{2}} S_{red}^{(g\ast)}(u+il,v) \, , \quad
  S_{m}^{(g\ast)}(u,v)=\prod _{l=-\frac{m-1}{2}}^{\frac{m-1}{2}} S_{m}^{(g\ast)}(u+il,v) \, .
 \label {S-gm}
\ee
It is now immediate to recognize the scattering factors between a bound state of $F_{+\perp}$ with center $u$ and 'length' $m$ and a fermion or a scalar:
\ba
&& S^{(gF)}_m (u,v)=\frac{u-v+\frac{im}{2}}{u-v-\frac{im}{2}} S_{red,m}^{(gF)}(u,v) \, , \quad S^{(gf)}_m (u,v)=\frac{u-v+\frac{im}{2}}{u-v-\frac{im}{2}} S_{red,m}^{(gf)}(u,v) \, , \nonumber \\
&& S^{(g\bar F)}_m (u,v)=S_{red,m}^{(g\bar F)}(u,v) \, , \quad S^{(g\bar f)}_m (u,v)=S_{red,m}^{(g\bar f)}(u,v) \, , \quad S^{(gs)}_m (u,v)= S_{red,m}^{(gs)}(u,v) \, . \label {Sgm-oth}
\ea
By means of (\ref {chimaster}) one shows that these factors equal the ones appearing in (\ref {glu-finm}).
Equations (\ref {eq-gm}) are then completed by taking into account that rapidities $u_j^g$, $u_j^{\bar g}$ appear into strings (\ref {rap-str1}, \ref  {rap-str2}). Because of the properties
\ba
&& \prod _{k'=0}^{l-1} \frac{u_k^{g,m}-u_j^{g,l}-\frac{i}{2}(l-1-2k')+\frac{i}{2}(m+1)}{u_k^{g,m}-u_j^{g,l}-\frac{i}{2}(l-1-2k')-\frac{i}{2}(m+1)}
\, \frac{u_k^{g,m}-u_j^{g,l}-\frac{i}{2}(l-1-2k')+\frac{i}{2}(m-1)}{u_k^{g,m}-u_j^{g,l}-\frac{i}{2}(l-1-2k')-\frac{i}{2}(m-1)} \cdot \nonumber \\
&& \cdot S_{red,m}^{(gg)}(u_k^{g,m},u_j^{g,l}+\frac{i}{2}(l-1-2k')) =S^{(gg)}_{ml}(u_k^{g,m},u_j^{g,l}) \, , \\
&& \prod _{k'=0}^{l-1} S_{m}^{(g\bar g)}(u_k^{g,m},u_j^{\bar g,l}+\frac{i}{2}(l-1-2k')) =S^{(g\bar g)}_{ml}(u_k^{g,m},u_j^{\bar g,l})
\ea
which follow from  (\ref {chimaster}), one finally finds that equations (\ref {glu-finm}) are reproduced. In analogous fashion, other expressions, derived from the BMN vacuum, can be reproduced, as well, starting from equations on the GKP vacuum.

\medskip
\noindent
{\bf The energy.}

We end this part by spending some words on the energy of the gluon and of its bound states (the same will apply to its barred companion). Immediately, we can put it in the general form at any $g$
\be
E^{(g)}_m(u)=2m- \int \frac{dv}{2\pi} \left [ \chi (v,u |m)+\Phi (v) \right ] \frac{d}{dv} \gamma ^{(s)} (v) \, ,
\ee
which we rewrite as
\ba
&&  E^{(g)}_m(u)=2m+ \int_{-\infty}^{\infty}\frac{dk}{4\pi^2}\,\frac{i\pi}{e^{\frac{k}{2}}-e^{-\frac{k}{2}}}\,
(\gamma _+^{\textrm{{\o}}}(\sqrt{2}gk)- \textrm{sgn}(k)\gamma _-^{\textrm{{\o}}}(\sqrt{2}gk))
\left[\frac{2\pi}{ik}\,e^{-|k|\frac{m+1}{2}}e^{-iku}- \right. \\
&&\ \ -\frac{2\pi}{ik}\left. \,e^{-\frac{|k|}{2}}\sum_{n=1}^{\infty}\left(\left(\frac{g}{\sqrt{2}i x(u+ \frac{im}{2})}\right)^n+\left(\frac{g}{\sqrt{2}i x(u- \frac{im}{2})}\right)^n\right)\,J_n(\sqrt{2}gk)
-\frac{2\pi}{ik}\,J_0(\sqrt{2}gk)e^{-\frac{|k|}{2}}
\right]  \nonumber \, .
\ea
By means of the relation (3.40) in \cite{Basso} and of the Fourier transform
\be
\int_0^\infty \frac{dk}{k}\,e^{-k\left(\frac{m}{2}\pm iu\right)}\, J_n(\sqrt{2}gk)=
\frac{(\pm 1)^n}{n}\left(\frac{g}{\sqrt{2}i x(u\mp \frac{im}{2})}\right)^n  \qquad ,
\ee
the expression above becomes
\be
E^{(g)}_m(u)=2m+\int_0^\infty\frac{dk}{k}\,\frac{\gamma _+^{\textrm{{\o}}}(\sqrt{2}gk)}
{1-e^{-k}}\,[\cos ku \,e^{-k\frac{m}{2}}-1]
-\int_0^\infty\frac{dk}{k}\,\frac{\gamma _-^{\textrm{{\o}}}(\sqrt{2}gk)}{e^{k}-1}\,[\cos ku \,e^{-k\frac{m}{2}}-1] \, ,
\label {E-g-m}
\ee
in agreement with \cite{Basso}.

\medskip

\noindent
{\bf Perturbative strong coupling regime.}

In this limit we can use the following results
\be
\textrm{exp} \left [ -i \tilde \chi (u,v|m,l) \right ] =\textrm{exp} \left [ \frac{\sqrt{2}m l}{g(\bar v-\bar u)}+ O(1/g^3) \right ]   \label {im}
\ee
and
\ba
\chi (w, u|m) +\Phi (w) &=& \frac{m}{\sqrt{2}g (\bar u -\bar w)} +O(1/g^2) \, , \quad \textrm{when} \ \ |\bar w|>1 \, , \\
\chi (w, u|m) +\Phi (w) &=& O(1/g)  \, , \quad \textrm{when} \ \ |\bar w|<1 \, . \label {chiphimper}
\ea
Repeating all the steps we did for one single gluon, we write the scattering factor between two bound states of $m$ and $l$ gluons, respectively, in the perturbative regime (up to terms $O(1/g^2)$) as
\be
S^{(gg)}_{ml}(u,v) = \textrm{exp} \left [ \frac {iml}{\sqrt{2}g (\bar u - \bar v)} \left ( 1+
\frac{1}{2}\left ( \frac{1+\bar u}{1- \bar u}\right )^{1/4} \left ( \frac{1-\bar v}{1+\bar v }\right )^{1/4}  + \frac{1}{2}\left ( \frac{1-\bar u}{1+ \bar u}\right )^{1/4} \left ( \frac{1+\bar v}{1-\bar v }\right )^{1/4} \right ) \right ] \label {Smlglu-str} \, .
\ee
For what concerns energy (\ref {E-g-m}) and momentum (\ref {P-glum}) of a bound state of $m$ gluons, we have the simple relativistic result
\be
E^{(g)}_m(\theta )=2m \cosh \theta \, , \quad P^{(g)}_m(\theta )=2m \sinh \theta \, , \label {EP-glum}
\ee
in terms of the hyperbolic variable $\theta $ defined by $u=\sqrt{2}g \tanh 2\theta $. As already written, the results above apply to the barred gluon as well.

\subsection{The meson and its bound states}

As hinted before, at infinite coupling in the perturbative regime a bound state of a small fermion and a small anti-fermion arises. As anticipated, we will call this state 'meson'.

Now, we elaborate a proposal for the complete set of ABA equations satisfied by mesons. One meson is described by the stack (\ref {prima0}) in which, of course, the index $F$ is replaced by $f$, i.e.
\ba\label{prima1}
u_{f,k} &=& u_{M,k}+i  \quad\quad
u_{a,k} = u_{M,k}+\frac{i}{2}  \quad\quad
u_{b,k} = u_{M,k}  \nonumber\\
u_{\bar f,k} &=& u_{M,k}-i  \quad\quad
u_{c,k} = u_{M,k}-\frac{i}{2} \, .
\ea
We already emphasised that this stack does not give rise to a genuine bound state for finite $g$, but describes a virtual state: in this perspective, we can formally write the ABA equations describing the scattering between virtual states (\ref {prima1}) with themselves and with other excitations at generic $g$. Of course, we should start by deducing -- in the usual manner -- the scattering between two virtual mesons:
\ba\label {SMM-virtual}
S^{(MM)}(u,v) &=& \frac{u-v+i}{u-v-i} S^{(ff)}(u+i,v+i) S^{(ff)}(u-i,v+i)S^{(ff)}(u+i,v-i)S^{(ff)}(u-i,v-i) \nn\\
&=&\frac{u-v+i}{u-v-i}\,
 \textrm{exp} \Bigl \{i \int\frac{dw}{2\pi} \chi_M(w,u|1) \frac{d}{dw} \chi_M(w,v|1) - \nonumber \\
&-&i \int \frac{dw}{2\pi}\frac{dz}{2\pi} \chi_M(w,u|1) \frac{d^2}{dwdz}\Theta (w,z) \chi_M(z,v|1) \Bigr \} \ \,\, ,
\ea
with the definitions
\ba
\chi_M(v,u|m) &= & \sum_{k=-\frac{m-1}{2}}^{\frac{m-1}{2}}\chi_M(v,u+ik|1)
\quad\quad \mbox{with}\ m\geq 1 \nn\\
\chi_M(v,u|1) &= & -\chi_H(v,u+i)-\chi_H(v,u-i) \ .
\label{chi-def}
\ea
Then, we can introduce also the other excitations and derive, at least formally, the ABA equations involving $N_M$ mesons
\ba
1 &=& e^{iR P^{(M)}(u_{M,k}) }\,e^{2i D^{(M)}(u_{M,k})}\prod_{h=1}^{H}{S}^{(Ms)}(u_{M,k},u_h)
\,\prod_{j=1}^{N_g}{S}^{(Mg)}(u_{M,k},u^g_{j})
\cdot  \label {aba-mes} \\
&\cdot & \,\prod_{j=1}^{N_{\bar g}}{S}^{(M\bar g)}(u_{M,k},u^{\bar g}_{j})
\,\prod_{j=1}^{N_f}{S}^{(Mf)}(u_{M,k},u_{f,j})
\prod_{j=1}^{N_{\bar f}}{S}^{(M\bar f)}(u_{M,k},u_{\bar f,j})
\prod_{j=1, j\neq k}^{N_M}{S}^{(MM)}(u_{M,k},u_{M,j}) \, , \nonumber
\ea
where $ P^{(M)}(u)=P^{(f)}(u+i)+P^{(f)}(u-i), D^{(M)}(u)=D^{(f)}(u+i)+D^{(f)}(u-i)$ and the scattering factors read
\ba
S^{(Ms)}(u,v) =
\frac{u-v+\frac{i}{2}}{u-v-\frac{i}{2}}\,
\textrm{exp} \Bigl \{ i\chi_M (v,u|1)-i \int \frac{dw}{2\pi}\frac{d\Theta}{dw}(v,w) \,\chi_M(w,u|1)\Bigr \} \ ,
\label{M-s}
\ea
\ba
S^{(Mg)}(u,v)&=&  \frac{u-v+\frac{3i}{2}}{u-v+\frac{i}{2}}
\,\textrm{exp} \Bigl \{i \int \frac{dw}{2\pi} \chi_M(w,u|1) \frac{d}{dw} \left[\chi(w,v|1)+\Phi(x)\right] - \nonumber \\
&-&i \int \frac{dw}{2\pi}\frac{dz}{2\pi} \chi_M(w,u|1) \frac{d^2}{dwdz}\Theta (w,z) \left[\chi(z,v|1)+\Phi(z)\right] \Bigr \} \ ,
\label{M-g}
\ea
\ba
S^{(M\bar g)}(u,v)&=& \frac{u-v-\frac{i}{2}}{u-v-\frac{3i}{2}}
\,\textrm{exp} \Bigl \{i \int \frac{dw}{2\pi} \chi_M(w,u|1) \frac{d}{dw} \left[\chi(w,v|1)+\Phi(x)\right] - \nonumber \\
&-&i \int \frac{dw}{2\pi}\frac{dz}{2\pi} \chi_M(w,u|1) \frac{d^2}{dwdz}\Theta (w,z) \left[\chi(z,v|1)+\Phi(z)\right] \Bigr \} \ ,
\label{M-gbar}
\ea
\ba
S^{(Mf)}(u,v)&=&  \frac{u-v+i}{u-v}
\,\textrm{exp} \Bigl \{-i \int \frac{dw}{2\pi} \chi_M(w,u|1) \frac{d}{dw} \chi_H(w,v) + \nonumber \\
&+&i \int \frac{dw}{2\pi}\frac{dz}{2\pi} \chi_M(w,u|1) \frac{d^2}{dwdz}\Theta (w,z) \chi_H(z,v) \Bigr \} \ ,
\label{M-f}
\ea
\ba
S^{(M\bar f)}(u,v)&=&  \frac{u-v}{u-v-i}
\,\textrm{exp} \Bigl \{-i \int \frac{dw}{2\pi} \chi_M(w,u|1) \frac{d}{dw} \chi_H(w,v) + \nonumber \\
&+&i \int \frac{dw}{2\pi}\frac{dz}{2\pi} \chi_M(w,u|1) \frac{d^2}{dwdz}\Theta (w,z) \chi_H(z,v) \Bigr \} \, .
 \label{M-fbar}
\ea
Eventually, these ABA equations (\ref{aba-mes}) \footnote{When writing them, we have deliberately neglected the multiplication by $S^{(MF)}(u,v)$ and $S^{(M\bar F)}(u,v)$ (which could be immediately obtained respectively from (\ref{M-f}) and (\ref{M-fbar}) via the usual substitution $-\chi_H(w,u)\longrightarrow \chi_F(w,u)+\Phi(w)$), since we know ({\it cf.} (\ref{SFFgh}) and (\ref{Sff-final})) that this contribution from the large fermions, forced to live in the giant hole regime at strong coupling, would be exponentially small.} will acquire the aforementioned physical meaning in the infinite coupling limit where it is possible to parametrise $u_M =\sqrt{2}g \coth 2\theta _M$ in terms of the hyperbolic rapidity, and then $P^{(M)}(\theta_M)=2\sinh \theta_M$, $D^{(M)}(\theta_M)=- P^{(M)}(\theta_M)\ln g$ and
\be
{S}^{(Ms)}(\theta_{M},u_{h}) =
		\exp \left [ \frac{i}{\sqrt{2}g} \frac{1}{\coth 2\theta _M -1} \right ]
		\ [ S^{(fs)}(\theta_{M},u_{h}) ]^2  \label {SMs} \, ,
\ee
\be
{S}^{(Mg)}(\theta_{M},\theta ^g) =
    \ S^{(fg)}(\theta_{M},\theta^{g})  \ S^{(\bar fg)}(\theta_{M},\theta^{g}) \, ,\label {SMg}
\ee
\be
{S}^{(M\bar g)}(\theta_{M},\theta ^{\bar g}) =
    \ S^{(f \bar g)}(\theta_{M} ,\theta^{\bar g})  \ S^{(\bar f \bar g)}(\theta_{M} ,\theta^{\bar g}) \, ,    \label {SMbarg}
\ee
\be
{S}^{(Mf)}(\theta_{M},\theta_{f}) =  \exp \left [ \frac{i}{\sqrt{2} g} \frac{1}{(\coth 2\theta _M -\coth 2\theta _f )} +O\left (\frac{1}{g^2} \right ) \right ]  [S^{(f f)}(\theta_{M},\theta_f)]^2 \, , \label {SMf}
\ee
\be
{S}^{(M\bar f)}(\theta_{M},\theta _{\bar f}) = \exp \left [ \frac{i}{\sqrt{2} g} \frac{1}{(\coth 2\theta _M -\coth 2\theta _{\bar f})} +O\left (\frac{1}{g^2} \right ) \right ] [ S^{(f f)}(\theta_{M},\theta_{\bar f})]^2  \, ,
\label {SMbarf}
\ee
\ba
{S}^{(MM)}(\theta_{M},\theta_{M}^{\prime}) &=&
   \exp \left [ \frac{i\sqrt{2}}{g} \frac{1}{(\coth 2\theta _M -\coth 2\theta _M ^{\prime})} +O\left (\frac{1}{g^2} \right )\right ] \ [ S^{(ff)}(\theta_{M} ,\theta _M^{\prime})]^4 \, . \label {sMM}
\ea
Explicit expressions for the scattering $S^{(fs)}(\theta_{M},u_{h})$ is given by (\ref {Ssfper}), with $\bar v=\coth 2\theta _M$ and for all the other $S$ factors (entering the previous r.h.ss.) in Appendix C.3.

Now, if we come back to (\ref {SMM-virtual}), we can notice that, in analogy to equations (\ref {pot-string}) for small fermion/antifermion (and to (\ref {eq-g}) for gluons), a rational pre-factor in front of the scattering factor is present.
This suggests that the same mechanism, which produces (virtual) mesons and bound states of gluons should also produce (virtual) bound states of mesons: in fact, these will be represented by strings with the customary form
\be
u_{M,k}=u_{M}^m+\frac{i}{2}(m-1-2k), \qquad k=0,\dots,m-1 \, ,
\ee
where all roots, once scaled, collapse towards the real axis at infinite coupling. At generic coupling, these virtual states enjoy scattering phases which, formally, do not distance themselves from the ones previously found for actual particles; indeed, when considering a process involving a bound state of $l$ mesons with center $u$ and a bound state of $m$ mesons with center $v$, the resulting scattering matrix emerges from the fusion of (\ref {SMM-virtual}):
\ba\label {SMMlm}
S^{(MM)}_{lm}(u,v)&=& \prod_{j=0}^{l-1}\prod_{k=0}^{m-1}\,S^{(MM)}\left(u+\frac{i}{2}(l-1-2j),v+\frac{i}{2}(m-1-2k)\right) =\\
&=& \textrm{exp} \Bigl \{ -i\tilde \chi (u,v|l,m)+i \int _{-\infty}^{+\infty} \frac{dw}{2\pi} \chi_M(w,u|l) \frac{d}{dw} \chi_M(w,v|m) - \nonumber \\
&-&i \int \frac{dw}{2\pi}\frac{dz}{2\pi} \chi_M(w,u|l) \frac{d^2}{dwdz}\Theta (w,z) \chi_M(z,v|m) \Bigr \} \nonumber \, ,
\ea
with the definitions (\ref{chi-def}). Similarly, from a generalisation of (\ref{M-s})-(\ref{M-gbar}) we can deduce the scattering phases concerning bound states of $m$ mesons against scalars or bound states of $l$ gluons:
\ba
S^{(Ms)}_{m}(u,v) =
\frac{u-v+\frac{im}{2}}{u-v-\frac{im}{2}}\,
\textrm{exp} \Bigl \{ i\chi_M (v,u|m)-i \int \frac{dw}{2\pi}\frac{d\Theta}{dw}(v,w) \,\chi_M(w,u|m)\Bigr \}
\ea
\ba
S^{(Mg)}_{ml}(u,v)&=& \prod_{k=-\frac{m-1}{2}}^{\frac{m-1}{2}}
\frac{u-v+\frac{i}{2}(l+2k+2)}{u-v-\frac{i}{2}(l+2k-2)}
\,\textrm{exp} \Bigl \{i \int \frac{dw}{2\pi} \chi_M(w,u|m) \frac{d}{dw} \left[\chi(w,v|l)+\Phi(x)\right] - \nonumber \\
&-&i \int \frac{dw}{2\pi}\frac{dz}{2\pi} \chi_M(w,u|m) \frac{d^2}{dwdz}\Theta (w,z) \left[\chi(z,v|l)+\Phi(z)\right] \Bigr \} ,
\ea
\ba
S^{(M\bar g)}_{ml}(u,v)&=& \prod_{k=-\frac{m-1}{2}}^{\frac{m-1}{2}}
\frac{u-v+\frac{i}{2}(l+2k-2)}{u-v-\frac{i}{2}(l+2k+2)}
\,\textrm{exp} \Bigl \{i \int \frac{dw}{2\pi} \chi_M(w,u|m) \frac{d}{dw} \left[\chi(w,v|l)+\Phi(x)\right] - \nonumber \\
&-&i \int \frac{dw}{2\pi}\frac{dz}{2\pi} \chi_M(w,u|m) \frac{d^2}{dwdz}\Theta (w,z) \left[\chi(z,v|l)+\Phi(z)\right] \Bigr \} .
\ea
It is crucial for the following (TBA) noticing the drastic simplification of the fusion into a multiplication when $g=\infty$ (zero shifts)
\be
S^{(MM)}_{lm}(\theta , \theta ') = \left[S^{(MM)}(\theta , \theta ') \right]^{lm}  \label {sMMlm-str} \, ,
\ee
along with the scattering phases between a bound state of $l$ mesons and a bound state of $m$ gluons which become
\be
S^{(Mg)}_{lm}(\theta , \theta ')= [S^{(gM)}_{ml}(\theta ',\theta)]^{-1}=[ S^{(Mg)}(\theta ,\theta ')]^{lm} \, ,\quad S^{(M\bar g)}_{lm}(\theta ,\theta ')=
[S^{(\bar gM)}_{ml}(\theta ', \theta )]^{-1}=[ S^{(M\bar g)}(\theta ,\theta ')]^{lm} \, .  \label {sMglm-str}
\ee
Explicit expressions for (\ref {sMMlm-str}, \ref {sMglm-str}) in terms of hyperbolic variables are given by formul{\ae} (\ref {sMM-str}, \ref {sMg-str}) in Appendix C.

We need to conclude this section with the dispersion relations. At least in principle, we can write the energy and momentum of the virtual meson (of mass $=2$) and of its $m$ bound state at finite $g$, albeit these are not stable particle: the energy as
\be
E^{M}_m(u)=2m+\gamma _m^{(M)}(u) \,\,\,\, ,
\ee
with
\ba
\gamma _m^{(M)}(u)&=&\int _{0}^{+\infty}  \frac{dk}{k} \frac{\cos ku}{\sinh \frac{k}{2}} \left ( \cosh k \ e^{-\frac{m}{2}k}-\cosh \frac{k}{2} \right ) \gamma _+^{\textrm{{\o}}}(\sqrt{2}gk)  \quad \textrm{for} \ \ m \ \ \textrm{odd} \, , \nonumber \\
\gamma _m^{(M)}(u)&=&\int _{0}^{+\infty}  \frac{dk}{k} \frac{\cos ku}{\sinh \frac{k}{2}} \left ( \cosh k \ e^{-\frac{m}{2}k}- 1 \right )  \gamma _+^{\textrm{{\o}}}(\sqrt{2}gk) \quad \textrm{for} \ \ m \ \ \textrm{even} \, ;
\label {gamma-mes}
\ea
and the momentum as
\ba
P_m^{(M)}(u)&=&\int _{0}^{+\infty}  \frac{dk}{k} \frac{\sin ku}{\sinh \frac{k}{2}} \left ( \cosh \frac{k}{2}-\cosh k \ e^{-\frac{m}{2}k} \right ) \gamma _-^{\textrm{{\o}}}(\sqrt{2}gk)  \quad \textrm{for} \ \ m \ \ \textrm{odd} \, , \nonumber \\
P_m^{(M)}(u)&=&\int _{0}^{+\infty}  \frac{dk}{k} \frac{\sin ku}{\sinh \frac{k}{2}} \left ( 1-\cosh k \ e^{-\frac{m}{2}k} \right )  \gamma _-^{\textrm{{\o}}}(\sqrt{2}gk) \quad \textrm{for} \ \ m \ \ \textrm{even} \, ,
\label {mom-mes}
\ea
both in terms of the functions $\gamma ^{\textrm{{\o}}}_{\pm}$ defined in (\ref {BES-BAS}, \ref {BES-BAS2}). The (rest) masses of these resonances are protected at all values of the coupling to be the purely additive values $E^{(M)}_m(u=\pm \infty)=2m$ (zero binding energy), which are indeed stable as long as the momentum is zero. Yet, these acquire full (any momentum) stability in the strong coupling limit where boost invariance is recovered. In fact, if we enter the perturbative regime $g \rightarrow +\infty$, $u=\sqrt{2}g \bar u$, with $|\bar u|>1$, we obtain the relativistic dispersion relations
\be
E_m^{(M)}(\theta )= 2m \cosh \theta \, , \quad P_m^{(M)}(\theta )=2m \sinh \theta \, , \label {E-mom-mes}
\ee
in terms of the hyperbolic rapidity $\theta$, defined via $\bar u =\coth 2\theta $.

\section {Pentagonal amplitudes at strong coupling (perturbative regime) and confinement}\label{sez10}
\setcounter{equation}{0}

An important application of the above scattering data, which implies a non-trivial check of them, is the construction of the so-called pentagonal amplitudes, $P$ \cite{BSV1, BSV2, BSV3, BSV4, BSV5}. The latter, in their turn, are the building blocks of an infinite expansion -- the BSV series -- of the gluonic (MHV) scattering amplitudes. In this section, we want to compute the pentagonal factors, $P$, relevant at large $g$, so to prepare the analysis of the BSV series (at strong coupling) in next section.

The BSV series is a sum over the (intermediate) multi-particle states, where the particles may be, -- at generic finite coupling --, scalars, fermions, gluons and bound states thereof, as analysed above. The simplest example is provided by the six-particle amplitude (or, in other terms, the equivalent hexagonal Wilson loop)
\ba\label{Wesa}
W_{hex} &=& \sum _{N=0}^{+\infty} \frac{1}{N!} \sum_{a_1}\cdots\sum_{a_N}\int \prod _{k=1}^N \left [ \frac{du_k}{2\pi} \mu_{a_k} (u_k) e^{-\tau E_{a_k}(u_k) +i\sigma p_{a_k}(u_k) +im_{a_k} \phi} \right ] \times \nn\\
&\times & P_{a_1...a_N}(0|u_1...u_N) P_{a_1...a_N}(-u_N...-u_1|0) \, ,
\ea
which is expressed by means of the measures $\mu_{a_i}(u_i)$ (corresponding to quadrangular amplitudes) and the multi-particle pentagonal amplitudes $P_{a_1...a_N}(0|u_1...u_N)$, representing the transition from the vacuum to an intermediate state with $N$ particles of the kinds listed above, each one associated to a label $a_i$. When we go to the strong coupling limit, we have to disentangle the integrations over internal rapidities by performing the limit $g\rightarrow +\infty$ in the integrand. This procedure means that we have to add different contributions.

The first one comes from performing the limit $g\rightarrow +\infty$
with integration variables fixed. This part depends on excitations
in the non-perturbative regime and is dominated by scalars, and may reserve very interesting surprises as anticipated in \cite{BSV4}. In fact, this contribution would come from a (genuinely) quantised string in $S^5$ and would elude the minimal area argument of the $AdS_5$ string. However, this regime misses contributions from regions in which rapidities are large: these are recovered by adding the integrals in which integration rapidities are scaled
before taking the limit $g\rightarrow +\infty$.

More precisely, if we scale the integration rapidity $u=\sqrt{2}g \bar u$, with $\bar u$ fixed, we have the two following regimes. If $|\bar u| >1$ we are in the giant hole regime. In this regime all the excitations behave in the same way.
In particular, as we showed in Section 4, scattering
phases $- i \ln S$ between any pair of excitations are all the same and are all proportional to the coupling $g$ (\ref{der2theta}). The same happens to energies and momenta  \cite {Basso}. This property is crucial, since it implies that contributions to scattering amplitudes coming from integrations in these regions (all scaled rapidities $|\bar u| >1$) are exponentially suppressed.

Instead, things are different if $|\bar u| <1$, i.e. in the (string) perturbative regime (for all particles except scalars: the rapidity of the latter in this regime does not scale, but instead $\bar u= u+\frac{2}{\pi} \ln m(g)$, where $m(g)\sim g^{1/4} e^{-\frac{\pi}{\sqrt{2}} g}$, as seen above). In this regime, energy, momentum and scattering factors are expanded in inverse powers of the coupling constant $g$. Additional structure is added when expressing the pentagonal transition between a $M$ particle state to a $N$ particle state in terms of the one-particle-to-one-particle transitions because of the matrix representation carried by the single particle (thus the singlets makes an exception to this). In this operation polynomials in the rapidity appear as denominators, taking into account the different representations to which the $S$ matrices can belong. For instance, in the case of the hexagon (\ref{Wesa}), extensively discussed below, as we start from the GKP vacuum, we need consider only pentagonal amplitudes to the other possible singlet states. In particular, this polynomial is a monomial in the case of the transition (from the vacuum) to a two particle state of a fermion and an antifermion (which, though, belong to the $4$ and $\bar 4$, respectively). This monomial  'squares' in the integrand of the amplitude contribution to (\ref{Wesa})
\be
P^{(f\bar f)}(0|u,v)P^{(\bar f f)}(-v,-u|0)=\frac{1}{(u-v)^2+4}\,\frac{1}{P^{(f\bar f)}(u|v)P^{(\bar f f)}(v|u)} \, .
\label{ff-ampl}
\ee
Instead, the transition from the vacuum into the two scalar singlet is even more depressed, albeit the rapidity does not scale (for a scalar, but is added a $g$-depending constant). In fact, the $P$ factor contains at the denominator a polynomial of degree $2$ multiplied by $g^2$ ({\it cf.} \cite{BSV2, BSV3} for details), and then the two scalar contribution to the hexagonal amplitude writes down:
\be
W_{hex}^{(ss)}=3\int\frac{du dv}{(2\pi)^2}\,\frac{\mu^s(u)\mu^s(v)}{g^4[(u-v)^2+4][(u-v)^2+1]}
\,\frac{e^{-\tau[E^s(u)+E^s(v)]+i\sigma[p^s(u)+p^s(v)]}}
{P^{(ss)}(u|v)P^{(ss)}(v|u)} \, ,
\ee
where we ought to consider that $\mu^s(u)=O(g)$ and $P^{(ss)}(u|v)=O(1/g)$. Hence, this integral turns out to be of order $W_{hex}^{(ss)}=O(g^0)$, then subdominant with respects to semi-classical approximation (contributed by the gluons, for instance). Actually, while the scalar contributions are really subdominant in the perturbative regime, on the contrary fermion ones behave in a subtle manner: in fact, the lorentzian function in front of (\ref{ff-ampl}) would entail a contribution from the singularity $\bar u - \bar v=\pm \sqrt{2} i/g$ pinching the real axis when $g\rightarrow +\infty$ \cite{BSV3}. But in our picture this is is the contribution given by their bound state, the meson indeed. Moreover, also the greater multi-fermion coalescence are taken into account by the multi-meson and meson-bound-state contributions, {\it cf.} below. In summary, we are in the presence of a phenomenon in which the fermions coalesces at least in a fermion and antifermion couple and disappear from the spectrum as free particles: as anticipated, this is a sort of {\it confinement} typical of MHV gluon scattering amplitudes/Wilson loops at strong coupling, not evident at first glance from the 2D scattering factors. To be fully precise, although the string theory minimal surface confirms this disappearance, and the appearance of the meson \cite{AM, FTT} and its bound states \cite{TBA-amp1, TBA-amp2, TBA-amp3}, nevertheless a detailed multi-fermion description is missing so far \cite{35}.

Concluding this preamble, these polynomials in the denominator produce in the general case negative powers of the coupling constant after scaling\footnote{Rapidity of scalars do not need to scale.} the rapidities and thus 'depress' the amplitude. Of course, these polynomials are absent if the excitations belong to a $SU(4)$ singlet (see also \cite {BSV2, BSV3} for a detailed analysis of the two particle case). Therefore, we can argue that the leading contributions in the perturbative regime are due to particles behaving as singlets under $SU(4)$, indeed. They are gluons and their bound states, as already proven by the detailed two particle analysis of \cite{BSV3}. But, at strong coupling, we have shown necessary to add mesons and their bound states to the spectrum, as well.

Now, the pentagonal amplitudes $P$ enjoy at general coupling a series of axioms depending on the $S$-matrix entries. Therefore for the latter we need to use our previous (strong coupling) perturbative expansions at leading order and 'solve' the axioms. For exposition's sake, we give in the following the complete list of the $P$ factors (gluon-gluon, gluon-meson, meson-meson and bound states, contributing at leading order), leaving the details of their derivation in the Appendix C\footnote {We have to say that the expansion of the gluon-gluon $P$ factors -- formul{\ae} (\ref {Pgg},\ref {Pgbarg}) -- previously appeared in \cite {BSV1}.}.

We start from the gluon and then the bound states of $\ell$ of them. In this gluonic sector the rapidity enjoys (at perturbative strong coupling) the parametrisation $u=\sqrt{2}g \tanh 2\theta$. Thanks to this, the three axioms (6-8) in \cite{BSV1} for the gluon ($g$) and its barred companion ($\bar g$, the other component of the massless spin $1$ field) simplify their arguments:
\be
P^{(gg)}(-\theta|-\theta ')=P^{(gg)}(\theta '|\theta),  \,\,\,\,\, P^{(g\bar g)}(-\theta|-\theta ')=P^{(g\bar g)}(\theta '|\theta) \, ,
\ee
\be
P^{(gg)}(\theta |\theta ')=S^{(gg)}(\theta ,\theta ') P^{(gg)}(\theta '|\theta ), \,\,\,\,\, P^{(g\bar g)}(\theta |\theta ')=S^{(g\bar g)}(\theta ,\theta ') P^{(g\bar g)}(\theta '|\theta ) \, ,
\ee
\be
P^{(gg)}(\theta -i\pi/2|\theta ')= P ^{(g\bar g)} (\theta '|\theta ) \, ,
\label{g-axioms}
\ee
and we can solve them with input the leading order expansion of the gluon-gluon scattering matrix (\ref {Sggtheta}, \ref {Sgbartheta}). We obtain
\be
\alpha P^{(gg)}(\theta, \theta ')=1+\frac{i}{2\sqrt{2}g} \frac{\cosh 2\theta \cosh 2\theta '}{\sinh (2\theta -2\theta ')}\left [1+\cosh (\theta -\theta ')-i\sinh (\theta -\theta ') \right ] + O(1/g^2) \label {Pgg} \, ,
\ee
\be
\alpha P^{(g\bar g)}(\theta, \theta ')=1+\frac{i}{2\sqrt{2}g} \frac{\cosh 2\theta \cosh 2\theta '}{\sinh (2\theta -2\theta ')}\left [-1+\cosh (\theta -\theta ')-i\sinh (\theta -\theta ') \right ] + O(1/g^2) \label {Pgbarg} \, ,
\ee
and the symmetric channels $P^{(\bar g\bar g)}(\theta , \theta ')=P^{(gg)}(\theta , \theta ')$, $P^{(\bar g g)}(\theta , \theta ')=P^{(g\bar g)}(\theta , \theta ')$. The constant $\alpha $ may equal $\pm 1$: its precise value is not fixed by the axioms, but by the comparison with data derived from the Thermodynamic Bubble Ansatz (TBA) in \cite {TBA-amp2}. As we wrote above, formul{\ae} (\ref {Pgg}, \ref {Pgbarg}) with $\alpha =1$ have been already reported in \cite {BSV1}.

For what concerns $P$ factors of gluon bound states, we may conjecture, along the lines of the previous equations (\ref{g-axioms}) for the single gluons, the following functional relations as axioms: $P^{(gg)}_{ml}(\theta , \theta ')=P^{(\bar g\bar g)}_{ml}(\theta , \theta ')$, $P^{(g\bar g)}_{ml}(\theta , \theta ')=P^{(\bar g g)}_{lm}(\theta , \theta ')$ and moreover
\ba
&& P_{lm}^{(gg)}(-\theta , -\theta ')=P_{ml}^{(gg)}(\theta ', \theta ) \, , \quad P_{lm}^{(gg)}(\theta -i\pi/2,\theta ')= P_{ml}^{(g\bar g)}(\theta ', \theta ) \, , \nonumber \\
&& P^{(gg)}_{lm}(\theta , \theta ')=S^{(gg)}_{lm}(\theta , \theta ')P^{(gg)}_{ml}(\theta ', \theta ) \, , \quad
P^{(g\bar g)}_{lm}(\theta , \theta ')=S^{(g\bar g)}_{lm}(\theta , \theta ')P^{(\bar gg)}_{ml}(\theta ', \theta )
\label {Pgglm-post} \, .
\ea
Moreover, we recall that the $S$-matrix factors are simply multiplicative at perturbative strong coupling: $S^{(gg)}_{ml}(\theta, \theta ')=[S^{(gg)}(\theta, \theta ')]^{ml}$
and $S^{(g\bar g)}_{ml}(\theta, \theta ')=[S^{(g\bar g)}(\theta, \theta ')]^{ml}$. Therefore, solutions to (\ref {Pgglm-post}) should enjoy the same property, which entails upon expansion for large $g$
\be
\alpha _{ml}P^{(gg)}_{ml}(\theta, \theta ')=1+\frac{iml}{2\sqrt{2}g} \frac{\cosh 2\theta \cosh 2\theta '}{\sinh (2\theta -2\theta ')}\left [1+\cosh (\theta -\theta ')-i\sinh (\theta -\theta ') \right ] + O(1/g^2) \label {Pmgg} \, ,
\ee
\be
\alpha _{ml}P^{(g\bar g)}_{ml}(\theta, \theta ')=1-\frac{iml}{2\sqrt{2}g} \frac{\cosh 2\theta \cosh 2\theta '}{\sinh (2\theta -2\theta ')}\left [1-\cosh (\theta -\theta ')+i\sinh (\theta -\theta ') \right ] + O(1/g^2) \label {Pmgbarg} \, ,
\ee
or in barred rapidities $\bar u= \tanh 2\theta$
\ba
&&\alpha _{ml} P_{ml}^{(gg)}(\bar u, \bar u')=1+\frac{iml}{2\sqrt{2}g} \frac{1}{\bar u -\bar u'} \Bigl [ 1+ \nn \\
&& + \frac{1}{2}(1-i) \left ( \frac{1+\bar u}{1-\bar u} \right )^{1/4} \left ( \frac{1-\bar u '}{1+\bar u'} \right )^{1/4} + \frac{1}{2}(1+i) \left ( \frac{1+\bar u'}{1-\bar u'} \right )^{1/4} \left ( \frac{1-\bar u }{1+\bar u} \right )^{1/4} \Bigr ]
\ea
\ba
&&\alpha _{ml} P_{ml}^{(g\bar g)}(\bar u, \bar u')=1-\frac{iml}{2\sqrt{2}g} \frac{1}{\bar u -\bar u'} \Bigl [ 1-  \nn\\
&& - \frac{1}{2}(1-i) \left ( \frac{1+\bar u}{1-\bar u} \right )^{1/4} \left ( \frac{1-\bar u '}{1+\bar u'} \right )^{1/4} - \frac{1}{2}(1+i) \left ( \frac{1+\bar u'}{1-\bar u'} \right )^{1/4} \left ( \frac{1-\bar u }{1+\bar u} \right )^{1/4} \Bigr ]
\ea
Overall constants $\alpha _{ml}=\alpha_{lm}$ can be equal to $\pm 1$ and are constrained by the comparison with the TBA of \cite{TBA-amp2}.

\medskip

Let us now consider the meson and its bound states, and in particular recall that for all of them the rapidity enjoys the perturbative parametrisation $u=\sqrt{2}g \coth 2\theta$. As we discussed before, these are
self-conjugate particles and this property allows us to postulate the following set of functional relations (which now will be meaningful only in the perturbative strong coupling regime, where the particle does exist) for the single meson $P$ factor:
\ba
\label{axiomMeson}
&& P^{(MM)}(\theta,\theta')=P^{(MM)}(-\theta ',-\theta) \, , \quad \nn \\
&& P^{(MM)}(\theta,\theta')=S^{(MM)}(\theta,\theta')P^{(MM)}(\theta ',\theta) \,  , \nn \\
&& P^{(MM)}(\theta -i\pi/2,\theta')=P^{(MM)}(\theta ',\theta ) \ , \label {PMM-ax}
\ea
where $S^{(MM)}$ is given by (\ref {sMM-str}). We write the solution of (\ref {axiomMeson}) as
\be
\beta P^{(MM)}(\theta, \theta')=1-\frac{1}{\sqrt{2}g}\frac{i\sinh 2\theta \sinh 2\theta '}{ \sinh (2\theta -2\theta ')}
\sqrt{2}\cosh \left (\theta -\theta ' -i\frac{\pi}{4} \right ) +O(1/g^2)  \, ,  \label {KMMstrongsmall}
\ee
where $\beta =\pm 1$.
For mesons bound states, we have anew the multiplicativity of the scattering factors in the perturbative regime, namely $S^{(MM)}_{ml}(\theta, \theta ')=[S^{(MM)}(\theta, \theta ')]^{ml}$. Which, in its turn, imply the same property on $P$ factors, i.e. upon expanding at large $g$
\be
\beta _{ml} P^{(MM)}_{ml}(\theta, \theta ')=1-\frac{ ml}{\sqrt{2}g}\frac{i \sinh 2\theta \sinh 2\theta '}{ \sinh (2\theta -2\theta ')}
\sqrt{2}\cosh \left (\theta -\theta ' -i\frac{\pi}{4} \right )  + O(1/g^2) \label {PmMM} \, ,
\ee
or in barred variables $\bar u=\coth 2\theta$
\ba
&&\beta _{ml} P_{ml}^{(MM)}(\bar u, \bar u')=1+\frac{iml}{2\sqrt{2}g} \frac{1}{\bar u -\bar u'} \cdot  \nn \\
&& \cdot \Bigl [ (1-i) \left ( \frac{\bar u +1}{\bar u -1} \right )^{1/4} \left ( \frac{\bar u '-1}{\bar u' +1} \right )^{1/4} + (1+i) \left ( \frac{\bar u -1}{\bar u +1} \right )^{1/4} \left ( \frac{\bar u '+1}{\bar u '-1} \right )^{1/4} \Bigr ] \, ,
\ea
where $\beta _{ml}=\beta _{lm}$ can equal $\pm 1$.

\medskip

Eventually\footnote {Even if contributions of small fermions to amplitudes is suppressed at strong coupling with respect to gluons and mesons, we give also the strong coupling limit of their P factors.
We refer to formul{\ae} (38) of \cite {BSV3} and use
formul{\ae} (\ref {Sfftheta}, \ref {Sffmir}, \ref {Sfbarfmir}) for the (strong coupling) perturbative regime of the fermion-(anti)fermion scattering factor and its mirror, respectively.
We eventually obtain
\ba
&&[P^{(ff)}(\theta, \theta ')]^2 = -\frac{\sinh \theta \sinh \theta ' \sinh 2\theta \sinh 2\theta '}{g^2 \sinh (\theta -\theta ') \sinh (2\theta -2\theta ')} \Bigl [ 1+\frac{i}{2\sqrt{2}g}\frac{\sinh 2\theta \sinh 2\theta '}{\sinh (2\theta -2\theta ')} \left ( 1-\cosh (\theta -\theta ')+i\sinh (\theta -\theta ')\right ) \Bigr ] \nn \\
&&[P^{(f\bar f)}(\theta, \theta ')]^2 = \frac{\cosh \theta \cosh \theta '}{\cosh (\theta -\theta ')}
\Bigl [ 1+\frac{i}{2\sqrt{2}g}\frac{\sinh 2\theta \sinh 2\theta '}{\sinh (2\theta  -2\theta ')} \left ( 1-\cosh (\theta -\theta ')+i\sinh (\theta -\theta ')\right ) \Bigr ]
\label {Pfbarf} \, .
\ea}, we consider the scattering between (bound states of) mesons and (bound states of) gluons.
We are now looking for functions $P^{(Mg)}_{ml}$, $P^{(M\bar g)}_{ml}$, $P^{(gM)}_{ml}$, $P^{(\bar g M)}_{ml}$ which may conjecturally satisfy the functional properties (meaningful only at perturbative strong coupling)
\ba
&& P^{(ab)}_{ml}(-\theta , -\theta ')=P^{(ba)}_{lm}(\theta ', \theta ) \, , \nn \\
&& P^{(Mg)}_{ml}(\theta , \theta ')=S^{(Mg)}_{ml}(\theta , \theta ') P^{(gM)}_{lm}(\theta ', \theta ) \, , \quad
P^{(M\bar g)}_{ml}(\theta , \theta ')=S^{(M\bar g)}_{ml}(\theta , \theta ') P^{(\bar gM)}_{lm}(\theta ', \theta ) \, , \nn \\
&& P^{(Mg)}_{ml}\left (\theta - \frac{i\pi}{2} , \theta '\right )=P^{(\bar g M)}_{lm}(\theta ', \theta ) \, , \quad
P^{(M\bar g)}_{ml}\left (\theta - \frac{i\pi}{2} , \theta '\right )=P^{(g M)}_{lm}(\theta ', \theta )  \, .
\label {Pmg-ax}
\ea
We write solutions to these equations in the form $\gamma ^{(ab)}_{ml}P^{(ab)}_{ml}=1+\frac{ml}{\sqrt{2}g}K^{(ab)}+O\left (\frac{1}{\lambda} \right )$, where
$\gamma ^{(ab)}_{ml}$ can equal $\pm 1$, $\gamma ^{(Mg)}_{ml}=\gamma ^{(gM)}_{lm}=\gamma ^{(M\bar g)}_{ml}=\gamma ^{(\bar gM)}_{lm}$ and
\ba
&& K^{(Mg)}(\theta , \theta ')=K^{(M\bar g)}(\theta , \theta ')= \frac{\sinh 2\theta \cosh 2\theta '}{\sqrt{2} \cosh (2\theta -2\theta ')}[ \sinh (\theta -\theta ')+i \cosh (\theta -\theta ') ]  \, , \label {KMg}\\
&& K^{(gM)}(\theta ', \theta )=K^{(\bar g M)}(\theta ' , \theta )= \frac{\sinh 2\theta \cosh 2\theta '}{\sqrt{2} \cosh (2\theta -2\theta ')}[ \sinh (\theta -\theta ')-i \cosh (\theta -\theta ') ] \label {KgM}\, ,
\ea
or alternatively in the barred variables
 \be
K^{(Mg)}(\bar u, \bar u')=\frac{1}{2\sqrt{2}} \frac{1}{\bar u -\bar u'}\Bigl [ (1+i) \left ( \frac{\bar u +1}{\bar u -1} \right )^{1/4} \left ( \frac{1-\bar u '}{1+\bar u'} \right )^{1/4} - (1-i) \left ( \frac{\bar u -1}{\bar u +1} \right )^{1/4} \left ( \frac{1+\bar u '}{1-\bar u '} \right )^{1/4} \Bigr ]
\ee
\be
K^{(gM)}(\bar u ', \bar u)=\frac{1}{2\sqrt{2}} \frac{1}{\bar u -\bar u'}\Bigl [ (1-i) \left ( \frac{\bar u +1}{\bar u -1} \right )^{1/4} \left ( \frac{1-\bar u '}{1+\bar u'} \right )^{1/4} - (1+i) \left ( \frac{\bar u -1}{\bar u +1} \right )^{1/4} \left ( \frac{1+\bar u '}{1-\bar u '} \right )^{1/4} \Bigr ] .
\ee

\medskip

\section{Hexagon at strong coupling}\label{sez11}
\setcounter{equation}{0}

\subsection{Aim and assumptions}

Now, we want to compute the hexagonal Wilson loop as the series proposed by \cite {BSV1, BSV2, BSV3}
\ba
W_{hex}&=&\sum _{N=0}^{+\infty} \frac{1}{N!} \sum_{a_1}\cdots\sum_{a_N} \int \prod _{k=1}^N \left [ \frac{du_k}{2\pi} \mu_{a_k} (u_k) e^{-\tau E_{a_k}(u_k) +i\sigma p_{a_k}(u_k) +im_{a_k} \phi} \right ] \times \nonumber \\ &\times & P_{a_1...a_N}(0|u_1...u_N) P_{a_1...a_N}(-u_N...-u_1|0)
\label {Whex-initial}
\ea
at strong coupling. As argued above, in this regime intermediate states which contribute are gluons and their bound states, together with mesons and their bound states. All of them are singlets and then for their pentagonal amplitudes a simple product and inversions hold when changing a rapidity from in to out:
\be
P_{a_1...a_N}(0|u_1...u_N) P_{a_1...a_N}(-u_N....-u_1|0) =
\prod_{i<j}^N\frac{1}{ P_{a_i,a_j}(u_i|u_j)\,P_{a_j,a_i}(u_j|u_i)}.
\label{P-inversion}
\ee
This formula entails an easy product to appear inside the hexagonal amplitude:
\be\label{Whex}
W_{hex}=\sum _{N=0}^{+\infty} \frac{1}{N!} \sum_{a_1}\cdots\sum_{a_N} \int \prod _{k=1}^N \left [ \frac{du_k}{2\pi} \mu_{a_k} (u_k) e^{-\tau E_{a_k}(u_k) +i\sigma p_{a_k}(u_k) +im_{a_k} \phi} \right ]
\displaystyle\prod_{i<j}^N\frac{1}{ P_{a_i,a_j}(u_i|u_j)\,P_{a_j,a_i}(u_j|u_i)} \, ,
\ee
where the indices $a_k$ label the species of different particles (including bound states): this is the formula we want first to match with initially, and then to sum up.

For the gluon and the bound states of $\ell$ of them, rapidity may be parametrised as $u=\sqrt{2}g \tanh 2\theta$. Then, we can recall their relativistic energy and momentum (\ref {EP-glum}) and notice how they are purely additive
\be
E_{\ell}^g(u)= \sqrt{2} \ell \cosh \theta = \ell E_1^g(u) \, , \qquad p_{\ell}^g(u)=\sqrt{2} \ell \sinh \theta= \ell p_1^g(u) \label {glu-Ep} \, .
\ee
In the following, the expression will be written in terms of
$$
\frac{\sqrt{\lambda}}{2\pi}=\sqrt{2}g
$$
in order to facilitate the comparison with the existing literature on this subject.
Gluonic measures appearing in (\ref {Whex}) is given by ($\ell=1$ is the gluon)
\be
\frac{du}{2\pi} \mu _{\ell}^g (u)=\frac{d\theta}{2\pi} \mu _{\ell}^g (\theta)=\frac{i}{\lim \limits_{\theta '\rightarrow \theta} (\theta '-\theta )P_{\ell \ell}^{(gg)}(\theta , \theta ') } \, \frac{d\theta}{2\pi} \, ,
\ee
with $P_{\ell \ell}^{(gg)}$ given by (\ref {Pmgg}). In order to have agreement with TBA it is enough to choose $\alpha _{\ell \ell}=(-1)^{\ell-1}$. Then, in terms of the single gluon measure (at strong coupling)
\be
\mu^g (u)=-1+\dots
\ee
we can express the $\ell$ gluon bound-state ones
\be
\frac{du}{2\pi} \mu _{\ell}^g (u)=\frac{\sqrt{\lambda}}{2\pi} (-1)^{\ell}\frac{d\theta}{\pi \ell ^2 \cosh ^2 2\theta}=\frac{du}{2\pi} \frac{(-1)^{\ell-1}}{\ell^2}\mu^g (u)=\frac{d\theta}{2\pi} \frac{(-1)^{\ell-1}}{\ell^2}\mu^g (\theta)\label {glu-meas} \, .
\ee
On the other hand, let us remind that the meson and its bound-states enjoy the rapidity parametrisation $u=\sqrt{2}g \coth 2\theta$ with relativistic energy and momentum (\ref {E-mom-mes}) which are purely additive (zero binding energy) as well
\be
E_{m}^M(u)=2 m  \cosh \theta = m E_1^M(u) \, , \qquad p_{m}^M(u)= 2 m  \sinh \theta= m p_1^M(u)
\label {mes-Ep} \, .
\ee
The measure for bound states of $m$ mesons is
\be
\frac{du}{2\pi} \mu _{m}^M (u)=\frac{d\theta}{2\pi} \mu _{m}^M (\theta)=\frac{i}{\lim \limits_{\theta '\rightarrow \theta} (\theta '-\theta )P_{mm}^{(MM)}(\theta , \theta ') } \, \frac{d\theta}{2\pi}  \, ,
\ee
with $P_{mm}^{(MM)}$ given by (\ref {PmMM}). We choose $\beta _{mm}=(-1)^{m-1}$ in order to have agreement with TBA.
Then, similarly to gluons, we obtain the meson bound-state measures
\be
\frac{du}{2\pi} \mu _{m}^M (u)=\frac{\sqrt{\lambda}}{2\pi} (-1)^{m-1}\frac{ d\theta}{\pi m^2 \sinh ^2 2\theta} =\frac{du}{2\pi} \frac{(-1)^{m-1}}{m^2}\mu^M (u) =\frac{d\theta}{2\pi} \frac{(-1)^{m-1}}{m^2}\mu^M (\theta) \label {mes-meas} \, ,
\ee
where the single meson measure (at strong coupling)
\be
\mu^M (u)=-1+\dots \, .
\ee

We shall point out that, for compactness sake, the paths of integration are not explicitly written in what follows, so we need to describe now the choices we adopt throughout the rest of the section. The integrations over gluon rapidity $u$ can be computed along the interval $[-\frac{\sqrt{\lambda}}{2\pi}, +\frac{\sqrt{\lambda}}{2\pi}]$, which turns to the whole real axis when the map to hyperbolic rapidities $\q$ is performed
$u=\frac{\sqrt{\lambda}}{2\pi}\tanh (2\q)$. Otherwise, the integrations for meson $u$-rapidity is intended on a section of a straight line below the real axis, lying in the small fermion sheet \cite{TBA-amp2} \cite{BSV3} (we remind that large fermions in this regime behave as giant holes, hence their contributions are exponentially suppressed because of their dispersion laws \cite{Basso} and scattering factors (\ref{SFFgh})\,), namely
$(+\infty-i\epsilon,+\frac{\sqrt{\lambda}}{2\pi})\cup(-\frac{\sqrt{\lambda}}{2\pi},-\infty-i\epsilon)$, where $0<\epsilon<2$. When switching to meson $\q$ rapidity according to the map $u=\frac{\sqrt{\lambda}}{2\pi}\coth (2\q)$, the integration may be set to run along the straight line $(-\infty+i\hat\varphi,+\infty+i\hat\varphi)$, where $\hat\varphi$ (according to \cite{TBA-amp2}) can be fixed in terms of conformal ratios, as shown below (\ref{hatvarphi}). Eventually, the same straight line is chosen for the integrations on the gluonic $\q$ rapidity.

\subsection{One particle}

Let us start from one particle contribution. With 'one particle contribution' we mean that in (\ref {Whex})
we consider only one insertion, which can be a gluon, a meson or bound states of thereof. Therefore, we can write down
\be
W_{hex}^{(1)}=\sum _{\ell=1}^{+\infty} \int \frac{du}{2\pi} \mu _{\ell}^g (u) e^{-\tau E_{\ell}^g(u) +i\sigma p_{\ell}^g(u)}
\left ( e^{i\ell \phi} + e^{-i\ell \phi} \right )+\sum _{m=1}^{+\infty} \int \frac{du}{2\pi} \mu _{m}^M (u) e^{-\tau E_{m}^M(u) +i\sigma p_{m}^M(u)}
\ee
which at strong coupling reads
\ba
W_{hex}^{(1)}&=&+\frac{\sqrt{\lambda}}{2\pi} \sum _{\ell=1}^{+\infty} \int
\frac{d\theta}{\pi \ell ^2 \cosh ^2 2\theta} (-1)^{\ell}  e^{- \sqrt{2} \tau \ell \cosh \theta +i \sqrt{2} \sigma \ell \sinh \theta}
\left ( e^{i\ell \phi} + e^{-i\ell \phi} \right )- \nonumber \\
&-& \frac{\sqrt{\lambda}}{2\pi} \sum _{m=1}^{+\infty} \int
\frac{d\theta}{\pi m^2 \sinh ^2 2\theta} (-1)^m e^{-2 \tau m \cosh \theta +2 i \sigma m  \sinh \theta} = \nonumber \\
&=& \frac{\sqrt{\lambda}}{2\pi} \int
\frac{d\theta}{\pi \cosh ^2 2\theta} \left [ \textrm{Li}_2 \left (-e^{- \sqrt{2} \tau  \cosh \theta +i \sqrt{2} \sigma \sinh \theta +i\phi} \right ) + \textrm{Li}_2 \left (-e^{-\sqrt{2} \tau \cosh \theta +i \sqrt{2} \sigma \sinh \theta -i\phi} \right )
\right ]- \nonumber \\
&-& \frac{\sqrt{\lambda}}{2\pi} \int
\frac{d\theta}{\pi \sinh ^2 2\theta}  \textrm{Li}_2 \left (-e^{- 2 \tau \cosh \theta +2i \sigma  \sinh \theta} \right )
\equiv W_{hex}^{(g)} + W_{hex}^{(M)}\, , \label {1-pt}
\ea
where $\textrm{Li}_2(z)=\sum\limits _{m=1}^{+\infty} \frac{z^m}{m^2}$ is the dilogarithm function.

\subsection{Two particles}

Let us pass to two particle terms $W_{hex}^{(2)}=W_{hex}^{(gg)}+W_{hex}^{(MM)}+W_{hex}^{(Mg)}$, in which we distinguish three contributions: gluon-gluon, meson-meson and gluon-meson.
\medskip

{\bf Gluon-gluon}

\medskip

Let us start from gluon-gluon:
\ba
W_{hex}^{(gg)}&=&\frac{1}{2}\sum _{\ell _1=1}^{+\infty} \sum _{\ell _2=1}^{+\infty}\int \frac{du_1}{2\pi} \mu _{\ell _1}^g (u_1)
\frac{du_2}{2\pi} \mu _{\ell _2}^g (u_2) e^{-\tau E_{\ell _1}^g(u_1) +i\sigma p_{\ell _1}^g(u_1) }e^{-\tau E_{\ell _2}^g(u_2) +i\sigma p_{\ell _2}^g(u_2) }  \cdot \label {Wgg}
 \\
&\cdot & \left \{ \frac{e^{i(\ell _1+\ell _2) \phi} +  e^{-i(\ell _1+\ell _2) \phi}}{P_{\ell _1 \ell _2}^{(gg)}(u_1|u_2) P_{\ell _2 \ell _1}^{(gg)}(u_2|u_1)} +
\frac{e^{i(\ell _1-\ell _2) \phi}+e^{-i(\ell _1-\ell _2) \phi}}{P_{\ell _1 \ell _2}^{(g\bar g)}(u_1|u_2) P_{\ell _2 \ell _1}^{(\bar gg)}(u_2|u_1)} \right \}
\nonumber \, .
\ea
At strong coupling the symmetric product of $P$ factors (\ref {Pgg}, \ref {Pgbarg}), entering (\ref {Wgg}), enjoys the property
\be\label{Kbound}
\frac{1}{P_{\ell _1 \ell _2}^{(gg)}(u_1|u_2) P_{\ell _2 \ell _1}^{(gg)}(u_2|u_1)}=
\frac{1}{P_{\ell _1 \ell _2}^{(g\bar g)}(u_1|u_2) P_{\ell _2 \ell _1}^{(\bar gg)}(u_2|u_1)}=
1-\frac{2\pi}{\sqrt{\lambda}} \ell _1 \ell _2
K_{sym} ^{(gg)}(\theta _1, \theta _2) \, ,
\ee
where
\be\label{kernel}
K_{sym} ^{(gg)}(\theta _1, \theta _2)=  \frac{\cosh 2\theta _1 \cosh 2\theta _2}{2\cosh (\theta _1-\theta _2)} \, .
\ee
We get
\ba
W_{hex}^{(gg)}&=& \left ( -\frac{\sqrt{\lambda}}{2\pi} \right )^2 \frac{1}{2}\sum _{\ell _1=1}^{+\infty} \sum _{\ell _2=1}^{+\infty}
 \int \frac{d\theta _1}{\pi \ell _1 ^2\cosh ^2 2\theta_1} \frac{d\theta _2}{\pi \ell _2 ^2\cosh ^2 2\theta_2}  (-1)^{\ell _1 +\ell _2}
 \cdot \nonumber \\
&\cdot & 4 \cos \ell _1 \phi \, \cos \ell _2 \phi  \, e^{-\sqrt{2} \tau \ell _1 \cosh \theta _1 +i \sqrt{2} \sigma \ell _1 \sinh \theta _1} e^{-\sqrt{2} \tau \ell _2 \cosh \theta _2 +i \sqrt{2} \sigma \ell _2 \sinh \theta _2} + \\
&+& \left ( -\frac{\sqrt{\lambda}}{2\pi} \right ) \frac{1}{2}\sum _{\ell _1=1}^{+\infty} \sum _{\ell _2=1}^{+\infty}
 \int  \frac{d\theta _1}{\pi \ell _1 \cosh ^2 2\theta_1} \frac{d\theta _2}{\pi \ell _2 \cosh ^2 2\theta_2}
K_{sym}^{(gg)} (\theta _1, \theta _2) (-1)^{\ell _1 +\ell _2} \cdot \nonumber \\
&\cdot & 4 \cos \ell _1 \phi \, \cos \ell _2 \phi  \, e^{-\sqrt{2} \tau \ell _1 \cosh \theta _1 +i \sqrt{2} \sigma \ell _1 \sinh \theta _1} e^{-\sqrt{2} \tau \ell _2  \cosh \theta _2 +i \sqrt{2} \sigma \ell _2 \sinh \theta _2} \, .
\ea
We can now perform the sums over $\ell _1$, $\ell _2$ and get
\be
W_{hex}^{(gg)}=\frac{1}{2} \left [ W_{hex}^{(g)} \right ]^2 - \frac{1}{2} \frac{\sqrt{\lambda}}{2\pi}
\int  \frac{d\theta _1}{\pi \cosh ^2 2\theta_1} \frac{d\theta _2}{\pi  \cosh ^2 2\theta_2} K_{sym}^{(gg)} (\theta _1, \theta _2) L^{g}(\theta _1) L^{g}(\theta _2)   \, , \label {2-pt}
\ee
where we used the short notation
\be
L^{g}(\theta)=\ln \left [ \left (1+e^{i\phi-\sqrt{2} \tau \cosh \theta+i\sqrt{2} \sigma \sinh \theta } \right )
\left (1+e^{-i\phi-\sqrt{2} \tau \cosh \theta+i\sqrt{2} \sigma \sinh \theta } \right )\right ] \, .
\ee

\medskip

{\bf Meson-meson}

\medskip

The meson-meson contribution is written in a completely analogous way:
\ba
W_{hex}^{(MM)}&=&\frac{1}{2}\sum _{m_1=1}^{+\infty} \sum _{m_2=1}^{+\infty}\int \frac{du_1}{2\pi} \mu _{m _1}^M (u_1)
\frac{du_2}{2\pi} \mu _{m _2}^M (u_2) \frac{1}{P_{m _1 m _2}^{(MM)}(u_1|u_2) P_{m _2 m _1}^{(MM)}(u_2|u_1)} \cdot
\nonumber \\
&\cdot & e^{-\tau E_{m _1}^M(u_1) +i\sigma p_{m _1}^M(u_1) }e^{-\tau E_{m _2}^M(u_2) +i\sigma p_{m _2}^M(u_2) }
\ea
Expression (\ref {PmMM}) for the strong coupling limit of mesonic P factor implies the property
\be\label{KboundM}
\frac{1}{P_{m_1 m_2}^{(MM)}(u_1|u_2) P_{m_2 m_1}^{(MM)}(u_2|u_1)}=1-\frac{2\pi}{\sqrt{\lambda}} m_1 m_2
K_{sym} ^{(MM)}(\theta _1, \theta _2) \, ,
\ee
where
\be\label{kernelM}
K_{sym} ^{(MM)}(\theta _1, \theta _2)= -\frac{\sinh 2\theta _1 \sinh 2\theta _2}{\cosh (\theta _1-\theta _2)} \, .
\ee
We get
\ba
W_{hex}^{(MM)}&=& \left ( \frac{\sqrt{\lambda}}{2\pi} \right )^2 \frac{1}{2}\sum _{m _1=1}^{+\infty} \sum _{m _2=1}^{+\infty}
 \int \frac{d\theta _1}{\pi m _1 ^2\sinh^2 2\theta_1} \frac{d\theta _2}{\pi m_2 ^2\sinh ^2 2\theta_2}(-1)^{m_1+m_2}
 \cdot \nonumber \\
&\cdot & e^{-2\tau m_1  \cosh \theta _1 +2i \sigma m_1  \sinh \theta _1} e^{-2\tau m_2  \cosh \theta _2 +2i \sigma m_2  \sinh \theta _2} + \nonumber\\
&+& \left ( -\frac{\sqrt{\lambda}}{2\pi} \right ) \frac{1}{2}\sum _{m_1=1}^{+\infty} \sum _{m_2=1}^{+\infty}
 \int \frac{d\theta _1}{\pi m_1 \sinh ^2 2\theta_1} \frac{d\theta _2}{\pi m_2 \sinh ^2 2\theta_2} (-1)^{m_1+m_2}
K_{sym}^{(MM)} (\theta _1, \theta _2)  \cdot \nonumber \\
&\cdot &  e^{-2\tau m_1 \cosh \theta _1 +2i \sigma m_1  \sinh \theta _1} e^{-2\tau m _2  \cosh \theta _2 +2i \sigma m_2 \sinh \theta _2} \nonumber \, .
\ea
We can now perform the sums over $m_1$, $m_2$ and get
\be
W_{hex}^{(MM)}=\frac{1}{2} \left [ W_{hex}^{(M)} \right ]^2 - \frac{1}{2} \frac{\sqrt{\lambda}}{2\pi}
\int   \frac{d\theta _1}{\pi \sinh ^2 2\theta_1} \frac{d\theta _2}{\pi  \sinh ^2 2\theta_2} K_{sym}^{(MM)} (\theta _1, \theta _2) L^{M}(\theta _1) L^{M}(\theta _2)   \label {2-ptM} \, ,
\ee
where we used the short notation
\be
L^{M}(\theta)=\ln \left (1+e^{-2\tau \cosh \theta+2i\sigma \sinh \theta } \right ) \, .
\ee
\medskip

{\bf Meson-gluon}

\medskip

Next step is to consider the meson-gluon contribution
\ba
W_{hex}^{(Mg)}&=&\frac{1}{2}\sum _{m=1}^{+\infty} \sum _{\ell=1}^{+\infty}\int \frac{du_1}{2\pi} \mu _{m }^M (u_1)
\frac{du_2}{2\pi} \mu _{\ell}^g (u_2)
 e^{-\tau E_{m }^M(u_1) +i\sigma p_{m}^M(u_1) }e^{-\tau E_{\ell}^g(u_2) +i\sigma p_{\ell}^g(u_2) } \cdot \nonumber \\
 &\cdot & 2 \left [ \frac{e^{i\ell \phi }}{P_{\ell m }^{(gM)}(u_2|u_1) P_{m  \ell}^{(Mg)}(u_1|u_2)} +
\frac{e^{-i\ell \phi }}{P_{\ell m }^{(\bar gM)}(u_2|u_1) P_{m  \ell}^{(M\bar g)}(u_1|u_2)} \right ] \, .
\ea
Now, at strong coupling, with the redefinitions $u_1=\sqrt{2}g \coth 2\theta_1$, $u_2=\sqrt{2}g \tanh 2\theta_2$
expressions (\ref {KMg}, \ref {KgM}) imply the property
\be
\frac{1}{P_{\ell m }^{(gM)}(u_2|u_1) P_{m  \ell}^{(Mg)}(u_1|u_2)}=
\frac{1}{P_{\ell m }^{(\bar gM)}(u_2|u_1) P_{m  \ell}^{(M\bar g)}(u_1|u_2)}=1-\frac{2\pi}{\sqrt{\lambda}} \ell m K_{sym}^{(Mg)}
(\theta _1, \theta _2) \, ,
\ee
where
\be
K_{sym}^{(Mg)} (\theta _1, \theta _2)= K^{(Mg)} (\theta _1, \theta _2)+K^{(gM)} (\theta _2, \theta _1)=K_{sym}^{(gM)} (\theta _2, \theta _1)=\sqrt{2} \frac{\cosh 2\theta _2 \sinh 2\theta _1 \sinh (\theta _1-\theta _2)}{\cosh (2\theta _2 -2\theta _1)} \, .
\ee
Remembering the measures (\ref {glu-meas}, \ref {mes-meas}) and the forms of energies and momenta (\ref {glu-Ep}, \ref {mes-Ep}) and performing the sums, we arrive at the expression
\be
W_{hex}^{(Mg)}=W_{hex}^{(g)} W_{hex}^{(M)}+ \frac{\sqrt{\lambda}}{2\pi} \int  \frac{d\theta _1}{\pi \sinh ^2 2\theta_1} \frac{d\theta _2}{\pi  \cosh ^2 2\theta_2} K_{sym}^{(Mg)} (\theta _1, \theta _2) L^{M}(\theta _1) L^{g}(\theta _2)
\ee

\subsection{Comparison and checks with TBA}

We now compare our previous predictions at strong coupling with TBA outcome. We use (F.42-F.46) of \cite {TBA-amp2}: these expressions
depend on the functions $\epsilon (\theta - i\hat \varphi)$, $\tilde \epsilon (\theta - i\hat \varphi)$,
which satisfy integral equations
\ba
\epsilon (\theta - i\hat \varphi)&=&E(\theta)  - \int \frac{d\theta '}{\pi \cosh ^2 2\theta '}K_{sym}^{(gg)} (\theta , \theta ') L(\theta ') + \int \frac{d\theta '}{\pi \sinh ^2 2\theta '}K_{sym}^{(Mg)} (\theta ', \theta ) \tilde L(\theta ') \nonumber  \\
&& \label {eps-int} \\
\tilde \epsilon (\theta - i\hat \varphi)&=&\sqrt{2}E(\theta)  - \int \frac{d\theta '}{\pi \cosh ^2 2\theta '}K_{sym}^{(Mg)} (\theta , \theta ') L(\theta ') + \int \frac{d\theta '}{\pi \sinh ^2 2\theta '}K_{sym}^{(MM)} (\theta , \theta ') \tilde L(\theta ') \nonumber \, ,
\ea
where the integrations are performed along the straight line $\mathbb{R}+i\hat\varphi$, and the functions $\ti L(\q),\,L(\q)$ are defined as
$$
\tilde L(\theta )=\ln \left (1+e^{-\ti\epsilon(\theta-i\hat\varphi)} \right ) \, , \quad L(\theta)=\ln \left [ \left (1+\mu e^{-\epsilon(\theta-i\hat\varphi)} \right )  \left (1+\mu ^{-1}e^{-\epsilon(\theta-i\hat\varphi)} \right ) \right ]
$$
while \cite {TBA-amp2}
\be
E(\theta )=-\frac{i}{\sqrt{2}}\sinh\q \ln \left(\frac{u_1}{u_2 u_3}\right)
+i\sinh\left(\q-\frac{i\pi}{4}\right)\ln\left(\frac{1}{u_2}-1\right) \ .
\ee
The relation between $\hat\varphi$ and the cross-ratios can now be clarified by means of the functions $\epsilon,\,\ti\epsilon$, since
\be
\ln \sqrt{\frac{u_1}{u_2 u_3}} = \epsilon(i\pi/4-i\hat\varphi)\ , \ \qquad
\ln\left(\frac{1}{u_2}-1\right)= \ti\epsilon(-i\hat\varphi)\ .
\label{hatvarphi}
\ee
We expand (\ref {eps-int}) at large $E(\theta )$ and plug such expansions in (F.45, F.46)
of \cite {TBA-amp2}, by using the Taylor expansion ($|f|\ll 1$)
\be
\textrm{Li}_2  (-e^{-F-f}) = \textrm{Li}_2  (-e^{-F})+f \ln (1+e^{-F})+ O(f^2) \, .
\ee
 We obtain
\ba
(F.45)+(F.46)&=&-\int
\frac{d\theta}{\pi \cosh ^2 2\theta} \left [ \textrm{Li}_2 \left (-\mu e^{-E(\theta) } \right ) + \textrm{Li}_2 \left (-\mu ^{-1}e^{-E(\theta) } \right ) \right ] + \nonumber \\
&+& \int
\frac{d\theta}{\pi \sinh ^2 2\theta} \textrm{Li}_2 \left (-e^{-\sqrt{2}E(\theta) } \right )+
\nonumber \\
&+& \int \frac{d\theta _1}{\pi \sinh ^2 2\theta _1} \int \frac{d\theta _2}{\pi \sinh ^2 2\theta _2}
K_{sym}^{(MM)} (\theta _1 , \theta _2)  \tilde L _E(\theta _1) \tilde L _E(\theta _2)  + \nonumber \\
&+&  \int \frac{d\theta _1}{\pi \cosh ^2 2\theta _1} \int \frac{d\theta _2}{\pi \cosh ^2 2\theta _2}
K_{sym}^{(gg)} (\theta _1 , \theta _2)   L_E(\theta _1) L_E(\theta _2) - \nonumber \\
&-&  2 \int \frac{d\theta _1}{\pi \sinh ^2 2\theta _1} \int \frac{d\theta _2}{\pi \cosh ^2 2\theta _2}
K_{sym}^{(Mg)} (\theta _1 , \theta _2) \tilde L_E(\theta _1)
L_E(\theta _2) \, ,
\ea
where we used the notations
\be
\tilde L _E(\theta )=\ln \left (1+e^{-\sqrt{2}E(\theta)} \right ) \, , \quad L_E(\theta)=\ln \left [ \left (1+\mu e^{-E(\theta)} \right )  \left (1+\mu ^{-1}e^{-E(\theta)} \right ) \right ] \, .
\ee

Summing this result with (F.42), (F.43), (F.44) we get
\ba
\sum _{k=42}^{46}(F.k)&=&-\int
\frac{d\theta}{\pi \cosh ^2 2\theta} \left [ \textrm{Li}_2 \left (-\mu e^{-E(\theta) } \right ) + \textrm{Li}_2 \left (-\mu ^{-1}e^{-E(\theta) } \right ) \right ] + \nonumber \\
&+& \int
\frac{d\theta}{\pi \sinh ^2 2\theta} \textrm{Li}_2 \left (-e^{-\sqrt{2}E(\theta) } \right )+
\nonumber \\
&+& \frac{1}{2}\int \frac{d\theta _1}{\pi \sinh ^2 2\theta _1} \int \frac{d\theta _2}{\pi \sinh ^2 2\theta _2}
K_{sym}^{(MM)} (\theta _1 , \theta _2) \tilde L_E(\theta _1) \tilde L_E(\theta _2) + \nonumber \\
&+&   \frac{1}{2} \int \frac{d\theta _1}{\pi \cosh ^2 2\theta _1} \int \frac{d\theta _2}{\pi \cosh ^2 2\theta _2}
K_{sym}^{(gg)} (\theta _1 , \theta _2)  L_E(\theta _1) L_E(\theta _2) - \nonumber \\
&-&   \int \frac{d\theta _1}{\pi \sinh ^2 2\theta _1} \int \frac{d\theta _2}{\pi \cosh ^2 2\theta _2}
K_{sym}^{(Mg)} (\theta _1 , \theta _2) \tilde L_E(\theta _1) L_E(\theta _2) \, . \label {tba-res}
\ea
Now, in order to compare with our results, we parametrise the cross-ratios $u_1$, $u_2$, $u_3$
as in \cite {BSV2} (formula (157)), i.e.
\be
\frac{1}{u_2}=1+e^{2\tau} \, , \quad \frac{1}{u_3}=1+(e^{-\tau}+e^{\sigma +i\phi})(e^{-\tau}+e^{\sigma -i\phi}) \, ,
\quad \frac{u_1}{u_2u_3}=e^{2\sigma +2\tau} \label {crossR} \, .\
\ee
Consequently, if we define $\mu=e^{i\phi}$, we obtain the relation
\be
\mu + \mu ^{-1}=\frac{1-u_1-u_2-u_3}{\sqrt{u_1u_2u_3}} \ . \label {mu-phi}
\ee
In addition, the function $E(\theta )$ may be written in a more transparent form as
\be
E(\theta)= \sqrt{2}\tau \cosh \theta -i \sqrt{2} \sigma \sinh \theta \, . \label {E-bsv}
\ee
Plugging (\ref {mu-phi}, \ref {E-bsv}) into (\ref {tba-res}),
the following relation
\be
W_{hex}=W_{hex}^{(1)}+W_{hex}^{(2)}+....=\exp \left ( -\frac{\sqrt{\lambda}}{2\pi} [ \sum _{k=42}^{46}(F.k) ] \right ) \cong 1- \frac{\sqrt{\lambda}}{2\pi}
[ \sum _{k=42}^{46}(F.k) ] + \frac{1}{2} \left ( \frac{\sqrt{\lambda}}{2\pi} \right ) ^{2} [\sum _{k=42}^{46}(F.k) ]^2+.....
\ee
is in agreement with expressions for $W_{hex}^{(1)}, W_{hex}^{(2)}$ computed in last subsection.

\subsection{Re-summation of the BSV series}

The agreement displayed above between the series written in \cite{BSV1} for hexagonal Wilson loops and the TBA for scattering amplitudes \cite{TBA-amp1, TBA-amp2, TBA-amp3} can be made even tighter, since it is not restricted to one and two particle contributions, but instead it does also extend to any number of particles. Even better, the BSV series for the hexagon (\ref{Whex}) can be fully re-summed by exploiting some standard techniques: eventually we will reproduce (as for the strong coupling regime) the TBA (in the form elaborated in) \cite{TBA-amp2}. In the following we will produce the main steps, but leave some further details and generalisations for an incoming publication \cite{35}.

The expression to sum up is the simple manipulation of the initial formula, (\ref{Whex}), strictly speaking valid for singlets, which we re-call here for practical reasons
\be\label{W6}
W_{hex}=\sum _{N=0}^{+\infty} \frac{1}{N!} \sum_{a_1}\cdots\sum_{a_N}\int \prod _{k=1}^N \left [ \frac{du_k}{2\pi} \mu_{a_k} (u_k) e^{-\tau E_{a_k}(u_k) +i\sigma p_{a_k}(u_k) +i m_{a_k} \phi} \right ]
\displaystyle\prod_{i<j}^N\frac{1}{ P_{a_i,a_j}(u_i|u_j)\,P_{a_j,a_i}(u_j|u_i)} \, ,
\ee
where the indices $a_k$ label the species of different particles (including bound states). \\

In general, not only at strong coupling, we may use a path integral trick of the type as in \cite{Meneghelli:2013tia, Bourgine:2014yha}, but then we should integrate eventually the extra $\rho$-field(s) \cite{35}. Thus, we better perform on the above series (\ref{W6}) a similar trick without the $\rho$ field(s), the Hubbard-Stratonovich transform \cite{Henkel}. The latter makes use of the well know identity of functional gaussian integration in the presence of a linear source term \footnote{The following formula is the infinite dimensional $d\rightarrow \infty$ version of
\be
\langle e^{X_1 J_1} e^{X_2 J_2} \cdots e^{X_d J_d} \rangle =\sqrt{\textrm{det}T} \int \prod_{i=1}^d \frac{d X_i}{\sqrt{2\pi}} e^{-\frac{1}{2} \sum\limits _{i,j=1}^d X_i T_{ij} X_j}  e^{ \sum\limits_{i=1}^d X_i J_i}= e^{\frac{1}{2} \sum\limits_{i,j=1}^d J_i G_{ij} J_j}   \, ,
\nn
\ee
with propagator $G=T^{-1}$, where we choose, for the continuum limit of the external field $J_i\rightarrow J(u)$, the configuration of point-like sources $J(u)=\sum_{i=1}^N a_i \delta(u-u_i)$, {\it cf.} \cite{35} for details.}
\be
\prod_{i<j}^N e^{\langle X_{(a_i)}(u_i) \, X_{(a_j)}(u_j)\rangle}=\langle e^{X_{(a_1)}(u_1)}\cdots e^{X_{(a_N)}(u_N)} \rangle \, ,
\ee
($e^{\langle X_{(a_i)}(u_i) \, X_{(a_i)}(u_i)\rangle} \simeq 1$) for allowing the summation to act on the single exponential of the r.h.s.\footnote{We mention the talk held by B. Basso at IGST 2013 in Utrecht concerning one gluon (without bound states) and \cite{Bettelheim:2014gma} for useful suggestions. We wish also to notice the possibility of interpreting the free boson $c=1$ 2D CFT (Coulomb gas) correlation function formul{\ae} by means of this one.}. This means that we need also to relate the pentagonal amplitudes $P_{a,b}(u|v)$ to correlators (and then to the kinetic part) of the gaussian field $X_{(a)}$ in this way
\be
\frac{1}{P_{a,b}(u|v)\,P_{b,a}(v|u)}=
e^{\langle X_{(a)}(u)\,X_{(b)}(v) \rangle} \ .
\label{P-corr}
\ee
For instance, we associate gluons to the (fluctuating) field $X^g_{(1)}(u)=X^g(u)$, whereas we denote their bound states as $X^g_{(\ell)}(u)$, where $\ell$ stands for the number of components. Thus, the 'linearisation' of the exponent is complete, namely we can recast the hexagonal Wilson loop (\ref{W6}) into a shape aiming at re-summing the different contributions:
\ba
W_{hex} =
\langle\, \mbox{exp}\left\{ \int \frac{du}{2\pi} \sum_{a} \left [ \mu_a(u) \, e^{-\tau E_{a}(u) +i\sigma p_{a}(u)+im_a\phi}
\,e^{X_{(a)}(u)} \right ] \right\}
\,\rangle \, .
\label{Wg-allcoupling}
\ea
With this series (on any particle kind, $a$) inside, (\ref{Wg-allcoupling}) can be interpreted as a Kac-Feynman path integral (partition function) for any value of the coupling, nevertheless it is at strong coupling $\lambda\longrightarrow\infty$ that a tangible simplification occurs since it can be summed up. For simplicity's sake, we will initially include only gluons and their bound states, then adapt our derivation easily to mesons and their bound states. Eventually, we will consider the general system (at strong coupling only), composed of gluons, meson and bound states. For, in this regime, the bound states enjoy a series of simple, useful properties:  their energies and momenta (\ref{glu-Ep}) are simply additive, as so is the relation $X^g_{(a)}=a\,X^g_{(1)}$ implied by the peculiar limit form $\ell _1 \ell _2 K_{sym}^{(gg)}(u,v)$ of the (bound state) gluonic kernels in (\ref{Kbound}) via (\ref{P-corr}) on the bound state fields $X^g_{(a)}$; and finally the measures $\mu^{g}_\ell(u)$ (\ref{glu-meas}) exhibit a peculiar square at denominator. Altogether these properties turn out to be crucial to re-sum the gluonic part of the hexagonal Wilson loop (\ref{Wg-allcoupling}), $W^{(g)}_{hex}$, in a handy shape, and they bring up the dilogarithm function Li$_2(x)$ (tuned by the third property, {\it cf.} \cite{35} for more details):
\be\label{Wsch}
W^{(g)}_{hex} =
\langle\,\mbox{exp}\left\{ -\int\frac{du}{2\pi}\,\mu^g(u)\left[
\mbox{Li}_2 (-e^{-\tau E^g_1(u) +i\sigma p^g_1(u)+i\phi}\,e^{X^g(u)})+\mbox{Li}_2 (-e^{-\tau E^g_1(u) +i\sigma p^g_1(u)-i\phi}\,e^{X^g(u)})
\right] \right\}\,\rangle \, .
\ee
Now, we can make explicit the gaussian measure in (\ref{Wsch}) as a kinetic term so to read $W^{(g)}_{hex}$ as a quantum mechanics partition function for the field $X^g(u)$
\be
\displaystyle W^{(g)}_{hex}=Z^{(g)}[X^g]=\int \m{D}X^g \,e^{-\m{S}^{(g)}[X^g]} \, ,
\ee
where the action $\m{S}^{(g)}[X^g]$, directly expressed in terms of the hyperbolic rapidity $\theta$, has the form
\ba\label{azione}
\m{S}^{(g)}[X^g] &=& \frac{1}{2}\int d\q\,d\q'\, X^g(\q)T^g(\q,\q')X^g(\q')+ \nn\\
&+& \int \frac{d\q'}{2\pi}\,\mu^{g}(\q')\left[
\mbox{Li}_2 (-e^{-E(\q')+i\phi}\,e^{X^g(\q')})+\mbox{Li}_2 (-e^{-E(\q')-i\phi}\,e^{X^g(\q')})
\right] \,,
\ea
with $E(\q)$ coinciding with the derived (\ref{E-bsv}). Of course the kinetic kernel $T^g(\q',\q'')$ is the inverse
\be
\int d\q' \,G^g(\q,\q')T^g(\q',\q'')=\delta(\q-\q'')
\ee
of the Green function
\be
G^g(\theta,\theta')=\langle X^g(\theta)X^g(\theta') \rangle \ \footnote{We could realise that directly $G^g(\q,\q')$ (instead of $T^g(\q',\q'')$) appears in the action by means of an Hubbard-Stratonovich transform which introduces some gaussian field $\rho^g$ coupled to $X^g$: in this way we will end up with the usual form of the Yang-Yang potential of Nekrasov-Shatashvili \cite{Nekrasov:2009rc} as it would be following {\it ab initio} the path integral trick contained in \cite{Meneghelli:2013tia, Bourgine:2014yha}, {\it cf.} \cite{35}.}.
\ee
Remarkably, the action (\ref{azione}) is proportional to $g$, which is going to $+\infty$, so making possibile the applicability of the saddle point with classical equation of motion:
\be
X^g(\q)-\int \frac{d\q'}{2\pi}\,G^g(\q,\q')\mu^{g}(\q')\,\log\left[(1+e^{X^g(\q')}e^{-E(\q')+i\phi})(1+e^{X^g(\q')}e^{-E(\q')-i\phi})\right]=0 \ ,
\ee
where the Green function, at strong coupling, can be easily related to (the symmetric part of) the gluonic pentagonal amplitude
\be
G^g(\theta,\theta')= -\frac{2\pi}{\sqrt{\lambda}}\,K^{(gg)}_{sym}(\theta,\theta')+O(1/\lambda) \ .
\ee
The introduction of the 'pseudo-energy' $\epsilon(\q)$ via the relation
$\epsilon(\q-i\h{\varphi})=E(\q)-X^g(\q)$, leads us to the special version of the TBA equations for gluons only, the first of (\ref{eps-int}) in which we fully neglect $\tilde \epsilon (\theta - i\hat \varphi)$, {\it i.e} the meson contribution.
In other words we have found an action (a Yang-Yang functional) whose differentiation give rise to equations in TBA form \cite{TBA-Y, TBA-Z1}, without thermodynamics. In this respect the generation of the Li$_2 (x)$ function via summation on bound states is of fundamental importance.

As for the mesonic sector, the reasonings outlined above can be easily repeated by substituting the gluon and bound states thereof with the meson and bound states thereof, respectively \footnote{As anticipated about the bound state analysis in section \ref{sez9}, a more mathematical understanding of the contributions of the mesons, as small fermion-antifermion state, and their bound states should be given in future \cite{35} with a mechanism where the poles pinch the integration axis \cite{Nekrasov:2009rc, Meneghelli:2013tia, Bourgine:2014yha, Bettelheim:2014gma, BSV3}.}. In first place, we associate the fields $X^M_{(\ell)}(\q)$ to bound states of mesons, each represented by the single meson $X^M_{(1)}(\q)=X^M(\q)$ by means of the relation $X^M_{(\ell)}(\q)=\ell\,X^M_{(1)}(\q)$. From the identification
\be
\frac{1}{P^M_{a,b}(u|v)\,P^M_{b,a}(v|u)}=
e^{\langle X^M_{(a)}(\q)X^M_{(b)}(\q') \rangle}
\ee
it follows that the meson-only hexagonal Wilson loop $W^{(M)}_{hex}$ assumes a shape analogous to (\ref{Wg-allcoupling}) and can be re-summed at all coupling, even though a remarkable simplification occur at strong coupling, owing to the properties of the mesonic kernel:
\be
W^{(M)}_{hex} = \langle\,\mbox{exp}\left\{ - \int\frac{du}{2\pi}\,\mu^M(u)
\mbox{Li}_2 (-e^{-\tau E_1^{M}(u) +i\sigma p_1^{M}(u)}\,e^{X^M(u)})
\right\}\,\rangle
\ee
Again, the meson hexagonal Wilson loop can be associated to a partition function, defined via the
action $\m{S}^{(M)}[X^M]$
\be\label{azioneM}
\m{S}^{(M)}[X^M]=\frac{1}{2}\int d\q'\,d\q''\, X^M(\q')T^{M}(\q',\q'')X^M(\q'')+\int \frac{d\q''}{2\pi}\,\mu^M(\q'')
\mbox{Li}_2 (-e^{-\sqrt{2}E(\q'')}\,e^{X^M(\q'')})
\ee
which, under extremisation, gives the equation of motion:
\be
X^M(\q)-\int \frac{d\q'}{2\pi}\,G^M(\q,\q')\,\mu^M(\q')\,\log\left[1+e^{X^M(\q')}e^{-\sqrt{2}E(\q')}\right]=0
\ee
where the mesonic Green function has been introduced
\be
\displaystyle\int d\q'\,G^M(\q,\q')T^M(\q',\q'')=\delta(\q-\q'') \ .
\ee
If we define the function $\ti\epsilon(\q-i\h\varphi)=\sqrt{2}E(\q)-X^M(\q)$, we obtain uniquely the mesonic TBA equation, {\it i.e.} the second of (\ref{eps-int}) where we discard the gluonic contribution (considered above).
\medskip\\
\textbf{Complete system}\\
After the considerations outlined above for incomplete systems, made of a single type of particle (and relative bound states) at one time, we can now cope with the complete system, including gluons and mesons together, by arranging the gluonic and mesonic fields into a vector, and the measures as well:
\be
X^a(\q)=
\left(
  \begin{array}{cccccc}
     X^g(\q) \\
     X^M(\q) \\
  \end{array}
\right)
\qquad\qquad
\mu_a(\q)=
\left(
  \begin{array}{cccccc}
     \mu_1(\q) \\
     \mu_2(\q) \\
  \end{array}
\right)  \equiv
\left(
  \begin{array}{cccccc}
     \ \mu^g(\q) \\
     \mu^M(\q) \\
  \end{array}
\right)
\ee
(the label $a$ takes the values $a=1,2$; the sum convention on repeated indices is assumed). The complete hexagonal amplitude can thus be expressed as
\be\label{W=partiz}
W_{hex}=\int \m{D}X_1\,\m{D}X_2\, e^{-\m{S}[X]} \, ,
\ee
where the action reads
\ba\label{action}
&&\m{S}[X] = \frac{1}{2}\int d\q\,d\q'\,X^a(\q)\, T_{ab}(\q,\q') \, X^b(\q') + \\
&&+ \int \frac{d\q}{2\pi}\,\left[\mu_1(\q)\,
\mbox{Li}_2 (-e^{-E(\q)+i\phi}\,e^{X_1(\q)})+\mu_1(\q)\,\mbox{Li}_2 (-e^{-E(\q)-i\phi}\,e^{X_1(\q)})
+\mu_2(\q)\,\mbox{Li}_2 (-e^{-\sqrt{2}E(\q)}\,e^{X_2(\q)})
\right] \,,\nn
\ea
and the Green functions, now arranged in a matrix, can anew be associated to the pentagonal amplitudes
\be\label{green}
G^{ab}(\q,\q')=-\frac{2\pi}{\sqrt{\lambda}}
\left(
  \begin{array}{cccccc}
     K_{sym}^{(gg)}(\q,\q') & K_{sym}^{(Mg)}(\q',\q) \\
     K_{sym}^{(Mg)}(\q,\q') & K_{sym}^{(MM)}(\q,\q') \\
  \end{array}
\right)  \, ,
\ee
and allow us to define the kinetic $2\times 2$ matrix as its inverse:
\be
\int d\q' \,G^{ab}(\q,\q')\,T_{bc}(\q',\q'')=\delta^a_c\,\delta(\q-\q'') \, .
\ee
Upon extremising the action $\m{S}[X]$, we obtain
\be
\label{eqmoto2}
X^a(\q)- \int \frac{d\q'}{2\pi}\,G^{ab}(\q ,\q')\,\mu_b(\q')\,L^b(\q') = 0 \ ,
\ee
where we have assumed the definitions
\be
L^1(\q) = \log\left[(1+e^{X^1(\q)}e^{-E(\q)+i\phi})(1+e^{X^1(\q)}e^{-E(\q)-i\phi})\right] ,\,\,\,\,  L^2(\q)\equiv\log\left[1+e^{X^2(\q)}e^{-\sqrt{2}E(\q)}\right] \,\, .
\ee
These equations of motions (\ref{eqmoto2}) match the TB(ubble)A equations (\ref{eps-int}) of \cite{TBA-amp2}, provided we identify the pseudo-energies as $\epsilon(\q-i\h\varphi)=E(\q)-X^1(\q)$ and $\ti\epsilon(\q-i\h\varphi)=\sqrt{2}E(\q)-X^2(\q)$, which simply translates into
\be
L(\q)=L^1(\q), \,\,\,\, \ti L(\q)=L^2(\q).
\ee
Since the action (\ref{action}) possesses the diverging prefactor $\m{S}[X]\propto \sqrt{\lambda}$, the hexagonal Wilson loop $W_{hex}$ (\ref{W=partiz}) is dominated by the classical configuration, achieved by imposing the equations of motion on the fields, and therefore with the aid of (\ref{eqmoto2}) we can rewrite the kinetic term in the action (\ref{action}) as
\ba\label{cinetico}
&& \frac{1}{2}\,\int d\q\,d\q'\,X^a(\q)\,T_{ab}(\q,\q')\,X^b(\q')= \nn\\
&& =-\frac{\sqrt{\lambda}}{2\pi}\int\frac{d\q\,d\q'}{(2\pi)^2}\,\frac{L^1(\q)\,L^1(\q')}
{\cosh2\q\,\cosh2\q'\,\cosh(\q-\q')}
+ \frac{\sqrt{\lambda}}{2\pi}\int\frac{d\q\,d\q'}{2\pi^2}\,\frac{L^2(\q)\,L^2(\q')}
{\sinh2\q\,\sinh2\q'\,\cosh(\q-\q')} + \nn\\
&& +\frac{\sqrt{\lambda}}{2\pi}\,\sqrt{2}\int\frac{d\q\,d\q'}{\pi^2}\,\frac{L^2(\q)\,L^1(\q')}
{\sinh2\q\,\cosh2\q'}\,\frac{\sinh(\q-\q')}{\cosh(2\q-2\q')} \ .
\ea
Eventually, the sum of the kinetic term (\ref{cinetico}) and the potential part, given by the second line of (\ref{action}) computed on the solution on the saddle point equations, amounts to the critical Yang-Yang functional $\frac{\sqrt{\lambda}}{2\pi}\,YY_{cr}$\,, which has been computed in \cite{TBA-amp2} by adding together the right hand sides of the formulae from (F.42) to (F.46).

\section {Conclusions in perspective}\label{sez12}
\setcounter{equation}{0}
We have derived the complete set of Asymptotic Bethe Ansatz (ABA) equations referring to the GKP vacuum instead that to the half-BPS state (Beisert-Staudacher equations \cite{BS}). These describe the dynamics of all the elementary excitations over the GKP vacuum (gluons, fermions and scalars), but  they also admit solutions in the form of complexes of Bethe and/or auxiliary roots, the so-called strings or stacks. The latter are the bound states, among whose the most important are the bound states of the elementary particles (the other are bound states of the auxiliary or isotopic roots, yet important for the spectrum TBA and so on). In this way, we have performed the 'fusion' of the fundamental (elementary and isotopic) excitations, which is in its whole an alternative way to perform the bootstrap of $S$-matrices ({\it cf.} for instance \cite{giuseppe} for a review).

Moreover, we outlined this system of algebraic equations at all coupling values, also including weak and strong (in different dynamical regimes) coupling.  Above all, we have mainly focused on the scattering phases between all kind of particles at any coupling, but also the new feature of two defects has arisen in the form of new scattering phases for any flavour. Then, we have devoted a meticulous care to the behaviour of the scattering factors in the three possible, -- non-perturbative, perturbative and giant hole --, regimes which allow different large $g$ expansions: for all these three, we obtained explicit expressions of all the scattering factors.

If the momentum of any particle enters the ABA equations, the energy/anomalous dimension is the final object expressed via a solution of these equations. And we could confirm for these first two conserved charges the achievements by \cite{Basso}, but also have been led to consider all the higher integrals of motion (which do play a so important r\^ole in the construction of the dressing factor in the usual ABA on the BMN vacuum).

A deeper look at the form of these new ABA equations brought to our attention an interesting property or identification for them: the $su(4)$ residual R-symmetry constraint the elementary particles to have as rapidities the inhomogeneities of a $su(4)$ symmetric spin chain of $S$-matrices which belong at any lattice site to the characteristic representation of the particle, {\it i.e.} $\bf 1$, $\bf 4$, $\bf \bar 4$, $\bf 6$ (for gluons, fermions, antifermions, scalars, respectively). Thus, as anticipated in \cite{FPR}, the matrix structure of the ABA equations could be inferred from the $SU(4)$ symmetry, but the specific form of the scalar factors and its $g$-dependence must be computed explicitly. For instance, in this perspective, the two defects are simply two purely transmitting impurities which still respect the $SU(4)$ symmetry. Moreover, the particular $g$-dependence shows explicitly the decoupling of the six scalars in the non-perturbative regime and their approach to the $O(6)$ non-linear sigma model $S$-matrix in \cite{Zam^2}, being, besides, the defects of no importance in this limit. More importantly, we have seen from the fusion of a fermion and an antifermion the formation of a new particle in the $g\rightarrow +\infty$ perturbative regime: a meson. Then we also identified bound states thereof.

At last, but not least we have been looking for confirmation and deep comparison of our careful strong coupling outcomes with the scattering amplitude/WL TBA \cite{TBA-amp1}-\cite{TBA-amp3} via the OPE or flux tube (BSV) series \cite{TBA-amp2}, \cite{BSV1}-\cite{BSV5} (see, also, {\it e.g.} \cite{Hatsuda:2014oza} for the multi-Regge limit). In fact, the basic object of the latter, the so-called pentagon amplitude, can be expressed via the aforementioned scattering factors as proposed for the gluons in \cite{BSV1}. The bound states of the latter, the meson and its bound states appear to be the only other relevant particles at leading order (the minimal area  of classical string). Therefore, we have checked explicitly those features by re-summing the BSV series \cite{BSV1} in case of a null hexagonal Wilson loop: we have used the saddle point method at large $g$ to obtain the critical equations coinciding with the TBA equations of \cite{TBA-amp1, TBA-amp2, TBA-amp3}. Then, we have computed the action on them and obtained the same (critical) Yang-Yang functional (or free energy) as in \cite{TBA-amp1, TBA-amp2, TBA-amp3}. Interestingly, the same set-up should be easily applicable to the computation of the heptagon WL. Nevertheless, it would desirable to have a more direct understanding of the phase we dubbed {\it confinement} of the fermions, which disappear as free particles, inside the mesons and their bound states.

\medskip
\medskip
\medskip
{\bf Acknowledgements}

It is a pleasure to thank G. Papathanasiou for explaining us the content of his \cite{P2} concerning the present topic, and also J. Balog, B. Basso, A. Belitsky, D. Bombardelli, A. Bonini, J.-E. Bourgine, L. Griguolo, I. Kostov, A. Sagnotti, Y. Satoh, D. Seminara, R. Tateo, P. Vieira for useful discussions. Hospitality (of DF) at APCTP during the Focus Program on Solving AdS/CFT 2 and Scuola Normale Superiore (Pisa) is kindly acknowledged. This project was  partially supported by INFN grants GAST and FTECP, the Italian
MIUR-PRIN contract 2009KHZKRX-007, the UniTo-SanPaolo research  grant Nr TO-Call3-2012-0088 {\it ``Modern Applications of String Theory'' (MAST)}, the ESF Network {\it ``Holographic methods for strongly coupled systems'' (HoloGrav)} (09-RNP-092 (PESC)) and the MPNS--COST Action MP1210.

{\bf Addendum} A special thank is addressed to L. Bianchi and M. S. Bianchi for pointing us factor $2$ typos in front of some $\theta$s in formul{\ae} (\ref{SMf}-\ref{sMM}) {\it after} this paper was published: this oversight is corrected here, though {\it not} in the journal.  

\appendix

\section{Functions} \label {functions}
\setcounter{equation}{0}

This appendix is devoted to the introduction of the functions we used throughout the paper.

\medskip

In the study of scalars we found convenient to use the following shorthand notations:
\be
\Phi (u)=\Phi _0(u)+\Phi _H (u) \, , \quad \phi (u,v)=\phi _0(u-v)+\phi _H (u,v) \, ,
\ee
with
\ba
\Phi _0(u) &=&-i\ln \frac{i+2u}{i-2u}\, , \quad \Phi _H(u)=-i \ln \left ( \frac {1+\frac {g^2}{2{x^-(u)}^2}}{1+\frac {g^2}{2{x^+(u)}^2}} \right )\, ,  \label {Phi} \\
\phi _0(u-v)&=& i\ln \frac{i+u-v}{i-u+v} \, ,\quad \phi _H(u,v)=-2i \left [ \ln \left ( \frac {1-\frac {g^2}{2x^+(u)x^-(v)} }{1-\frac {g^2}{2x^-(u)x^+(v)}} \right )+i\theta (u,v)\right] \, , \label {phi}
\ea
$\theta (u,v)$ being the dressing phase \cite {BES} and $ x(u)=\frac{u}{2}\left[1+\sqrt{1-\frac{2g^2}{u^2}}\right]$, $ x^\pm(u)=x(u\pm\frac{i}{2})$.
We used also
\be
\varphi(u,v)=\frac{1}{2\pi} \frac{d}{dv} \phi (u,v) \label {fphidef} \, .
\ee

\medskip

For what concerns gluon and their bound states, we used the function
\be
\chi (v,u|l)=\chi _0 (v-u|l+1) +\chi _H \left (v, u-\frac{il}{2}\right )+\chi _H \left (v, u+\frac{il}{2}\right ) \, ,
\nonumber
\ee
where
\be
\chi _0(u|l) = i \ln \frac{il+2u}{il-2u}=2 \arctan \frac{2u}{l} \, , \quad \chi _H(v,u)= i \ln \left ( \frac{1-\frac{g^2}{2x^-(v)x(u)}}{1-\frac{g^2}{2x^+(v)x(u)}} \right ) \, ,
\ee
which enjoys the expression
\be
\chi (v,u|l)=i \ln \left ( \frac{x^+(v)-x\left (u-\frac{il}{2}\right )}{x\left (u+\frac{il}{2}\right )-x^-(v)} \right ) +i\ln \left ( \frac{1-\frac{g^2}{2x^-(v)x\left (u-\frac{il}{2}\right )}}{1-\frac{g^2}{2x^+(v)x\left (u+\frac{il}{2}\right )}} \right ) \, .
\ee
Scattering factors involving gluons and their bound states are expressed in terms of the function
\ba
\chi (v,u|l)+\Phi (v)&=& i \ln \left ( \frac{x^+(v)-x\left (u-\frac{il}{2}\right )}{x\left (u+\frac{il}{2}\right )-x^-(v)} \right ) +i\ln
\left ( \frac{\frac{g^2}{2x\left (u-\frac{il}{2}\right )}-x^-(v)}{x^+(v)-\frac{g^2}{2x\left (u+\frac{il}{2}\right )}} \right )
\label {chiphi} \, .
\ea

\medskip

Finally, for large fermions we introduced the function
\be
\chi _F(u,v)=  \chi _0(u-v|1) +\chi _H(u,v) = i \ln \left ( \frac{x^+(u)-x(v)}{x(v)-x^-(u)} \right )  \, .
\ee
Scattering factors involving large fermions depend on the function
\be
\chi _F(u,v)+\Phi (u)=i\ln \frac{x^+(u)-x(v)}{x(v)-x^-(u)}+i\ln \left ( - \frac{x^-(u)}{x^+(u)} \right )
\, . \label {chiF2}
\ee
Scattering factors for small fermions are obtained from scattering factors for large fermions
after the substitution
\be
\chi _F(u,v)+\Phi (u) \rightarrow -\chi _H (u,v)= i \ln \left ( \frac{1-\frac{g^2}{2x^+(u)x(v)}}{1-\frac{g^2}{2x^-(u)x(v)}} \right )=
i \ln \left ( \frac{1-\frac{x_f(v)}{x^+(u)}}{1-\frac{x_f(v)}{x^-(u)}} \right ) \, ,
\ee
where
\be
x_f(v)=\frac{g^2}{2x(v)}= \frac{v}{2}\left [ 1-\sqrt{1-\frac{2g^2}{v^2}} \right ] \, .
\ee

\section{Useful formul{\ae}} \label {formuli}
\setcounter{equation}{0}

\subsection{Fourier transforms}

We collect here some of the Fourier transforms
\be
\hat f(k)=\int _{-\infty}^{+\infty} du e^{-iku} f(u)
\ee
of functions $f(u)$ we use in the main text.

\medskip

For scalar we used
\ba
\Phi _0(u) &=& -i\ln \frac{i+2u}{i-2u}\quad \Rightarrow \quad \hat \Phi _0(k)= \int _{-\infty}^{+\infty} du e^{-iku} \Phi _0(u)=-\frac{2\pi}{ik}
e^{-\frac{|k|}{2}} \, , \label {Phi0ft} \\
\Phi _H(u)&=&-i \ln \left ( \frac {1+\frac {g^2}{2{x^-(u)}^2}}{1+\frac {g^2}{2{x^+(u)}^2}} \right )
 \quad \Rightarrow \quad \hat \Phi _H(k)= \frac{2\pi }{ik} e ^{-\frac{|k|}{2}} [1-J_0(\sqrt{2}gk)]  \label {PhiHft}
\ea
and also
\ba
\phi _0(u-v)&=&  i\ln \frac{i+u-v}{i-u+v}  \quad \Rightarrow \quad \hat \phi _0(k)=\frac{2\pi e^{-|k|}}{ik} \, , \label {fou3} \\
\varphi _0(u-v)&=&\frac{1}{2\pi}\frac{d}{dv} \phi _0(u-v)=-\frac{1}{\pi}\frac{1}{1+(u-v)^2} \quad \Rightarrow \quad
\hat \varphi _0(k)= -e^{-|k|} \, , \\
\phi _H(u,v)&=&-2i \left [ \ln \left ( \frac {1-\frac {g^2}{2x^+(u)x^-(v)} }{1-\frac {g^2}{2x^-(u)x^+(v)}} \right )+i\theta (u,v)\right] \, , \nonumber \\
\hat \phi _H(k,t)&=& \int _{-\infty}^{+\infty} du e^{-iku} \int _{-\infty}^{+\infty} dv e^{-itv} \phi _H(u,v)= \nonumber \\
&-& 8i \pi ^2 \frac{e^{-\frac {|t|+|k|}{2}}}{k|t|}\Bigl [ \sum _{r=1}^{\infty} r (-1)^{r+1}J_r({\sqrt
{2}}gk) J_r({\sqrt {2}}gt)\frac {1-{\mbox {sgn}}(kt)}{2}
 + \nonumber \\
&+&{\mbox {sgn}} (t) \sum _{r=2}^{\infty}\sum _{\nu =0}^{\infty}
c_{r,r+1+2\nu}(g)(-1)^{r+\nu} \Bigl (
J_{r-1}({\sqrt {2}}gk) J_{r+2\nu}({\sqrt {2}}gt)-  \label {fou4}  \\
&-& J_{r-1}({\sqrt {2}}gt) J_{r+2\nu}({\sqrt {2}}gk)\Bigr ) \Bigr ] \, . \nonumber
\ea
We remark that in previous literature integral equations concerning the scalar sector are often written by using the 'magic kernel' $\hat K$ \cite {BES}, related to $\hat \phi _H$ by
\be
\hat \phi _H(k,t)+\hat \phi _H(k,-t)=8i\pi ^2 g^2 e ^{-\frac{t+k}{2}} \hat K (\sqrt{2}gk, \sqrt{2}gt) \, , \quad t,k >0 \label {magic}\, .
\ee

\medskip

For what concerns gluon bound states, we introduced
\be
\chi _0(u|l)=i \ln \frac{il+2u}{il-2u} =2 \arctan \frac{2u}{l} \quad \Rightarrow \hat \chi _0(k|l)= \int _{-\infty}^{+\infty} du e^{-iku} \chi _0(u|l)=\frac{2\pi}{ik}
e^{-|k|\frac{l}{2}} \label {chi0ft}
\ee
and for higher loops the function
\be
\chi (v,u|l)=\chi _0 (v-u|l+1) +\chi _H \left (v, u-\frac{il}{2}\right )+\chi _H \left (v, u+\frac{il}{2}\right ) \, ,
\nonumber
\ee
where
\be
\chi _H(v,u)= i \ln \left ( \frac{1-\frac{g^2}{2x^-(v)x(u)}}{1-\frac{g^2}{2x^+(v)x(u)}} \right ) \, ,
\ee
whose Fourier transform reads
\ba
&& \int _{-\infty}^{+\infty} du \int _{-\infty}^{+\infty} dv e^{-ikv} e^{-itu} \chi (v, u|l) = 2\pi \delta (t+k) \frac{2\pi}{ik}
e^{-|k|\frac{l+1}{2}}+ \nonumber \\
&+& i \sum _{n=1}^{+\infty} n(-1)^n \frac{2\pi}{k} \frac{2\pi}{|t|} e^{-\frac{|k|}{2}} e^{-\frac{|t|l}{2}}J_n (\sqrt{2}gk)J_n (\sqrt{2}gt) \label {chift} \, .
\ea
In getting (\ref {chift}) we used the Fourier transforms
\be
\int _{-\infty} ^{+\infty} du e^{-iku} \frac{1}{x\left ( u \pm i \frac{l}{2} \right )^n}=\pm n \left ( \frac{\sqrt{2}}{ig} \right )^n \theta (\pm k) \frac{2\pi}{k} e^{\mp \frac{l}{2}k} J_n(\sqrt{2}gk) \, .
\ee
It is useful to Fourier transform $\chi (v,u|l)$ and $\chi _H(v,u)$  with respect only to the variable $v$:
\ba
&& \int _{-\infty}^{+\infty} dv e^{-ikv} \chi (v, u|l)= e^{-iku} \frac{2\pi}{ik}e^{-|k|\frac{l+1}{2}}
+ \label  {chiftv}\\
&+& i \sum _{n=1}^{+\infty} \left ( \frac{g}{\sqrt{2}i \ x\left (u-\frac{il}{2}\right ) } \right )^n
 \frac{2\pi}{k} e^{-\frac{|k|}{2}}  J_n (\sqrt{2}gk) +  i \sum _{n=1}^{+\infty} \left ( \frac{g}{\sqrt{2}i \ x\left (u+\frac{il}{2}\right ) } \right )^n
 \frac{2\pi}{k} e^{-\frac{|k|}{2}}  J_n (\sqrt{2}gk) \, , \nonumber \\
&& \int _{-\infty}^{+\infty} dv e^{-ikv} \chi _H(v,u)= i  \frac{2\pi}{k} e^{-\frac{|k|}{2}}  \sum _{n=1}^{+\infty} \left ( \frac{g}{\sqrt{2}i \ x (u) } \right )^n J_n (\sqrt{2}gk) \label  {chiHftv} \, .
\ea

\medskip

Finally, for large fermions we introduced the function
\be
\chi _F(v,u)= \chi _0(v-u|1) +\chi _H(v,u) = i \ln \left ( \frac{x^+(v)-x(u)}{x(u)-x^-(v)} \right ) \, ,
\ee
whose Fourier transform with respect to $v$ is easily extracted from (\ref {chiftv}, \ref {chiHftv}):
\be
 \int _{-\infty}^{+\infty} dv e^{-ikv} \chi _F(v,u)= e^{-iku} \frac{2\pi}{ik}e^{-\frac{|k|}{2}} + i  \frac{2\pi}{k} e^{-\frac{|k|}{2}}  \sum _{n=1}^{+\infty} \left ( \frac{g}{\sqrt{2}i \ x (u) } \right )^n J_n (\sqrt{2}gk) \label  {chiFftv} \, .
\ee

\subsection{BES and BES-like integral equations}

The BES integral equation for the density $\hat \sigma _{BES} (k)$ in Fourier space reads as
\be
\hat \sigma _{BES} (k)= - \frac{2ik}{1-e^{-|k|}} \frac{\hat \phi _H(k,0)}{\pi} + \frac{ik}{4(1-e^{-|k|})} \int \frac{dt} {\pi ^2} \hat \phi _H(k,t) \hat \sigma _{BES} (t) \label {BES} \, .
\ee
Owing to the parity properties $\hat \sigma _{BES} (k)=\hat \sigma _{BES} (-k)$, we can restrict this equation in the region $k>0$. Introducing the kernel $\hat K$
\be
\hat \phi _H(k,t)+\hat \phi _H(k,-t)=8i\pi ^2 g^2 e ^{-\frac{t+k}{2}} \hat K (\sqrt{2}gk, \sqrt{2}gt) \, , \quad t,k >0 \, , \label {kernels}
\ee
we have
\be
\hat \sigma _{BES} (k)= \frac{4\pi g^2 k}{\sinh \frac{k}{2}}\hat K(\sqrt{2}gk,0)- \frac{g^2 k}{\sinh \frac{k}{2}}\int _{0}^{+\infty} dt e^{-\frac{t}{2}} \hat K(\sqrt{2}gk,\sqrt{2}gt) \hat \sigma _{BES} (t) \label {BES2} \, .
\ee
We can connect to quantities used in \cite {Basso} by means of
\be
\hat \sigma _{BES} (k)= \frac{\pi}{\sinh \frac{k}{2}} \left [ \gamma _+^{\textrm{\o}}(\sqrt{2}gk)+ \gamma _-^{\textrm {\o}}(\sqrt{2}gk) \right ] \, , \quad k>0 \label {BES-BAS}
\ee
and
\be
 \gamma _+^{\textrm {\o}}(\sqrt{2}gk)=2 \sum _{n\geq 1} 2n \gamma _{2n}^{\textrm {\o}} J_{2n} (\sqrt{2}gk) \, , \quad
  \gamma _-^{\textrm {\o}}(\sqrt{2}gk)=2 \sum _{n\geq 1} (2n-1) \gamma _{2n-1}^{\textrm {\o}} J_{2n-1} (\sqrt{2}gk)
  \label {BES-BAS2} \, .
\ee
The total density at order $\ln s$ is $\hat \sigma _{\ln s}(k)=-8\pi \delta (k) + \hat \sigma _{BES}(k)$ which satisfy the equation
\be
\hat \sigma _{\ln s} (k)= - 8\pi \delta (k) + \frac{ik}{4(1-e^{-|k|})} \int \frac{dt} {\pi ^2} \hat \phi _H(k,t) \hat \sigma _{\ln s} (t)
\label {sigmalns} \, .
\ee
The Fourier transform of the density associated to the first generalised scaling function \cite {FRS} satisfies the equation
\be
\hat \sigma ^{(1)}(k)=\frac{\pi }{\sinh \frac{|k|}{2}}[e^{-\frac{|k|}{2}}-J_0(\sqrt{2} gk) ]
+\frac{i k}{1-e^{-|k|}} \int _{-\infty}^{+\infty} \frac{dt}{4\pi ^2}
\hat \phi _H (k,t) \Bigl [ 2\pi + \hat \sigma ^{(1)}(t)  \Bigr ] \, .  \label {sigma(1)}
\ee
Eventually, the density 'all internal holes', which satisfies equation (3.8) of \cite {FRO6} with $L=3$
is solution of
\be
\hat \sigma (k;x)= \frac{2\pi e^{-|k|}}{1-e^{-|k|}} \left (\cos k x  -1 \right ) +  \frac{ik}{1-e^{-|k|}} \int _{-\infty}^{+\infty} \frac{dt}{4\pi ^2}
\hat \phi _H (k,t) \Bigl [ 2\pi (\cos t x -1 )+ \hat \sigma (t;x) \Bigr ] \, . \label {tilSigeqA}
\ee

\subsection{Integrals}

In the one loop case we make use of the following integrals
\be
- \int _{-\infty} ^{+\infty}\frac{dv}{2\pi}\ln \frac{ib+v}{ib-v} \ \frac{d}{dv} \ln \frac{\Gamma (a+iv-iu)}{\Gamma (a-iv+iu)}=
i\ln \frac{\Gamma (a+b+iu)}{\Gamma (a+b-iu)} \, , \quad a,b >0 \,  \label {integ0}
\ee
and
\be
\int  _{-\infty} ^{+\infty} \frac{dv}{2\pi}\ln \frac{ib-v+w}{ib+v-w} \ \frac{d}{dv} \ln \frac{ic-v+u}{ic+v-u} =i \ln \frac{i(c+b)-u+w}{i(c+b)+u-w}
 \, , \quad b,c > 0 \, \label {integ0bis} \, .
\ee
In order to show that (bound states of) gluons do not couple to (type $b$) isotopic roots, we used the following results
\be
\ \bullet  \int _{-\infty} ^{+\infty}\frac{dv}{2\pi} \ln \frac{x^+(v)-x\left (u-\frac{il}{2}\right )}{x\left (u+\frac{il}{2}\right )-x^-(v)} \ \frac{d}{dv} \ln
\frac{\frac{i}{2}+v-u'}{\frac{i}{2}-v+u'}= i \ln \frac{x\left (u+\frac{il}{2}\right )-x(u'-i)}{x(u'+i)-x\left (u-\frac{il}{2}\right )} \, , \quad
l \geq 1 \, ,
 \label {integlu}
\ee
\ba
&\bullet &\int _{-\infty} ^{+\infty}\frac{dv}{2\pi} \ln \frac{x\left (u+\frac{il}{2}\right )-x^+(v)}{x^-(v)-x\left (u-\frac{il}{2}\right )} \ \frac{d}{dv} \ln
\frac{\frac{i}{2}+v-u'}{\frac{i}{2}-v+u'}= i \ln \frac{x(u'-i)-x\left (u-\frac{il}{2}\right )}{x\left (u+\frac{il}{2}\right )-x(u'+i)} + \nonumber \\
&& i \ln \frac{u-u'+\frac{il}{2}-i}{u-u'+\frac{il}{2}} \frac{u-u'-\frac{il}{2}}{u-u'-\frac{il}{2}+i} \, , \quad  l \geq 2 \,
 \label {integlu2} \\
&\bullet & \int _{-\infty} ^{+\infty}\frac{dv}{2\pi} \ln \frac{x^+(u)-x^+(v)}{x^-(v)-x^-(u)} \ \frac{d}{dv} \ln
\frac{\frac{i}{2}+v-u'}{\frac{i}{2}-v+u'}= i \ln \frac{x(u'-i)-x^-(u)}{x^+(u)-x(u'+i)} + \nonumber \\
&& i \ln \frac{\frac{i}{2}-u+u'}{\frac{i}{2}+u-u'} \, .
 \label {integlu3}
\ea
In calculations for the strong coupling limit of scattering factors, we used the following integrals
\ba
&\bullet & \int _{-1}^{1}dk \frac{1}{u-k} \left ( \frac{1+k}{1-k} \right ) ^{\frac{1}{4}} = -\pi \sqrt{2} \left [ 1- \left ( \frac{u+1}{u-1} \right )^{\frac{1}{4}} \right ]  \, , \quad |u| >1 \label {integ1}\\
&\bullet & \int _{-1}^{1}dk P \frac{1}{u-k} \left ( \frac{1+k}{1-k} \right )^{\frac{1}{4}}=
-\pi \sqrt{2} + \pi \left ( \frac{1+u}{1-u} \right )^{\frac{1}{4}}  \, , \quad |u| <1
\label {integ2} \\
&\bullet & \int_{-1}^1 dz\, \frac{1}{z-\bar v}\,\frac{1}{\sqrt{1-z^2}}=-\frac{\pi \textrm{sgn}(\bar v)}{\sqrt{\bar v^2-1}} \, ,
\quad |\bar v|>1, \label {integ2bis} \\
&\bullet & \int_{-1}^1 dz\, PV \frac{1}{z-\bar v}\,\frac{1}{\sqrt{1-z^2}}=0 \, ,
\quad |\bar v|<1, \label {integ2ter} \\
&\bullet & \int _{|\bar w| \geq 1} \frac{d\bar w}{2\pi} \frac{1}{\bar w -\bar u} PV \frac{1}{\bar w -\bar z}
\left ( \frac{\bar w +1}{\bar w -1} \right ) ^{\frac{1}{4}}=
\frac {\frac{1}{2} \left ( \frac{\bar z +1}{\bar z -1} \right ) ^{\frac{1}{4}}
- \frac{1}{\sqrt{2}} \left ( \frac{1+\bar u}{1-\bar u} \right ) ^{\frac{1}{4}}} {\bar z -\bar u} \, , \quad
|\bar u|\leq 1 \, , \ |\bar z|\geq 1 \, , \label {integ3} \\
&\bullet & \int _{|\bar z|\geq 1} \frac{d\bar z}{2\pi} \frac{1}{\bar z \sqrt{1-\frac{1}{\bar z ^2}}} \frac{1}{\bar x_f(\bar v)-\bar x (\bar z)}\left ( \frac{\bar z-1}{\bar z+1} \right )^{\frac{1}{4}}\frac{1}{\bar u -\bar z}= \label {integ4} \\ &=& \frac{\sqrt{\frac{1-2\bar x_f (\bar v)}{1+2\bar x_f (\bar v)}} + \frac{1}{\sqrt{2}}\left (\bar x_f(\bar v) - \frac{1}{2} \right ) \left [  \left ( \frac{1+\bar u}{1-\bar u} \right ) ^{\frac{1}{4}}+  \left ( \frac{1-\bar u}{1+\bar u} \right ) ^{\frac{1}{4}} \right ] +  \frac{1}{\sqrt{2}} \left (\bar x_f(\bar v) + \frac{1}{2} \right )
\sqrt{\frac{1-\bar u}{1+\bar u}} \left [    \left ( \frac{1-\bar u}{1+\bar u} \right ) ^{\frac{1}{4}}- \left ( \frac{1+\bar u}{1-\bar u} \right ) ^{\frac{1}{4}} \right ] }{2\bar x_f(\bar u)(\bar u -\bar v)} \nonumber
\ea

\section{Collection of scattering factors} \label {scatt-fact}
\setcounter{equation}{0}

\subsection{One loop: explicit expressions}

We list here the scattering factors at one loop:

$\bullet$ Scalar - Scalar
\be
S^{(ss)}_0(u_h,u_{h'})= -\frac{\Gamma \left (\frac{1}{2}-iu_h \right ) \Gamma \left (\frac{1}{2}+iu_{h'} \right ) \Gamma (1+iu_h-iu_{h'})}{\Gamma \left (\frac{1}{2}+iu_h \right )
\Gamma \left (\frac{1}{2}-iu_{h'} \right )\Gamma (1-iu_h+iu_{h'})}    \, . \label {S_0}
\ee
Formula (\ref {S_0}) does agree with result (3.8) of Basso-Belitsky \cite{BB}, but seems to be the inverse of (2.13) of Dorey-Zhao \cite{DoreyZhao}.

\medskip

$\bullet $ Gluon - Gluon
\be
S^{(gg)}_0(u,v)=-
\frac{\Gamma\left(1+iu-iv)\right)}{\Gamma\left(1-iu+iv)\right)}
\frac{\Gamma\left(\frac{3}{2}-iu\right)}{\Gamma\left(\frac{3}{2}+iu\right)}
\frac{\Gamma\left(\frac{3}{2}+iv\right)}{\Gamma\left(\frac{3}{2}-iv\right)}  \ . \label {Sgg21l}
\ee
In addition, we have
\be
S^{(gg)}_0(u,v)=S^{(\bar g\bar g)}_0(u,v)=S^{(g\bar g)}_0(u,v)\frac{u-v+i}{u-v-i} \, , \quad S^{(\bar g g)}_0(u,v)=
[S^{(g\bar g)}_0(v,u)]^{-1} \, .
\ee
\medskip

$\bullet $ (Large) Fermion - (Large) Fermion
\be
S^{(FF)}_0(u,v)=\frac{\Gamma(1+iu-iv)}{\Gamma(1-iu+iv)}\,
\frac{\Gamma(1-iu)}{\Gamma(1+iu)}\,\frac{\Gamma(1+iv)}{\Gamma(1-iv)}
\label {SFF1l}
\ee
and when antifermions get involved
\be
S^{(FF)}_0(u,v)=S^{(F\bar F)}_0(u,v)=S^{(\bar FF)}_0(u,v)=S^{(\bar F\bar F)}_0(u,v) \, .
\ee
\medskip

$\bullet $ Gluon - Scalar
\ba
&& S^{(gs)}_0(u,u_h)=[S^{(sg)}_0(u_h,u)]^{-1}=S^{(\bar g s)}_0(u,u_h)=[S^{(s\bar g)}_0(u_h,u)]^{-1}= \nonumber \\
&=& \frac{\Gamma\left(1+iu-iu_h\right)}{\Gamma\left(1-iu+iu_h)\right)}
\frac{\Gamma\left(\frac{1}{2}+iu_h\right)}{\Gamma\left(\frac{1}{2}-iu_h\right)}
\frac{\Gamma\left(\frac{3}{2}-iu\right)}{\Gamma\left(\frac{3}{2}+iu\right)} \, .
\label {Sgs1l}
\ea

\medskip

$\bullet $ (Large) Fermion - Scalar
\ba
&& S^{(Fs)}_0(u,u_h)=S^{(\bar F s)}_0(u,u_h)=[S^{(sF)}_0(u_h,u)]^{-1}= [S^{(s\bar F)}_0(u_h,u)]^{-1}=\nonumber \\
&=& \frac{\Gamma(\frac{1}{2}+iu-iu_h)}{\Gamma(\frac{1}{2}-iu+iu_h)}\,\frac{\Gamma(1-iu)}{\Gamma(1+iu)}\,
\frac{\Gamma(\frac{1}{2}+iu_h)}{\Gamma(\frac{1}{2}-iu_h)} \, .
\label {SFs1l}
\ea

\medskip

$\bullet $ Gluon - (Large) Fermion
\ba
&& S^{(gF)}_0(u,v)=[S^{(Fg)}_0(v,u)]^{-1}=S^{(\bar g \bar F}_0(u,v)= [S^{(\bar F \bar g)}_0(v,u)]^{-1}= \nonumber \\
&=& -\frac{\Gamma\left(\frac{1}{2}+iu-iv\right)}{\Gamma\left(\frac{1}{2}-iu+iv\right)}
\frac{\Gamma\left(\frac{3}{2}-iu\right)}{\Gamma\left(\frac{3}{2}+iu\right)}
\frac{\Gamma\left(1+iv \right)}{\Gamma\left(1-iv\right)}   \label {SgF21l}
\ea
and
\be
S^{(\bar g F)}_0(u,v)=[S^{(F \bar g)}_0(v,u)]^{-1}=
S^{(g \bar F)}_0(u,v)=[S^{(\bar F g)}_0(v,u)]^{-1}=S^{(gF)}_0(u,v)\frac{u-v-i/2}{u-v+i/2} \, .
\ee

\subsection{All loops: expressions in terms of solutions of integral equations}

We list here the factors found in \cite {FPR}. We start from the 'direct' S factors:
\ba
{S}^{(ss)}(u,v)&=& -\textrm{exp} [ -i\Theta (u,v) ] \, , \label {Sss} \\
{S}^{(FF)}(u,v)&=& \textrm{exp} \Bigl \{ i \int _{-\infty}^{+\infty}\frac{dw}{2\pi} [\chi _F (w,u)+\Phi (w)] \frac{d}{dw} [\chi _F (w,v)+\Phi (w)] - \nonumber \\
&-& i \int \frac{dw}{2\pi} \frac{dz}{2\pi} [\chi _F (w,u)+\Phi (w)] \frac{d^2}{dwdz}\Theta (w,z) [\chi _F(z,v)+\Phi (z)] \Bigr \}  \, ,  \label {SFF} \\
{S}^{(FF)}(u,v)&=&{S}^{(F\bar F)}(u,v)={S}^{(\bar FF)}(u,v)={S}^{(\bar F\bar F)}(u,v) \, , \label {SFF2} \\
{S}^{(gg)}(u,v)&=& -\textrm{exp} \Bigl \{ -i\chi _0(u-v|2)+i \int _{-\infty}^{+\infty} \frac{dw}{2\pi} [\chi  (w,u|1)+\Phi (w)] \frac{d}{dw} [\chi (w,v|1)+\Phi (w)] - \nonumber \\
&-&i \int \frac{dw}{2\pi}\frac{dz}{2\pi} [\chi  (w,u|1)+\Phi (w)] \frac{d^2}{dwdz}\Theta (w,z) [\chi (z,v|1)+\Phi (z)] \Bigr \} \label {Sgg} = \\
&=& \frac{u-v+i}{u-v-i} \ {S}^{(gg)}_{red}(u,v) =-\textrm{exp} \{ -i\chi_0(u-v|2) \} \ {S}^{(gg)}_{red}(u,v) \, , \nonumber \\
{S}^{(gg)}_{red}(u,v)&=& \textrm{exp} \Bigl \{i \int _{-\infty}^{+\infty} \frac{dw}{2\pi} [\chi  (w,u|1)+\Phi (w)] \frac{d}{dw} [\chi (w,v|1)+\Phi (w)] - \nonumber \\
&-&i \int \frac{dw}{2\pi}\frac{dz}{2\pi} [\chi  (w,u|1)+\Phi (w)] \frac{d^2}{dwdz}\Theta (w,z) [\chi (z,v|1)+\Phi (z)] \Bigr \} \, , \\
S^{g\bar g} (u,v)&=&[S^{\bar g g} (v,u)]^{-1}= {S}^{(gg)}_{red}(u,v) \label {Sgg2} \, .
\ea

The 'mixed' S factors are:
\ba
S^{(sF)}(u,v)&=&[S^{(Fs)}(v,u)]^{-1}=\textrm{exp} \Bigl \{ -i [ \chi _F(u,v) +\Phi (u)]+ \nonumber \\
&+& i \int \frac{dw}{2\pi} \frac{d\Theta}{dw}(u,w) [\chi _F(w,v)+\Phi (w)] \Bigr \} \, , \label {SsF} \\
S^{(sF)}(u,v)&=&S^{(s\bar F)}(u,v) \, , \quad S^{(Fs)}(u,v)=S^{(\bar F s)}(u,v) \, , \label {SFs} \\
S^{(gs)}(u,v)&=&[S^{(sg)}(v,u)]^{-1}=S^{(\bar gs)}(u,v)=[S^{(s\bar g)}(v,u)]^{-1}= \nonumber \\
&=& \textrm{exp} \Bigl \{ i [ \chi (v,u|1) +\Phi (v)]- i \int \frac{dw}{2\pi} \frac{d\Theta}{dw}(v,w) [\chi (w,u|1)+\Phi (w)] \Bigr \} \, , \label {Sgs} \\
{S}^{(gF)}(u,v)&=& [{S}^{(Fg)}(v,u)]^{-1}=-\textrm{exp} \Bigl \{ -i\chi_0(u-v|1)+ \nonumber \\
&+& i \int _{-\infty}^{+\infty} \frac{dw}{2\pi} [\chi  (w,u|1)+\Phi (w)] \frac{d}{dw} [\chi _F (w,v)+\Phi (w)] - \nonumber \\
&-& i \int \frac{dw}{2\pi}\frac{dz}{2\pi} [\chi  (w,u|1)+\Phi (w)] \frac{d^2}{dwdz}\Theta (w,z) [\chi _F(z,v)+\Phi (z)] \Bigr \}= \, , \label {SgF} \\
&=& \frac{u-v+\frac{i}{2}}{u-v-\frac{i}{2}}\ S^{(gF)}_{red}(u,v)=-\textrm{exp}  \{ -i\chi_0(u-v|1) \}
\ S^{(gF)}_{red}(u,v) \nonumber \\
S^{(gF)}(u,v)&=&S^{(\bar g\bar F)}(u,v) \, , \label {SgF2} \\
S^{(\bar g F)}(u,v)&=& [S^{( F \bar g)}(v,u)]^{-1}=\textrm{exp} \Bigl \{ i \int _{-\infty}^{+\infty} \frac{dw}{2\pi} [\chi  (w,u|1)+\Phi (w)] \frac{d}{dw} [\chi _F (w,v)+\Phi (w)] - \nonumber \\
&-& i \int \frac{dw}{2\pi}\frac{dz}{2\pi} [\chi  (w,u|1)+\Phi (w)] \frac{d^2}{dwdz}\Theta (w,z) [\chi _F(z,v)+\Phi (z)] \Bigr \} \, , \label {SgF3} \\
S^{(g \bar F)}(u,v)&=& [S^{( \bar F g)}(v,u)]^{-1}=S^{(\bar g F)}(u,v) \, . \label {SgF4}
\ea
The S matrices involving small fermions are obtained from the corresponding ones for large fermions by means of the replacement
\be
\chi_F (v,u)+\Phi (v) \, \longrightarrow\, -\chi_H (v,u) \, . \label {repla}
\ee
All the scalar factors are expressed in terms of known functions listed in Appendix \ref {functions} and the 'dynamical' function $\Theta (u,v)$ \cite {FPR}, which equals
\be
\Theta (u,v)=\Theta '(u,v) +\tilde P(v) \, , \label {theta-app}
\ee
where $\Theta '(u,v)$ and $\tilde P(v)$ are found as solutions of the linear integral equations
\ba
\Theta '(u,v)&=&\phi (u,v)+\Phi (u) - \int _{-\infty}^{+\infty} dw   \varphi (u,w) \Theta '(w,v) \, , \label {theta'-app} \\
\tilde P(v)&=&-\Phi (v) - \int _{-\infty}^{+\infty} \frac{dw}{2} [ \varphi (v,w)- \varphi (v,-w) ] \tilde P(w)
\label {tildeP-app} \, .
\ea

\medskip

{\bf Scattering factors involving bound states of gluons}

If bound states of gluons are present, the factors involving the gauge field should be generalised as follows.

The scattering factor between a bound state of gluons with length $m$ and center $u$ and a scalar is
\ba
S^{(gs)}_m(u,v)&=&[S^{(sg)}_m(v,u)]^{-1}=S^{(\bar gs)}_m(u,v)=[S^{(s\bar g)}_m(v,u)]^{-1}= \nonumber \\
&=& \textrm{exp} \Bigl \{ i [ \chi (v,u|m) +\Phi (v)]- i \int \frac{dw}{2\pi} \frac{d\Theta}{dw}(v,w) [\chi (w,u|m)+\Phi (w)] \Bigr \} \, . \label {Sgms}
\ea
The scattering factor between a bound state of gluons with length $m$ and center $u$ and large fermions is
\ba
{S}^{(gF)}_m(u,v)&=& [{S}^{(Fg)}_m (v,u)]^{-1}=-\textrm{exp} \Bigl \{ -i\chi_0(u-v|m)+ \nonumber \\
&+& i \int _{-\infty}^{+\infty} \frac{dw}{2\pi} [\chi  (w,u|m)+\Phi (w)] \frac{d}{dw} [\chi _F (w,v)+\Phi (w)] -  \label {SgFm} \\
&-& i \int \frac{dw}{2\pi}\frac{dz}{2\pi} [\chi  (w,u|m)+\Phi (w)] \frac{d^2}{dwdz}\Theta (w,z) [\chi _F(z,v)+\Phi (z)] \Bigr \} \, , \nonumber \\
S^{(gF)}_m(u,v)&=&S^{(\bar g\bar F)}_m(u,v) \, , \label {SgF2m} \\
S^{(\bar g F)}_m(u,v)&=& [S^{( F \bar g)}_m(v,u)]^{-1}=\textrm{exp} \Bigl \{ i \int _{-\infty}^{+\infty} \frac{dw}{2\pi} [\chi  (w,u|m)+\Phi (w)] \frac{d}{dw} [\chi _F (w,v)+\Phi (w)] - \nonumber \\
&-& i \int \frac{dw}{2\pi}\frac{dz}{2\pi} [\chi  (w,u|m)+\Phi (w)] \frac{d^2}{dwdz}\Theta (w,z) [\chi _F(z,v)+\Phi (z)] \Bigr \} \, , \label {SgF3m} \\
S^{(g \bar F)}_m(u,v)&=& [S^{( \bar F g)}_m(v,u)]^{-1}=S^{(\bar g F)}_m(u,v) \, . \label {SgF4m}
\ea
The replacement $\chi _F (w,v)+\Phi (w) \rightarrow -\chi _H(w,v)$ gives the corresponding quantities for small fermions.

The scattering factors between a bound state of gluons with length $m$ and center $u$ and a bound state of gluons with length $l$ and center $v$ are
\ba
{S}^{(gg)}_{ml}(u,v)&=& \textrm{exp} \Bigl \{ -i\tilde \chi (u,v|m,l)+i \int _{-\infty}^{+\infty} \frac{dw}{2\pi} [\chi  (w,u|m)+\Phi (w)] \frac{d}{dw} [\chi (w,v|l)+\Phi (w)] - \nonumber \\
&-&i \int \frac{dw}{2\pi}\frac{dz}{2\pi} [\chi  (w,u|m)+\Phi (w)] \frac{d^2}{dwdz}\Theta (w,z) [\chi (z,v|l)+\Phi (z)] \Bigr \} \, , \label {Sgglm}
\ea
\ba
{S}^{(g\bar g)}_{ml}(u,v)&=& \textrm{exp} \Bigl \{i \int _{-\infty}^{+\infty} \frac{dw}{2\pi} [\chi  (w,u|m)+\Phi (w)] \frac{d}{dw} [\chi (w,v|l)+\Phi (w)] - \nonumber \\
&-&i \int \frac{dw}{2\pi}\frac{dz}{2\pi} [\chi  (w,u|m)+\Phi (w)] \frac{d^2}{dwdz}\Theta (w,z) [\chi (z,v|l)+\Phi (z)] \Bigr \} \label {Sgbarglm} \, ,
\ea
where
\be
\tilde \chi (u,v|m,l)=\chi _0 (u-v|m+l) - \chi _0 (u-v|m-l) +2\sum _{\gamma =1}^{m-1} \chi _0 (u-v|m+l-2\gamma) \, .
\label {tilde-chi}
\ee
In the particular case $m=l=1$, since $e^{i\chi _0(u-v|0)}=-1$, one recovers from (\ref {Sgglm})
the gluon-gluon scattering factor (\ref {Sgg}).

\medskip

{\bf Remark} The following relations hold, for $u$ real and $u^2<2g^2$,
\be
\chi (v,u|m)+\Phi (v)=\sum _{l=-\frac{m-1}{2}}^{\frac{m-1}{2}} [\chi (v,u+il|1) +\Phi (v) ] \, , \quad m \geq 1 \label {chimaster} \, ,
\ee
where $\chi (v,u+il|1)$ has to be understood as analytical continuation of $\chi (v,u|1)$.

To prove this statement, we refer to (\ref {chiphi}) and remember the following properties
\be
\lim _{\epsilon \rightarrow 0^+} x(u-i\epsilon)=\lim _{\epsilon \rightarrow 0^+} \frac{g^2}{2 x(u+i\epsilon)} \, ,
\ee
which are valid for $u$ real and $u^2<2g^2$.
Therefore, when the complex variable $u$ crosses the real axis in the region $-\sqrt{2}g <\textrm{Re} u < \sqrt{2}g$, the function $x(u)$ is analytically continued in $g^2/2 x(u)$. With the help of this property, relation (\ref {chimaster}) is easily shown.

\medskip

\subsection{Strong coupling and mirror in hyperbolic rapidities}

In the strong coupling perturbative regime, the scattering matrices for gluons, fermions and mesons can be suitably recast in term of hyperbolic rapidities, according to the following identities (written up to $O(1/g^2)$ corrections):\\
\textbf{Gluons:}
\be
u^g=\sqrt{2}g \bar u^g \, , \quad \bar u^g= \tanh (2\theta) \, ,
\ee
\textbf{Fermions:}
\be
u_f=\sqrt{2}g \bar u_f \, , \quad \bar u_f=\coth (2\theta) \qquad\qquad \mbox{or else} \qquad\qquad 2\bar x_f= \tanh \theta \, ,
\ee
\textbf{Mesons:}
\be
u_M=\sqrt{2}g \bar u_M \, , \quad \bar u_M=\coth (2\theta) \, .
\ee
Explicitly, we obtain:
\be
S^{(gg)}(\theta,\theta') = \exp\left\{\frac{i}{\sqrt{2}g}\left[\frac{1}{\tanh2\theta-\tanh2\theta'}+\frac{\cosh2\theta \cosh2\theta'}{2\sinh(\theta-\theta')}\right]
+O\left(\frac{1}{g^2}\right)\right\} \, , \label {Sggtheta}
\ee

\be
S^{(g\bar g)}(\theta,\theta') = \left(1-\frac{1}{\sqrt{2}g}\,\frac{2i}{\tanh2\theta-\tanh2\theta'}+O\left(\frac{1}{g^2}\right)\right)
\,S^{(gg)}(\theta,\theta') \, , \label {Sgbartheta}
\ee

\be
S^{(ff)}(\theta,\theta') =
\exp\left\{-\frac{i}{2\sqrt{2}g}\frac{\sinh2\theta \sinh2\theta'}{\sinh(2\theta-2\theta')}\,(\cosh(\theta-\theta')-1)
+O\left(\frac{1}{g^2}\right)\right\} \, , \label {Sfftheta}
\ee

\be
S^{(gf)}(\theta,\theta_f) = \exp\left\{\frac{i}{4g}\,\frac{2\cosh(\theta_f-\theta)+\sqrt{2}}{\tanh2\theta-\coth2\theta_f}
+O\left(\frac{1}{g^2}\right)\right\} \, ,
\label {Sgftheta}
\ee

\be
S^{(\bar gf)}(\theta,\theta_f) = \exp\left\{\frac{i}{4g}\,\frac{2\cosh(\theta_f-\theta)-\sqrt{2}}{\tanh2\theta-\coth2\theta_f}
+O\left(\frac{1}{g^2}\right)\right\} \, ,
\label {Sbargftheta}
\ee

\be
S^{(MM)}(\theta,\theta')=\textrm{exp} \left [ -\frac{i}{\sqrt{2}g} \frac{\sinh 2\theta \sinh 2\theta '}{\sinh (\theta -\theta ')} \right ] \, , \label {sMM-str}
\ee

\be
S^{(Mg)}(\theta,\theta')=S^{(M\bar g)}(\theta,\theta')=\textrm{exp} \left [ -\frac{i}{g} \frac{\cosh (\theta -\theta ')}{\tanh 2\theta '-\coth 2\theta } \right ] \, . \label {sMg-str}
\ee

\textbf{Mirror transformations:}\\
The mirror rotation should be implemented in different ways on the scattering phases, depending on the kind of particle the transformation is acting on. For instance, in the scalar case it is achieved by means of a shift
$u \longrightarrow u^\gamma=u+i$. For gluons, the mirror transform is performed via a closed path across the complex rapidity plane ($u^\gamma=u$), passing through a cut, so that actually the initial and final points do not lie on the same sheet. Defining a procedure for the mirror rotation on fermions is more involved, and for this purpose we refer to \cite{BSV3}. A complete all coupling study of all the mirror S matrices can be found in \cite {BEL}.
Nevertheless, as long as the perturbative strong coupling regime is concerned, the mirror rotation gets simplified: for gluons, scalars and mesons it amounts to an imaginary shift in the hyperbolic rapidities $\theta ^{\gamma}=\theta +i\frac{\pi}{2}$. For instance, we have:
\be
S^{(gg)}_{mir}(\theta ,\theta') = S^{(gg)}(\theta+i\frac{\pi}{2},\theta')
= \exp\left\{\frac{i}{\sqrt{2}g}\left[\frac{1}{\tanh2\theta-\tanh2\theta'}-\frac{\cosh2\theta \cosh2\theta'}{2i\cosh(\theta-\theta')}\right]
+O\left(\frac{1}{g^2}\right)\right\} \, ,
\label {Sggmir}
\ee

\ba
S^{(g\bar g)}_{mir}(\theta ,\theta') &=& S^{(g\bar g)}(\theta+i\frac{\pi}{2},\theta') =
\left(1-\frac{1}{\sqrt{2}g}\,\frac{2i}{\tanh2\theta-\tanh2\theta'}+O\left(\frac{1}{g^2}\right)\right)\cdot
\nonumber \\
&\cdot & \exp\left\{\frac{i}{\sqrt{2}g}\left[\frac{1}{\tanh2\theta-\tanh2\theta'}-\frac{\cosh2\theta \cosh2\theta'}{2i\cosh(\theta-\theta')}\right]
+O\left(\frac{1}{g^2}\right)\right\}  \, , \label {Sgbargmir}
\ea

\be
S^{(ff)}_{mir}(\theta ,\theta') =
\exp\left\{-\frac{i}{2\sqrt{2}g}\frac{\sinh2\theta \sinh2\theta'}{\sinh(2\theta-2\theta')}\,(i\sinh(\theta-\theta')-2)
+O\left(\frac{1}{g^2}\right)\right\} \, ,
\label {Sffmir}
\ee

\be
S^{(f\bar f)}_{mir}(\theta ,\theta') =
\exp\left\{\frac{1}{4\sqrt{2}g}\frac{\sinh2\theta \sinh2\theta'}{\cosh(\theta-\theta')}\,
+O\left(\frac{1}{g^2}\right)\right\} \, ,
\label {Sfbarfmir}
\ee

\be
S^{(MM)}_{mir}(\theta ,\theta')=\textrm{exp} \left [ \frac{1}{\sqrt{2}g} \frac{\sinh 2\theta \sinh 2\theta '}{\cosh (\theta -\theta ')} \right ] \, ,  \label {sMM-str-mir}
\ee

\be
S^{(Mg)}_{mir}(\theta ,\theta')=S^{(M\bar g)}_{mir}(\theta ,\theta')=\textrm{exp} \left [ \frac{1}{g} \frac{\sinh (\theta -\theta ')}{\tanh 2\theta '-\coth 2\theta } \right ] \, . \label {sMg-str-mir}
\ee

\subsection{On the factor $\Theta$}

Since the scalar-scalar factor $\Theta $ is the building block for all the scattering factors, we give here some alternative constructions for it. Following what was done in \cite {FPR}, we define the function
\be
M(u,v)=\frac{\Theta (u,v)+\Theta (u,-v)}{2} \, ,
\ee
which stores all the information on $\Theta $, since (see (2.23) of \cite {FPR})
\be
\Theta (u,v)=M(u,v)-M(v,u) \, .
\ee
Then, we use (2.21) of \cite {FPR} to express $M(u,v)$ in terms of densities $\sigma ^{(1)}(u)$ and $\sigma (u;v)$.
Formul{\ae} (\ref {S1-def}, \ref {Su-def}) and Neumann expansions (\ref {S-Neu}) allow to arrive at
\ba
&& M(u,v)=\frac{i}{2} \ln \frac{\Gamma (1+iu+iv) \Gamma (1+iu-iv) }{\Gamma (1-iu+iv) \Gamma (1-iu-iv) }+i\ln \frac{\Gamma \left (\frac{1}{2}-iu \right )}{\Gamma \left (\frac{1}{2}+iu \right )}+ \label {Mfinal} \\
&+& \sum _{n=0}^{\infty} \sum _{p=1}^{\infty} \frac{1}{ip} \left ( \frac{ig}{\sqrt{2}} \right )^p [S_p^{(1)}+S_p^{\prime} (v)] \left [
\left (\frac{1}{x\left (u+\frac{i}{2}(1+2n)\right )} \right )^p - \left (-\frac{1}{x\left (u-\frac{i}{2}(1+2n)\right )} \right )^p \right ] \, . \nonumber
\ea
The first line of (\ref {Mfinal}) is the one loop contribution; the second line is the higher than one loop correction. We can manipulate (\ref {Mfinal}) in order to get alternative expressions for $M(u,v)$: for instance,
\ba
&& M(u,v)=\frac{i}{2} \ln \frac{\Gamma (1+iu+iv) \Gamma (1+iu-iv) }{\Gamma (1-iu+iv) \Gamma (1-iu-iv) }+i\ln \frac{\Gamma \left (\frac{1}{2}-iu \right )}{\Gamma \left (\frac{1}{2}+iu \right )}+ \label {Mfinal3} \\
&+& \sum _{p=1}^{\infty} [S_p^{(1)}+S_p^{\prime} (v)] \int _{-\pi}^{+\pi} \frac{d\theta }{2\pi i}
e^{-ip\theta} \ln \frac {\Gamma \left (\frac{1}{2} -i \sqrt{2}g \sin \theta -iu \right )}{\Gamma \left (\frac{1}{2} -i \sqrt{2}g \sin \theta +iu \right )}  \,  \nonumber
\ea
and
\ba
&& M(u,v)=\frac{i}{2} \ln \frac{\Gamma (1+iu+iv) \Gamma (1+iu-iv) }{\Gamma (1-iu+iv) \Gamma (1-iu-iv) }+i\ln \frac{\Gamma \left (\frac{1}{2}-iu \right )}{\Gamma \left (\frac{1}{2}+iu \right )}+ \label {Mfinal2} \\
&+& \sum _{n=0}^{\infty} \sum _{p=1}^{\infty} \frac{(-1)^{n+p}}{i n! (n+p)!} \left ( \frac{g}{\sqrt{2}} \right )^{2n+p} [S_p^{(1)}+S_p^{\prime} (v)] \left [\psi ^{(2n+p-1)} \left (\frac{1}{2}-iu \right ) - \psi ^{(2n+p-1)}\left (\frac{1}{2} +iu \right ) \right ] \, . \nonumber
\ea

\section{Some calculations on how to get scattering factors} \label {some-calc}
\setcounter{equation}{0}

\subsection{Scalars vs others}

The 'fermionic' contribution to $Z_4(v)-P$ reads
\be
Z_4(v)-P|_{F}=F^F (v,u_{F,j})+ \int \frac{dw}{2\pi} \frac{d}{dw} \Phi (w) F^F(w,u_{F,j}) \, .  \label {d1}
\ee
Inserting the integral equation (\ref {F-F}) for $F^F$, we arrive at
\be
Z_4(v)-P|_{F}=\chi _F (v,u_{F,j})+ \Phi (v)+\int \frac{dw}{2\pi} \frac{d}{dw}\left [ \phi (w,v)+\Phi (w) \right ] F^F(w,u_{F,j}) \, .  \label {d2}
\ee
Remembering the integral equation (\ref {Theta-prime}) for $\Theta '$ and then, after an integration by parts, the equation (\ref {F-F}) for $F^F$, we get
\ba
Z_4(v)-P|_{F}&=&\chi _F (v,u_{F,j})+ \Phi (v)+\int \frac{dw}{2\pi} \frac{d}{dw}\Theta ' (w,v) [ \chi _F (w,u_{F,j})+ \Phi (w)] \label {d3} \\
&=&\chi _F (v,u_{F,j})+ \Phi (v)-\int \frac{dw}{2\pi} \frac{d}{dw}\Theta  (v,w) [ \chi _F (w,u_{F,j})+ \Phi (w)]=
i \ln S ^{(sf)}(v,u_{F,j}) \, .  \nonumber
\ea
The same calculation can be done for large antifermions. For gluons we can repeat the same reasonings after the substitution
$\chi _F (v,u) \rightarrow \chi (v,u|1)$. For small (anti)fermions we perform the substitution $\chi _F(v,u)+\Phi (v) \rightarrow
-\chi _H(v,u)$.

For scalar-scalar factor we have to consider
\be
Z_4(v)-P|_{s}=\Theta '(v,u_h)-\Phi (u_h)+ \int \frac{dw}{2\pi} \frac{d}{dw} \Phi (w) \Theta ' (w,u_h) \, .
\label {d4}
\ee
Now, we remember the integral equation (\ref {Theta-prime}) for $\Theta '$, which allows to write the identity
\be
\int \frac{dw}{2\pi} \frac{d}{dw} \Phi (w) \Theta ' (w,u_h) = -\int \frac{dw}{2\pi} \frac{\varphi (u_h,w)-\varphi (u_h,-w)}{2} \tilde P(w) \, . \label {d5}
\ee
Therefore, using the integral equation (\ref {tildePeq}) for $\tilde P$ we get
\be
Z_4(v)-P|_{s}=\Theta '(v,u_h)+\tilde P(u_h)=\Theta (v,u_h)=i\ln [-S^{(ss)}(v,u_h)] \, . \label {d6}
\ee

\subsection{Non scalars vs others}

Taking as a prototypical example the case of fermions, after multiplication by $e^{iP}$ we have to cope with
\be
\textrm{exp} \left [ i \int \frac{dv}{2\pi} [ \chi _F (v,u_{F,k})+\Phi (v) ] \frac{d}{dv} Z_4(v) \right ] \, , \label {e1}
\ee
which is the master relation from which the various scattering factors originate.

For instance, the fermion-fermion factor is
\be
\textrm{exp} \left [ i \int \frac{dv}{2\pi} [ \chi _F (v,u_{F,k})+\Phi (v) ] \frac{d}{dv} F^{F}(v,u_{F,j}) \right ]  \, . \label {e2}
\ee
Mixing equations (\ref {Theta-prime}) and (\ref {F-F}), we get
\be
F^F(v,u)=\chi _F(v,u)+\Phi (v) - \int \frac{dw}{2\pi} \frac{d}{dw}\Theta '(v,w) [\chi _F(w,u)+\Phi (w) ] \, ,
\ee
which, inserted in (\ref {e2}) gives the fermion-fermion factor as reported in Appendix \ref {scatt-fact}:
\ba
&& \textrm{exp} \Bigl \{ i \int _{-\infty}^{+\infty}\frac{dw}{2\pi} [\chi _F (w,u_{F,k})+\Phi (w)] \frac{d}{dw} [\chi _F (w,u_{F,j})+\Phi (w)] - \nonumber \\
&-& i \int \frac{dw}{2\pi} \frac{dx}{2\pi} [\chi _F (w,u_{F,k})+\Phi (w)] \frac{d^2}{dwdx}\Theta (w,x) [\chi _F(x,u_{F,j})+\Phi (x)]  \Bigr \}  \, .
\ea
For what concerns the fermion-scalar factor, it receives contribution from
\be
\textrm{exp} \left [ i \int \frac{dv}{2\pi} [ \chi _F (v,u_{F,k})+\Phi (v) ] \frac{d}{dv} \Theta '(v,u_h) \right ] =
\textrm{exp} \left [ i \int \frac{dv}{2\pi} [ \chi _F (v,u_{F,k})+\Phi (v) ] \frac{d}{dv} \Theta (v,u_h) \right ]  \label {e3}
\ee
Expression (\ref {e3}) has to be multiplied to the factor
\be
\frac{x_{F,k}-x_h^-}{x_h^+-x_{F,k}}=e^{i\chi _F (u_h, u_{F,k})}
\ee
present in equation (\ref {ferm1}) and to the factor $e^{i\Phi (u_h)}$ present in $e^{iP}$ to get the full fermion-scalar factor.

\section{Bethe equations} \label {Bethe-eqs}
\setcounter{equation}{0}
We list the complete set of Bethe Ansatz equations we found in this paper:

$\bullet $ Scalars
\ba
1&=&e^{iRP^{(s)}(u_h)+2iD^{(s)}(u_h)}
 \prod _{j=1}^{K_b} \frac{u_h-u_{b,j}+\frac{i}{2}}{u_h-u_{b,j}-\frac{i}{2}}  \prod _{\stackrel {h'=1}{h'\not=h}}^{H} S^{(ss)}(u_h,u_{h'})\prod _{j=1}^{N_{g}} S^{(sg)}(u_h,u_j^g)  \prod _{j=1}^{N_{\bar g}} S^{(s\bar g)}(u_h,u_j^{\bar g}) \cdot  \nonumber\\
&\cdot & \prod _{j=1}^{N_F} S^{(sF)}(u_h,u_{F,j})   \prod _{j=1}^{N_{\bar F}} S^{(s\bar F)}(u_h, u_{\bar F,j})
\prod _{j=1}^{N_f} S^{(sf)}(u_h,u_{f,j})   \prod _{j=1}^{N_{\bar f}} S^{(s\bar f)}(u_h, u_{\bar f,j})
\label {scal-final}
\ea

$\bullet $ Large fermions
\ba
1&=& e^{iRP^{(F)}(u_{F,k}) + 2i D^{(F)}(u_{F,k})}\prod _{j=1}^{K_a} \frac{u_{F,k}-u_{a,j}+i/2}{u_{F,k}-u_{a,j}-i/2} \prod _{h=1}^{H} S^{(F s)} (u_{F,k},u_h) \cdot  \nonumber \\
& \cdot &  \prod _{j=1}^{N_{F}} {S}^{(FF)}(u_{F,k},u_{F,j})  \prod _{j=1}^{N_{\bar F}} {S}^{(F\bar F)}(u_{ F,k},u_{\bar F,j})
\prod _{j=1}^{N_{f}} {S}^{(Ff)}(u_{F,k},u_{f,j})  \prod _{j=1}^{N_{\bar f}} {S}^{(F\bar f)}(u_{ F,k},u_{\bar f,j})
\cdot \nonumber \\
&\cdot & \prod _{j=1}^{N_g}  {S}^{(Fg)}(u_{F,k},u_{j}^g) \prod _{j=1}^{N_{\bar g}}
{S}^{(F\bar g)}(u_{F,k},u_{j}^{\bar g})  \label  {Fer-final}
\ea

$\bullet $ Large antifermions
\ba
1&=& e^{iRP^{(F)}(u_{\bar F,k}) + 2i D^{(F)}(u_{\bar F,k})} \prod _{j=1}^{K_c} \frac{u_{\bar F,k}-u_{c,j}+i/2}{u_{\bar F,k}-u_{c,j}-i/2} \prod _{h=1}^{H} S^{(\bar F s)} (u_{\bar F,k},u_h) \cdot  \nonumber \\
& \cdot &  \prod _{j=1}^{N_{F}} {S}^{(\bar FF)}(u_{\bar F,k},u_{F,j})  \prod _{j=1}^{N_{\bar F}} {S}^{(\bar F\bar F)}(u_{\bar F,k},u_{\bar F,j})
\prod _{j=1}^{N_{f}} {S}^{(\bar Ff)}(u_{\bar F,k},u_{f,j})  \prod _{j=1}^{N_{\bar f}} {S}^{(\bar F\bar f)}(u_{\bar F,k},u_{\bar f,j}) \nonumber \\
&\cdot & \prod _{j=1}^{N_g}  {S}^{(\bar Fg)}(u_{\bar F,k},u_{j}^g) \prod _{j=1}^{N_{\bar g}}
{S}^{(\bar F\bar g)}(u_{\bar F,k},u_{j}^{\bar g})
\label  {barFer-fin}
\ea

$\bullet $ Small fermions
\ba
1&=& e^{iRP^{(f)}(u_{f,k}) + 2i D^{(f)}(u_{f,k})} \prod _{j=1}^{K_a} \frac{u_{f,k}-u_{a,j}+i/2}{u_{f,k}-u_{a,j}-i/2} \prod _{h=1}^{H} S^{(f s)} (u_{f,k},u_h) \cdot  \nonumber \\
& \cdot &  \prod _{j=1}^{N_{F}} {S}^{(fF)}(u_{f,k},u_{F,j})  \prod _{j=1}^{N_{\bar F}} {S}^{(f\bar F)}(u_{f,k},u_{\bar F,j}) \prod _{j=1}^{N_f}  {S}^{(ff)}(u_{f,k},u_{f,j}) \prod _{j=1}^{N_{\bar f}}
{S}^{(f\bar f)}(u_{f,k},u_{\bar f,j}) \nonumber \\
&\cdot & \prod _{j=1}^{N_g}  {S}^{(fg)}(u_{f,k},u_{j}^g) \prod _{j=1}^{N_{\bar g}}
{S}^{(f\bar g)}(u_{f,k},u_{j}^{\bar g})
\label  {fer-fin}
\ea

$\bullet $ Small antifermions
\ba
1&=&  e^{iRP^{(f)}(u_{\bar f,k}) +2 i D^{(f)}(u_{\bar f,k})}  \prod _{j=1}^{K_c} \frac{u_{\bar f,k}-u_{c,j}+i/2}{u_{\bar f,k}-u_{c,j}-i/2} \prod _{h=1}^{H} S^{(\bar f s)} (u_{\bar f,k},u_h) \cdot  \nonumber\\
& \cdot &  \prod _{j=1}^{N_{F}} {S}^{(\bar f F)}(u_{\bar f,k},u_{F,j})  \prod _{j=1}^{N_{\bar F}} {S}^{(\bar f \bar F)}(u_{\bar f,k},u_{\bar F,j})
\prod _{j=1}^{N_{f}} {S}^{(\bar f f)}(u_{\bar f,k},u_{f,j})  \prod _{j=1}^{N_{\bar f}} {S}^{(\bar f \bar f)}(u_{\bar f,k},u_{\bar f,j}) \nonumber \\
&\cdot & \prod _{j=1}^{N_g}  {S}^{(\bar f g)}(u_{\bar f,k},u_{j}^g) \prod _{j=1}^{N_{\bar g}}
{S}^{(\bar f\bar g)}(u_{\bar f,k},u_{j}^{\bar g})
\label  {barfer-fin}
\ea

$\bullet $ Gluons
\ba
1&=&  e^{iRP^{(g)}(u_{k}^g) + 2i D^{(g)}(u_{k}^g)} \prod _{j=1, j\not=k}^{N_{g}}
 {S}^{(gg)}(u_{k}^{g},u_{j}^{g}) \prod _{j=1}^{N_{\bar g}} {S}^{(g\bar g)}(u_{k}^{g},u_{j}^{\bar g})
  \prod _{h=1}^H S^{(gs)}(u_k^g,u_h) \cdot  \nonumber \\
&\cdot &  \prod _{j=1}^{N_F} {S}^{(gF)}(u_{k}^{g},u_{F,j})
 \prod _{j=1}^{N_{\bar F}} {S}^{(g\bar F)}(u_{k}^{g},u_{\bar F,j})  \prod _{j=1}^{N_f} {S}^{(gf)}(u_{k}^{g},u_{f,j})
 \prod _{j=1}^{N_{\bar f}} {S}^{(g\bar f)}(u_{k}^{g},u_{\bar f,j})
\label  {glu-fin}
\ea

$\bullet $ Barred gluons
\ba
1&=& e^{iRP^{(g)}(u_{k}^{\bar g}) + 2i D^{(g)}(u_{k}^{\bar g})} \prod _{j=1}^{N_{g}}
 {S}^{(\bar gg)}(u_{k}^{{\bar g}},u_{j}^{g}) \prod _{j=1, j\not=k}^{N_{\bar g}} {S}^{(\bar g\bar g)}(u_{k}^{{\bar g}},u_{j}^{\bar g}) \prod _{h=1}^H S^{(\bar gs)}(u_k^{\bar g},u_h) \cdot \nonumber \\
&\cdot &  \prod _{j=1}^{N_F} {S}^{(\bar gF)}(u_{k}^{{\bar g}},u_{F,j})
 \prod _{j=1}^{N_{\bar F}} {S}^{(\bar g\bar F)}(u_{k}^{{\bar g}},u_{\bar F,j})  \prod _{j=1}^{N_f} {S}^{(\bar gf)}(u_{k}^{{\bar g}},u_{f,j})
 \prod _{j=1}^{N_{\bar f}} {S}^{(\bar g\bar f)}(u_{k}^{{\bar g}},u_{\bar f,j})
 \label {barglu-fin}
\ea

$\bullet $ Isotopic roots
\ba
&& \prod _{j=1}^{N_F}
\frac{u_{a,k}-u_{F,j}+i/2}{u_{a,k}-u_{F,j}-i/2} \prod _{j=1}^{N_f}
\frac{u_{a,k}-u_{f,j}+i/2}{u_{a,k}-u_{f,j}-i/2} = \prod _{j\not=k}^{K_a} \frac{u_{a,k}-u_{a,j}+i}{u_{a,k}-u_{a,j}-i} \prod _{j=1}^{K_b}
\frac{u_{a,k}-u_{b,j}-i/2}{u_{a,k}-u_{b,j}+i/2} \label {iso-1} \\
&& \prod _{h=1}^{H} \frac{u_{b,k}-u_h+i/2}{u_{b,k}-u_h-i/2}=\prod _{j=1}^{K_a} \frac{u_{b,k}-u_{a,j}-i/2}{u_{b,k}-u_{a,j}+i/2} \prod _{j=1}^{K_c} \frac{u_{b,k}-u_{c,j}-i/2}{u_{b,k}-u_{c,j}+i/2} \prod _{\stackrel {j=1}{j\not= k}}^{K_b} \frac{u_{b,k}-u_{b,j}+i}{u_{b,k}-u_{b,j}-i}
 \label {iso-3} \\
&& \prod _{j=1}^{N_{\bar F}}
\frac{u_{c,k}-u_{\bar F,j}+i/2}{u_{c,k}-u_{\bar F,j}-i/2} \prod _{j=1}^{N_{\bar f}}
\frac{u_{c,k}-u_{\bar f,j}+i/2}{u_{c,k}-u_{\bar f,j}-i/2}=\prod _{j\not=k}^{K_c} \frac{u_{c,k}-u_{c,j}+i}{u_{c,k}-u_{c,j}-i}   \prod _{j=1}^{K_b}
\frac{u_{c,k}-u_{b,j}-i/2}{u_{c,k}-u_{b,j}+i/2} \label {iso-2}
\ea



\begin{thebibliography}{xx}

\bibitem{Bethe}
H. Bethe,
{\sl On the theory of metals. 1. Eigenvalues and eigenfunctions for the linear atomic chain},
Z. Phys. {\bf 71} (1931) 205;

\bibitem{BS}
N. Beisert, M. Staudacher,
{\sl Long-Range $psu(2, 2|4)$ Bethe Ans�atze
for Gauge Theory and Strings},
Nucl. Phys. {\bf B727} (2005) 1 and
hep-th/0504190;

\bibitem{WRA}
C. Sieg, A. Torrielli, {\sl Wrapping interactions and the genus expansion of the 2-point function of composite operators}, Nucl. Phys. {\bf B723} (2005) 3 and hep-th/0505071
$\bullet $
J. Ambjorn, R. Janik, C. Kristjansen,
{\sl Wrapping interactions and a new source of
corrections to the spin chain/string duality},
Nucl. Phys. {\bf B736} (2006) 288
and hep-th/0510171
$\bullet $
A. Kotikov, L. Lipatov, A. Rej, M. Staudacher, V. Velizhanin,
{\sl Dressing and wrapping},
J. Stat. Mech. {\bf 10} (2007) P003 and
arXiv:0704.3586 [hep-th];

\bibitem{GKP}
S. Gubser, I. Klebanov, A. Polyakov,
{\sl  A Semiclassical limit of the gauge / string correspondence},
Nucl. Phys. {\bf B636 } (2002)  99  and hep-th/0204051;

\bibitem{MGKPW}
J.M. Maldacena, {\sl The large N limit of superconformal field
theories and supergravity}, Adv. Theor. Math. Phys.
{\bf 2} (1998) 231 and hep-th/9711200
$\bullet $
S. Gubser, I. Klebanov, A. Polyakov, {\sl Gauge theory
correlators from non-critical string theory},
Phys. Lett. {\bf B428} (1998) 105 and hep-th/9802109
$\bullet$
E. Witten, {\sl Anti-de Sitter space and holography}, Adv. Theor.
Math. Phys. {\bf 2} (1998) 253 and hep-th/9802150;

\bibitem{FTT}
S. Frolov, A. Tirziu, A. Tseytlin,
{\sl Logarithmic corrections to higher twist scaling at strong coupling from AdS/CFT},
Nucl. Phys. {\bf B766} (2007)
and hep-th/0611269;

\bibitem{BGK}
A. Belitsky, A. Gorsky, G. Korchemsky,
{\sl Logarithmic scaling in gauge/string correspondence},
Nucl. Phys. {\bf B748} (2006) 24 and hep-th/0601112;

\bibitem{Poly-cusp}
A. Polyakov,
{\sl Gauge fields as rings of glue},
Nucl. Phys. {\bf B164} (1980) 171;


\bibitem{Korchemsky:1987wg}
G. Korchemsky and A. Radyushkin, {\sl Renormalization of the Wilson Loops Beyond the Leading Order}, Nucl. Phys. {\bf B283} (1987) 342;

\bibitem{BES}
N. Beisert, B. Eden, M. Staudacher,
{\sl Transcendentality and
crossing}, J. Stat. Mech. {\bf 07} (2007) P01021 and hep-th/0610251;

\bibitem{FGR5}
D. Fioravanti, P. Grinza, M. Rossi, {\sl On the logarithmic powers of $sl(2)$ SYM$_4$},
Phys. Lett. {\bf B684} (2010) 52 and arXiv:0911.2425 [hep-th];

\bibitem{FRS}
L. Freyhult, A. Rej, M. Staudacher,
{\sl A Generalized Scaling Function for AdS/CFT},
J. Stat. Mech. {\bf 07} (2008) P015 and
arXiv:0712.2743 [hep-th];

\bibitem{BFR}
D. Bombardelli, D. Fioravanti, M. Rossi, {\sl Large spin
corrections in ${\cal N}=4$ SYM $sl(2)$: still a linear integral
equation}, Nucl. Phys. {\bf B810} (2009) 460 and arXiv:0802.0027 [hep-th];

\bibitem{FZ}
L. Freyhult, S. Zieme,
{\sl The virtual scaling function of AdS/CFT},
Phys. Rev. {\bf D79} (2009) 105009 and
arXiv:0901.2749 [hep-th]
$\bullet$
D. Fioravanti, P. Grinza, M. Rossi,
{\sl Beyond cusp anomalous dimension from integrability},
Phys. Lett. {\bf B675} (2009) 137
and arXiv:0901.3161 [hep-th];

\bibitem{BKK}
B. Basso, G. Korchemsky, J. Kotanski,
{\sl Cusp anomalous dimension in maximally supersymmetric Yang-Mills theory at strong coupling},
Phys. Rev. Lett. {\bf 100} (2008) 091601
and arXiv:0708.3933 [hep-th];

\bibitem{KM}
G. Korchemsky, {\sl Asymptotics of the Altarelli-Parisi-Lipatov evolution kernels of parton distributions}, Mod. Phys. Lett. {\bf A4} (1989) 1257;


\bibitem {reciprocity}
D. Fioravanti, G. Infusino, M. Rossi,
{\sl Reciprocity and self-tuning relations without wrapping}, arXiv:1510.02445 [hep-th];


\bibitem{BJL}
Z. Bajnok, R. Janik, T. Lukowski,
{\sl Four loops twist two, BFKL, wrapping and strings},
Nucl. Phys. {\bf B816} (2009) 376 and
arXiv:0811.4448 [hep-th]
$\bullet$
T. Lukowski, A. Rej, V. Velizhanin,
{\sl Five-loops anomalous dimension of twist-two operators},
Nucl. Phys. {\bf B831} (2010) 105 and
arXiv:0912.1624 [hep-th]
$\bullet$
V. Velizhanin,
{\sl Six-Loop Anomalous Dimension of Twist-Three Operators in N=4 SYM},
JHEP{\bf 11} (2010) 129 and arXiv:1003.4717 [hep-th];

\bibitem{Freyhult:2009fc}
L. Freyhult, A. Rej and S. Zieme, {\sl From weak coupling to spinning strings}, JHEP{\bf 02} (2010) 050 and arXiv:0911.2458 [hep-th];

\bibitem{Basso}
B. Basso,
{\sl Exciting the GKP String at Any Coupling},
Nucl. Phys. {\bf B857} (2012) 254
and arXiv:1010.5237 [hep-th];

\bibitem{FRO6}
D. Fioravanti, M. Rossi,
{\sl TBA-like equations and Casimir effect in (non-)perturbative AdS/CFT},
JHEP{\bf 12} (2012) 013 and arXiv:1112.5668 [hep-th];

\bibitem{BREJ}
B. Basso and A. Rej,
{\sl Bethe Ansaetze for GKP strings},
Nucl.Phys. {\bf B879} (2014) 162
and arXiv:1306.1741 [hep-th];

\bibitem{FPR}
D. Fioravanti, S. Piscaglia, M. Rossi,
{\sl On the scattering over the GKP vacuum},
Phys. Lett. {\bf B728} (2014) 288-295
arXiv:1306.2292 [hep-th];

\bibitem{BMN}
D. Berenstein, J.M. Maldacena, H. Nastase,
{\sl Strings in flat space and pp waves from $\m{N}=4$ Super Yang Mills},
JHEP{\bf 04} (2002) 013
and arXiv:hep-th/0202021;

\bibitem{DDV}
A. Kluemper, M.T. Batchelor, P. Pearce,
{\sl Central charges of the 6- and 19- vertex models with twisted boundary conditions},
J. Phys. {\bf A24} (1991) 3111
$\bullet $
C. Destri, H.J. de Vega,
{\sl New thermodynamic Bethe ansatz equations without strings},
Phys. Rev. Lett. {\bf 69} (1992) 2313;


\bibitem{FMQR}
D. Fioravanti, A. Mariottini, E. Quattrini, F. Ravanini, {\sl
Excited state Destri-de Vega equation for sine-Gordon and restricted
sine-Gordon models}, Phys. Lett. {\bf B390} (1997) 243
and hep-th/9608091;

\bibitem{FR}
D. Fioravanti, M. Rossi,
{\sl On the commuting charges for the highest dimension SU(2) operators in planar ${\cal N}=4$ SYM},
JHEP{\bf 08} (2007) 089 and arXiv:0706.3936 [hep/th];

\bibitem{TBA-Z1}
A.B. Zamolodchikov, {\sl Thermodynamic Bethe ansatz in
relativistic models. Scaling three state Potts and Lee-Yang
models}, Nucl. Phys. {\bf B342} (1990) 695;

\bibitem{TBA-Y}
C.N. Yang and C.P. Yang, {\sl Thermodynamics of one-dimensional system of bosons with repulsive delta function interaction}, J. Math. Phys. {\bf 10} (1969) 1115;


\bibitem{TBA}
D. Bombardelli, D. Fioravanti and R. Tateo,
{\sl Thermodynamic Bethe Ansatz for planar AdS/CFT: a proposal},
 J. Phys. A  {\bf 42} (2009) 375401
and arXiv:0902.3930 [hep-th]
$\bullet$
N. Gromov, V. Kazakov, A. Kozak and P. Vieira,
{\sl Exact Spectrum of Anomalous Dimensions of Planar N = 4 Supersymmetric Yang-Mills Theory: TBA and excited states},
Lett. Math. Phys. {\bf 91} (2010) 265, cf. also
arXiv:0902.4458 [hep-th]
%
$\bullet$
G. Arutyunov and S. Frolov,
{\sl Thermodynamic Bethe Ansatz for the $AdS_5 \times  S^5$ Mirror Model},
JHEP{\bf 05} (2009) 068
and arXiv:0903.0141 [hep-th]
$\bullet$
D. Bombardelli, D. Fioravanti and R. Tateo,
{\sl TBA and Y-system for planar $AdS_4/CFT_3$},
Nucl. Phys. {\bf B834} (2010) 543 and arXiv:0912.4715 [hep-th]
$\bullet$
N. Gromov and F. Levkovich-Maslyuk,
{\sl Y-system, TBA and Quasi-Classical strings in AdS(4) x CP3},
JHEP{\bf 06} (2010) 88 and
arXiv:0912.4911 [hep-th];


\bibitem{AM-amp}
L.F. Alday and J.M. Maldacena, {\sl  Gluon scattering amplitudes at strong coupling},
  JHEP{\bf 06} (2007) 064 and
  arXiv:0705.0303 [hep-th];


\bibitem{DKS}
J. Drummond, G. Korchemsky and E. Sokatchev,
   {\sl  Conformal properties of four-gluon planar amplitudes and Wilson loops},
  Nucl. Phys. {\bf B795} (2008) 385 and
  arXiv:0707.0243 [hep-th];

\bibitem{BHT}
A. Brandhuber, P. Heslop and G. Travaglini, {\sl MHV amplitudes in N=4 super Yang-Mills and Wilson loops},
  Nucl.\ Phys.\ B {\bf 794} (2008) 231 and
  arXiv:0707.1153 [hep-th];



\bibitem{BSV1}
B. Basso, A. Sever, P. Vieira,
{\sl Space-time S-matrix and Flux-tube S-matrix at Finite Coupling},
Phys. Rev. Lett. {\bf 111} (2013) 091602 and
arXiv:1303.1396 [hep-th];
\bibitem{BSV2}
B. Basso, A. Sever, P. Vieira,
{\sl  Space-time S-matrix and Flux tube S-matrix II. Extracting and Matching Data},
JHEP{\bf 01} (2014) 008 and arXiv:1306.2058 [hep-th];
\bibitem{BSV3}
B. Basso, A. Sever, P. Vieira,
{\sl  Space-time S-matrix and Flux tube S-matrix III. The two-particle contributions},
JHEP{\bf 08} (2014) 085 and
arXiv:1402.3307 [hep-th];
\bibitem{BSV4}
B. Basso, A. Sever, P. Vieira,
{\sl  Collinear limit of scattering amplitudes at strong coupling},
Phys. Rev. Lett. {\bf 113} (2014) 26, 261604 and arXiv:1405.6350 [hep-th];
\bibitem{BSV5}
B. Basso, A. Sever, P. Vieira,
{\sl  Space-time S-matrix and Flux tube S-matrix IV. Gluons and fusion},
 JHEP{\bf 09} (2014) 149 and
arXiv:1407.1736 [hep-th];



 \bibitem{TBA-amp1}
L.F. Alday, D. Gaiotto, J.M. Maldacena,
{\sl Thermodynamic Bubble Ansatz},
JHEP{\bf 09} (2011) 032
and arXiv:0911.4708[hep-th];

\bibitem{TBA-amp3}
L.F. Alday, D. Gaiotto, J.M. Maldacena, A. Sever, P. Vieira
{\sl Y-system for Scattering Amplitudes},
J.Phys. {\bf A43} (2010) 485401
and arXiv:1002.2459 [hep-th];

\bibitem{TBA-amp2}
L.F. Alday, D. Gaiotto, J.M. Maldacena, A. Sever, P. Vieira,
{\sl An Operator Product Expansion for Polygonal null Wilson Loops},
JHEP{\bf 04} (2011) 088
and arXiv:1006.2788 [hep-th];


\bibitem{Hatsuda:2010cc}
Y.~Hatsuda, K.~Ito, K.~Sakai and Y.~Satoh, {\sl Thermodynamic Bethe Ansatz Equations for Minimal Surfaces in $AdS_3$}, JHEP {\bf 1004} (2010) 108 and arXiv:1002.2941 [hep-th].

\bibitem{TBA-Z2}
Al. Zamolodchikov, {\sl On the thermodynamic Bethe ansatz equations for reflectionless ADE scattering theories},
Phys. Lett.  {\bf B253} (1991) 391;


\bibitem{KOR}
G. Korchemsky, {\sl Quasiclassical QCD pomeron}, Nucl. Phys. {\bf B462} (1996) 333 and hep-th/9508025;

\bibitem{OW}
E. Ogievetsky, P. Wiegmann,
{\sl Factorised S-matrix and the Bethe Ansatz for simple Lie groups},
Phys. Lett. {\bf B168} (1986) 360;

\bibitem{Baj-G}
Z. Bajnok and A. George, {\sl From defects to boundaries}, Int. J. Mod. Phys. {\bf A21} (2006) 1063 and hep-th/0404199;


\bibitem{DRE}
G. Arutyunov, S. Frolov, M. Staudacher,
{\sl  Bethe ansatz for quantum strings},
JHEP{\bf 10} (2004) 016 and hep-th/0406256
$\bullet $
N. Beisert, R. Hernandez, E. Lopez,
{\sl A Crossing-symmetric phase for AdS(5) x S**5 strings},
JHEP{\bf 11} (2006) 070 and hep-th/0609044;

\bibitem{FGR1}
D. Fioravanti, P. Grinza, M. Rossi, {\sl Strong coupling for planar
 ${\cal N}=4$ SYM: an all-order result},  Nucl. Phys. {\bf B810} (2009) 563
and arXiv:0804.2893 [hep-th];

\bibitem{FGR3}
D. Fioravanti, P. Grinza, M. Rossi,
{\sl The generalised scaling function: a systematic study},
JHEP{\bf 11} (2009) 037 and arXiv:0808.1886 [hep-th];

\bibitem{AM}
L.F. Alday, J.M. Maldacena, {\sl Comments on operators with large
spin}, JHEP{\bf 11} (2007) 019 and arXiv:0708.0672 [hep-th];

\bibitem{BK}
  B. Basso and G. Korchemsky,
{\sl Embedding nonlinear O(6) sigma model into N=4 super-Yang-Mills theory},
  Nucl. Phys. {\bf B807} (2009) 397 and
  arXiv:0805.4194 [hep-th];

\bibitem{Zarembo:2011ag}
K. Zarembo and S. Zieme, {\sl Fine Structure of String Spectrum in $AdS_5$ x $S^5$}, JETP Lett. {\bf 95} (2012) 219  [Erratum-ibid.  {\bf 97} (2013) 8,  504] and arXiv:1110.6146 [hep-th];


\bibitem{Zam^2}
A. Zamolodchikov and Al. Zamolodchikov, {\sl Relativistic Factorized S Matrix in Two-Dimensions Having O(N) Isotopic Symmetry}, Nucl. Phys. {\bf B133} (1978) 525 [JETP Lett.  {\bf 26} (1977) 457].



\bibitem{DoreyZhao}
N. Dorey, P. Zhao,
{\sl Scattering of giant holes},
JHEP{\bf 08} (2011) 134 and arXiv:1105.4596 [hep-th];

\bibitem{BB}
B. Basso, A. Belitsky, {\sl Luescher formula for GKP string},
Nucl. Phys. {\bf B860} (2012) 1
and arXiv:1108.0999 [hep-th];


\bibitem{strings}
M. Takahashi,  M. Suzuki,
{\sl One-dimensional anisotropic Heisenberg model at finite temperatures},
Prog. Theor. Phys. {\bf 48} (1972) 2187;
L. Faddeev,
{\sl How Algebraic Bethe Ansatz works for integrable model},
arXiv:hep-th/9605187;

\bibitem{Zamolodchikov:2013ama}
  A. Zamolodchikov,
{\sl Ising Spectroscopy II: Particles and poles at $T > T_c$},
  arXiv:1310.4821 [hep-th];

  \bibitem{Zarembo:2009au}
  K. Zarembo,
{\sl Worldsheet spectrum in AdS(4)/CFT(3) correspondence},
  JHEP{\bf 04} (2009) 135, arXiv:0903.1747 [hep-th];

  \bibitem{35}
A. Bonini, D. Fioravanti, S. Piscaglia, M. Rossi, {\sl Strong Wilson polygons from the lodge of free and bound mesons}, arXiv:1511.05851 [hep-th];



\bibitem{Nekrasov:2009rc}
  N. Nekrasov and S. Shatashvili,
{\sl Quantization of Integrable Systems and Four Dimensional Gauge Theories},
  arXiv:0908.4052 [hep-th];

  \bibitem{Meneghelli:2013tia}
  C. Meneghelli and G. Yang,
{\sl Mayer-Cluster Expansion of Instanton Partition Functions and Thermodynamic Bethe Ansatz},
  JHEP{\bf 05} (2014) 112 and
  arXiv:1312.4537 [hep-th];

  \bibitem{Bourgine:2014yha}
  J.-E. Bourgine,
{\sl Confinement and Mayer cluster expansions},
  Int. J. Mod. Phys. {\bf A29} (2014) 15,  1450077
  and arXiv:1402.1626 [hep-th];


\bibitem{Bettelheim:2014gma}
E. Bettelheim, I. Kostov, {\sl Semi-classical analysis of the inner product of Bethe states},
J. Phys. {\bf A47} (2014) 245401 and
arXiv:1403.0358 [hep-th];

\bibitem{Henkel}
M. Henkel, {\sl Conformal invariance and critical phenomena}, Springer 1999;

\bibitem{giuseppe}
G. Mussardo, {\sl Off critical statistical models: Factorized scattering theories and bootstrap program}, Phys.\ Rept.\  {\bf 218} (1992) 215;

\bibitem{BEL}
A. Belitsky,
{\sl Nonsinglet pentagons and NHMV amplitudes},
arXiv:1407.2853 [hep-th]
$\bullet $
A. Belitsky,
{\sl Fermionic pentagons and NMHV hexagon},
Nucl. Phys. {\bf B894} (2015) 108 and
arXiv:1410.2534 [hep-th];

\bibitem{Hatsuda:2014oza}
Y.Hatsuda, {\sl Wilson loop OPE, analytic continuation and multi-Regge limit}, JHEP {\bf 1410} (2014) 38, arXiv:1404.6506 [hep-th].


\bibitem{P2}
G. Papathanasiou,
{\sl Hexagon Wilson Loop OPE and Harmonic Polylogarithms},
JHEP{\bf 11} (2013) 150 and arXiv:1310.5735 [hep-th];


\end{thebibliography}
\end{document}